\documentclass[preprint]{aastex}
\usepackage{subfigure}

\shorttitle{ATLBS Extended source sample}
\shortauthors{Saripalli et al.}

\begin{document}

\title{ATLBS Extended Source Sample: The evolution in radio source morphology with flux density}

\author{L. Saripalli$^{1*}$,  R. Subrahmanyan$^{1}$,  K. Thorat$^{1,2}$, R. D. Ekers$^{1,3}$, 
R. W. Hunstead$^{4}$, H. M. Johnston$^{4}$ \& E. M. Sadler$^{4}$}
\affil{$^{1}$Raman Research Institute, C. V. Raman Avenue, Sadashivanagar, Bangalore 560080, India}
\affil{$^{2}$Joint Astronomy Programme, Indian Institute of Science, Bangalore 560012, India}
\affil{$^{3}$CSIRO Astronomy \& Space Sciences, Epping, NSW 2121, Australia}
\affil{$^{4}$Sydney Institute for Astronomy, School of Physics, University of Sydney, NSW 2006, Australia}
\email{$^{*}$lsaripal@rri.res.in}

\begin{abstract}

Based on the ATLBS survey we present a sample of extended radio sources and derive 
morphological properties of faint radio sources. 119 radio galaxies form the 
ATLBS-Extended Source Sample (ATLBS-ESS) consisting of all sources exceeding 30" 
in extent and integrated flux densities exceeding 1 mJy. We give structural details 
along with information on galaxy identifications and source classifications. 
The ATLBS-ESS, unlike samples with higher flux-density limits, has almost equal 
fractions of FR-I and FR-II radio galaxies with a large fraction of the FR-I population 
exhibiting 3C31-type structures. Significant asymmetry in lobe extents appears to be 
a common occurrence in the ATLBS-ESS FR-I sources compared to FR-II sources. We present 
a sample of 22 FR-Is at $z>0.5$ with good structural information. The detection of 
several giant radio sources, with size exceeding 0.7 Mpc, at $z>1$ suggests that giant 
radio sources are not less common at high redshifts. The ESS also includes a sample of 
28 restarted radio galaxies. The relative abundance of dying and restarting sources is 
indicative of a model where radio sources undergo episodic activity in which an active 
phase is followed by a brief dying phase that terminates with restarting of the central 
activity; in any massive elliptical a few such activity cycles wherein adjacent events 
blend may constitute the lifetime of a radio source and such bursts of blended activity 
cycles may be repeated over the age of the host. The ATLBS-ESS includes a 2-Mpc giant 
radio galaxy with the lowest surface brightness lobes known to date.

\end{abstract}


\section{Introduction}

Sensitive radio surveys have a key role in cosmology,  resting on their ability to detect faint radio
sources reaching down to sub-milliJansky flux densities and thereby representing a chronicle of the 
evolution in active galaxies across Hubble time. Radio 
source structures are rich in information: the extended structures serve as indicators of 
relativistic jet-type activity in AGNs; additionally,  their morphology is related to 
the temporal behavior of the jet activity, and the ambient interstellar and intergalactic environment 
in which they deposit their energy. 
AGN activity triggered by mergers and the accretion of intergalactic baryons may well be a phase in 
the lives of most luminous galaxies. Radio galaxies and quasars with extended structures represent 
circumstances in which there is obvious kinetic mode energetic impact of the AGN on the environment
and, therefore, radio surveys represent 
an opportunity for examining the role of such AGN feedback on galaxy evolution.

In most sensitive radio surveys there is often incomplete structural information because of 
limited surface brightness sensitivity,  and 
consequently the relatively faint and diffuse emission components are often unrepresented. 
In view of the potential missing structural information on faint radio sources as well as the puzzling
paucity of relic radio sources we embarked on a sensitive radio survey that was designed to achieve
high radio surface-brightness sensitivity. The Australia Telescope Low Brightness Survey (ATLBS; 
\citet{sub10}) was made using the Australia Telescope Compact Array (ATCA) at 1.4~GHz adopting an 
observing strategy that guarantees imaging over a wide field of view with complete spatial frequency 
coverage; this ensured the data were not limited by incomplete uv-coverage,  dynamic
range issues and that the theoretical rms noise sensitivity 
was achieved. The ATLBS achieved a surface brightness sensitivity that is nearly
an order of magnitude lower than previous wide-field surveys. 

The ATLBS survey was followed up with imaging in multiple array 
configurations at 1.4~GHz with the aim of obtaining good structural information on the sources.  These 
data will be presented in \citet{tho11} along with a discussion of the source counts.

Herein we discuss the radio structures of the ATLBS radio sources. We present an 
extended ATLBS radio source sample (which we refer to as the ATLBS-ESS sample) in which
all sources with angular extent larger than about 30 arcsec have been included. 
A sensitive survey such as the ATLBS that combines high surface brightness sensitivity and angular 
resolution is expected to reveal a variety of source structures,  and provides a sample that reveals 
the structural variety that is present in the relatively faint radio source population. 
We present different classes of radio sources as revealed by the ATLBS; with its rich haul of 
different structural types,  our sample is a valuable resource in the study of the life cycles of 
radio galaxies.

It is well recognized that flux limited radio surveys include powerful radio sources (that
invariably have Fanaroff-Riley type-II morphologies; \citep{fan74}) to relatively higher redshifts whereas 
the low power radio sources 
identified in such surveys are confined to lower redshifts. The dominance of 
low power sources in the low redshift population is owing to the slope of the radio luminosity function. 
A consequence of this bias in radio surveys is that population studies of low-luminosity sources have been
limited to low redshifts with little known of the relative abundance of this dominant group or of their
structures at higher redshifts. A few studies have attempted to extend our knowledge into this space 
and the general approach has been through radio-optical cross-correlation studies 
involving deep radio and optical surveys \citep{cle04,  sne03,  sad07,  chi09}. ATLBS has achieved 
relatively high radio surface brightness sensitivity and this is
important in detecting the faint extended structures in FR-I radio galaxies. Together with excellent 
imaging of the diffuse structures,  we have herein used a more traditional
method (using radio source morphologies) to extend the FR-I population study to higher redshifts. 

A unique and valuable aspect of the ATLBS survey is that the sky regions have been observed with 
excellent surface brightness sensitivity at low resolutions,  and the wide survey fields have also 
been followed up with high resolution radio observations. Reliable measures of integrated 
flux density provided by the low resolution images along with reliable images of structure at
higher angular resolution together yield useful estimates of the truly diffuse emission associated with
radio sources.  For example,  if at higher resolution there is a clustering of weak compact components 
and at lower resolution there is detected a source with substantially greater integrated flux density,  
a good estimate of the associated diffuse emission is possible.  This allows compilation
of samples of such radio sources that have substantial associated extended emission components with little 
compact structure: high resolution interferometer surveys might not only miss the extended 
emission,  but may also misclassify associated compact components as independent sources.

In Section~2 we present an overview of the observations followed by the method adopted for the 
creation of the ATLBS-ESS. We devote Sections~3 to 4 to a discussion of the relative abundances of different radio 
source types present in the sample. This characterization of the morphologies of the faint radio 
sources is followed by discussion and conclusions in Section~5.

We adopt a flat cosmology with Hubble constant H$_{o}$=71~km s$^{-1}$ Mpc$^{-1}$ and matter density parameter
$\Omega_{m}$=0.27. 

\section{The ATLBS Extended source sample: ATLBS-ESS}

The ATLBS radio images were made at 1388~MHz using the Australia Telescope Compact Array (ATCA) by mosaic 
observing 38 adjacent pointing positions covering about 8.4 square degrees of sky area.  The observations
were made in a set of array configurations and with sufficiently small dwell time in each pointing 
position so that the entire survey region is imaged with almost complete visibility coverage 
up to 750~m. The ATLBS images,  with beam FWHM of 50~arcsec and with high surface brightness 
sensitivity,  together with associated high resolution data that were simultaneously available on
long baselines to the 6-km antenna,  provided a measure of the extended emission associated with 
the radio source population down to about 0.4~mJy flux density \citep{sub10}. Important for the 
present work is the listing provided therein of the integrated flux density in the ATLBS sources along with 
listings of the flux density associated with compact components. 

\subsection{Radio and optical data for the ATLBS survey fields}

Detailed radio structures of the sources detected in the ATLBS survey were subsequently imaged by combining the 
visibilities along with additional data obtained using the ATCA 
in a set of extended 6-km array configurations.  Despite the relatively sparse visibility coverage
in the outer realms of the visibility plane,  the imaging of extended structures was accomplished in  
a convergent and reliable manner by aiding decovolution in initial self-calibration cycles 
by restricting the sky area to regions where sources were reliably detected in the 
high-surface-brightness ATLBS images.  The high resolution images of the ATLBS survey regions will  
be presented in \citet{tho11},  along with derived source counts. The synthesized images have a 
beam of FWHM 6~arcsecs and rms noise of 72~$\mu$Jy~beam$^{-1}$. 
The image quality is excellent in that they have no artifacts apparent above thermal noise.

The survey regions were also imaged in the Ks-band using the IRIS2 imager on the 3.9-m 
Anglo Australian Telescope (AAT). The ATLBS survey sky area was covered as a mosaic
of nearly 700 pointings with median seeing about 1~arcsec. The field centred at RA: 00h35m00s, 
Dec.: -67 00 00 (J2000.0; field~A) was completed barring a few pointings where as the field centred at RA: 00 59 17, 
Dec.: -67 00 00 (J2000.0; field~B) was only partially completed. 
A completeness down to 19 mag was achieved. 
The Ks-band images will be presented in an accompanying manuscript (Johnston et al.,  in preparation). 

In addition to observing in Ks-band, the ATLBS survey regions were also observed in optical g,  r and z 
band with the MOSAICII imager on the CTIO NOAO 4-m Blanco telescope. The larger field of view of 37'x37' 
meant that the two regions could be covered in fewer pointings.  The data reduction of the r-band imaging has 
been completed and the images have a $90 \%$ completeness 
level to an apparent magnitude of 22.5. Details of the grz-band optical imaging will be 
presented in a followup publication (Thorat et al.,  in preparation). 

Optical identifications (Id) for the ATLBS-ESS sources were made by examining the r-band images. Where
there was no identification in r (either because the area was not covered in the r-band imaging 
or because there was no object seen at the location of the radio core) or it was uncertain, we examined
the Ks band images. In situations where the area was not covered in either r or Ks images we used
the B and R-band images from the SuperCOSMOS Sky Survey \citep{ham01}. These identifications are based 
on all the material available and not just the images shown in the paper. In general the process of 
identifications ultimately involved examination of the vicinity of the centroid of the 
radio source in all four optical bands; the information related to this exercise is provided below. 

We have an ongoing program for obtaining spectra of the optical
identifications for all radio sources in the ATLBS fields using the AAOmega instrument on the AAT. 
To date 24 sources in the ATLBS-ESS have measured redshifts. 
Herein we give estimates of redshifts for the remaining sources using the derived  
r-band magnitude vs redshift relationship when spectra are not available 
(the details of the methods used for redshift estimates will be given in an accompanying 
paper: Thorat et al., in preparation). We give these estimated redshifts in brackets to distinguish 
them from measured values.

\subsection{Selection criteria for ATLBS-ESS}

ATLBS images made with the synthesized beam of 50~arcsec were used as the basic resource for compiling 
the ATLBS-ESS sample. In the two, ~2-deg mosaic images,  
only `islands' of image pixels with peaks exceeding five times the image rms noise were considered.
As described in \citet{sub10},  the integrated flux density in compact components within these 
source islands were computed from images made with 4-arcsec FWHM beam 
using exclusively interferometer baselines to the 6~km antenna
(so that extended emission was intentionally filtered out).  A first list of candidate sources was
created by including only those `islands' that had a ratio of integrated flux density to that in
compact components larger than or equal to 1.5. These sources potentially have extended 
emission that is more than half the flux density in the compact components. At relatively
higher values for the ratio,  selected sources have substantial low surface brightness diffuse
emission with little fine structure.  This list was generated by an automated algorithm that 
scanned the 50 and 4~arcsec images. It may be noted here that this automated procedure was only
applied to the sky regions in the mosaic image that were observed with effective primary beam 
attenuation within 0.5,  which excluded the outer parts of the images.

The radio structures of the ATLBS sources are best represented in 6~arcsec beam FWHM images that
were made by \citet{tho11}.  Separately,  these images, including their outer parts, were carefully 
examined by eye to identify 
extended radio sources.  This `second pass' over the survey regions was aimed at including any 
additional sources that may have been missed by the automated search algorithm.  Within the regions
scanned by the automated algorithm,  the examination by eye did not yield any additional extended
radio sources,  giving confidence in the completeness of the selection process.  However in this step  
we recognized 58 extended radio sources in the outer regions 
of the ATLBS survey and included them as 
well for the analyses and considerations presented herein.  Additionally,  examination by eye 
identified sources that were blended together in the 50~arcsec resolution images: these
were thereafter considered as separate sources.

The images with beam FWHM of 6~arcsec were convolved to a final beam of 
10~arcsec,  and as a final filter 
only those extended radio sources that were clearly resolved with at least three beams across were
retained.  The convolution improved the surface brightness sensitivity of the images,  thereby improving
the ability to detect diffuse emission.  This criterion corresponds approximately to restricting 
the ATLBS-ESS sample to sources with angular size exceeding 30~arcsec.  

Table~1 presents this sub-sample of extended ATLBS sources.  It contains 119 sources and the minimum 
integrated flux density of sources in this sub-sample is 1~mJy at 1.4~GHz.  
The Table includes information on the radio structural properties based on
the examination of the radio images by eye, and estimates of the angular extents (Column~3). 
The angular sizes were measured along the maximum extent of the source between the 3-sigma contours 
in the 10" maps. We also give the total and fractional core flux densities (Column~5 and 6 respectively) 
as well as source classification (Column~7).
Details of the flux density measurements will be given in \citet{tho11}. In several FR-I sources
although cores are the brightest components the fractional core flux densities are upper limits because of 
blending with surrounding emission. 
Where possible, the radio source position is taken from the position of the 
r-band optical identification. Where the r-band identification is not available or doubtful the coordinates assigned 
to the radio source is that of the identification in Ks-band, or B or R band; in cases where there is no
optical identification available, on the radio core position and in the absence of radio
cores we use the radio centroid position from the original 50" image. 
Additionally, the Table includes
information on the optical identifications,  measured r-band magnitudes, measured and inferred redshifts (given in
brackets) as well as any information on their galaxy 
environments based on examination of the optical 
fields.  As the Ks band imaging has covered field~B only partially the information on K-band identifications
for sources in this field is given only where available. 

The images of the 119 ATLBS-ESS sources are presented in Figs 1---119. The choice of the contours and image 
resolutions was made with a view to reveal structures of the faint emission regions present in several of the sources. 

The remaining extended
radio sources that have no recognizable extended structure and are mostly composed of well resolved 
diffuse emission have not been included in this sub-sample. These 23 sources 
are discussed separately in Section~4.6.

There are 11 radio galaxies in the ATLBS-ESS that have estimated projected 
linear sizes in excess of 700~kpc and hence are classified as giant radio sources. Nine of these giant radio galaxies 
are at photometric redshifts $z>1$.  Additionally,  there are
another 3 candidate giant radio sources that have large angular sizes 
(exceeding 84~arcsec). These have faint optical hosts and their
r-band magnitudes could not be reliably determined; nevertheless, the faintness of the host galaxy
suggests that they are likely to be located at redshifts $z>1$. The ATLBS-ESS giant radio galaxies are
listed in Table~3.

\begin{deluxetable}{lccccccl} 
\tabletypesize\scriptsize 
\tablecaption{ATLBS extended source sample} 
\tablehead{ 
\colhead{Name} & \colhead{RA} & \colhead{LAS(\arcsec)} & \colhead{Fig.} & \colhead{$S_{t}$}& \colhead{$S_{c}$} & \colhead{Type} &
\colhead{Comments} \\
               & \colhead{Dec}& \colhead{Morph.} & \colhead{No.} & \colhead{mJy} & \colhead{/$S_{t}$} & & \\}
\startdata 

J0022.1-6705*& 00:22:09.60 & 46  & 1  & 9.7 &0.26  &           & Bright B, R, K id. \\
            & -67:05:10.3 & HT   &    &     &     &          & r-band NA. In cluster.\\
J0022.7-6652& 00:22:44.98 & 102  & 2  & 4.9&0.29  & RS        & z=0.234, r=18.48. Extended core and,  \\
            & -66:53:05.2 & FR-II&    &     &     &          & faint detached outer extensions.\\
            &             &      &    &     &     &          & Bright B, R, K, r id. Neighbors.\\ 
J0023.6-6710& 00:23:44.51 & 84   & 3  & 28.1&0.09 & RS       & z=(1.41), r=22.67. Triple. Faint K,  r id. \\ 
            & -67:11:07.9 & FR-II&    &     &     & GRG      & Prominent core. Galaxy group \\ 
            &             &      &    &     &     &          & associated with NW lobe; \\
            &             &      &    &     &     &          & Lobe confused with group emission.\\
J0024.4-6636*& 00:24:26.77 & 150 & 5  & 13.8&0.16 & RS       & z=(0.21), r=17.92. Bright core with two \\ 
            & -66:36:12.8 & FR-I &    &     &     &          & extensions. Weak outer lobes. Bright B, \\  
            &             &      &    &     &     &          & R, K, r id. Close companions. Cluster.\\
J0024.6-6751*& 00:24:41.77 & 36  & 4  &21.1 & --- &          & K, r-band NA. No core. \\ 
            & -67:51:10.8 & FR-II&    &     &     &          & No B, R id. 50" centroid position given.\\
J0025.0-6658& 00:24:59.87 & 50   & 6  & 15.8&0.15 & HzFR-I    & z=(0.64), r=20.69. Id not the brightest. \\
            & -66:58:08.1 & WAT  &    &     &     &          & B, R, K, r id. No separate core seen.\\
J0025.2-6701& 00:25:16.11 & 34   & 7  &1.2  &0.14 &          & Weak,  diffuse lobes.  \\ 
            & -67:01:32.5 & FR-II&    &     &     &          & Faint K-only object 3" E of offset \\
            &             &      &    &     &     &          & core(?) Core position given.\\   
J0025.9-6621*& 00:25:58.59 & 60  & 8  &215.3&0.49 & HzFR-I    & Strong core, 2 short \\
            & -66:21:20.5 & FR-I &    &     &     &          & extensions. Very faint id only in K,  r. \\ 
J0026.4-6721& 00:26:28.43 & 30   & 9  &1.7  &0.36 &          & z=0.274, r=18.15.  Core and 2 extensions. \\ 
            & -67:21:48.9 & FR-I &    &     &     &          & Bright id, close companion.\\
J0026.8-6643& 00:26:49.23 & 72   & 10 &8.8  &0.41 &          & z=(0.21), r=17.96. Core with 2 jets and  \\
            & -66:44:01.1 & FR-I &    &     &     &          & fainter extensions. Id in B, R, K, r.\\
J0027.2-6624& 00:27:15.59 & 252  & 11 & 24.9&0.13 & RS       & z=0.073, r=15.06.  Bright id.  \\
            & -66:24:18.5 & FR-II&    &     &     &          & Core with jet. Source in cluster.\\ 
J0028.4-6733*& 00:28:26.47 & 42  & 12 &18.0 &0.17 &          & z=(0.47), r=19.93. B, R, K, r id\\ 
            & -67:33:48.6 & FR-I?&    &     &     &          & at centre. Lobe peaks well recessed.\\
J0028.9-6809*& 00:28:59.73 & 36  & 13 &34.9 &0.23 &          & z=(0.22), r=18. Extended core.\\ 
            & -68:09:25.6 & FR-I &    &     &     &          & Bright B, R, K, r id. \\      
J0029.0-6755*& 00:29:00.09 & 38  & 14 &134.0 &0.73& HzFR-I    & z=1.748, r=20.03.  Strong core, 2 short \\ 
            & -67:55:50.2 & FR-I &    &     &     &          & extensions. B, R, K, r id. Quasar.\\
J0028.9-6631& 00:29:01.89 & 105  & 15 &9.6  & 0.15& RS       & z=(1.78), r=23.26. Bright core. Detached  \\
            & -66:31:52.8 & FR-II&    &     &     & GRG      & lobes. One relic lobe. Faint K,  r id  \\
            &             &      &    &     &     &          & in cluster. Galaxy group in NW lobe. \\
J0030.0-6628& 00:30:00.67 & 50   & 16 &3.2  &0.36 & RS       & z=(2.19), r=23.78.  Triple,  strong core. \\
            & -66:28:08.9 & FR-II&    &     &     &          & Relic lobes. K, r id.\\
J0030.0-6604*& 00:30:04.03 & 57  & 17 &8.9  &0.08 &          & Triple. Faint K, r id. \\
            & -66:04:43.6 & FR-II&    &     &     &          &                                          \\
J0030.4-6723& 00:30:23.26 & 32   & 18 & 6.6 & 0.11& HzFR-I    & Weak core. NE lobe with strong \\ 
            & -67:23:17.7 & FR-I &    &     &     &          & peak. Highly asymmetric source.\\
            &             &      &    &     &     &          & Very faint r id.\\  
J0030.7-6714*& 00:30:44.92 & 43  & 19 & 12.1&0.6  & HzFR-I    & z=(0.97), r=21.76. Bright core, \\
            & -67:14:36.4 & FR-I &    &     &     &          & 2 extensions. B, R, K, r id.\\ 
J0030.9-6626*& 00:30:54.12 & 40  & 20 & 36.0&0.65 & HzFR-I    & z=(2.07), r=23.64. Core, 2 short,   \\ 
            & -66:26:35.5 & FR-I &    &     &     &          & broad extensions. Very faint K id. \\
            &             &      &    &     &     &          & Clearly seen in r.\\
J0031.0-6744*& 00:31:01.60 & 84  & 21 &5.51 & --- & GRG      & z=(1.31), r=22.51. No core. K-band NA.  \\ 
            & -67:44:42.8 & FR-II&    &     &     &          & Faint r id. Close companion to N. \\ 
            &             &      &    &     &     &          & 2 galaxies along N lobe.\\
J0031.1-6713& 00:31:06.63 & 55   & 22 & 54.8&0.05 &          & z=(1.14), r=22.15.   \\  
            & -67:13:49.9 & FR-II&    &     &     &          & Id in K, r only. Companion in \\
            &             &      &    &     &     &          & B, R, K, r.\\
J0031.1-6642*& 00:31:09.01 & 66  & 23 & 7.0 & 0.19& RS       & z=(0.77), r=21.16.  Asymmetric, bright  \\
            & -66:42:28.6 & FR-II&    &     &     &          & compact N lobe and relic to South.\\ 
            &             &      &    &     &     &          & Core with B, R, K, r id. \\
            &             &      &    &     &     &          & LAS=1.5' if relic component included.\\
J0031.3-6657*& 00:31:21.21 & 66  & 24 &17.7 &0.02 &          & Triple. \\
            & -66:57:22.8 & FR-II&    &     &     &          & Faint K, r id at core (?). \\
J0031.5-6748& 00:31:32.44 & 50   & 25 & 2.2 &0.2  & RS       & z=0.355, r=19.10.  Weak core and  \\
            & -67:49:00.5 & FR-I &    &     &     &          & 2 weak extensions. B, R, K, r id.\\
J0031.8-6727*& 00:31:48.23 & 102& 26 & 12.7&0.18 & RS       & z=1.156, r=19.24.  Triple. 2.3' if NE \\
            & -67:27:17.1 &  FR-II&    &     &     & GRG      & relic extension included. B, R, K id.\\         
J0032.9-6614*& 00:32:57.64 & 32  & 27 & 17.0&0.73 & HzFR-I    & z=0.915, r=20.79.  Bright core and 2 \\
            & -66:14:18.7 & FR-I &    &     &     &          & diffuse,  short extensions. \\
            &             &      &    &     &     &          & Bright B, R, K, r id. Quasar.\\ 
J0033.4-6714& 00:33:29.35 & 78   &28  &9.5  &0.21 &          & z=0.407, r=18.38.  Bright core \\
            & -67:14:19.3 & FR-I &    &     &     &          & and jets. Elongated source with  \\
            &             &      &    &     &     &          & constant width.Bright B, R, K, r id.\\
J0034.0-6639& 00:34:05.61 & 1020 &29  &86.1 &0.05 & RS       & z=0.110, r=16.79. Triple.  \\
            & -66:39:35.2 & FR-II&    &     &     & GRG      & Bright B, R, K, r id. Inner N lobe; \\
            &             &      &    &     &     &          & Amorphous outer lobes.\\
J0035.0-6612*& 00:35:01.87 & 37  &30  & 10.5&0.55 & RS       & z=0.465, r=19.10.  Strong elongated   \\
            & -66:12:52.3 & FR-I &    &     &     &          & core. B, R, K, r id. Offset from peak.\\
            &             &      &    &     &     &          & Linear feature to S (no K, r id).\\
J0035.1-6748& 00:35:07.91 & 32   &31  & 1.44&0.3  & HzFR-I    & z=(1.82), r=23.32.  Weak source.   \\
            & -67:48:40.7 & FR-I &    &     &     & Relic    & Elongated core. Faint K, r id only.\\
J0035.2-6638*& 00:35:18.00 & 45  &32  & 11.9&---  &          & No B, R, r, K id. \\
            & -66:38:35.10& FR-II&    &     &     &          & No separate core seen. Centroid position. \\
J0035.4-6636*& 00:35:25.28 & 90  &33  &11.7 &0.14 & GRG      & z=(2.38), r=23.98.  Triple. Faint K, r id.\\
            & -66:36:09.0 & FR-II&    &     &     &          &                                         \\
J0036.9-6645& 00:36:54.19 & 30   &34  &1.5 &0.39  &          & z=(0.23), r=18.16. Core+twin jets. Bright  \\
            & -66:45:13.9 & FR-I &    &     &     &          & B, R, K, r id at core. Two neighbors.\\    
J0037.1-6649*& 00:37:08.60 & 52  & 35 & 39.5&0.07 &         & z=(2.11), r=23.69. Id in r, very faint in K. \\
            & -66:49:39.1 & FR-II&    &     &     &          &                                 \\
J0037.3-6647*& 00:37:19.05 & 72  & 36 &2.5 &0.43  & RS        & z=(1.30), r=22.48. Bright core. Asymmetric  \\
            & -66:47:05.6 & FR-II&    &     &     &          & lobe flux. Relic lobe to South. Id in K, r.\\
J0037.7-6807*& 00:37:45.16 & 72  &37  & 19.1&0.08 & RS       & Bright core, 2 jets.   \\
            & -68:07:41.6 & FR-II&    &     &     &          & Asymmetric lobe flux. Relic lobe to SW.\\
            &             &      &    &     &     &          & Faint K-only id. K id position given.\\
J0038.6-6732*& 00:38:36.11 & 90  &38  &47.1 &0.01 & GRG      & Faint core. Faint  \\
            & -67:32:01.6 & FR-II&    &     &     &          & K-only id. K id position is given.\\
J0039.4-6601& 00:39:27.23 & 72   &39  &21.7 &0.03 & RS       & z=(1.06), r=21.97.  Asymmetric. Faint core.  \\
            & -66:01:06.9 & FR-II&    &     &     &          & Id in R, K, r. Companion 4" to north.\\
J0039.8-6624*& 00:39:53.82 & 90  &41  &42.7 &0.03 & GRG      & No id in B,  R, or K  \\
            & -66:24:51.4 & FR-II&    &     &     &          & at peak. Faint r id position given.\\
J0040.2-6553& 00:40:14.21 & 66   &40  &3.2 &0.09  &           & z=0.512, r=20.28.  Faint,  detatched lobes. \\
            & -65:53:25.07& FR-II&    &     &     &          & Bright B, R, K, r id. Core?\\
J0040.2-6729& 00:40:17.83 & 45   &42  &3.4  &0.38 & HzFR-I    & Bright core, 2  \\
            & -67:29:56.9 & FR-I &    &     &     & RS       & extensions. Faint r id. 3 blue galaxies close. \\
J0040.3-6703*& 00:40:23.64 & 75  &43  &2.9 &0.06  &           & z=(3.65), r=25.06. Inversion symmetric  \\
            & -67:02:40.1 & FR-II&    &     &     &          & lobes. Neighboring strong source, no id.\\ 
            &             &      &    &     &     &          & Weak radio core at centre. \\
            &             &      &    &     &     &          & Faint r id,  brighter in K. \\
J0040.7-6724& 00:40:46.59 & 35   &44  & 3.1 &0.23 &          & z=0.296, r=19.07.  Bright B, R, K, \\
            & -67:24:35.8 & HT   &    &     &     &          & r id at head. Cluster.\\
J0040.9-6638*& 00:40:57.43 & *60&45  & 5.5 &0.05 &          & No obvious id in B,  \\
            & -66:38:37.7 &  FR-II &    &     &     &          & R, K. Faint r id at weak core. Galaxy \\
            &             &      &    &     &     &          & in N lobe. Asymmetric lobe separation.\\ 
J0041.7-6726*& 00:41:47.35 & 40  &46  &78.6 &0.19 &          & z=0.292, r=17.34.  Includes halo emission.  \\
            & -67:26:26.5 & HT   &    &     &     &          & Bright B, R, K, r id. Complex region.\\
J0042.1-6728*& 00:42:09.27 & 45  &47  &30.0 &0.17 &          & z=(0.25), r=18.33.  Bright B, R, K, r id.\\
            & -67:27:56.4 & HT   &    &     &     &          &                            \\
J0043.2-6751*& 00:43:17.28 & 30  &48  &40.2 &0.59 &          & z=(0.38), r=19.42.  Id in B, R, K, r. \\
            & -67:51:45.7 & WAT  &    &     &     &          & Clear cluster. \\ 
J0043.4-6738& 00:43:28.83 & 50   &49  &7.1  &---  & Relic    & Faint N lobe. Faint\\
            & -67:38:45.8 & FR-II&    &     &     &          & K-only id? No core. K id position given.\\
J0043.6-6624& 00:43:37.04 & 42   &50  &5.9  &0.13  & HzFR-I    & z=(0.99), r=21.81.  Id in B, R, K, r\\
            & -66:24:47.2 & WAT  &    &     &     & Relic    &                       \\
J0043.8-6659& 00:43:53.16 & 45   &51  &9.2  &---  &          & z=(0.39), r=19.47.  B, R, K, r id (star-like).  \\
            & -66:59:25.4 & FR-I &    &     &     &          & Brighter in B. No core. Hybrid morphology.\\
J0044.3-6746*& 00:44:20.18 & 96  &52  &60.2 &0.05 &         & z=(0.29), r=18.78. B, R, K, r id at core.\\ 
            & -67:46:59.9 & FR-II&    &     &     &          &                             \\
J0044.7-6656& 00:44:47.63 & 108  &53  &9.3  &0.14 & RS       & z=(0.72), r=21.01. Bright core, inner double.\\ 
            & -66:56:39.5 & FR-II&    &     &     & GRG      & Faint K, r id. Lobes with hotspots.\\
J0045.0-6620*& 00:45:01.39 & 45  &54  &15.9 &0.09 &          & Id only in R and   \\
            & -66:20:31.9 & FR-II&    &     &     &          & K at core. r image NA.\\ 
J0045.5-6726*& 00:45:31.87 & 40  &55  &11.8 &0.12 &          & z=(0.27), r=18.57. Bright B, R, K, r id. \\  
            & -67:26:36.7 & FR-I &    &     &     &          & Weak core. Asymmetric extensions. \\ 
            &             &      &    &     &     &          & 2 close neighbors.\\
J0046.1-6630*& 00:46:09.88 & 50  &57  &6.0  &0.39 & HzFR-I    & Core and 2 weak extensions.  \\
            & -66:30:09.2 & FR-I &    &     &     & RS       &  No B, R, r id. K band NA. \\ 
            &             &      &    &     &     &          & Core position given.\\
J0046.2-6637*& 00:46:12.72 & 50  &56  &19.2 &0.04 &          & z=(0.37), r=19.32.  B, R, K, r id? \\
            & -66:37:07.5 & FR-II&    &     &     &          & Weak core? Fat lobes. Alternately, quasar?\\
            &             &      &    &     &     &          & Star-like id 10-arcsec NW has z=(0.81), m=21.29.\\
J0046.4-6730*& 00:46:27.26 & 78  &58  &71.7 &---  & RS       & Faint K, r only id(?). \\
            & -67:30:17.4 & FR-II&    &     &     &          & Galaxy chain along N lobe. No core.\\
            &             &      &    &     &     &          & Faint r object at SE lobe peak.\\  
J0047.1-6715*& 00:47:12.51 & 60  &59  &17.2 &---  &          & Weak core? No B, R id.\\
            & -67:14:45.7 & FR-II&    &     &     &          &  Faint K, r id.\\ 
J0049.3-6705& 00:49:18.94 & 48   &60  &17.2 &0.14 &         & No K id. Core, \\ 
            & -67:05:02.5 & FR-II&    &     &     &          & faint diffuse r id. Companions.\\
J0049.3-6703& 00:49:22.16 & 78   &61  &2.6  &0.16 & RS       & z=(0.47), r=19.94.  Bright B, R, K, r id.\\ 
            & -67:03:58.5 & FR-I &    &     &     &          & Extended core.\\
J0049.9-6639& 00:49:58.05 & 132  &62  &11.4 &0.04 & GRG      & No K id.  \\
            & -66:39:25.5 & FR-II&    &     &     &          & Faint r id at core. Asymmetric lobes.\\ 
J0050.0-6619*& 00:50:03.63 & 50  &63  &10.8 &0.07 &          & Core position. No R,  \\
            & -66:19:42.9 & FR-II&    &     &     &          & K id. r-image NA. Possible faint B id.\\
J0052.7-6636*& 00:52:44.08 & 45  &64  &58.9 &0.05 &          & z=(1.33), r=22.53. Clear r id  \\
            & -66:36:30.9 & FR-II&    &     &     &          & at centre. K-band NA. No B, R id.\\
J0052.7-6651& 00:52:48.69 & 50   &65  &41.7 &0.22 &          & z=(0.24), r=18.21.  Galaxy chain along  \\
            & -66:51:12.5 & HT   &    &     &     &          & axis. Brightest galaxy at head. B, R, r id.\\
J0052.8-6641*& 00:52:51.59 & 66  &66  &232.0 &0.01&          & Prominent wing to W   \\
            & -66:41:27.7 & FR-II&    &     &     &          & at core location. No B, R, K, r id.\\
J0053.5-6553*& 00:53:32.72 & 55  &67  &205.4 &0.68& HzFR-I    & z=(0.73), r=21.05.  Bright core, 2 jets.  \\
            & -65:53:00.7 & FR-I &    &     &     &          & Star-like B, R, r id,  brighter in B.\\
J0054.6-6650& 00:54:39.59 & 60   &68  &3.5 &0.41  & RS       & Core, 2 faint extensions. \\
            & -66:50:35.7 & FR-II&    &     &     &          & No B, R, K, r id. Faint r id 2" from \\ 
            &             &      &    &     &     &          &  radio peak? Radio peak position given.\\
J0055.7-6610& 00:55:44.53 & 72   &69  &28.0 &0.12 &          & z=(0.22), r=18.04. Bright B, R, r id has \\
            & -66:10:41.7 & HT   &    &     &     &          & close companions.\\
J0055.9-6802& 00:55:57.03 & 40   &70  &74.0 & 0.64&          & z=(0.37), r=19.36. Bright core and  \\
            & -68:02:22.3 & FR-I &    &     &     &          & 2 extensions. Cluster. B,  R, r id.\\
J0056.4-6651& 00:56:27.16 & 52   &71  &22.2 &0.08 &          & z=(0.19), r=17.75. Bright B, R, r id at \\
            & -66:51:21.7 & WAT  &    &     &     &          & centre. Cluster. \\
J0056.6-6743& 00:56:37.90 & 47   &72  &5.8  &0.11 &          & z=(0.93), r=21.64. Morphology/LAS in doubt.  \\
            & -67:43:42.1 & FR-II&    &     &     &          & 1.8' FR-II or 47" hybrid morphology. R, \\
            &             &      &    &     &     &          & r-only id. Id also for NE hotspot.\\
J0056.9-6632& 00:56:57.22 & 75   &73  &4.4  &0.05 &          & z=0.249, r=18.60. Bright B, R, r id. Weak core.\\  
            & -66:32:39.4 & FR-II&    &     &     &          &                                           \\
J0057.0-6734& 00:57:04.39 & 66   &74  &5.4  &0.10 &          & z=0.307, r=19.02. Triple. Id in B, R, r. \\
            & -67:34:13.2 & FR-II&    &     &     &          & Object in B in east lobe.\\
J0057.1-6633& 00:57:11.76 & 40   &75  & 5.9 &0.18 & HzFR-I    & Faint diffuse radio  \\
            & -66:33:38.7 & FR-I &    &     &     &          & emission 1.2' S. Faint R id (?),  not \\
            &             &      &    &     &     &          & seen in B. Artifact in r-image. Central \\
            &             &      &    &     &     &          & radio peak position given.\\
J0057.2-6651& 00:57:07.00 & 126  &76  &29.7 &0.05 &          & z=0.236, r=18.19.  Asymmetric in extent, flux, \\
            & -66:50:59.5 & FR-I &    &     &     &          & morphology. Bright R, B, r id at bright core.\\
J0057.4-6606& 00:57:24.85 & 90   &77  &8.9  &0.08 & GRG      & z=(1.39), r=22.65. Faint id at core in R, r,   \\
            & -66:06:30.3 & FR-II&    &     &     &          & fainter in B. Several neighboring galaxies.\\
J0057.4-6703& 00:57:27.20 & 42   &78  &10.7 &0.4  &           & z=0.260, r=18.64. Bright B, R, r id. Cluster.\\
            & -67:03:19.9 & WAT  &    &     &     &          &                                   \\   
J0057.7-6701& 00:57:43.54 & 40   &79  & 10.1&0.32 &          & z=0.261, r=18.13.  Bright id in B, R, r.\\
            & -67:01:36.7 & WAT  &    &     &     &          &                                    \\        
J0057.7-6655*& 00:57:45.17 & 40  &80  & 2.6 &0.24 &          & z=(0.66), r=20.79.  r id at core.\\
            & -66:55:07.3 & FR-II&    &     &     &          & Bright star nearby.\\
J0057.8-6711& 00:57:50.16 & 37   &81  &4.4  & --- &          & Asymmetric in flux.  \\
            & -67:11:35.6 & FR-II&    &     &     &          & No B, R, r id.  \\
            &             &      &    &     &     &          & 50" centroid position given.\\
J0057.9-6633& 00:57:54.56 & 45   &82  &29.8 &0.27 &          & Bright core. Bright star close.\\
            & -66:33:58.4 & WAT  &    &     &     &          &  Id in K-band?\\
            &             &      &    &     &     &          &  Radio core position given.\\
J0059.6-6712& 00:59:41.17 & 40   &83  &9.61 &0.23 & HzFR-I    & z=(0.50), r=20.09. Core and 2 jets. \\
            & -67:12:58.9 & FR-I &    &     &     &          & Cluster. B, R, r id. Overlapping,\\
            &             &      &    &     &     &          & unrelated NE source with id.\\ 
J0101.1-6600& 01:01:07.69 & 150  &84  &10.9 &0.12 & RS       & z=(0.24), r=18.20.  Bright core, two\\
            & -66:00:18.7 & FR-II&    &     &     &          & detached lobes. Weak,  diffuse S lobe.\\
            &             &      &    &     &     &          & Bright B, R, r id. No id in N lobe.\\
J0101.5-6742*& 01:01:34.12 & 40  &85  &21.7 &0.39 & HzFR-I    & z=(0.59), r=20.50. B, R, r id. Several neighbors\\
            & -67:42:13.5 & WAT  &    &     &     &          &                   \\ 
J0102.1-6552*& 01:02:12.14 & 66  &86  &12.5 &---  & Relic    & z=(1.11), r=22.09. No core. Emission gap  \\
            & -65:52:19.6 & FR-II&    &     &     &          & between lobes. B, r id (?) on radio axis.\\
J0102.3-6614*& 01:02:15.88 & 60  &87  &15.0 &0.06 &          & Triple. Faint B, R(?), r id.\\
            & -66:14:49.7 & FR-II&    &     &     &          &                       \\  
J0102.4-6632*& 01:02:27.66 & 55  &88  &27.0 &0.03 &          & Faint r id. Not seen in B.  \\
            & -66:32:13.4 & FR-II&    &     &     &          & Weak core. Bright star object close.\\
J0102.5-6621& 01:02:31.63 & 50   &89  &12.8 &---  &          & z=(1.50), r=22.84. Central r id. No core.\\
            & -66:21:21.3 & FR-II&    &     &     &          &                      \\
J0102.6-6658& 01:02:37.05 & 60   &90  &7.9  & 0.35& HzFR-I    & z=(0.61), r=20.59. Bright core and twin,  \\
            & -66:58:26.4 & FR-I &    &     &     & RS       & faint extensions. No jets. R, r id only. \\
            &             &      &    &     &     &          & Several nearby galaxies.\\
J0102.6-6734& 01:02:41.48 & 60   &91  &9.9  & 0.13&          & z=0.065, r=15.53. Bright B, R, r id at core.\\ 
            & -67:34:02.8 & WAT  &    &     &     &          &                  \\
J0102.6-6750& 01:02:42.43 & 204  &92  & 38.0&0.4  &           & z=(0.13), r=16.78. Bright B, R, r id. Core,  \\
            & -67:50:32.5 & FR-I &    &     &     &          & 2 jets. May be larger? Rich cluster.\\         
J0102.9-6722& 01:02:56.48 & 38   &93  &3.72 & 0.13& HzFR-I    & z=(0.84), r=21.38. Bright B, R, r id at centre.  \\ 
            & -67:22:20.2 & WAT? &    &     &     &          &                    \\
J0103.1-6632& 01:03:10.03 & 30   &94  &2.46 &0.5  &           & z=0.398, r=18.87. Core, 2 faint extensions.  \\
            & -66:32:21.5 & FR-I &    &     &     &          & B, R, r id. Close companions in r. Faint id \\
            &             &      &    &     &     &          & in E component? Extent less than 30"?\\
J0103.2-6614& 01:03:15.05 & 120  &95  &101.2&0.006&          & z=0.331, r=18.22. Low axial ratio lobes. \\
            & -66:14:25.4 & FR-II&    &     &     &          & Weak core? Bright B, R, r id. \\
J0103.7-6632& 01:03:44.49 & 78   &96  &3.9  &0.10 & RS       & z=(0.59), r=20.53. Triple. B, R, r id. \\
            & -66:32:27.1 & FR-II&    &     &     &          & Close companion. Cluster?\\
J0103.7-6747*& 01:03:44.55 & 35  &97  & 6.7 &0.50 &          & z=0.329, r=18.44. Core and 2 extensions. \\
            & -67:47:51.8 & FR-I &    &     &     &          & Bright id in B, R, r.\\
J0104.1-6719& 01:04:08.42 & 40   &98  &13.7 & 0.08&         & z=(1.44), r=22.74.  No id in B,  R. Id in r. \\
            & -67:19:16.2 & FR-II&    &     &     &          &                     \\   
J0104.3-6609& 01:04:21.26 & 90   &99  &5.22 &0.07 & RS       & z=(1.19), r=22.26.  Inner double between  \\
            & -66:09:17.3 & FR-II&    &     &     & GRG      & outer lobes. r-only id.\\
J0104.4-6704& 01:04:27.68 & 90   &101 &8.8  &0.15 & RS      & z=(1.39), r=22.64. Triple. Faint r id.\\
            & -67:04:23.5 & FR-II&    &     &     & GRG       &                          \\
J0105.0-6608*& 01:05:00.85 & 52  &100 &47.1 &---  &          & z=(0.85), r=21.42.  No core. R id? Not   \\
            & -66:08:56.1 & FR-II&    &     &     &          & seen in B. Clear r id. Neighbours.\\
J0105.1-6617*& 01:05:08.22 & 37  &102 &10.0 &0.24 & RS       & Extended core, one extension. \\
            & -66:17:47.8 & FR-II&    &     &     &          &  No B, R id. Offset r, K  object\\
            &             &      &    &     &     &          & position given. Related to NE (no id) \\
            &             &      &    &     &     &          & source? 1' source? Bright star near id.\\
J0105.3-6736& 01:05:19.04 & 66   &103 &20.4 &0.18 & RS       & z=(0.26), r=18.49.  Inner double. \\
            & -67:36:04.7 & FR-II&    &     &     &          & B, R, r id at core. Close neighbors.\\ 
J0105.7-6604*& 01:05:44.38 & 50  &104 &18.0 &---  &          & No core. No B,  R,  r, K id.  \\
            & -66:04:04.6 & FR-II&    &     &     &          & Bright star close. Centroid position.\\
J0105.7-6609& 01:05:45.73 & 84   &105 &14.5 &0.02 &          & z=(0.98), r=21.78.  r id at weak core. \\  
            & -66:09:42.8 & FR-II&    &     &     &          & Asymmetric lobes. r object at E lobe peak.\\
J0105.7-6643*& 01:05:47.18 & 50  &106 &185.5 &0.73& HzFR-I    & Strong core, 2 extensions. \\
            & -66:43:52.4 & FR-I &    &     &     &          & No B, R, r id. Radio peak position.\\
J0106.0-6653*& 01:06:01.90 & 34  &107 &3.8 &0.49  &           & z=0.262, r=17.59.  Extended core. Bright B, \\
            & -66:53:37.3 & FR-I &    &     &     &          & R, r id. Group,  close neighbours.\\
            &             &      &    &     &     &          & Likely $<30"$ if E source unrelated; \\
            &             &      &    &     &     &          & very faint r id for E source?\\
            &             &      &    &     &     &          & R-only galaxy along extension to E.\\ 
J0106.2-6645*& 01:06:13.78 & 40  &108 &5.4  &---  &          & No B, R, r id. Weak core? \\
            & -66:45:37.9 & FR-II&    &     &     &          & Faint K id? 50" centroid position given.\\ 
J0106.2-6545*& 01:06:24.17 & 66  &109 &457.3&0.47 &         & Bright core, 2 jets.  \\
            & -65:45:21.1 & FR-I &    &     &     &          & Core position given. B, R id. r-band NA.\\
J0106.8-6645*& 01:06:49.61 & 108 &110 &38.3 &0.02 & GRG      & z=(1.21), r=22.29.  Faint r id at core.  \\
            & -66:46:07.3 & FR-II&    &     &     &          & Lobes asymmetric in flux, separation.\\
            &             &      &    &     &     &          &  Closer NW lobe stronger.\\
J0107.3-6626& 01:07:20.57 & 35   &111 &1.4  &---  & HzFR1    & Core (?). No id. Faint \\ 
            & -66:26:46.8 & FR-I &    &     &     &          & galaxies within source. Centre peak position. \\
J0108.2-6727& 01:08:13.35 & 30   &112 & 9.5 &0.43 & HzFR-I    & z=(1.22), r=22.32.  Bright core, 2 short  \\
            & -67:27:04.0 & WAT? &    &     &     &          & extensions; group emission? Id in r.\\
J0108.6-6655*& 01:08:38.92 & 37  &113 &19.5 &0.28 & HzFR-I    & z=0.528, r=19.95.  Sharp boundaries like \\
            & -66:55:28.6 & FR-I &    &     &     &          & FR-II. Id in R, r at core, not seen in B.\\    
J0109.1-6743*& 01:09:07.41 & 90  &115 &16.5 &0.08 & RS       & Extended,  weak core.   \\
            & -67:43:13.9 & FR-II&    &     &     &          & Bright B, R id. r-band NA. R-id position. \\
            &             &      &    &     &     &          & Relic S lobe,  hotspot at N lobe end.\\
J0109.1-6653& 01:09:08.92 & 45   &114 &2.7  &0.13 &          & Extended core. B, R, r id. \\
            & -66:53:21.3 & WAT  &    &     &     &          & Bright star close. R-band position.\\
J0110.7-6727*& 01:10:46.81 & 60  &116 &431.4&0.54 & HzFR-I    & Strong core, two extensions. \\
            & -67:27:55.9 & FR-I &    &     &     &          & No R, B, r id. Radio peak position.\\
J0110.7-6705& 01:10:46.98 & 78   &117 & 32.8&0.01 &          & z =(0.80), r=21.27.  Bright hotspots. Wings. \\ 
            & -67:05:15.0 & FR-II&    &     &     &          & Faint B, R, rid at core.\\
J0112.3-6634*& 01:12:18.92 & 60  &118 &456.5& 0.36&         & Extended core; 2 extensions. \\
            & -66:34:44.7 & FR-I &    &     &     &          & Bright B,  R id. r image NA. B-band position.\\
J0113.2-6705*& 01:13:16.76 & 66  &119 & 117.2&0.01&         & Core? B, R id. Brighter   \\
            & -67:05:38.4 & FR-II&    &     &     &          & in B. Neighbor. r image NA. B-band position.\\
            
\enddata 
\tablecomments{Sources selected from regions that lie outside the effective primary beam 
attenuation of 0.5 are denoted by an asterisk. WAT: wide angle tail structures, HT: head-tail type structure, 
RS: restarted source, HzFR-I: $z>0.5$ FR-I, GRG: giant radio galaxy. Redshifts given in brackets refer to values 
estimated using the r-band magnitude-redshift relation.}  
\end{deluxetable}
\clearpage

\section{Morphological classification}

Herein we present a characterization of the extended radio sources detected in ATLBS. 
Radio sources are large scale manifestations of central nuclear activity and may in some cases exhibit
structures that also pre-date that activity. The variety in radio morphologies may be indicative of basic differences 
in central engine characteristics and environmental conditions in which they develop; the 
structural form may also evolve as the central engine ages and the properties of the beam change. 
It is therefore meaningful and insightful to examine the distribution in source structures in radio 
source populations and over cosmic time,  not only from the viewpoint of the gross type --- whether
the radio morphology is of FR-I or FR-II type --- but also whether the sources may be 
currently active,  inactive or starting a new activity. 

\subsection{Radio source morphologies: general considerations}

Sources with FR-I morphology invariably manifest a pair of 
bright jets or two-sided extensions to the core and lack well defined lobes;
the cores are the brightest components.
FR-II sources are usually observed to have a pair of distinct edge-brightened lobes with 
possibly embedded hotspots/warmspots. Jets are often not obvious,  unless observed with high
dynamic range and sensitivity.  
In the ATLBS-ESS sample, some FR-I sources have bright elongated cores,  which with higher
resolution appear as twin collimated jets on either side of an unresolved core component. 
More often the cores have weak,  continuous and elongated extensions that hardly resemble classic jets. 

A relic double radio source is one in which the beams have ceased. It is relatively easier to recognize
relicts of FR-II type radio galaxies. Presence of twin edge-brightened lobes that,  when observed with high 
resolution and sensitivity,  lack compact hotspots,  jets and a core (or only display a weak core), 
may with some confidence classify the source as a relic FR-II.   An edge-brightened 
double source that lacks compact features in the lobes,  which nevertheless possesses a bright core, 
might be classified as a restarting FR-II radio source.  A double-double morphology is a fairly 
unambiguous case of a restarted FR-II radio galaxy. Restarted FR-II sources
may sometimes appear to have the typical edge-brightened hot-spot radio structure with additional 
remnant low-surface brightness relic emission that is disjoint and outside of the main source; the 
relict lobe (or lobes) may be collinear with the inner double or sometimes offset in angle.

The appearance of an FR-II radio 
source in which the beam power drops substantially as it approaches the end of activity is not
clear: it may be that the beams become lossy and transform to bright FR-I type jets (which are believed
to have intrinsically low beam powers): such sources may be recognized by the joint appearance of 
edge brightened lobes,  without hotspots,  together with possibly twin bright jets.

Since the original FR-I/II classification scheme was proposed by \citet{fan74},  
there have been several attempts to refine the criteria for separation of extended radio sources
into the two major morphological types \citep{owe89,  owe94,  zir95,  par96,  har07}.
\citet{par96} have noted that the morphologies of FR-I types are varied: the bulk of low
power radio sources have low luminosity double-lobe type morphologies and only a small percentage ($4\%$) 
resemble the archetypal source 3C31. Since we will be referring to the 3C31-type morphology often we
give its definition: such a source structure is described as "naked-jet" type, where a bright core is
accompanied by twin rapidly expanding jets (e.g. see Fig. 10).
The double-lobe type FR-I sources are considered as 
those in which the jets are light but not powerful enough to end in bright compact hotspots 
and are affected by surface instabilities and,  therefore,  may be lossy and bright. 
Thus a source that has two lobes, 
a bright core and twin extensions is also to be classified as an FR-I: the lobe-type FR-I. 

An unsolved problem is,  therefore,  whether at least some of the sources considered to
be FR-I type may,  in fact,  be FR-II sources in which the beams are close to the ends of their activity phase.
The possibility of conversion between the two morphological types,  specifically from FR-II to FR-I type, 
has been suggested and invoked previously \citep{owe94,  zir95,  led97,  sar09}.  Nevertheless,  it 
is accepted that the central engine in any FR-II source (also FR-I) does switch off and that such 
beam activity is only a phase in the lives of luminous galaxies. At present there is no 
understanding of the manner in which a radio source structure evolves as the central engine turns off,  and
whether the timescale of the turning off is abrupt or comparable to the active phase. 
If the switch off happens over a significant period of time,  we may expect to see significant 
numbers of dying FR-II sources: perhaps they are the non-3C31 type FR-I sources? 

\subsection{Radio source morphologies: classification method}

While the primary separation of ATLBS-ESS sources into FR-I (radio sources with bright cores 
and twin bright large-scale jets or edge-darkened lobes) and FR-II sources 
(radio sources with twin edge-brightened lobes) follows the traditional classification 
scheme of \citet{fan74},  we have adopted additional criteria in an attempt to identify sources within 
relic and restarted sub classes: 

FR-II (relic) type: These sources are edge-brightened radio galaxies that lack hotspots,  
jets and cores. These sources appear simply as pairs of diffuse edge-brightened lobes.

FR-II (restarted) type: Sources of this sub-class have 
edge-brightened lobes that may or may not have emission
peaks at their ends ({\it i.e.},  which may or may not be relic lobes) but contain an inner (recessed) 
double source or an elongated radio core or an unusually (for FR-II sources) bright radio core.

FR-I (relic) type: These are sources with a core with twin 
extensions or trails that do not resemble typical jets 
or lobes. No other compact emission apart from the core is present. Although the core is often 
the brightest part of the source the contrast is not high for sources in this sub-class.

FR-I (restarted) type: Sources of this sub-class are required to have an extended core or a 
core with a pair of bright,  short extensions,  which are necessarily accompanied 
by emission regions further out that are detached from the core and substantially fainter. 
Membership of this sub-class requires that the outer emission regions are not edge-brightened. 

In classifying the radio sources based on the above criteria we have used all the radio data available to us
(the original 50" images, the 4" images sensitive to compact components and the 6" images that give detailed 
morphologies). 

\section{Radio morphology distribution and characteristics in the ATLBS-ESS}

Of the total of 119 radio sources that comprise the ATLBS-ESS,  64 ($54\%$) are 
FR-II sources and 55 ($46\%$) are FR-I type sources (the FR-I sources include 18 sources 
that have wide angle tail (WAT) structures or head-tail (HT) type structures). 
The ATLBS-ESS,  which has sources down to mJy 
flux density,  has nearly three times larger fraction of FR-I sources compared to the 
well-known 3CRR sample,  which has a much greater flux density cutoff.

A large fraction ($83\%$) of the sample has firm identifications.
In 20 FR-II sources,  either radio cores are not obvious (although a candidate host galaxy is identified
in 6) or optical identifications have not been possible (although a core component is present in 4) or the best
optical candidate was chosen with the available data. Five of the 55 FR-I
sources have no optical counterparts identified although cores are clearly observed. 

A large fraction of the FR-I population (not including WAT,  HT types) 
is of the asymmetric type where a core is accompanied by a pair of lobes that appear to 
extend to different distances on the sky.  The ratio of lobe lengths on the two sides 
exceeds 1.5 in $30\%$ (12/37) of the sources 
(Figs. 5, 18, 30, 31, 42, 55, 76, 90, 92, 94, 107, 111). Even if two of the sources 
01 03 10.03 -66 32 21.5 and 01 06 01.90 -66 53 37.26 (Figs. 94, 107) may not meet the LAS criterion
(see Table~1) and hence are excluded, close to a third of the FR-I sources have more than $50\%$ difference in the lobe extents 
on the two sides; therefore, significant asymmetry appears to be a common occurrence in the ATLBS FR-I sources. 
Additionally,  the FR-I sources in ATLBS-ESS are mostly of the 3C31-type and not
of the double-lobe type structure that dominates samples with relatively higher flux density cutoffs.

Morphological differences and flux asymmetries are quite prevalent in these strongly asymmetric FR-I sources,
occurring in nearly half of them. All five asymmetric FR-I sources (Figs. 18, 31, 55, 76, 92)
that have pronounced 
flux asymmetries also have substantial morphological asymmetry.  
Sources with the greatest asymmetry in lobe extents do not necessarily have the greatest asymmetry in 
flux density; however,  one of the most asymmetric FR-I sources (Fig. 92) does
indeed have the greatest asymmetry in flux density.


In contrast to these morphological characteristics of FR-I sources,  a significantly smaller fraction 
of FR-II sources display such pronounced asymmetry in lobe extents. If we omit from consideration 
those ten FR-II sources that have no radio core and no obvious optical IDs (because lack of a 
core or clear ID in FR-II sources 
results in ambiguity as regards the location of the center in contrast to the case of FR-I 
sources where bright cores makes knowledge of the position of an optical Id less necessary),  and those where
the total extent is unclear (owing to confusion as to whether outlying components are independent
sources or extensions),  7/54 sources or $13\%$ alone display pronounced asymmetry in which
the ratio of extents exceeds 1.5 (Figs. 23, 26, 49, 59, 84, 87, 110). 
The lobe-extent asymmetry in these FR-II sources 
appears to be frequently accompanied by asymmetries in flux density but  
less often by asymmetries in morphologies.

\subsection{The $z>0.5$ FR-I radio galaxies.}

As mentioned in the introduction little is known 
of the low power radio galaxy population at higher redshifts because imaging their low brightness 
structures required highly sensitive radio surveys. The high surface brightness sensitivity achieved 
in the ATLBS has the potential to detect and
identify the radio morphologies of low power radio galaxies at relatively high redshifts,  besides
the high redshift powerful extended radio sources. In Table~1 we have indicated the
22 FR-I type radio sources from the ATLBS-ESS that are estimated to have redshifts $z > 0.5$. 
In this section we have attempted to characterize their radio structures and make a comparison with the structures 
of the remaining, lower redshift ($z<0.5$) FR-I sources.

For three of these FR-I sources we have spectroscopically measured redshifts:
two have broad emission lines and hence classify as FR-I quasars. 
Seven sources have optical hosts that are either very faint in the r-band images or are not
seen in these images. These sources are likely to have redshifts $z$ exceeding 0.5. In the case of 
00 57 11.76 -66 33 38.7 (Fig. 75),  it was not possible to obtain 
the magnitude of the host galaxy in either r-band (due to an image artifact at the source location) 
or K-band (because of contamination from a bright star in the vicinity). 
In this case we used the FR-I/II break power to estimate 
the redshift using the measured total flux density of the source and the 
assumption that this break power does not evolve with redshift. 
Using the upper limit to the total 1.4~GHz powers of FR-I 
sources---the break power of $10^{25}$~W~Hz$^{-1}$---we use the total flux density of the FR-I 
source, 5.69~mJy, and derive an upper limit of $z < 0.63$ to the redshift. 

Five of the 22 FR-I sources are of WAT or probably WAT-type structures.  The remainder of high redshift FR-I sources 
have predominantly 3C31 type morphologies (14/17) with only 3 having lobe-type morphologies. 

The low redshift FR-I population in the ATLBS-ESS sample (excepting those with WAT/HT type and one
with hybrid structure) has a similarly low fraction (3/19) of lobe-type morphologies. In contrast to 
the findings of \citet{par96}, whose use the much brighter B2 sample, where only a small percentage ($4\%$) of FR-I sources 
are of the type resembling the archetypal source 3C31,  the ATLBS survey finds substantially larger fractions 
of 3C31-type sources.  More than a fourth (6/22) of the high redshift FR-I sources are of core-twin-jet type that 
appear to have the brightest cores in the entire sample of ATLBS-ESS sources. In these the extensions are weak 
or ill-formed with most of the source dominated by the bright core. The fractional abundance of 
this source type at $z<0.5$ is smaller (3/33).

The low and high redshift FR-I sources do not display differences in the incidence 
of pronounced asymmetry in lobe extents. If we omit the WAT and HT sources from consideration,  
of the 12 FR-I sources that have lobe extents differing 
by more than $50\%$, 7/19 are at $z<0.5$ and 5/17 are at $z>0.5$. 

Because of the edge-darkened nature of FR-I sources in which the lobe surface brightness progressively 
diminishes with distance from the core, it is more meaningful to examine the structures of  
FR-Is with relatively larger angular size to identify relic and restarted radio sources. We examined sources 
with angular size $\ge 1$~arcmin,  which ensured well resolved images from the ATLBS observations.
There are 4 sources that are identified as being at high redshift ($z > 0.5$)
and 11 that are identified as being at low redshift ($z < 0.5$). 
While $18\%$ (4/22) of high redshift FR-Is have angular sizes larger than or equal to an arc-minute 
nearly double that fraction ($33\%$,  11/33) is seen to have these large angular sizes in the low redshift FR-I 
population.  Two of the 11 low redshift FR-I sources are WAT type; of the remaining nine $z < 0.5$ FR-I sources, 
two are recognized as having restarted nuclear activity (Figs. 5 and 30). 
Among the four $z> 0.5$ FR-I sources with LAS $\ge 1$~arcmin,  one is recognized as potentially having a 
restarted activity (Fig. 90) and another as a source in which the nuclear activity has ceased
(Fig. 31). For these small sub-samples derived from the ATLBS-ESS, 
there is no evidence for a difference in the incidence of restarted and
relic FR-I sources at redshifts below and above $z=0.5$. 

In summary,  we have attempted to compare the morphologies of low and high redshift FR-I sources using the
ATLBS-ESS. There appear to be no major differences in morphological properties of the high and low 
redshift samples of FR-I sources; in their asymmetry properties as well as the fractions with structures
corresponding to relic or restarted sources. There is,  however,  indication that there is a higher abundance 
of FR-I sources with a bright-core and twin weak jets at higher redshifts.

\subsection{Relic sources}

Sources in which the core is weak or absent and no hotspots or jets are seen are candidates for 
relics. However,  as mentioned earlier,  if the central activity ceases gradually rather than 
abruptly,  there may be FR-II sources observed that are close to the end of their lifetime and at a time when
the lobes are devoid of hotspots,  the central region has a 
core and twin,  lossy,  bright jets,  the lobe has suffered substantial entrainment and may be  
buoyantly displaced by the ambient medium. Such sources might be mistaken for an FR-I or even a restarted FR-II. 
On the other hand,  a relic FR-I may continue to be centrally brightened 
as a consequence of weakening of the peripheral parts of the lobes as the energy injection ceases
and brightening at the center as weakened jets become increasingly lossy. 

Based on the definitions adopted in Section~3.2 for relic FR-I radio sources,  we have 
recognized two sources as being relic FR-Is: J0035.1-6748 (Fig.~31),
and possibly J0043.6-6624 (Fig.~50). In J0035.1-6748,  there is no compact emission (even 
at the core). The source J0043.6-6624 is a WAT-type relic FR-I in which a weak core 
appears disjoint from the two extensions. In J0035.1-6748, the central extended emission 
component is associated with a faint galaxy and to the south is a faint extended feature. There is no compact emission
seen at 4" resolution. When smoothed to 10" resolution a much larger source emerges particularly to the
north where a faint extension is seen extending nearly twice the extent of the feature to the south. The
bright source at the centre accompanied by weaker extensions together with the lack of any core or jets at 
the resolution of 4" are characteristics that suggest the source to be a relic FR-I.

As for relic FR-II radio sources,  only one source J0102.1-6552 (Fig.~86) qualifies the 
definition; however,  we also note J0043.4-6738 (Fig.~49) as a possible relic FR-II candidate. 
The first source has no core and no obvious hotspot at the ends of the lobes. The second candidate 
has a weak compact source (at 4", contours not shown in the figure) between the two lobe components but 
its identification as the radio core
is uncertain because it is located off the radio axis (about 3 arcsec west). Moreover,  although there is an 
extended emission peak in the southern lobe (also at 4", contours not shown) it is not at the end of 
the lobe as is expected in an archetypal FR-II 
source. In both sources the lobes have relatively small axial ratios as is expected for evolved relic lobes.

Most sources manifest central cores,  jets,  or hotspots that are key indicators of continuing AGN 
activity. However the sample of ATLBS-ESS relics have been selected on the basis that they 
lack most of these components. The poor abundance of relic radio sources even in a survey
such as the ATLBS that has exceptional surface brightness sensitivity adds weight to the 
well-known rarity of relic sources \citep{blu00}. 
It may be noted here that the identification of relic sources has been done herein 
solely on the basis of radio morphology and without considering additional indicators such as 
the spectral index \citep{par07,dwa09}.

To gauge the relative effectiveness of our survey in identifying relic radio sources,  we compared the ATLBS 
images with the only other comparable radio survey for the survey region: the 843~MHz SUMSS \citep{boc99, mau03}. 
The ATLBS sources were identified from images with beam FWHM 50~arcsec,  which is very similar to
the resolution in SUMSS. The observing frequencies of the two surveys are not too far 
apart, although the image rms noise in SUMSS is more than 10 times higher than that in ATLBS. 
Of the 4 relic sources that have potentially been identified in ATLBS,  
one (J0035.1-6748; Fig.~31) is not detected in SUMSS whereas it is detected at the
14$\sigma$ level in the ATLBS.  The remaining relic sources are detected in SUMSS; however,  they are 
observed with peak brightness of 3-5 times rms noise where as in the ATLBS images they are detected 
at 25-100 times the rms noise in the ATLBS. Their angular extents in the SUMSS image are 
in most cases significantly smaller than their angular sizes in the ATLBS. 
The ATLBS has,  therefore,  clear advantages in sensitivity to detecting relic radio sources. 

The FR-I and FR-II relics in our survey are not among the sources with the lowest surface brightness. 
Only J0035.1-6748,  which is the FR-I relic source,  
has a low surface brightness of 3.5~mJy~arcmin$^{-2}$.
Both the FR-I relic sources are at redshift $z>0.5$.

\subsection{Restarting radio sources}

Admittedly,  identification of relic or restarted activity amongst the lower-power FR-I radio 
sources may often be ambiguous. The properties of the ambient gas and the time elapsed since the cessation
of beam activity are some of the factors that influence the appearance of remnant synchrotron plasma. 
It is interesting to compare methods for recognizing sources that are in the restarting phase
among FR-I and FR-II types. In both types recognizing a source to be restarting hinges on 
recognizing relic features; however,  a problem specific to FR-I sources is that the lobes 
fade with distance from the core and this makes it difficult to 
cleanly separate renewed activity from past activity especially if the source is not well resolved.
The likelihood of detecting relic emission is greater if the host galaxy resides in a dense
cluster medium that may ram pressure limit expansion losses. 
An example of such a circumstance may be in the source J0024.4-6636 (Fig.~5). It has a central
core with twin extensions that are accompanied by weak,  diffuse emission regions farther out that are 
asymmetrically located
with respect to the central source. The host is in a rich cluster. The western outer lobe appears to be 
experiencing intracluster gas weather: it is relatively brighter,  closer to the core and 
also appears to be offset from the source axis where there is a concentration of galaxies (e.g. \citep{sub08, hot11}. 
The outer lobes of this source may be in a transition state fading to become radio-ghosts.

The components needed to classify an FR-I source as restarting are an active
core together with outer extensions that represent older activity. Morphological signature of an active
core is a compact central source accompanied by twin jets; at lower resolution the renewed activity
may appear as an elongated source or simply a bright core with a relatively steep spectral index.
Evidence for relict FR-I type activity may be in the form of substantially weaker 
extensions (single or twin) that take the form of broad (non edge-brightened) 
emission regions that may be detached from the core. Based on the criteria in Section~3.2 
we have recognized 7 FR-I restarting sources in the ATLBS-ESS. 
Of these three are likely to be at redshifts exceeding 0.5. 
On the other hand,  21 ATLBS-ESS sources are recognized as candidate restarted FR-II sources. 
They have the following defining characteristics: an edge-brightened
lobe morphology (with or without hotspots) accompanied by a central bright,  extended core or an inner-double source
straddling the core. We also include sources where only one relic lobe is observed along the radio axis. 
In Table~1 the candidate restarted sources have been indicated. 

In all,  $24\%$ (28/119 sources, including ) of ATLBS-ESS radio sources appear to have restarted their 
nuclear activity.  By source type,  $33\%$ (21/64) FR-II sources compared to only $13\%$ (7/55) FR-I
type have signatures corresponding to a restarting of nuclear activity: the fraction of FR-II 
sources that appear to have recently restarted is about twice that of FR-Is. It may be cautioned here that
these estimates are tentative in the light of above mentioned difficulties with reliable recognition
of restarting especially in the case of FR-I sources. At least 11 of the 21 restarted FR-II 
sources are estimated to have redshifts $z>0.5$. 

The radio sources identified as restarted radio galaxies exhibit non-classic FR-I and FR-II morphologies. 
For the larger sample of restarted FR-II radio sources we have compiled the following characteristics: the average 
fractional core flux density is 0.17 which is more than twice the value for normal FR-II sources in the ESS (including
seven upperlimits and excluding 
the 2 relics) and twice below the value for normal ESS FR-Is (that excludes relics, restarts and the hybrid morphology
source). Excluding the two quasars among the restarted FR-IIs does not change 
the fractional core flux density. Also if we exclude the 9 FR-Is with fractional core flux densities above
0.5 (to exclude possible quasars) the value still remains high at 0.27 for the FR-Is. Although this value is 
likely to be lower because several FR-I cores are blended with surrounding jet emission, the restarted FR-IIs 
appear to have rather high fractional core flux densities, intermediate between FR-IIs and FR-Is. Also as many 
as 10 of the 21 have at least one relic radio lobe and 4 out of the 5 with the most prominent cores have at 
least one relic lobe. The 7 sources that have either an inner double or extensions to the core have an
average fractional core power of 0.09. Only 5 out of 19 (excluding the two quasars) have lobe pairs with 
hotspots in them three of which also have an inner double. For the 8 restarted FR-II sources whose classification
hinges on having bright cores in association with edge-brightened lobes the average linear size is 606~kpc which 
is larger than the average linear size (514~kpc) for the 25 normal FR-IIs for which sizes could be estimated.
Hence projection-related effects may not be playing a significant role in boosting the core flux density.

Four of the seven restarting FR-I sources appear to have pronounced asymmetry in lobe extents; in these
the lobe extent on one side is more than one and a half times the extent on the opposite side. 
In contrast,  only 3 of the 20 restarted FR-II sources (Figs.~23, 26, 84) 
show pronounced side-to-side asymmetry in lobe extents 
(with ratio $>1.5$). There is an indication therefore that restarting FR-I sources are more likely to be 
asymmetric. 

In many restarted radio sources the renewed activity manifests as an inner double that is embedded
within a pair of outer relic lobes. 
J0031.1-6642 (FR-II; Fig.~23), J0031.8-6727 (FR-II; Fig.~26) and J0035.0-6612
 (FR-I; Fig.~30) 
are ATLBS-ESS sources where on one side there is an accompanying elongated diffuse emission,  
aligned but well separated from the main source. Such morphology may be indicative 
of restarting after a relatively longer time.  In both source types there may be considerable spread
in the time spent in a quiescent mode and it is remarkable that in all the cases identified the radio axis
appears to be unchanging over successive active phases.

These sources with detached but aligned relic-like emission indicate that the lobes,  once they cease 
to be active,  may not always expand and experience a drastic drop in surface brightness  
\citep{lea91}. We discuss each of these three sources below.
In all three cases the weak,  aligned and well separated extended emission is seen only on one side. 

J0031.1-6642 (Fig.~23): A 66~arcsec FR-II radio source with four components. 
The bright core
is identified with a star-like object. Towards the south is a hotspot-like lobe and to the north (at 
much larger separation than the southern hotspot) is another hotspot-like lobe. All three are
aligned. There is no bridge connecting the northern hotspot to the core. Interestingly,  to the south of
the closer,  southern hotspot is a relic-like extended feature (with no galaxy identification).  
This extended source is oriented at an angle with respect to the radio axis formed by the core and the 
two hotspots. 

J0031.8-6727 (Fig.~26): This 102~arcsec FR-II source has four components. 
This is a classic triple
with a bright core and two edge-brightened lobes straddling it. This triple is asymmetric in
lobe separation. The core is connected with the NE lobe by fainter narrow emission. The 
farther SW lobe is separated from the core by a large emission gap. There is the fourth 
component to the extreme NE separated from the NE lobe by a gap in emission. This component
is faint and lies exactly on the radio axis and on the side of the closer lobe. 
The core is identified with a star-like object. We
have obtained its optical spectrum. It shows at least two broad emission lines. Given its angular
size (102 arcsec),  its redshift (z=1.156) and morphology,  the source is a restarted,  845~kpc giant radio quasar.

J0035.0-6612 (Fig.~30): A bright extended core with a pair of bright jets. To the south
of the FR-I radio source is a narrow,  almost constant-width extension after a gap. 

In both the restarting FR-II sources J0031.1-6642 as well as
J0031.8-6727,  the distant tail-like faint emission that is plausibly a relict
lobe is on the side of the shorter lobe. The latter source has an optical host that appears star-like 
and has broad emission lines: this is an FR-II quasar and,  therefore,  the jet axis is likely inclined at a 
large angle to the sky plane.  If we assume that the side-to-side asymmetry is owing to light travel
time effects and consequently the shorter lobe is on the far side,  the one-sided appearance of the
relict emission may also arise from light travel time effects in which the relict emission on the 
near side,  which is being observed at a later time,  has faded from view where as the lobe on the far
side continues to be above the detection threshold. Such a possibility allows an estimate of the time 
over which a lobe may fade as a result of expansion losses.

Alternately,  it may be that the ambient gas density is 
greater on the side of the shorter lobe and that causes the jet advance speed on that side 
to be lower and also ram-pressure impedes the expansion loss in the relict lobe on the same side.
In the case of J0031.1-6642,  the optical host is once again star-like and the two close-in 
lobes appear compact with little diffuse structure; the phenomenology in this case may be the same. 
However,  it may be noted that neither source shows a jet on the side of the more distant lobe,  
which is expected owing to beaming effects if light-travel-time is responsible for the observed asymmetries. 

We describe two other interesting restarted sources below.
J0023.6-6710 (Fig.~3) is a source with a triple structure. 
It is a 1~arcmin source with three separate components all aligned along an axis. The strong core is 
accompanied by a weak extended feature to the south-east and to the north-west by a strong extended source. 
The NW source is the strongest and it has a jet-like short extension to the NW,  again aligned with the main 
radio axis. This stronger and more extended NW component is
identified with a galaxy that appears to be a member of a close group of four galaxies. 
We have a redshift measurement available for the central galaxy associated with
the NW component: $z=0.78$. The redshift we infer for the galaxy associated with the central core is $z \sim 0.72$.
Given their common axis and the extensions along that axis for two outer sources it appears likely they are all 
related. A likely scenario is that the triple is a 
restarted FR-II radio source where the NW lobe happens to be co-located with the cluster radio emission.

J0028.9-6631 (Fig.~15) is a source similar to J0023.6-6710. It is triple with
a bright core identified with a faint object. The core is straddled by two very dissimilar radio sources. 
The northern source is edge-brightened and extended towards the core and at high resolution shows a hotspot. 
This NW lobe is associated with a group of three bright galaxies. None of it is located at the hotspot peak.
The southern source is a weak elongated relic-like lobe extended towards the core. We have classified the 
source as a restarted FR-II radio galaxy where one lobe shows a hotspot whereas the opposite lobe is a relic.
Alternately, the northern lobe could be much fainter and below the survey sensitivity.

Three of the ATLBS-ESS restarted sources (J0037.3-6647 and J0104.3-6609 both 
of which are FR-II sources and J0049.3-6703,  which is an FR-I source) are not apparent 
at 843~MHz in the SUMSS survey. All of the other restarted sources detected in the ATLBS survey
are significantly larger in angular size in the ATLBS images compared to that in SUMSS,  in some cases
the angular extent detected in ATLBS is more than a factor 2 of the SUMSS extent. 
The diffuse, disjoint tail-like elongated emission observed in the two sources ---
J0035.0-6612 and J0031.8-6727 --- are not detected in 843~MHz SUMSS
images,  whereas in J0031.1-6642  the relic emission is detected in SUMSS at 5 times the image rms noise. 
The ATLBS survey,  with its order-of-magnitude higher surface brightness sensitivity, has indeed 
facilitated recognition of restarted radio galaxies.

The outer lobes in restarted FR-II radio galaxies are in several cases lobes where there are no hotspots and hence 
are relic lobes with relatively low surface brightness. The lobes associated with
the 17-arcmin giant radio galaxy, J0034.0-6639 (Fig.~29) are the faintest 
(1~mJy~arcmin$^{-2}$; also see next section) followed
by the lobes of J0022.7-6652 (Fig.~2) and the 4-arcmin cluster source, J0027.2-6624 (Fig.~11) 
both with surface brightness about 3.7~mJy~arcmin$^{-2}$. The median surface brightness of 
these faintest lobes of the restarted FR-II source sample is about 6~mJy~arcmin$^{-2}$; these are detected in ATLBS 
images a factor 50 above the image thermal noise.  With no other examples of large angular size faint
emission regions (other than the 17-arcmin giant radio galaxy or the 4-arcmin cluster radio source) there is indeed a 
dearth of large angular size low-surface-brightness radio source components.

In Table~2 we give a summary of the source categories found in the ATLBS-ESS.

\begin{deluxetable}{rr} 
\tablewidth{0pt} 
\tablecaption{Summary of source categories} 
\tablehead{ 
\colhead{Total} &  \colhead{119} \\}
\startdata
FR-I            &  55           \\ 
FR-II           &  64           \\
                &               \\
                &               \\
FR-I            &               \\
Relics          &   2           \\ 
Restarts        &   7           \\
$z>0.5$         &  22           \\
                &               \\
                &               \\
FR-II           &               \\
Relics          &   2           \\
Restarts        &  21           \\
                &               \\
                &               \\
GRGs            &  14           \\
FR-I            &   0           \\
FR-II           &  14           \\
                
\enddata 
\end{deluxetable} 

\begin{deluxetable}{rrrccc} 
\tablewidth{0pt} 
\tabletypesize\scriptsize 
\tablecaption{Candidate giant radio galaxies from the ATLBS-ESS } 
\tablehead{ 
\colhead{Name} & \colhead{RA} &  \colhead{DEC} & \colhead{redshift} & \colhead{LAS} & \colhead{Linear size}\\
\colhead{} & \colhead{J2000} &  \colhead{J2000}  & & \colhead{arcsec} &  \colhead{Kpc}\\    }
\startdata 

J0023.6-6710 & 00:23:44.51 & -67:11:07.9 & (1.41) & 84  & 714\\
J0028.9-6631 & 00:29:01.89 & -66:31:52.8 & (1.78) & 105 & 897\\
J0031.0-6744 & 00:31:01.60 & -67:44:42.8 & (1.31) & 84  & 710\\
J0031.8-6727 & 00:31:48.23 & -67:27:17.1 & 1.156 & 102 & 845\\
J0034.0-6639 & 00:34:05.61 & -66:39:35.2 & 0.110 &1020 & 2022\\
J0035.4-6636 & 00:35:25.28 & -66:36:09.0 & (2.4)  & 90  & 744\\
J0038.6-6732 & 00:38:36.11\tablenotemark{(a)} & -67:32:01.6 & ---     & 90  & ---   \\
J0039.8-6624 & 00:39:53.82\tablenotemark{(a)} & -66:24:51.4 & ---     & 90  & ---   \\
J0044.7-6656 & 00:44:47.63 & -66:56:39.5 & (0.72) & 108 & 781\\
J0049.9-6639 & 00:49:58.05\tablenotemark{(a)} & -66:39:25.5 & ---     & 132 & ---    \\
J0057.4-6606 & 00:57:24.85 & -66:06:30.3 & (1.39) & 90  & 764\\
J0104.3-6609 & 01:04:21.26 & -66:09:17.3 & (1.19) & 90  & 750\\
J0104.4-6704 & 01:04:27.68 & -67:04:23.5 & (1.39) & 90  & 764\\
J0106.8-6645 & 01:06:49.61 & -66:46:07.3 & (1.21) & 108 & 901\\
\enddata 
\tablenotetext{(a)} {These sources have identifications that are too faint to determine reliable 
r-magnitudes. Given their faint magnitudes they are expected to lie at relatively high z.}
\tablecomments{Redshifts given in brackets refer to values estimated using the r-band magnitude-redshift
relation thus the corresponding linear sizes are also very estimated.}
\end{deluxetable} 

\subsection{The lowest surface brightness radio sources}

The ATLBS has a 5$\sigma$ detection limit of $0.6$~mJy arcmin$^{-2}$,  which is deeper 
than that in wide-field surveys like the WENSS,  NVSS or the SUMSS which have 
detection limits equivalent to 4-9~mJy~arcmin$^{-2}$ at 1.4~GHz.
Examples of sources in the literature that have the lowest radio surface brightness are
the cluster-wide halo source in Coma \citep{kim90},  the 
unusual relic PKS B1400-33 \citep{sub03},  the relic 
Rood27 \citep{har93} and the giant radio source SGRS J0515$-$8100 \citep{sub06}. 
The lowest surface brightness source component observed in the ATLBS survey are the lobes of 
the giant radio galaxy J0034.0-6639 (Fig.~29),  which have a surface brightness of 1~mJy~arcmin$^{-2}$ 
that is three times lower than that of the faintest radio structures known: 
the 3~mJy~arcmin$^{-2}$ cluster-wide relic in Coma. The lobes of several of the restarted
FR-I and FR-II radio galaxies (see Section~4.3) have surface brightness
about 2.5-6~mJy~arcmin$^{-2}$.  The lack of more detections of low surface brightness extended radio
sources in the ATLBS,  along with indications based on X-ray imaging for the existence of expanded radio
galaxies with low pressure lobes \citep{fab09},  argues for rapid disappearance of relic radio lobes 
once the energy injection switches off (also see \citet{mac11}).  

J0034.0-6639 is the largest angular size source in the ATLBS. It
has giant radio lobes that are amorphous over their whole extent except for a relatively small embedded 
edge-brightened component in the northern lobe. The structure of this compact feature 
is best represented in the 6~arcsec resolution image. 
The location of this component is close to midway along the northern lobe
and its structure indicates that it might be
the leading head of a new jet propagating through the older radio lobe. 
Surprisingly,  there is no discernible
counterpart to this relatively compact feature in the southern lobe. At the lowest resolution
of 50" the large-scale source morphology becomes apparent; warm spots are seen towards the
leading edges of the giant lobes.

The radio axis is close to the minor axis of the optical host galaxy - a characteristic 
noted for giant radio galaxies \citep{sar09}. 
The source is barely detected in the 45~arcsec, 843 MHz SUMSS image.

\subsection{Unusual and interesting sources}

J0043.8-6659 (Fig.~51) is a source whose nature is difficult to understand. Its two 'lobes' are
dissimilar in morphology. The eastern component is strong and core-like but extended in a NE-SW direction 
whereas the western component has an EW jet-like linear structure. There is a bright star-like
bluish object seen at the centre between the two components. While it is observed in B and R it almost 
disappears in K band.
There is no radio core and the spectrum of the central object has no emission lines. While several
of the characteristics suggest a quasar the lack of emission lines is a problem for the 
interpretation.

An unusual source is J0056.6-6743 (Fig.~72) where the collinear components could all be related and forming
a very asymmetric (in lobe separation) edge-brightened radio source; however,  there is a 
clear optical identification for the eastern,  well-removed 'hotspot' and this 
casts doubt on the nature of the associations. There are,  however,  arguments in 
support of both possibilities - the eastern hotspot may be an unrelated source or it could be 
the bright end of a northern lobe. Its location on the radio axis of the source 
and its resemblance to the hotspot at the end of the 
southern lobe suggest that it is related to the source. Optical spectroscopy of the ID in the NE source 
could reveal whether it is a star or galaxy. 

J0057.1-6633 (Fig.~75): This is an interesting lobe-type,  35~arcsec FR-I radio galaxy
that is accompanied by a possible relic emission visible as a weak,  extended feature to the SW, aligned with the 
radio axis but separated by a large gap in emission. At 4" resolution it is completely resolved out 
except for a weak compact source. Several faint galaxies are seen within the extent of the relic feature.
The nature of this SE extended emission region is unclear nor its relation to the FR-I radio source. 
Its surface brightness is among the lowest known,  2.8~mJy~arcmin$^{-2}$. 
The FR-I source is bright at the centre and has
edge-darkened lobes. A core is seen at 4" resolution; it is associated with a faint object in R. This
object is not seen in B. Unfortunately,  at r-band,  an artifact extending east-west has 
prevented a good image of this region. 

J0102.6-6750 (Fig.~92) is a large angular size FR-I radio galaxy with a very asymmetric morphology. 
But for the short jet-like feature to the east and the detached weak source it points to the source
could be classified as a head-tail radio soource. Not only are the emission regions on the two sides
very different in morphology they are also different in angular extent and total flux. 

We have also noted sources,  not belonging to the 119 ATLBS-ESS,  whose morphologies
at 6" resolution are unusual and difficult to classify. At this resolution they appear as a collection
of weak 1---3 sigma sources. Several of these sources have recognizable 
structures in images smoothed to lower resolution, that appear to be at low surface brightness levels of 
about 5~mJy~arcmin$^{-2}$. Some however just appear as weak and diffuse elongated emission regions.  
They did not get selected 
in the procedure adopted for the ATLBS-ESS. There the procedure involved recognizable structures in
6~arcsec resolution images with angular extent $\ge~30$~arcsec. However,  these sources were observed to be
connected sources in the intermediate resolution images (images smoothed to 10 and 20~arcsec beam FWHM). 
At these resolutions,  at least one is of head-tail structure with a galaxy identification,  
one is a faint linear source with galaxy identification in the central regions,  
and the others are weak diffuse sources; all have angular extents between 30" and 1~arcmin. 
The appearence of the some of the sources suggest relic stages of head-tail type or FR-II type radio sources.

\subsection{ATLBS-ESS: Galaxy and group halos?}

Apart from the sample of 119 sources given in Table 1, we compile 
ATLBS-ESS sources that are largely devoid of compact 
components at 4'' resolution but are found to have significant flux on extended scales (more than 
$50\%$ of the flux at 4'').  These $\sim 1$ mJy radio sources are likely candidate halo-type radio sources that
may be associated with individual galaxies,  groups and clusters. These sources appear as a collection
of individual resolved sources, some with a dominant source among them. None has a recognizable radio
galaxy structure whether of FR-I and FR-II type. The sample of 23 sources is given in Table~4. 
We give the centroid positions from the original 50" images along with the ratio (Lres/Hres) of
the flux densities in the 50" and 4" images.
As we might expect,  most appear to be co-located with optical galaxies and galaxy groups. 20
of the 23 sources in this sub-sample were at artifact-free sky locations on SUMSS images; 
nevertheless,  SUMSS does not appear to have detected 16 of the sources. Table~4 provides 
information on potential associations with optical galaxies and clusters,  this
is based on examining the r-band images. 

\begin{deluxetable}{lrrll} 
\tablewidth{0pt} 
\tabletypesize\scriptsize 
\tablecaption{Candidate halo-type radio sources in the ATLBS-ESS} 
\tablehead{ 
\colhead{Name} & \colhead{RA } &  \colhead{DEC } &\colhead{Lres/} &\colhead{Comments}\\
               & \colhead{J2000} &  \colhead{J2000}  &\colhead{Hres} &  \\    }
\startdata 
J0025.5-6710& 00:25:30.65 & -67:10:41.5 & 3.17 & Bright galaxy group. Includes two \\ 
            &             &             &      & discrete but resolved sources.\\
J0025.6-6727& 00:25:38.57 & -67:27:56.3 & 1.88 & Few faint galaxies seen.\\
J0026.9-6706& 00:26:55.17 & -67:06:23.5 & 1.72 & Galaxy group.\\
J0027.6-6750& 00:27:36.62 & -67:50:55.2 & 2.23 & Few faint galaxies seen.\\
J0030.5-6635& 00:30:30.66 & -66:35:13.2 & 1.59 & Bright group of 4 galaxies. \\
J0030.9-6727& 00:30:56.63 & -67:27:32.8 & 1.57 & Cluster of faint galaxies. \\
            &             &             &      & Discrete source with faint Id.\\
J0031.5-6633& 00:31:31.22 & -66:33:35.7 & 2.37 & Several faint galaxies seen.\\
J0031.8-6617& 00:31:51.85 & -66:17:46.6 & 1.65 & Galaxies seen,  two of which \\
            &             &             &      & are relatively brighter. \\
J0032.6-6757& 00:32:39.23 & -67:57:59.5 & 1.60 & Emission associated with bright\\ 
            &             &             &      & galaxy with several satellites(?)\\
J0035.2-6730& 00:35:15.95 & -67:30:34.9 & 3.60 & Several bright galaxies. One \\
            &             &             &      & discrete radio source has Id.\\
J0035.9-6555& 00:35:56.28 & -65:55:05.2 & 2.55 & Several galaxies. Several radio \\
            &             &             &      & sources,  one with Id.\\
J0039.1-6717& 00:39:07.54 & -67:17:17.7 & 2.16 & Several galaxies seen. One offset\\ 
            &             &             &      & source with bright galaxy Id.   \\
J0040.6-6800& 00:40:36.30 & -68:00:13.8 & 1.78 & No galaxies within low resolution contour.\\
J0044.2-6713& 00:44:12.18 & -67:13:50.4 & 2.24 & Rich group of galaxies. Discrete \\
            &             &             &      & source at centre with bright Id.\\
J0055.9-6802& 00:55:55.61 & -68:02:49.4 & ---- & Outskirts of a large cluster of galaxies. \\
            &             &             &      & Faint galaxies only.\\
J0100.1-6711& 01:00:08.67 & -67:11:17.7 & 3.11 & Bright galaxy and several faint galaxies.\\
J0100.3-6749& 01:00:22.14 & -67:49:27.8 & 1.87 & Several galaxies present. \\
J0101.1-6606& 01:01:07.68 & -66:06:24.2 & 1.72 & Bright face-on spiral galaxy. \\
            &             &             &      & Emission centred on spiral.\\
J0101.7-6756& 01:01:45.88 & -67:56:34.1 & 1.53 & Elongated emission region. No obvious \\
            &             &             &      & galaxy concentration seen.\\
J0103.7-6755& 01:03:44.76 & -67:55:24.9 & 1.77 & Bright galaxy and a few other \\
            &             &             &      & fainter galaxies.\\
J0105.1-6648& 01:05:07.82 & -66:48:28.4 & 1.72 & Bright galaxies.\\
J0106.2-6719& 01:06:16.98 & -67:19:25.7 & 2.49 & Few very faint galaxies seen.\\
J0108.3-6637& 01:08:19.26 & -66:37:18.5 & 2.09 & Bright galaxy and few faint galaxies. \\

\enddata 
\end{deluxetable} 
\clearpage

\section{Discussion and Conclusions}

Close to half of the ATLBS-ESS sources are of FR-I type: compared to the 3CRR radio source sample that has significantly
higher cutoff in flux density the ATLBS-ESS has three times greater fraction of FR-I type.  This is consistent with a
morphological evolution wherein the FR-I fraction increases with decreasing flux density.

Significant asymmetry in lobe extents appears to be a common occurrence in the ATLBS FR-I sources
compared to FR-II sources ($30\%$ vs $13\%$). 
The FR-I sources in ATLBS-ESS are mostly of the 3C31-type and not
of the double-lobe type morphology that has been suggested in a previous study as dominant.
It follows that the increasing abundance of FR-I sources at mJy flux densities is accompanied by a
morphological change in the FR-I population towards 3C31-type sources.

We find a large number of relatively high-redshift FR-I sources. These 22 high-redshift FR-I sources form an 
important database for examining 
evolution of low power sources because of their robust radio morphological classification unlike previous
attempts. The properties of the low and high redshift FR-I sources have been compared. There appear to be no 
major differences in morphological properties of the high and low redshift samples of FR-I sources.

We identify as many as 14 of the ATLBS-ESS radio sources to be giants 
with projected linear size exceeding 700 kpc; at least 9 of them are likely to be at redshifts exceeding unity. 
Based on the 7C survey, \citet{cot96} revealed the prevalence of giant radio sources out to redshift of unity.
Compared to the 7C survey, 
the ATLBS-ESS has relatively higher sensitivity over smaller sky area, 
and the substantial number of high redshift giants detected in ATLBS-ESS
reveals that giant radio sources are not less common at redshifts exceeding unity.
The giant sources detected in the ATLBS-ESS have similar radio powers compared to the 7C giants.

We consider herein a model in which the lifetime of an extended
radio galaxy consists of (i) an active phase in which the source extent dynamically grows over time, (ii) a dying phase
in which the central engine has switched off and the beams cease feeding the lobes, and (iii) a restarting phase in which
new activity is seen along with relict lobes created in the previous activity phase.

The fraction of FR-Is and FR-IIs where there are no signatures of on-going nuclear activity remains
small,  about $3\%$: consistent with the reported rarity of dead radio galaxies \citep{blu00}. 
The finding indicates that even in surveys like the ATLBS in
which the surface brightness sensitivity has been substantially improved, a significantly larger number of
relict radio sources are not detected.  This indicates that the relict or dying phase is short, independent of 
the survey sensitivity, and perhaps ended by a restarting of the central engine.    This last phase ends when the relict 
lobes are no longer distinctly discernible either because they have faded away owing to losses arising from their emissivity 
or expansion, or because the new beams have rejuvenated the entire relict lobes.  

A new finding of the ATLBS-ESS is that observed morphology of a large number of FR-I and FR-II radio galaxies 
may suggest signatures of restarted nuclear activity.
Nearly one-third of FR-IIs have "non-classic" morphologies that are more easily associated with a renewed beam activity;
in contrast, only one-eighth of the ATLBS-ESS FR-Is show signatures of restarted activity.
In the model considered above, it appears that FR-II sources may spend two-thirds of their lifetime in the active phase, one-thirds
in the restarting phase, and only a few percent in the dying phase.  In the case of FR-I sources, the active phase may be larger 
and the restarting phase correspondingly smaller.  We speculate that such episodic activity may continue over multiple cycles during 
which the source never disappears from radio surveys, following which the entire episodic activity may cease and the host
reverts to a radio-quiet state. These cycles of episodic activity may perhaps be the time during which the source grows and total duration of these 
activity phases represents the dynamical age of the source.  In this scenario, since the active phase is the dominant duration,
the dynamical age is close to the true age and spectral ages may be confusing owing to the episodic activity. 

We have listed the few radio sources that are observed to have faint, elongated emission regions on one side
that are aligned with but separated from the radio galaxy. These are likely to be radio galaxies that have had a previous epoch
of activity. We point out that we may be seeing old emission regions that are preserved in a relatively dense ambient
medium on one side; alternately, light travel time effects may be enabling 
the old lobe on the side pointing away from us to be visible (J0031.1-6642 and J0031.8-6727).

The ATLBS has detected among the faintest 
radio sources known. The giant radio galaxy  J0034.0-6639 has 
lobes that have the lowest surface brightness known. 
In imaging the faintest relict radio lobes identified in the ATLBS-ESS, which
includes this giant radio galaxy and also the aligned but
separated faint 'tail' in the large quasar J0031.1-6642,  the survey has shown that relict lobes
may continue to maintain their symmetry about the radio axis.  
This is in contrast to the case in radio sources at centers of galaxy clusters,
and is consistent with the expectation that extended radio sources in the
intergalactic gas away from cluster environments experience relatively benign
weather. 

The discussion presented here of the structures in radio galaxies imaged in one
of the most sensitive radio surveys is
of considerable relevance in the view of the upcoming sensitive,  all-sky radio surveys with instruments such as 
LOFAR and ASKAP. By focusing on the larger of the radio sources imaged in the ATLBS-ESS
we have sought to highlight the variety of source structures and science that may be anticipated.

\acknowledgments

ATCA is part of the Australia Telescope,  
which is funded by the Commonwealth of Australia for operation as a national Facility managed by CSIRO.
Cerro Tololo Inter-American Observatory, National Optical Astronomy Observatory, is operated by 
the Association of Universities for Research in Astronomy, under contract with the National Science Foundation.
We thank the AAO staff for their help during our observations with the IRIS2 imager.
We acknowledge the use of SuperCosmos,  an advanced photographic plate digitizing machine at the Royal 
Observatory of Edinburgh,  in the use of digitized images for some of our optical identification work.
LS would like to thank the Helena Kluyver female visitor program at ASTRON (Netherlands Institute 
for Radio Astronomy) where some of the work was completed. We thank the referee for several useful suggestions.

\begin{figure*}
\centering
\begin{tabular}{cc}
\begin{minipage}{0.47\linewidth}
\begin{center}
\frame{\includegraphics[angle=-90,  width=2.8in]{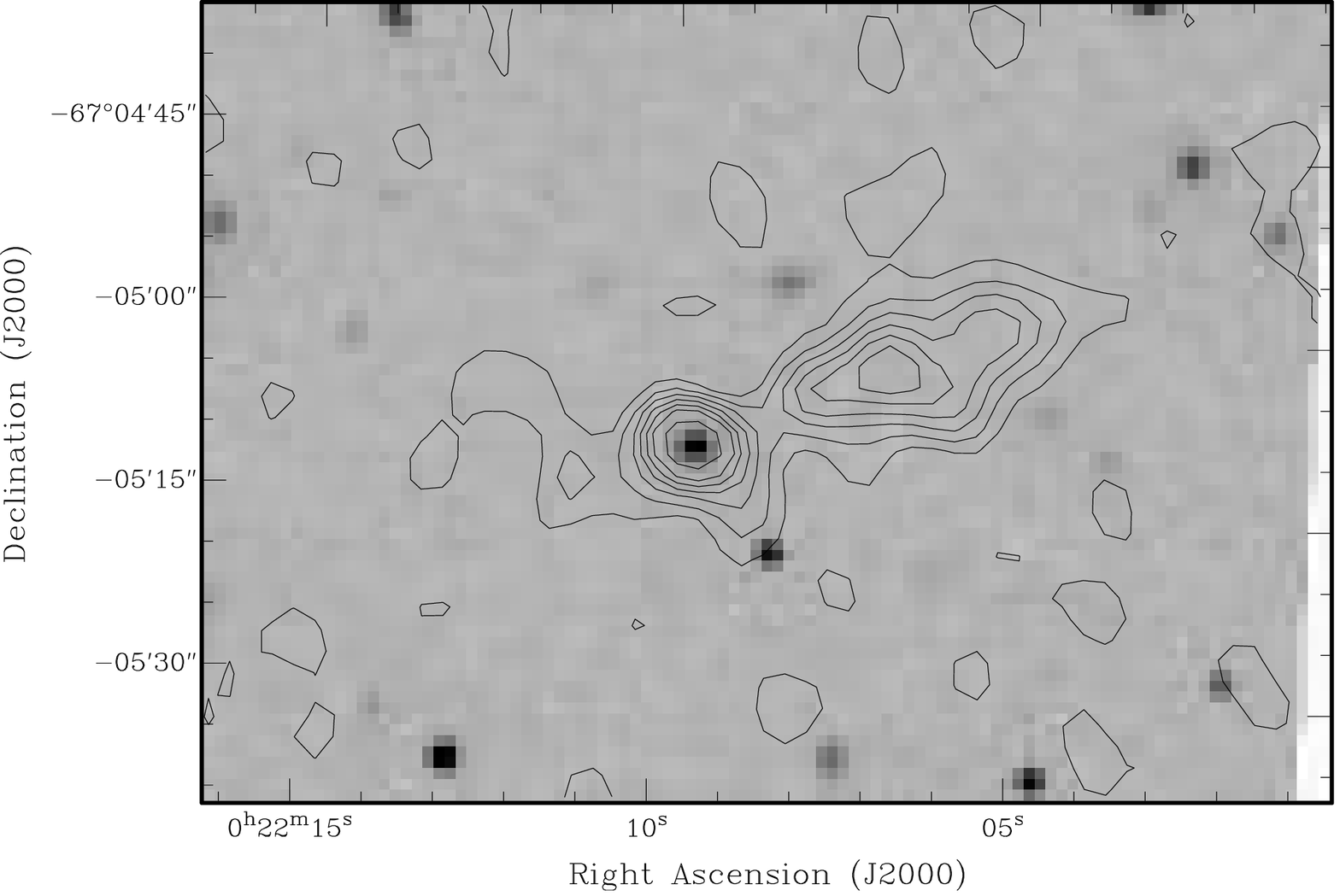}}
\end{center}
\caption{J0022.1-6705: $10^{-4}$ Jy x 1,   2,  3, 4, 5, 6, 8. All figures are
made with a beam having FWHM of 6"x6" and 1.4~GHz contours overlayed on r-band images unless indicated otherwise. 
In this image K band image is used.} 
\end{minipage}
&
\begin{minipage}{0.47\linewidth}
\begin{center}
\frame{\includegraphics[angle=-90, width=2.8in]{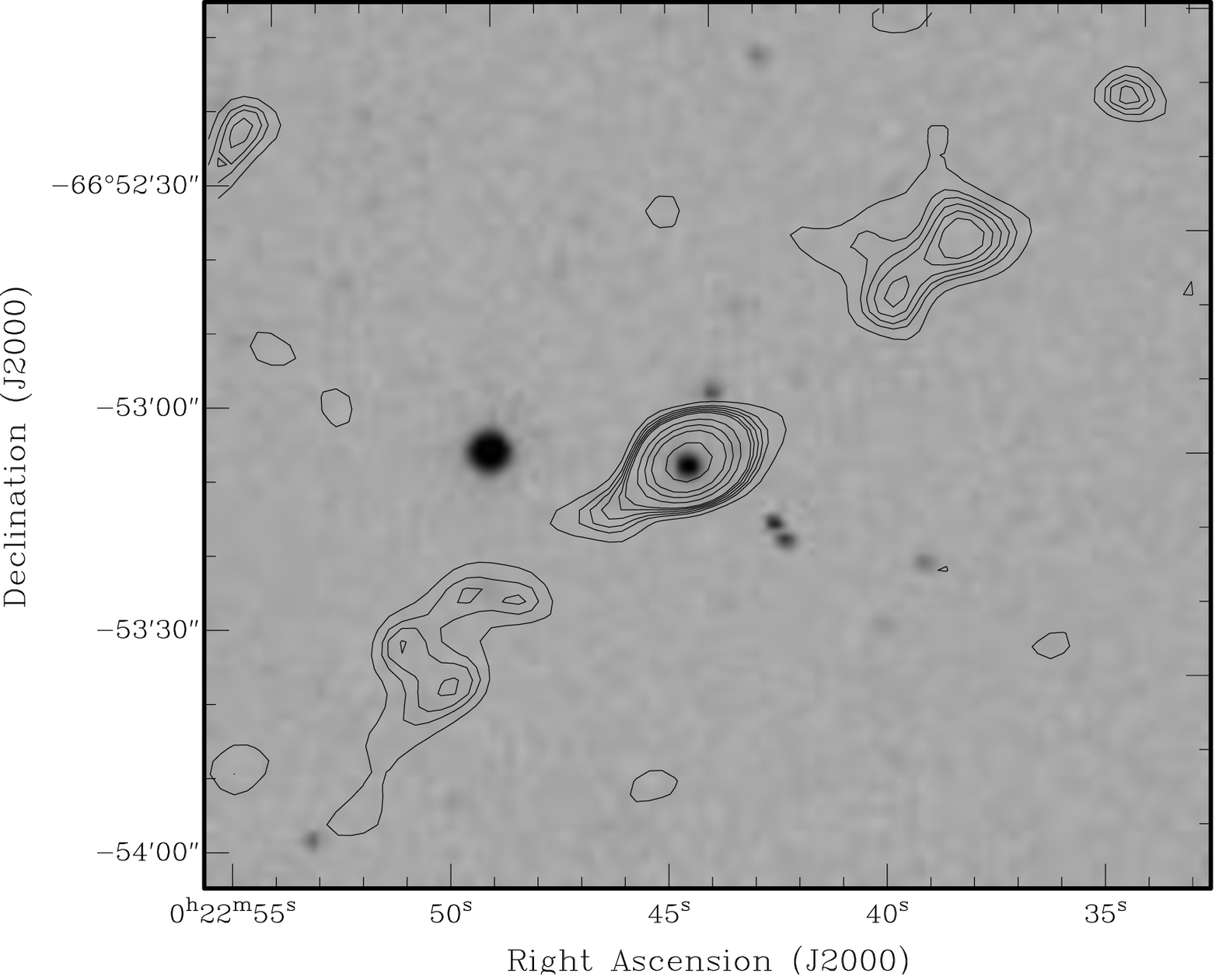}}
\end{center}
\caption{J0022.7-6652: Beam 10"; $10^{-4}$ Jy x 1.5, 2, 2.25, 2.5, 2.75, 3, 4, 5, 6, 8. K band image is
used here.}
\end{minipage}
\\
\end{tabular}
\end{figure*}

\begin{figure*}
\centering
\begin{tabular}{cc}
\begin{minipage}{0.47\linewidth}
\begin{center}
\frame{\includegraphics[angle=-90, width=2.8in]{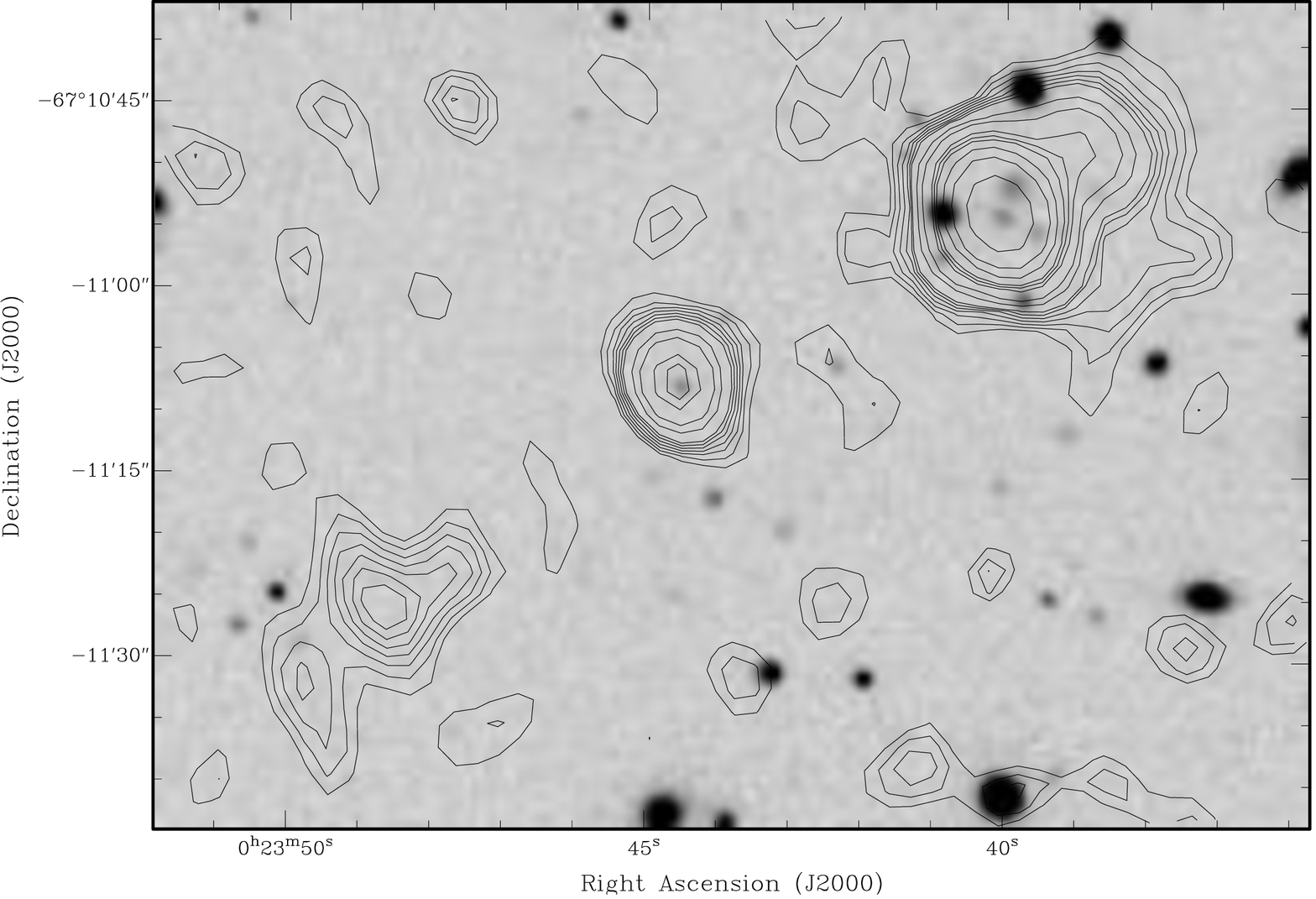}}
\end{center}
\caption{J0023.6-6710: $10^{-4}$ Jy x 1, 1.5, 2, 2.5, 3, 3.5, 4, 6, 8, 12, 14, 16, 22, 30, 50, 100.} 
\end{minipage}
&
\begin{minipage}{0.47\linewidth}
\begin{center}
\frame{\includegraphics[angle=-90, width=2.8in]{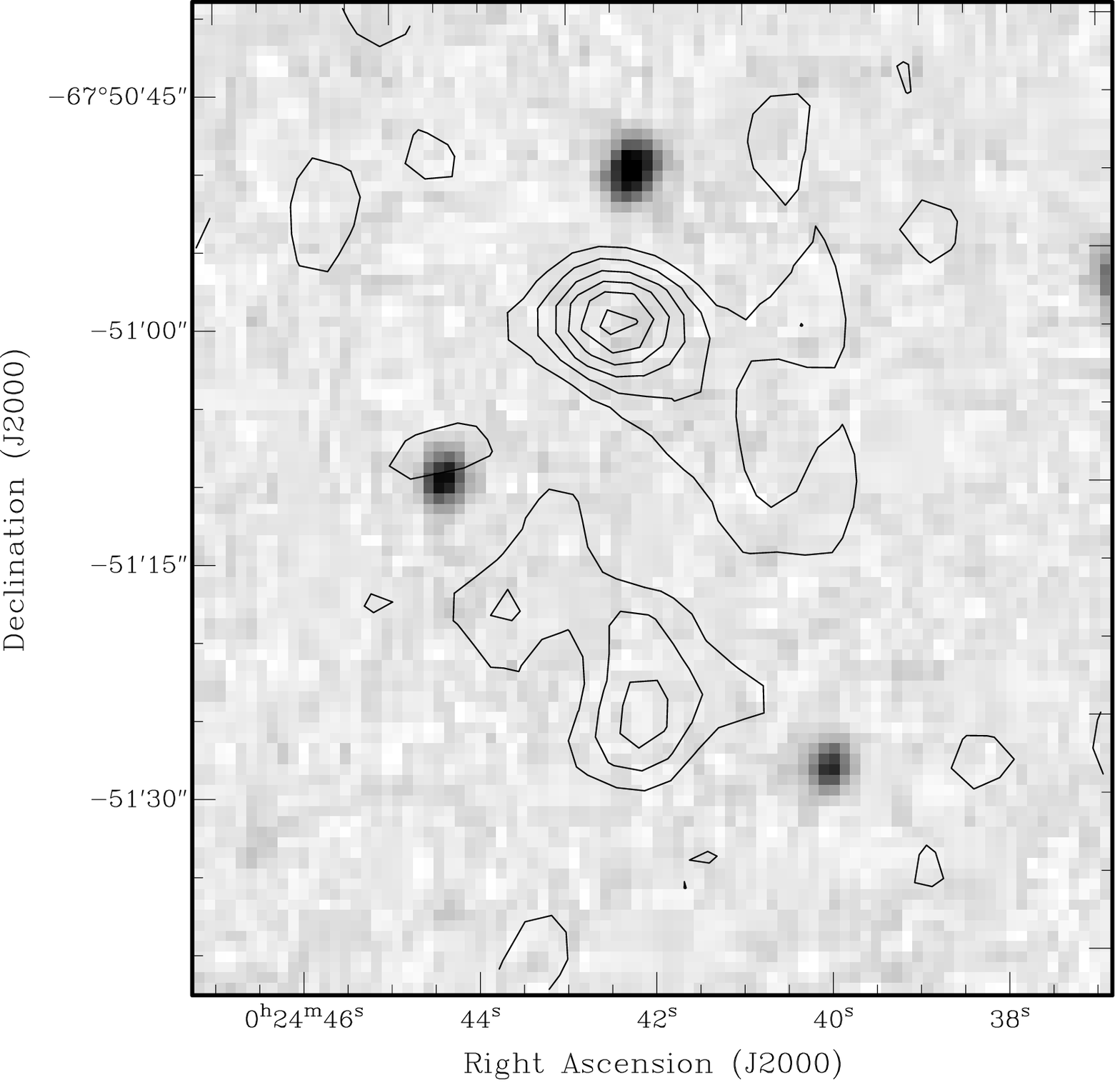}}
\end{center}
\caption{J0024.6-6751: $10^{-4}$ Jy x 1,   2,   3,   4,   5,   6. R band image is used here.} 
\end{minipage}
\\
\end{tabular}
\end{figure*}

\begin{figure*}
\centering
\begin{tabular}{cc}
\begin{minipage}{0.47\linewidth}
\begin{center}
\frame{\includegraphics[angle=-90, width=2.8in]{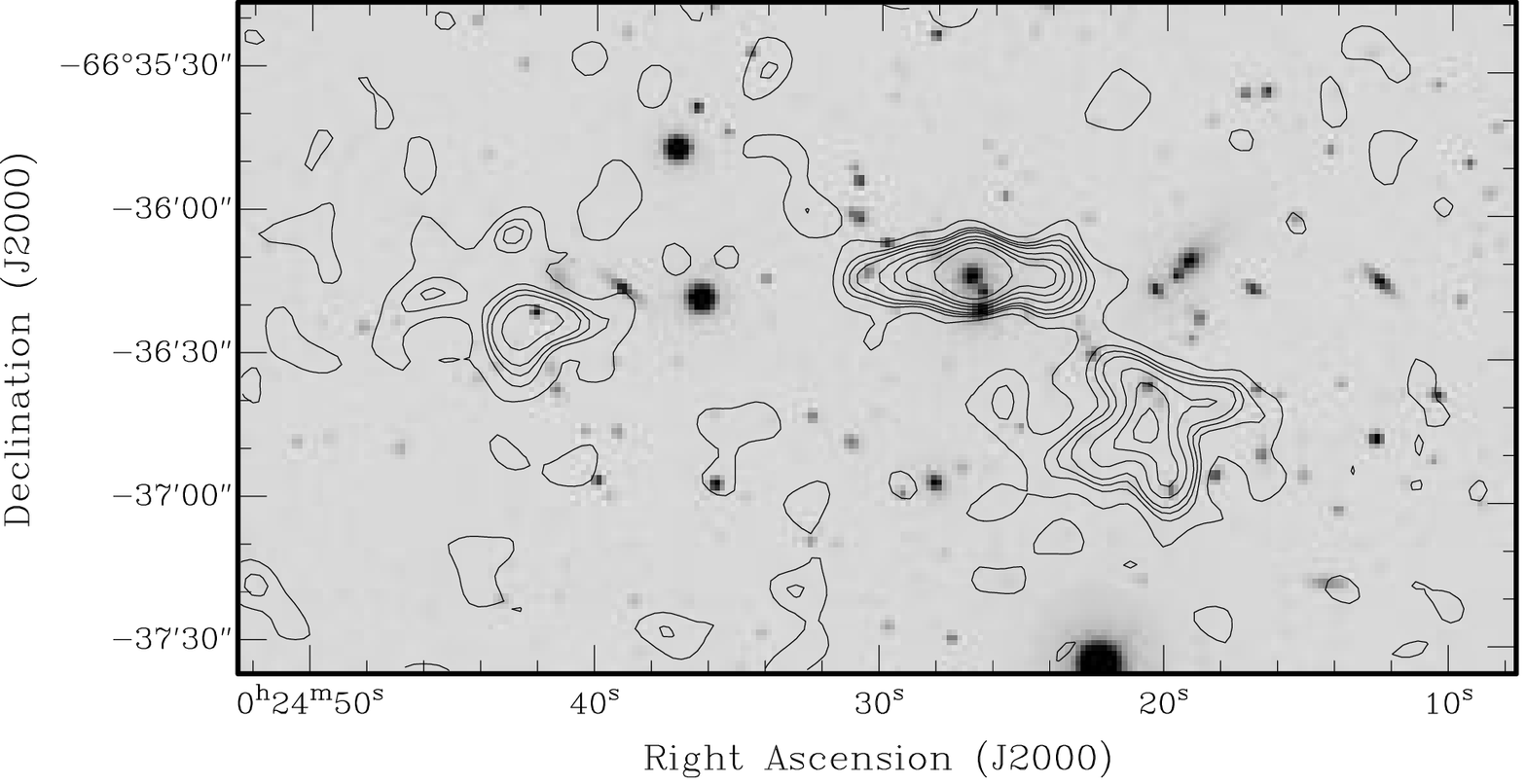}}
\end{center}
\caption{J0024.4-6636: Beam 10"; $10^{-4}$ Jy x 1, 2, 2.5, 3, 4, 5, 6, 8.} 
\end{minipage}
&
\begin{minipage}{0.47\linewidth}
\begin{center}
\frame{\includegraphics[angle=-90, width=2.8in]{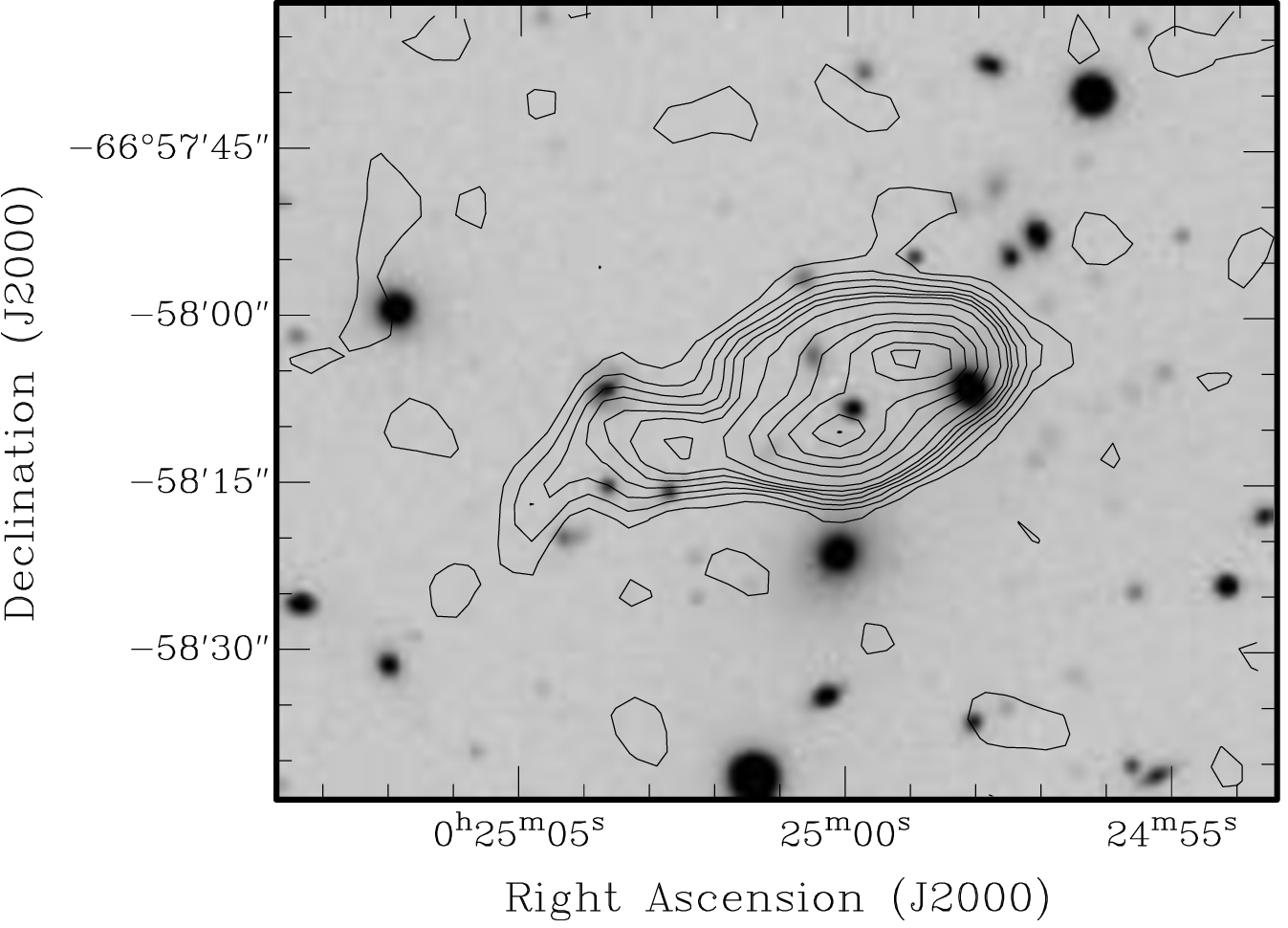}}
\end{center}
\caption{J0025.0-6658: $10^{-4}$ Jy x 1,   2,   3,   4,   5,   6,   8,   12,   16,   20,   24,   26.5.} 
\end{minipage}
\\
\end{tabular}
\end{figure*}



\begin{figure*}
\centering
\begin{tabular}{cc}
\begin{minipage}{0.47\linewidth}
\frame{\includegraphics[angle=-90, width=2.8in] {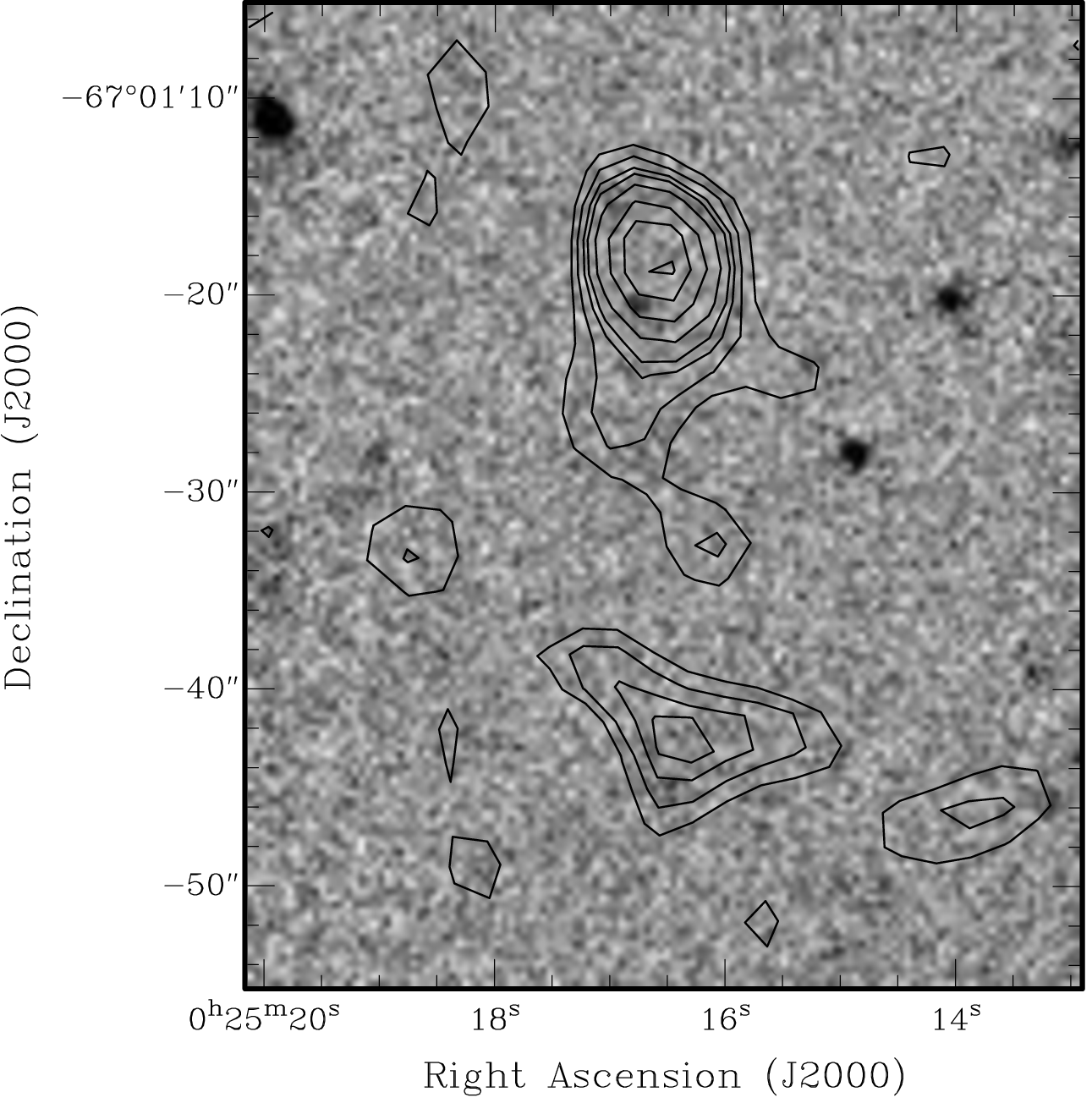}}
\caption{J0025.2-6701: $10^{-4}$ Jy x 1,   1.5,   2,   2.35,   3,   4,   5,   6.} 
\end{minipage}
&
\begin{minipage}{0.47\linewidth}
\frame{\includegraphics[angle=-90, width=2.8in]{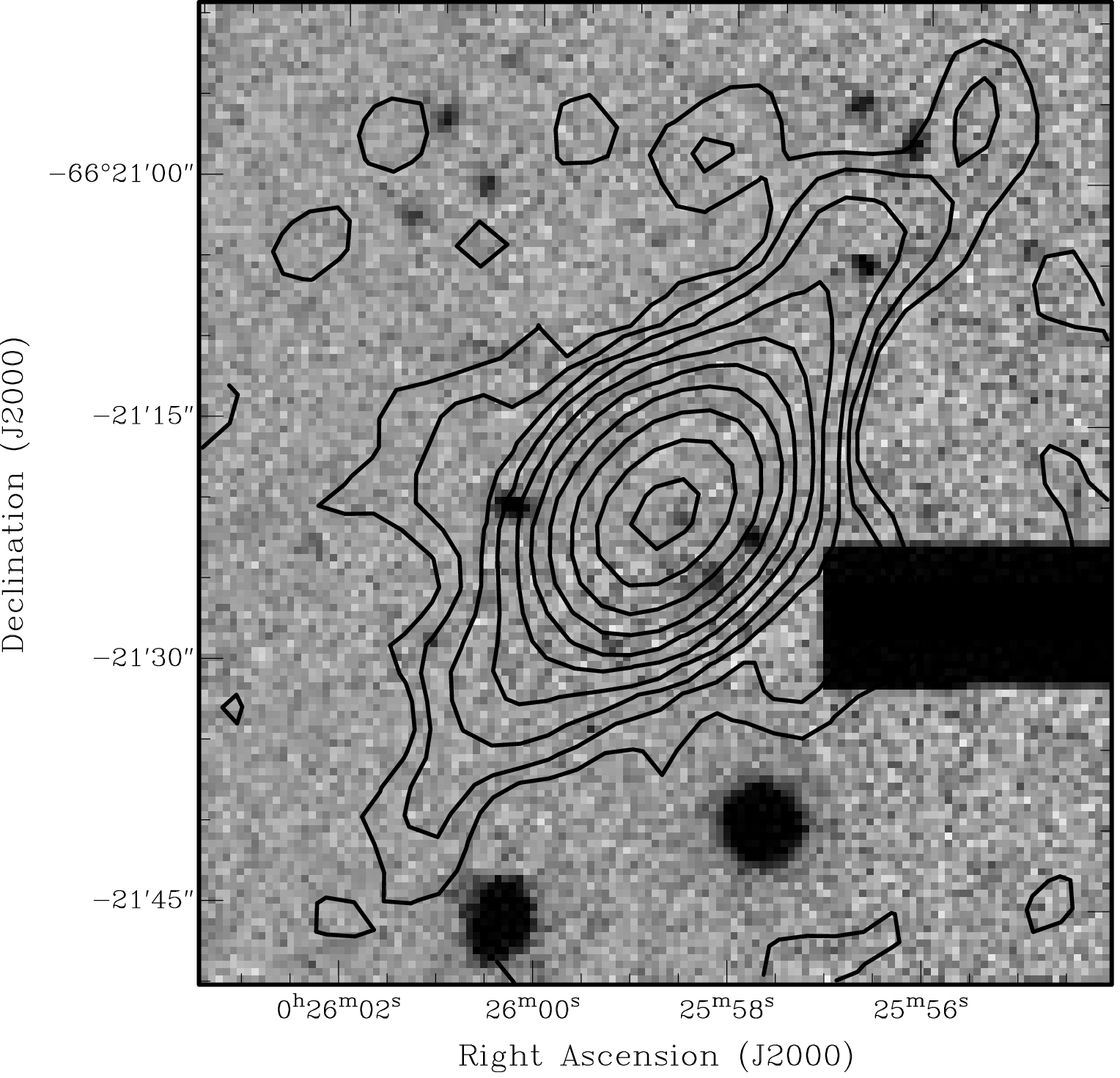}}
\caption{J0025.9-6621: $10^{-4}$ Jy x 1,   2,   4,   8,   16,   32,   64,  128,  256,  450.}
\end{minipage}
\\
\end{tabular}
\end{figure*}

\begin{figure*}
\centering
\begin{tabular}{cc}
\begin{minipage}{0.47\linewidth}
\frame{\includegraphics[angle=-90, width=2.8in]{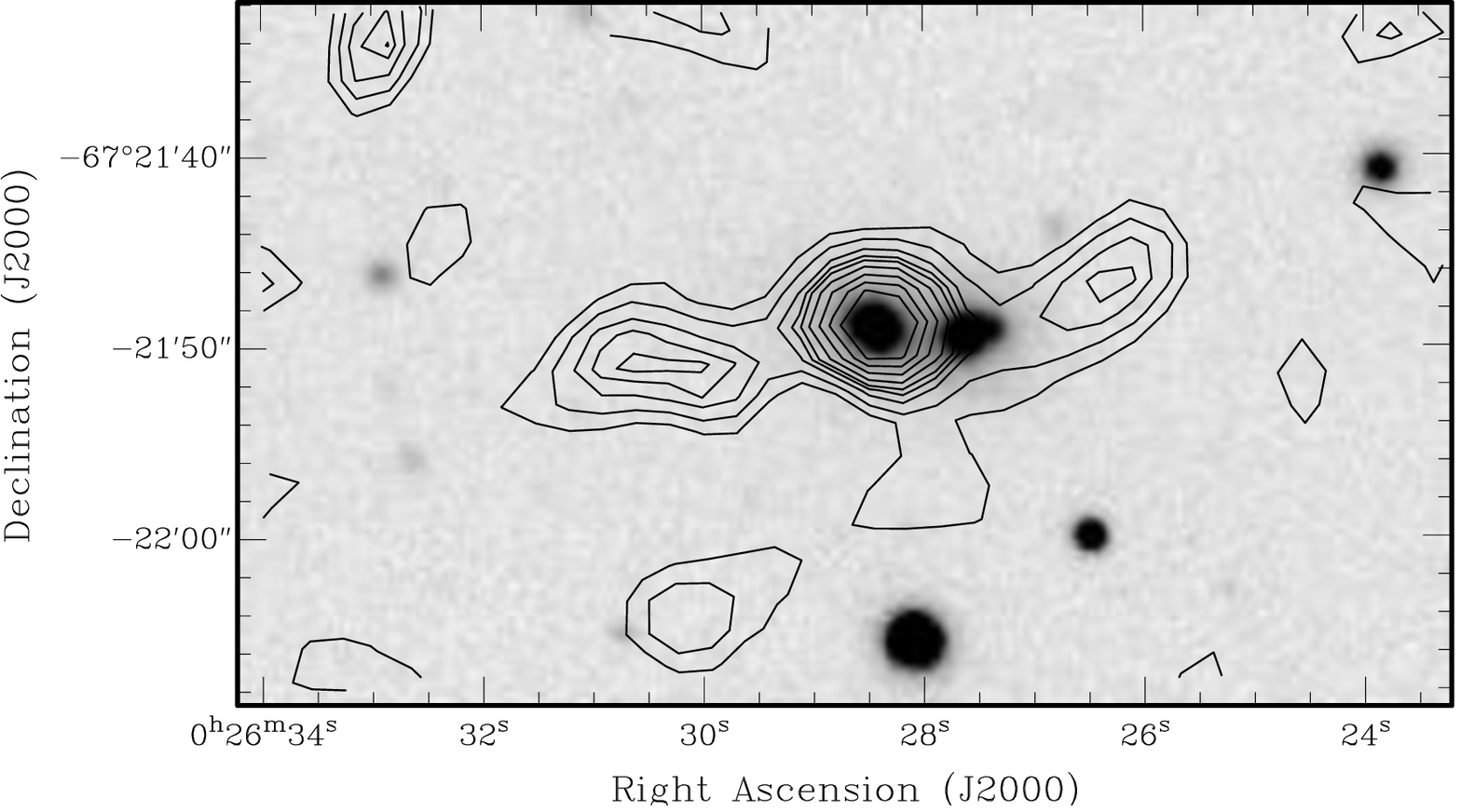}}
\caption{J0026.4-6721: $10^{-4}$ Jy x 1,   1.5,   2,   2.35,   2.6,   3,   3.5,   4,   4.5.} 
\end{minipage}
&
\begin{minipage}{0.47\linewidth}
\frame{\includegraphics[angle=-90, width=2.8in]{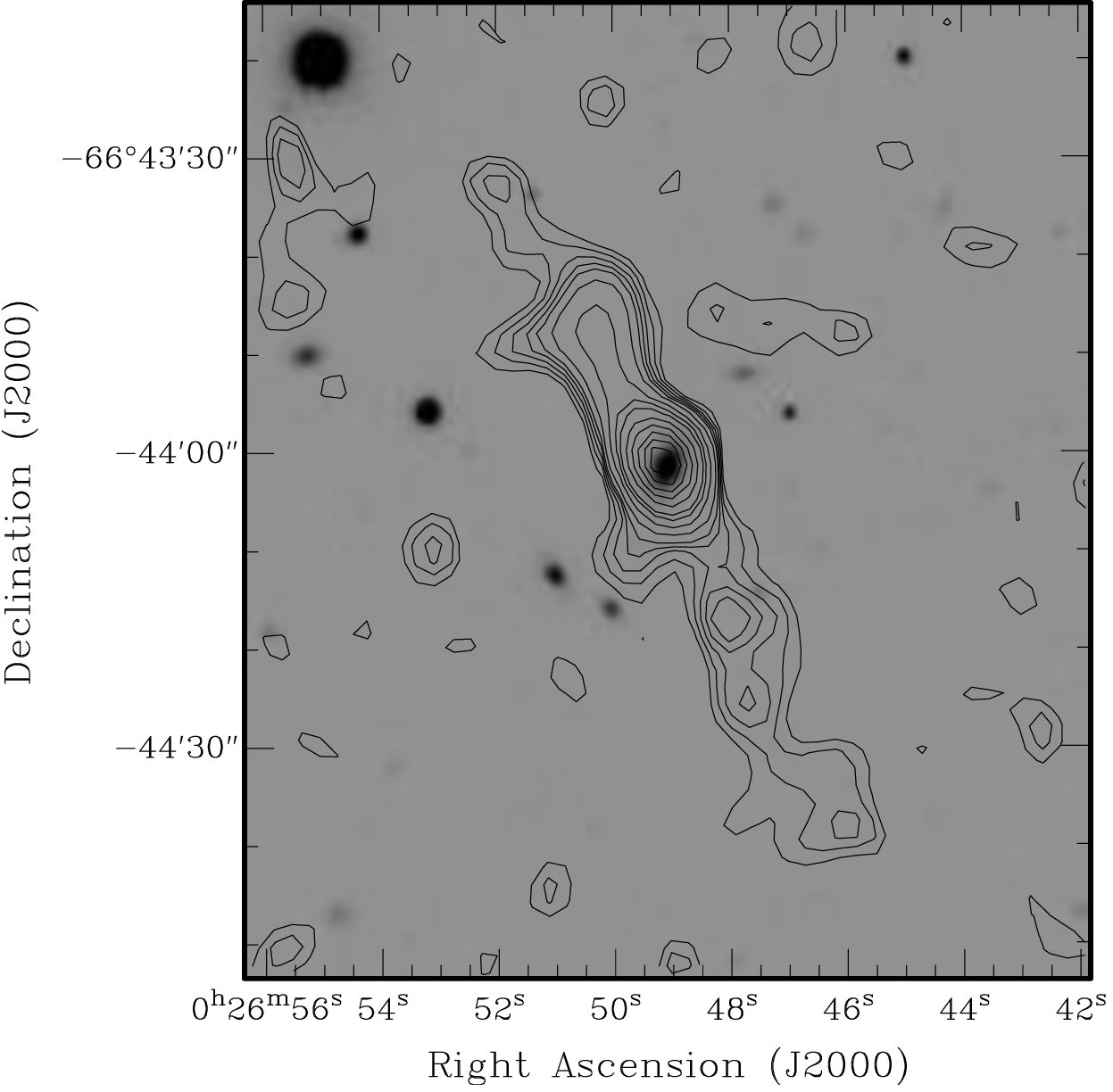}}
\caption{J0026.8-6643: $10^{-4}$ Jy x 1,   1.5,   2,   2.5,   3,   4,   6,   8,   10,   14,   18,  22,   26,  30.} 
\end{minipage}
\\
\end{tabular}
\end{figure*}

\begin{figure*}
\centering
\begin{tabular}{cc}
\begin{minipage}{0.47\linewidth}
\frame{\includegraphics[angle=-90, width=2.8in]{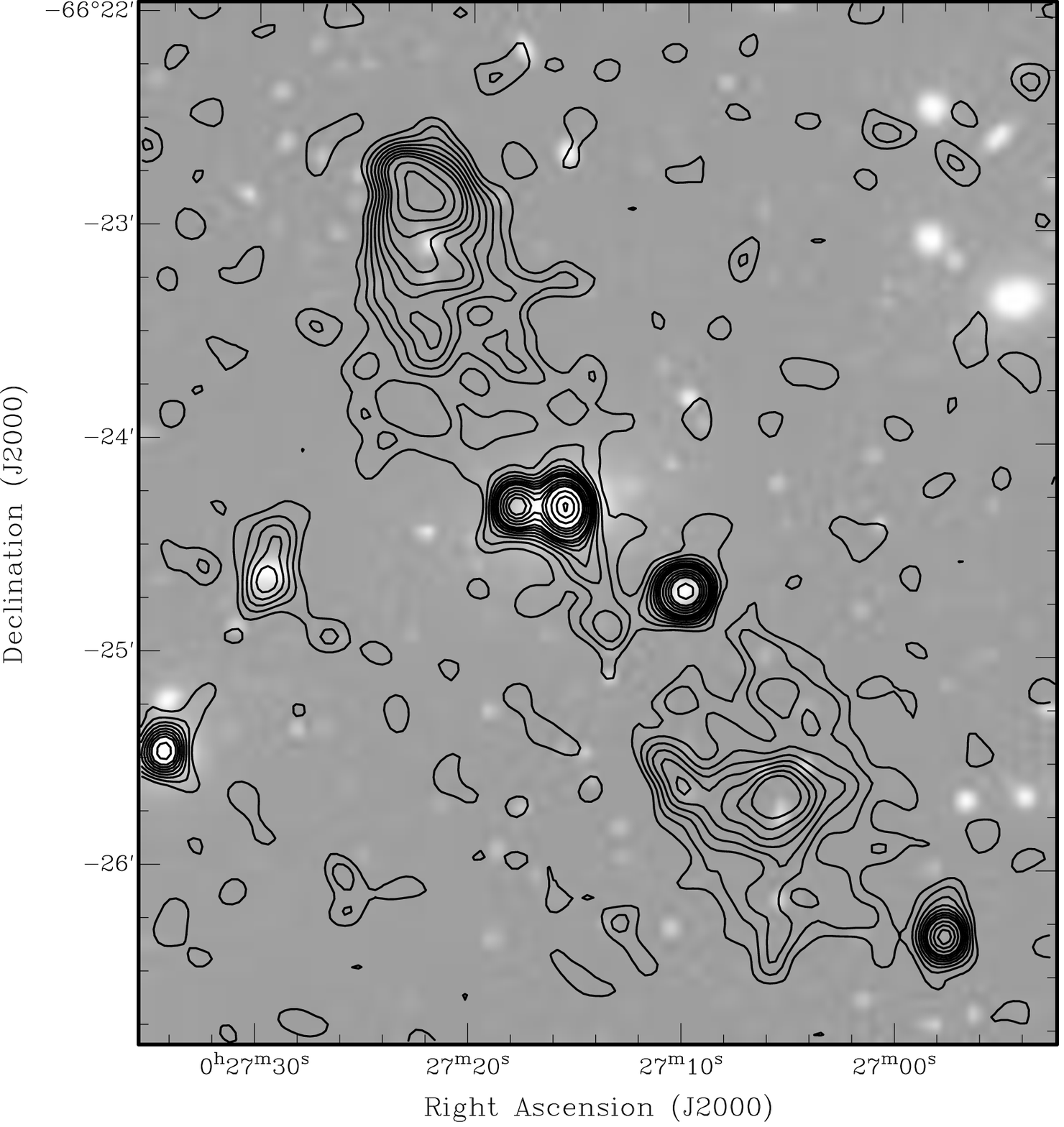}}
\caption{J0027.2-6624: Beam 10"; $10^{-4}$ Jy x 1,   2,   3,   4,   5,   6,   7,   8,   10,  12,   14,   16,   21,   26, 30.} 
\end{minipage}
&
\begin{minipage}{0.47\linewidth}
\frame{\includegraphics[angle=-90, width=2.8in]{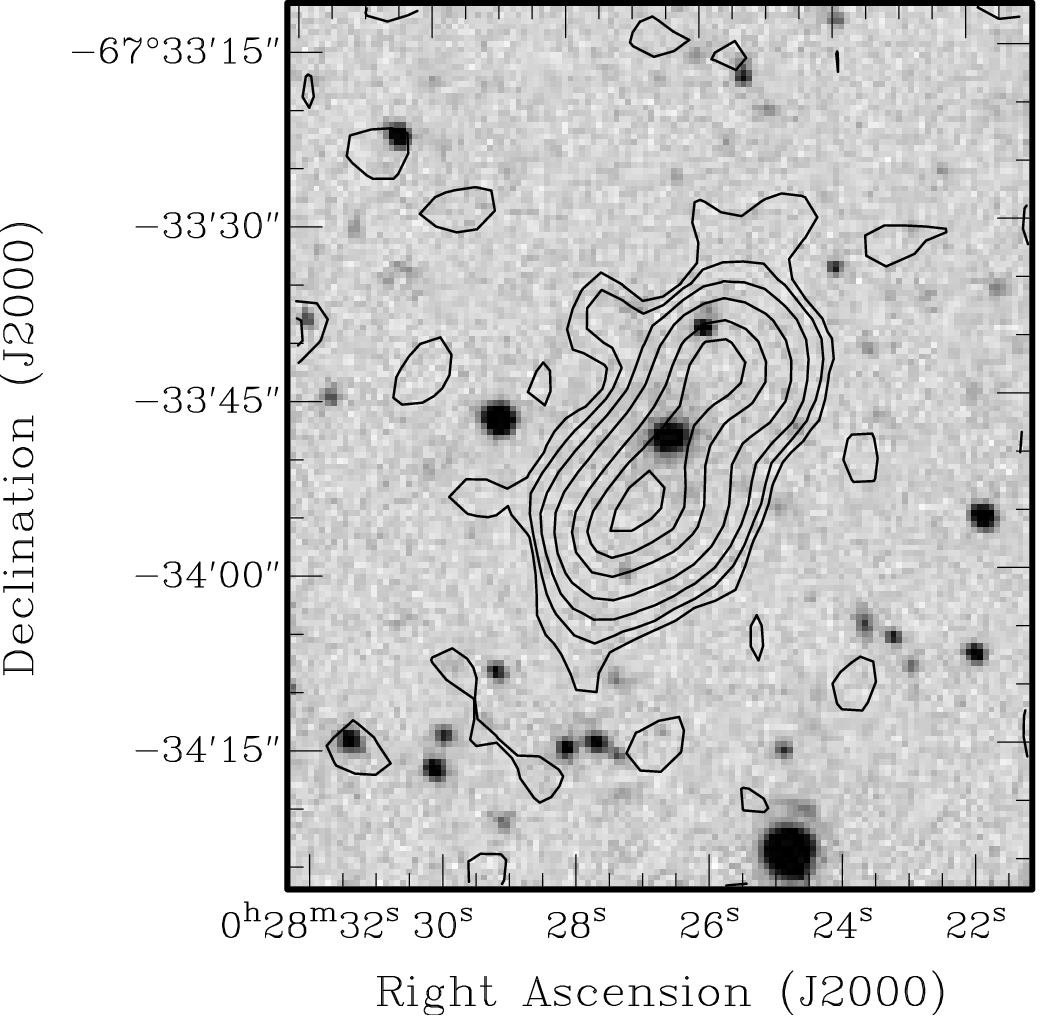}}
\caption{J0028.4-6733: $10^{-4}$ Jy x 1,   2,   4,   8,   16,   24,   32. } 
\end{minipage}
\\
\end{tabular}
\end{figure*}



\begin{figure*}
\centering
\begin{tabular}{cc}
\begin{minipage}{0.47\linewidth}
\frame{\includegraphics[angle=-90, width=2.8in]{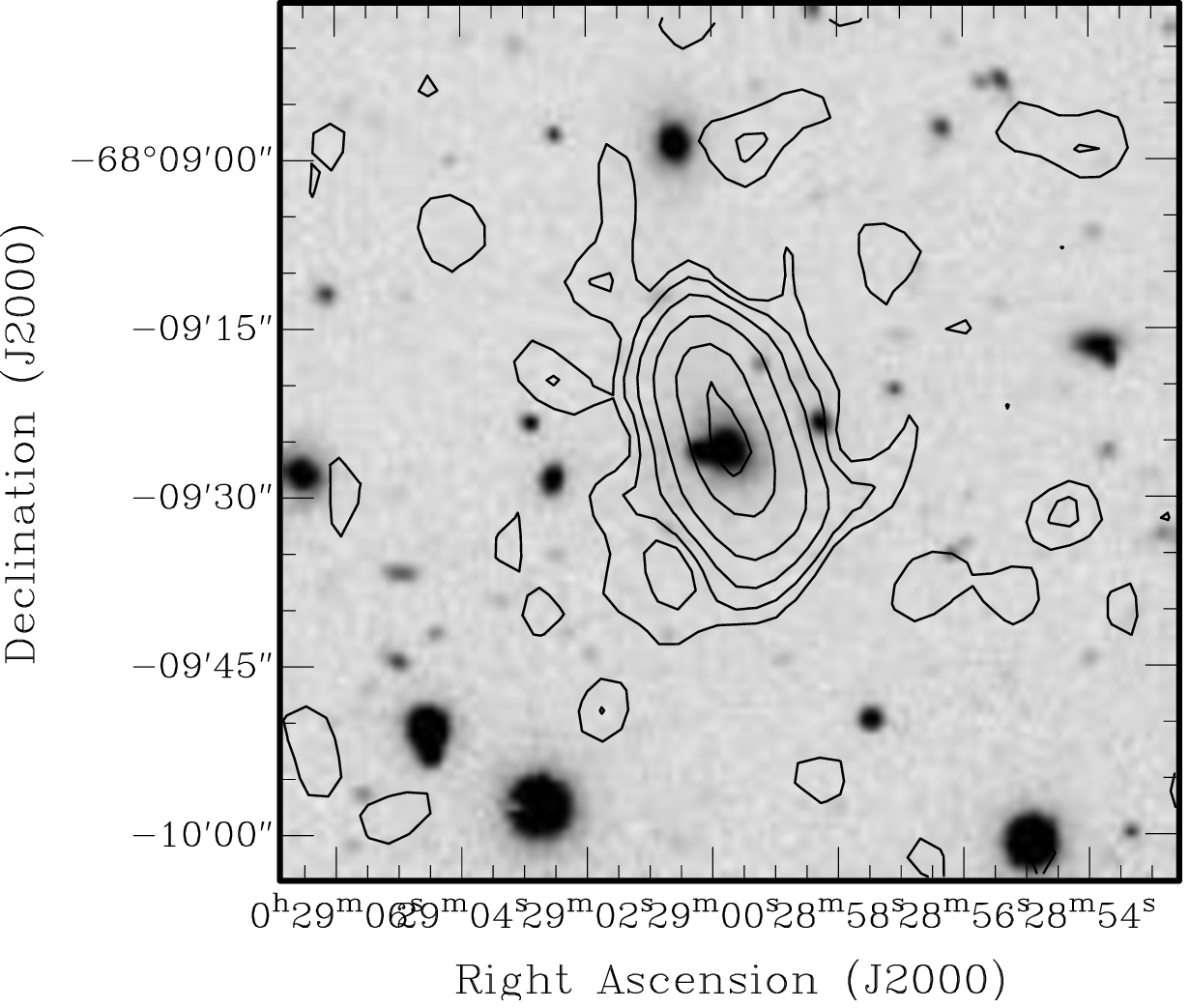}}
\caption{J0028.9-6809: $10^{-4}$ Jy x 1,   2,   4,   8,   16,   24.} 
\end{minipage}
&
\begin{minipage}{0.47\linewidth}
\frame{\includegraphics[angle=-90, width=2.8in]{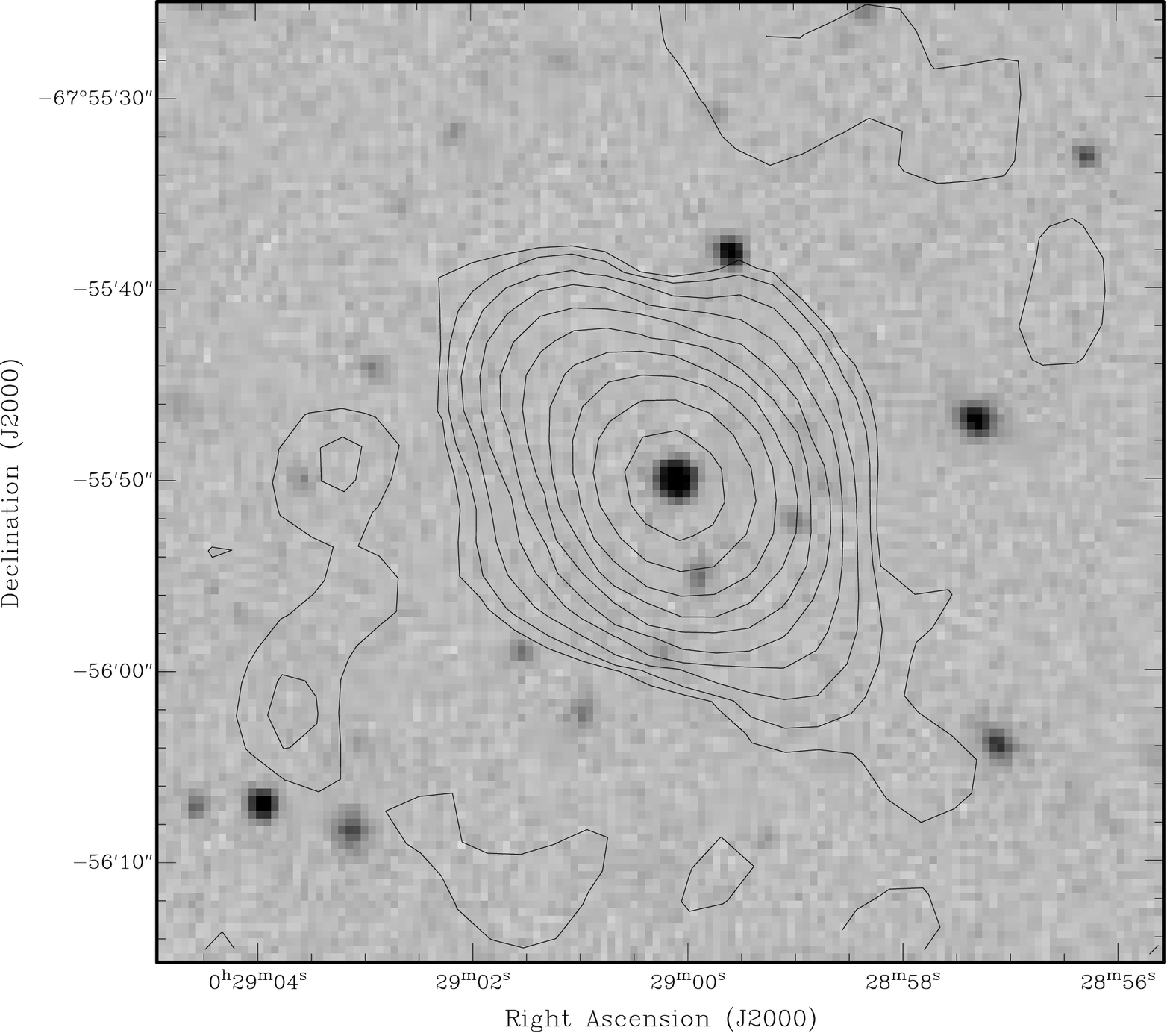}}
\caption{J0029.0-6755: $10^{-4}$ Jy x 1,   2,   4,   8,   16,   32,   64,  128,  256,  512}
\end{minipage}
\\
\end{tabular}
\end{figure*}

\begin{figure*}
\centering
\begin{tabular}{cc}
\begin{minipage}{0.47\linewidth}
\frame{\includegraphics[angle=-90, width=2.8in]{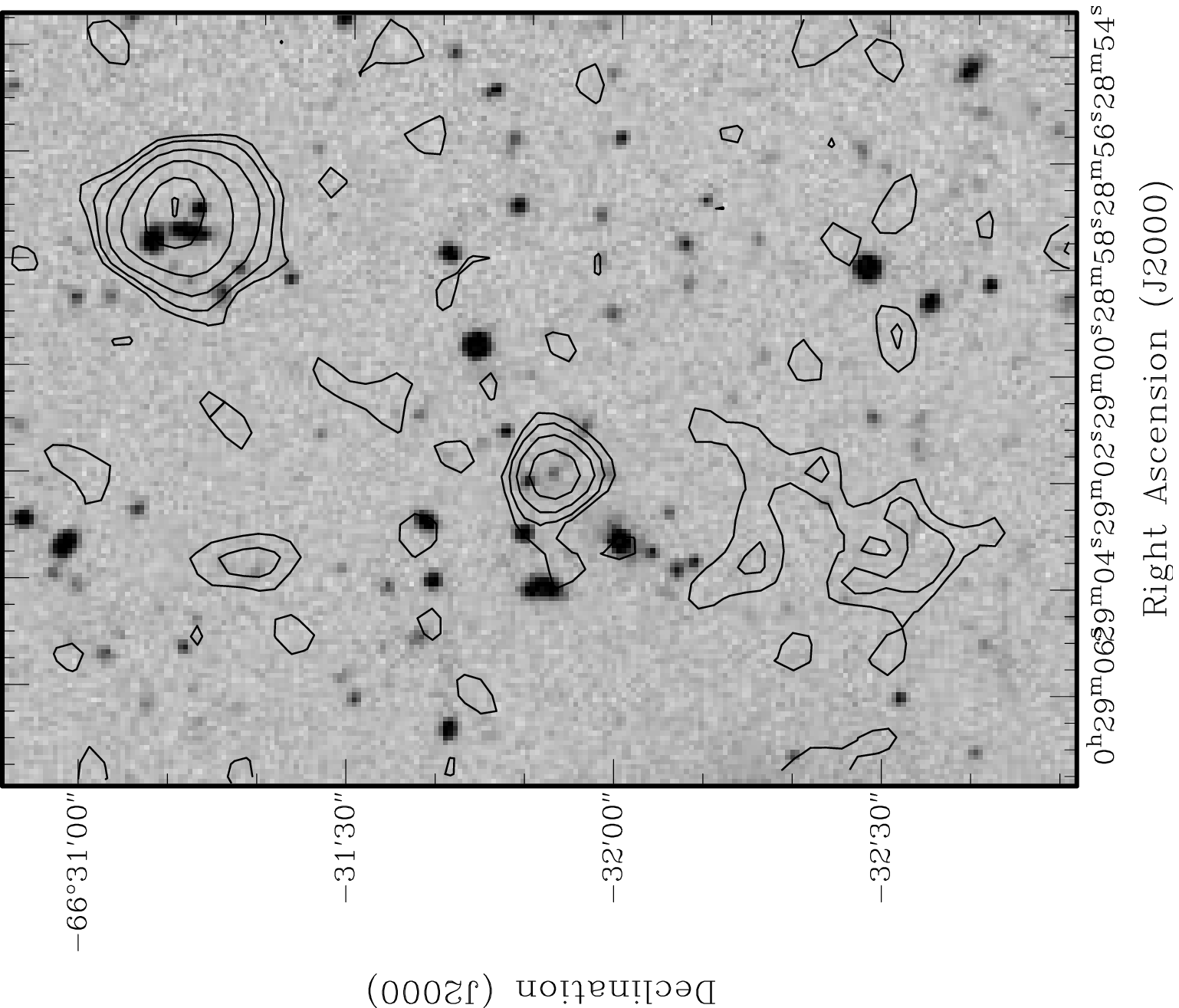}}
\caption{J0028.9-6631: $10^{-4}$ Jy x 1,   2,   4,   8,   16,   24.} 
\end{minipage}
&
\begin{minipage}{0.47\linewidth}
\frame{\includegraphics[angle=-90, width=2.8in]{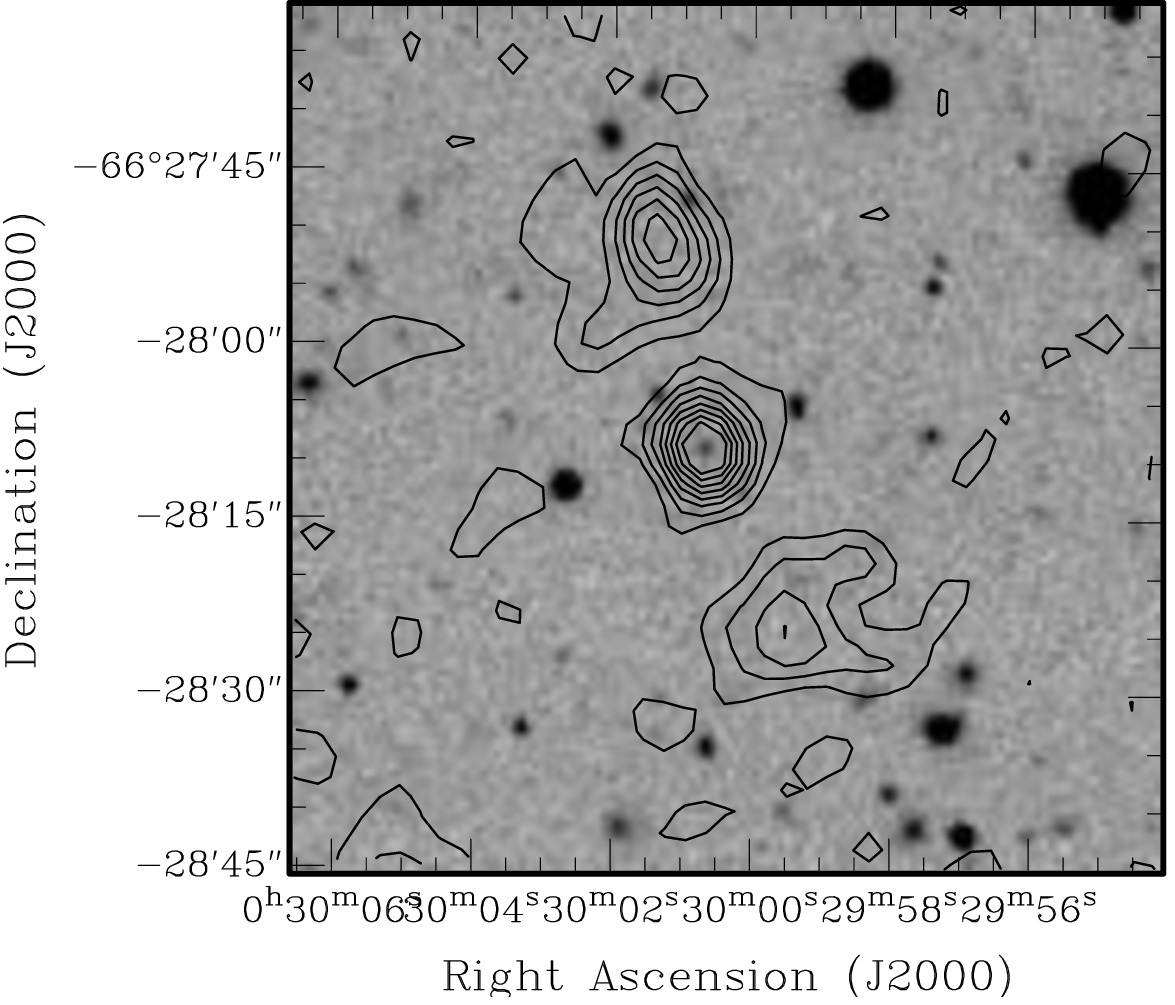}}
\caption{J0030.0-6628: $10^{-4}$ Jy x  1,  2,  3,  4,  5,  6,  7,  8.} 
\end{minipage}
\\
\end{tabular}
\end{figure*}

\begin{figure*}
\centering
\begin{tabular}{cc}
\begin{minipage}{0.47\linewidth}
\frame{\includegraphics[angle=-90, width=2.8in] {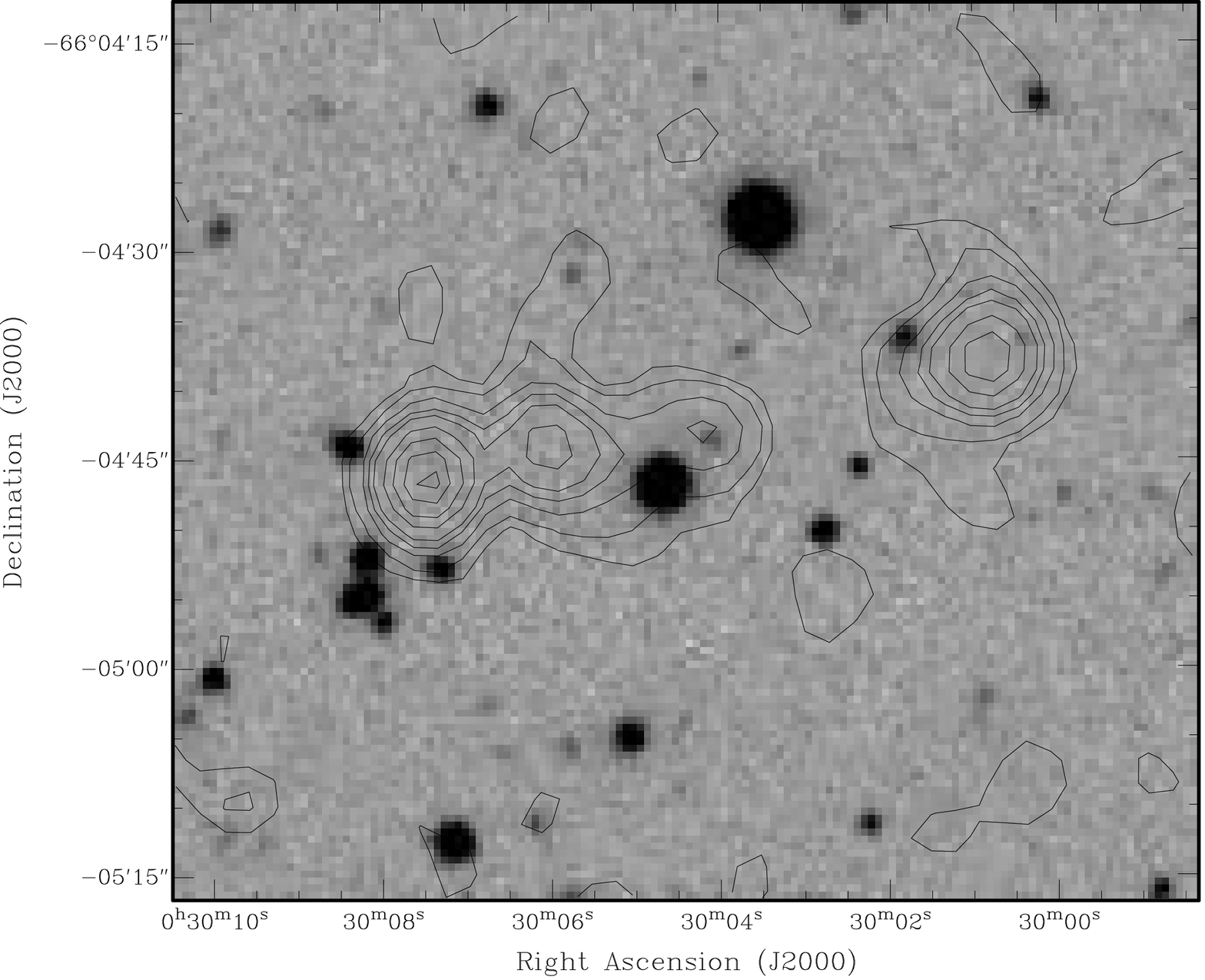}}
\caption{J0030.0-6604: $10^{-4}$ Jy x 1,  2,  4,  6,  8,  12,  16,  21,  25.} 
\end{minipage}
&
\begin{minipage}{0.47\linewidth}
\frame{\includegraphics[angle=-90, width=2.8in]{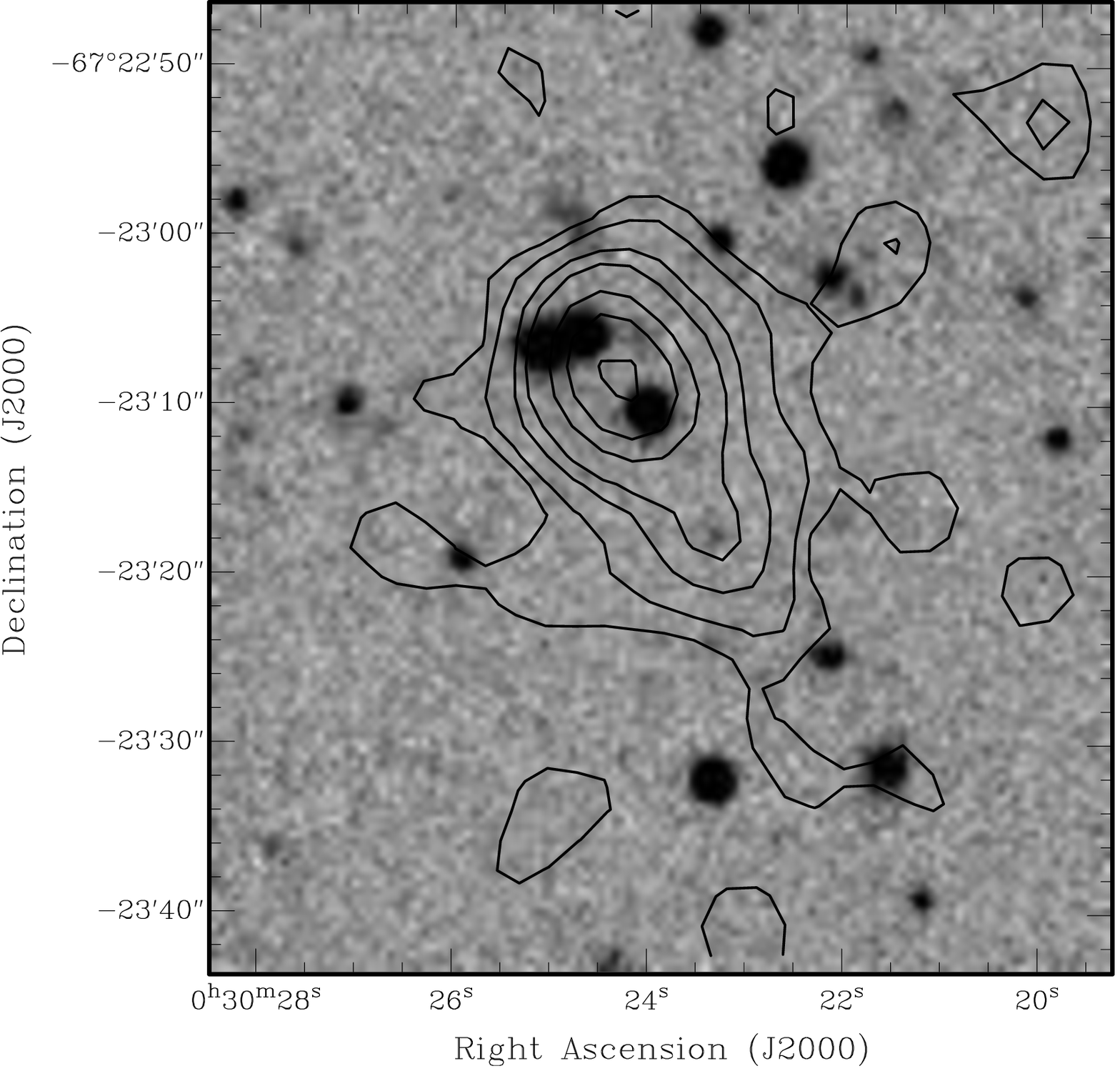}}
\caption{J0030.4-6723: $10^{-4}$ Jy x 1,  2,  4,  6,  11,  16,  24.}
\end{minipage}
\\
\end{tabular}
\end{figure*}

\begin{figure*}
\centering
\begin{tabular}{cc}
\begin{minipage}{0.47\linewidth}
\frame{\includegraphics[angle=-90, width=2.8in]{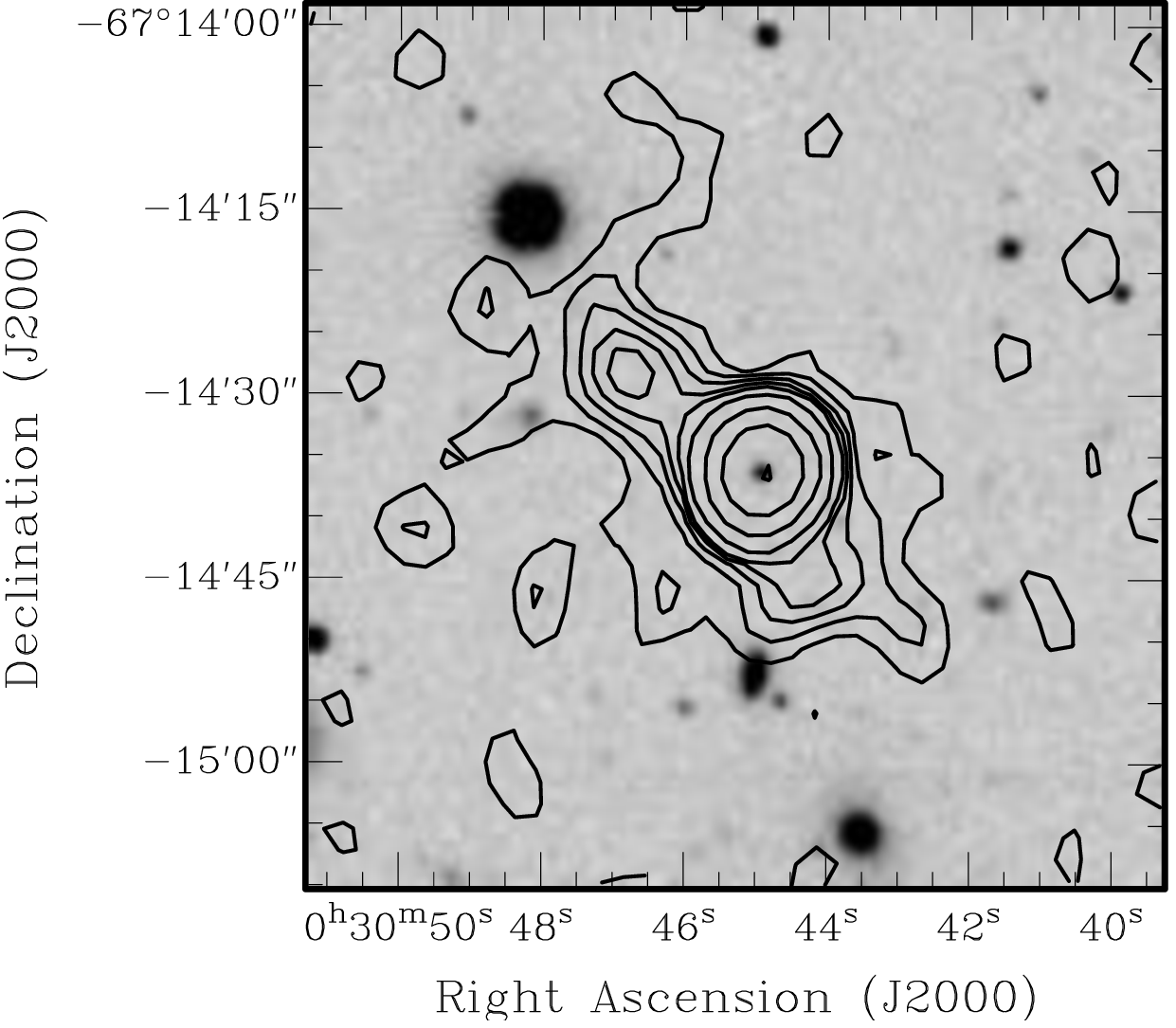}}
\caption{J0030.7-6714: $10^{-4}$ Jy x 1,  2,  3,  4,  5,  8,  16,  32.} 
\end{minipage}
&
\begin{minipage}{0.47\linewidth}
\frame{\includegraphics[angle=-90, width=2.8in]{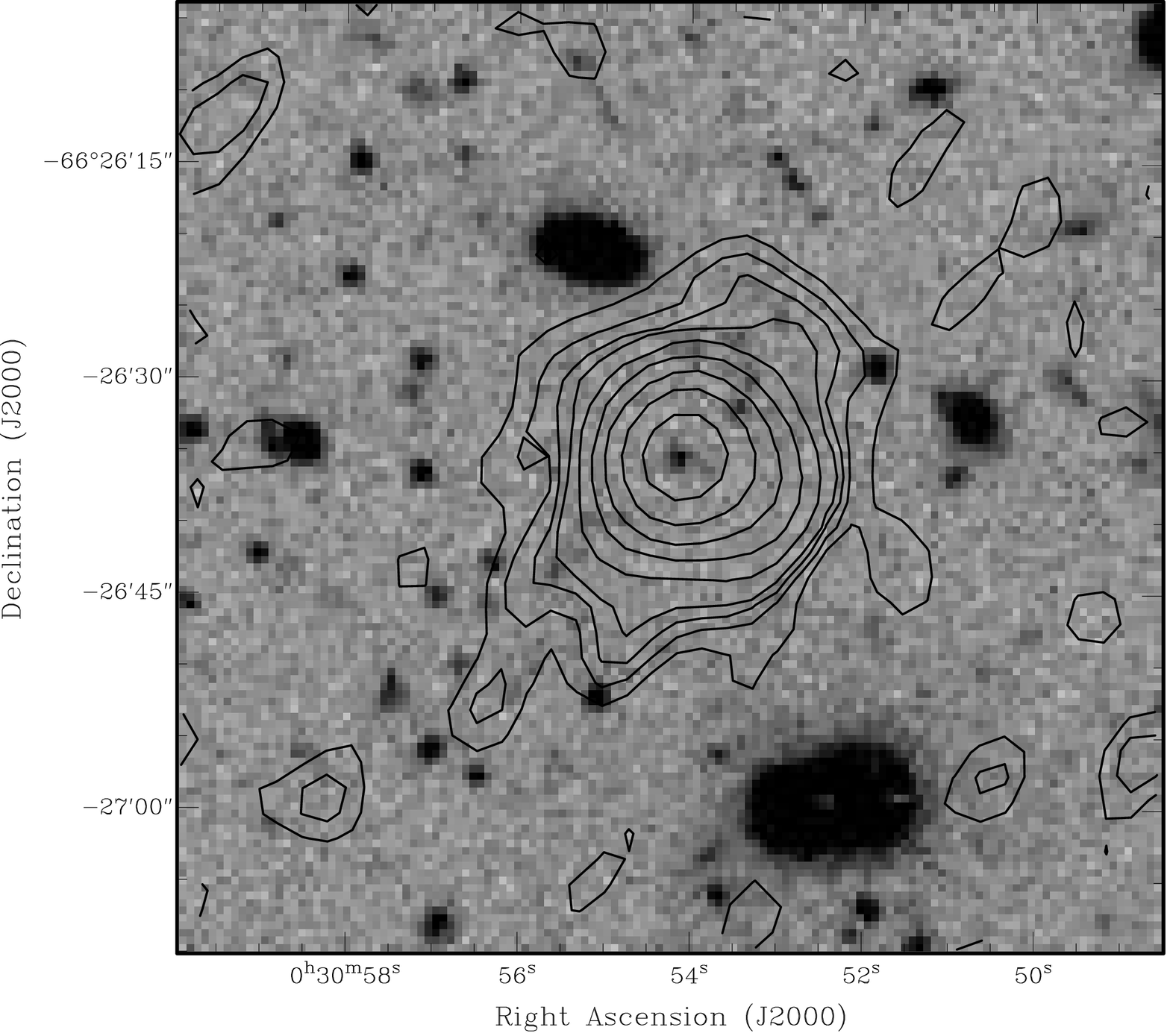}}
\caption{J0030.9-6626: $10^{-4}$ Jy x 1,  2,  3,  4,  8,  16,  32,  64, 128.} 
\end{minipage}
\\
\end{tabular}
\end{figure*}

\begin{figure*}
\centering
\begin{tabular}{cc}
\begin{minipage}{0.47\linewidth}
\frame{\includegraphics[angle=-90, width=2.8in] {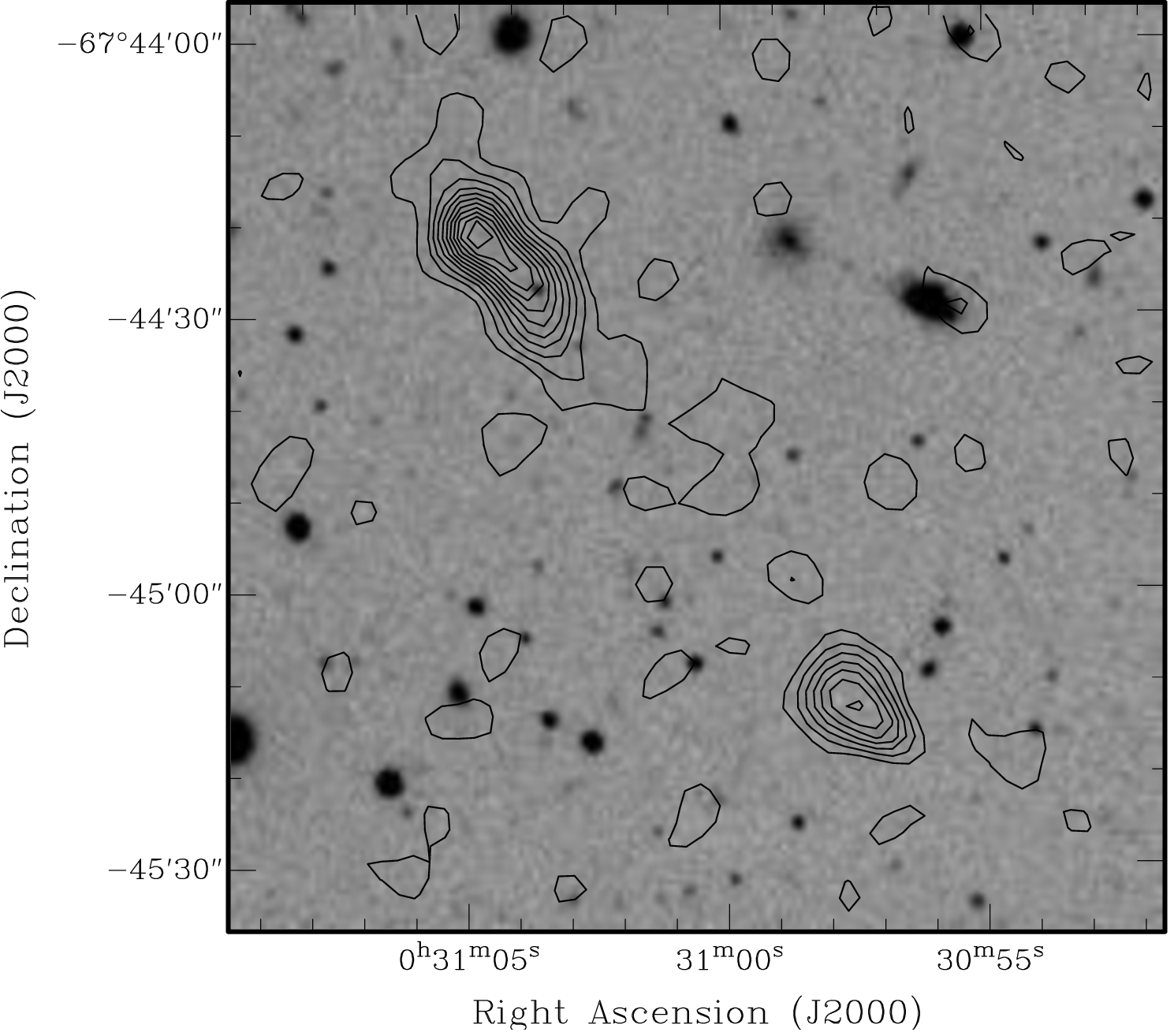}}
\caption{J0031.0-6744: $10^{-4}$ Jy x 1, 2, 3, 4, 5, 6, 7, 8, 9, 10, 11.} 
\end{minipage}
&
\begin{minipage}{0.47\linewidth}
\frame{\includegraphics[angle=-90, width=2.8in]{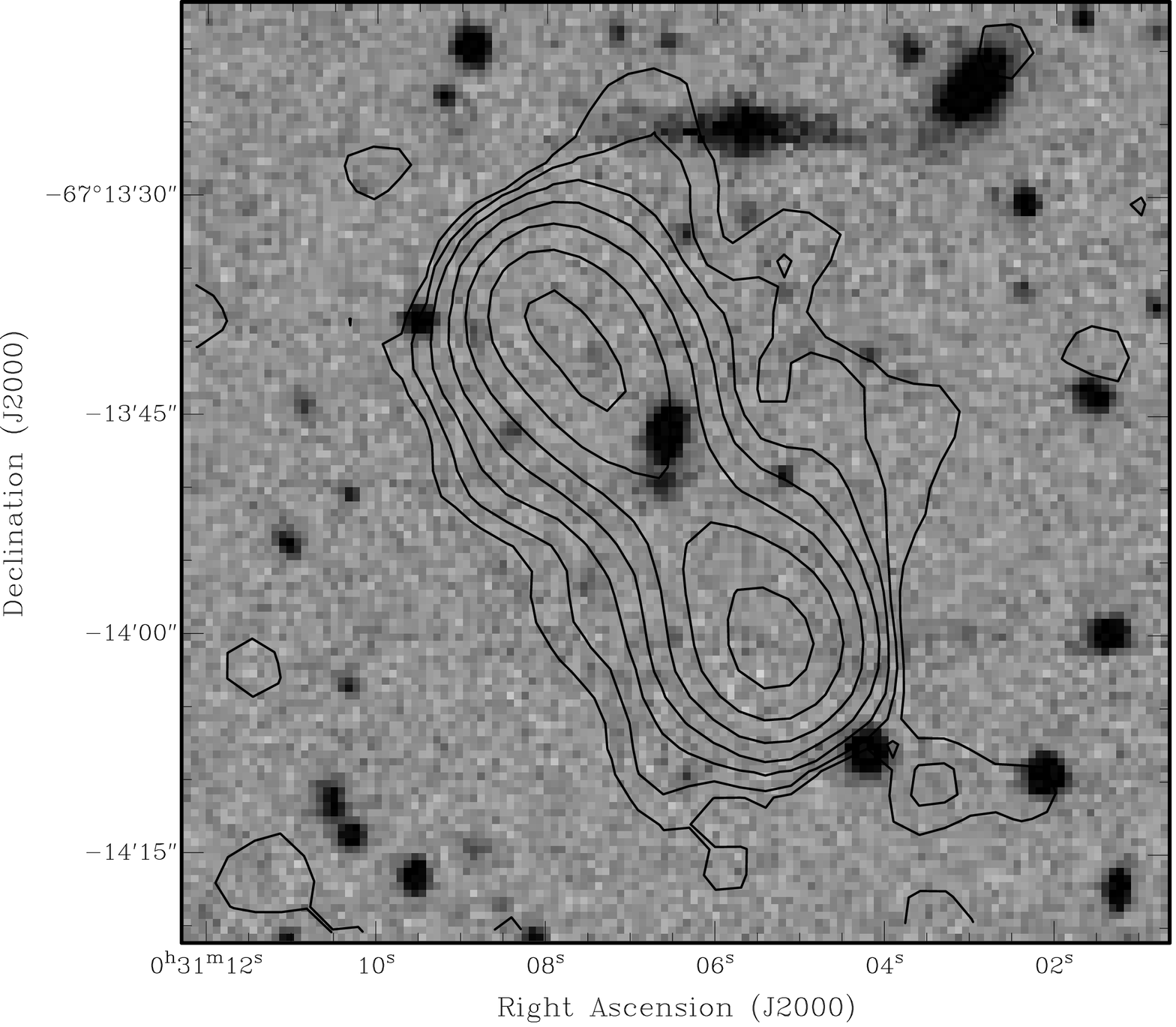}}
\caption{J0031.1-6713: $10^{-4}$ Jy x 1,  2,  4,  8,  16,  32,  64.}
\end{minipage}
\\
\end{tabular}
\end{figure*}

\begin{figure*}
\centering
\begin{tabular}{cc}
\begin{minipage}{0.47\linewidth}
\frame{\includegraphics[angle=-90, width=2.8in]{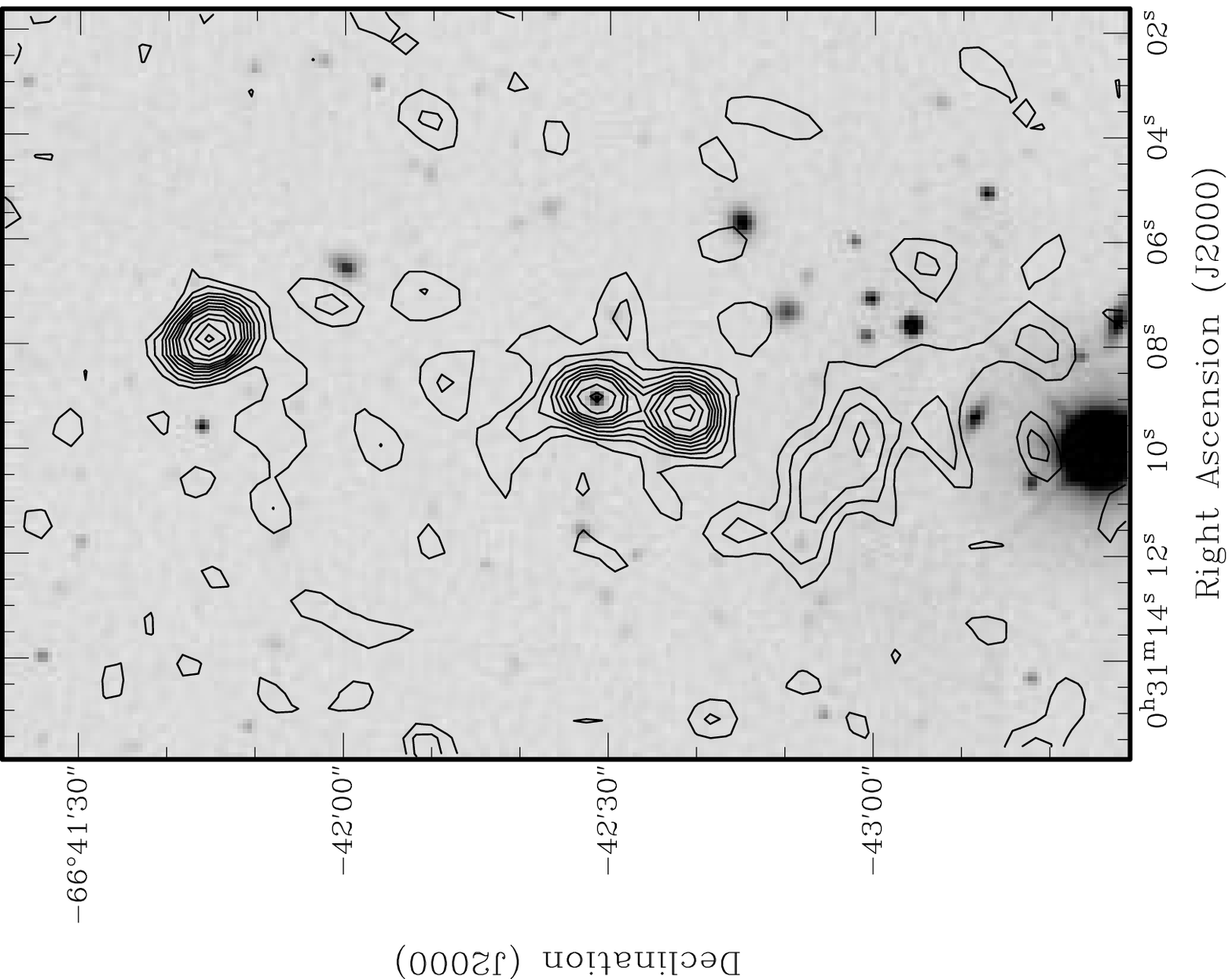}}
\caption{J0031.1-6642: $10^{-4}$ Jy x 1,  2,  3,  4,  5,  6,  7,  8,  10,  12,  14,  16.} 
\end{minipage}
&
\begin{minipage}{0.47\linewidth}
\frame{\includegraphics[angle=-90, width=2.8in]{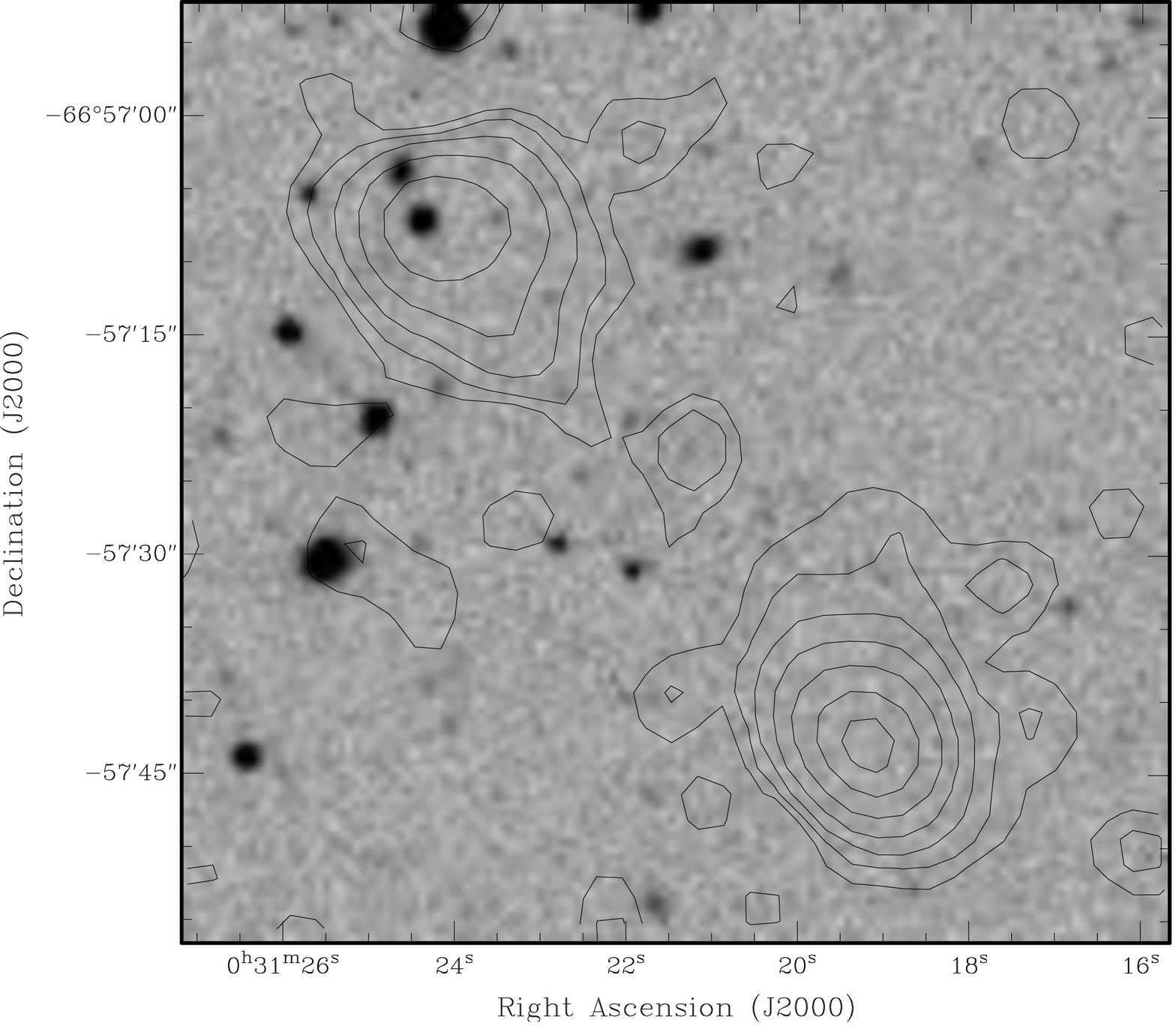}}
\caption{J0031.3-6657: $10^{-4}$ Jy x 1,  2,  4,  8,  16,  32,  52.} 
\end{minipage}
\\
\end{tabular}
\end{figure*}

\begin{figure*}
\centering
\begin{tabular}{cc}
\begin{minipage}{0.47\linewidth}
\frame{\includegraphics[angle=-90, width=2.8in] {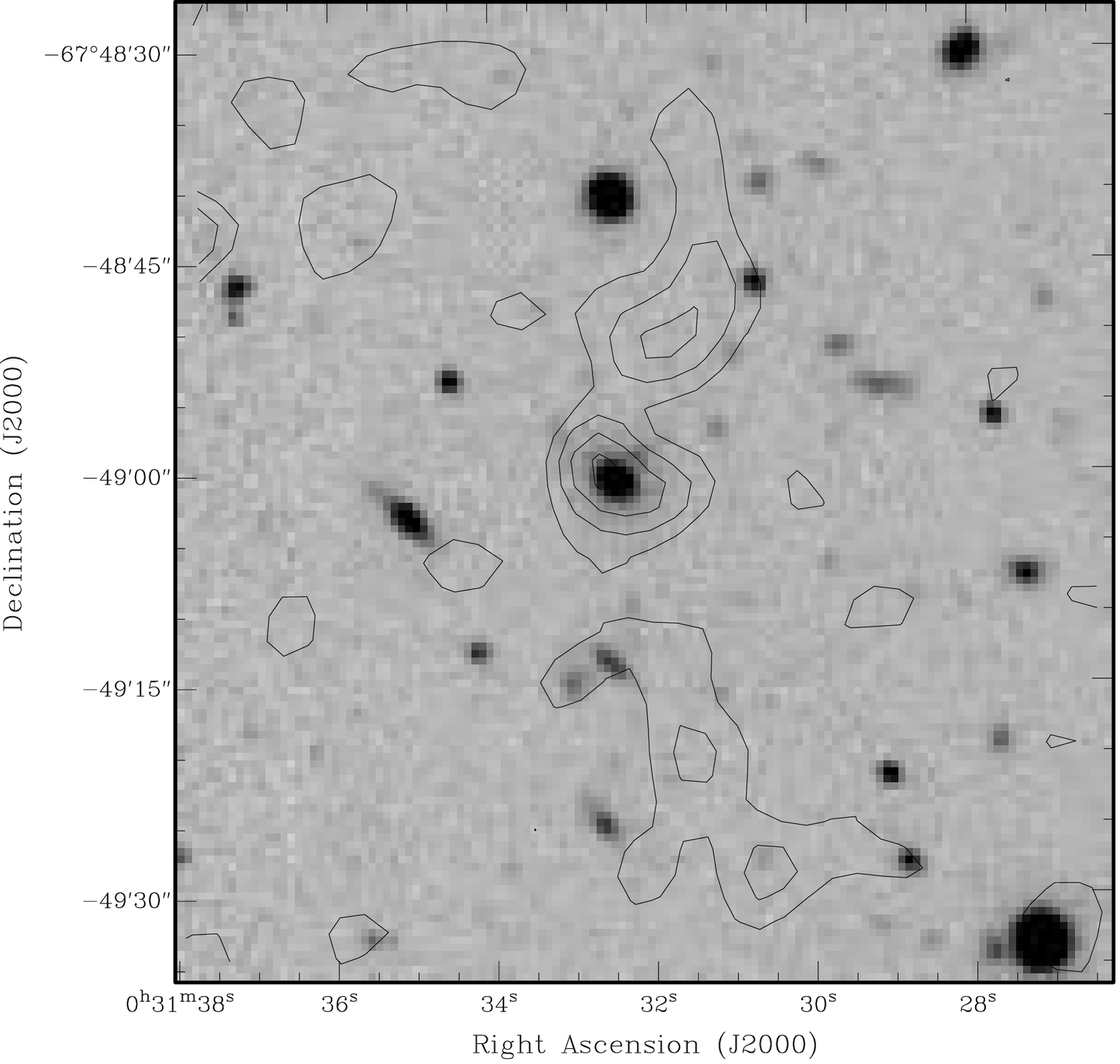}}
\caption{J0031.5-6748: $10^{-4}$ Jy x 1,  2,  3,  4.} 
\end{minipage}
&
\begin{minipage}{0.47\linewidth}
\frame{\includegraphics[angle=-90, width=2.8in]{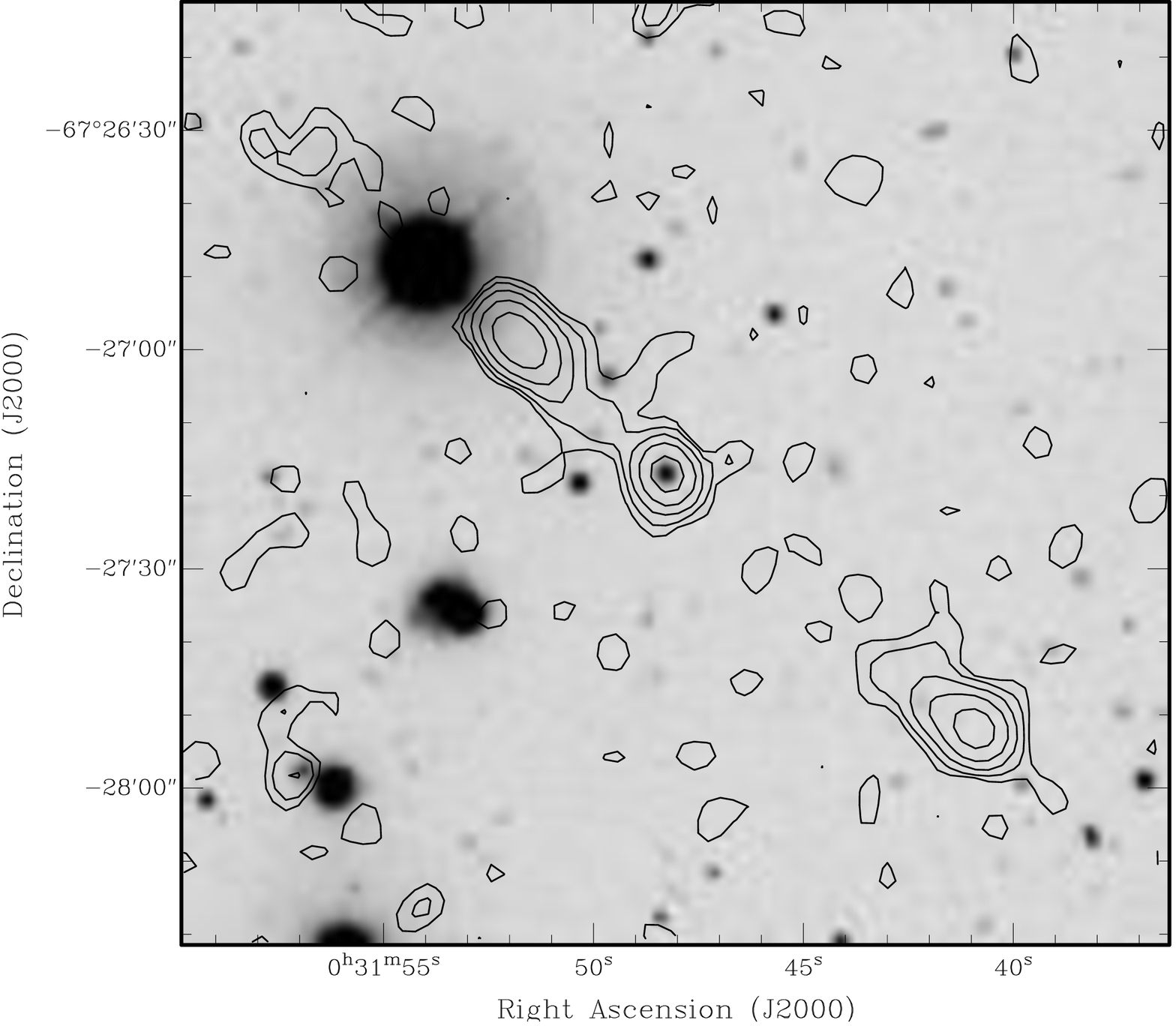}}
\caption{J0031.8-6727: $10^{-4}$ Jy x 1,  2,  4,  8,  16.}
\end{minipage}
\\
\end{tabular}
\end{figure*}

\begin{figure*}
\centering
\begin{tabular}{cc}
\begin{minipage}{0.47\linewidth}
\frame{\includegraphics[angle=-90, width=2.8in]{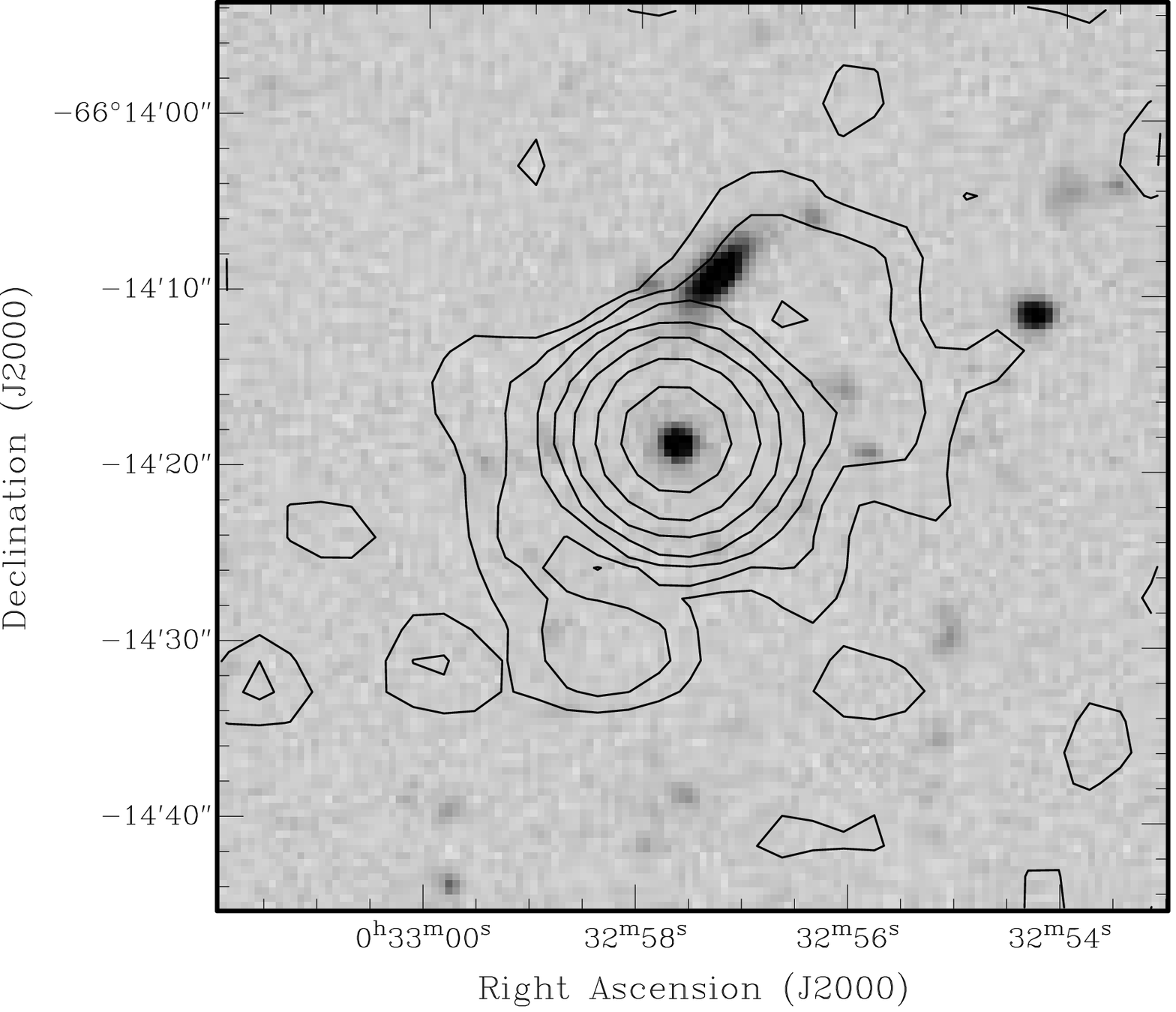}}
\caption{J0032.9-6614: $10^{-4}$ Jy x 1,  2,  4,  8,  16,  32,  64.} 
\end{minipage}
&
\begin{minipage}{0.47\linewidth}
\frame{\includegraphics[angle=-90, width=2.8in]{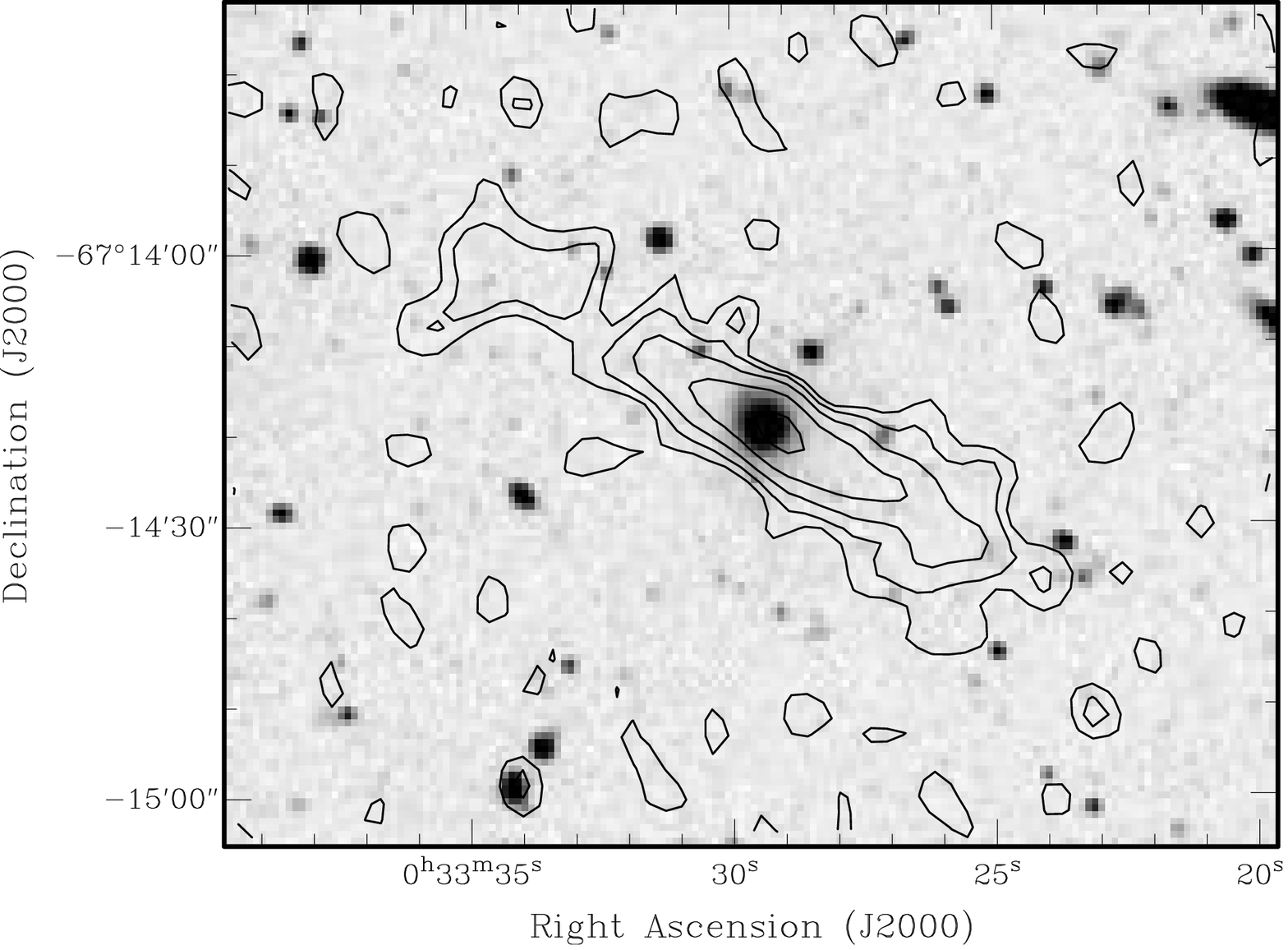}}
\caption{J0033.4-6714: $10^{-4}$ Jy x 1,  2,  4,  8,  16.} 
\end{minipage}
\\
\end{tabular}
\end{figure*}

\begin{figure*}[ht]
\centering
\mbox{\subfigure{\includegraphics[angle=-90, width=2.8in]{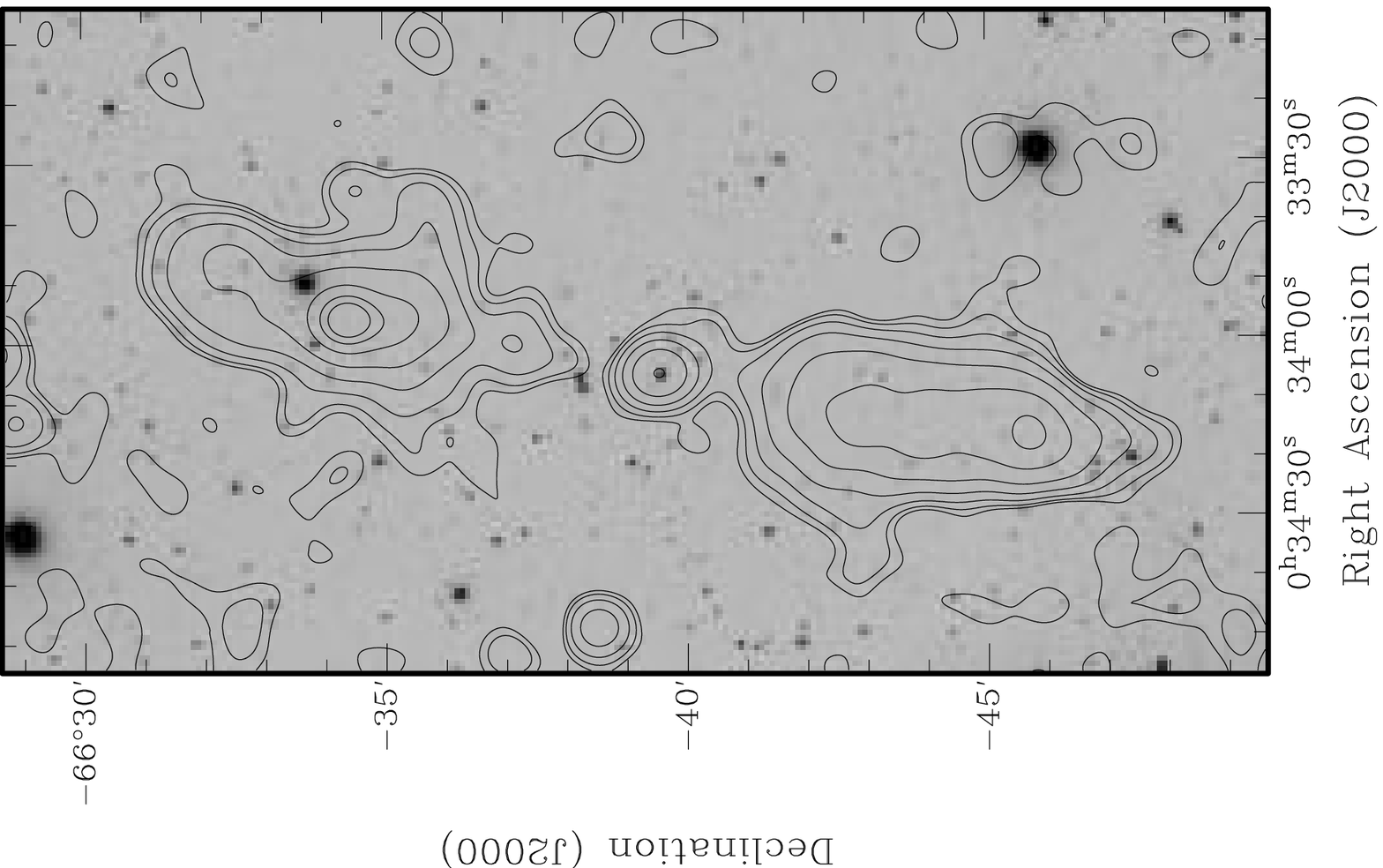}}\quad
\subfigure{\includegraphics[angle=-90, width=2.8in]{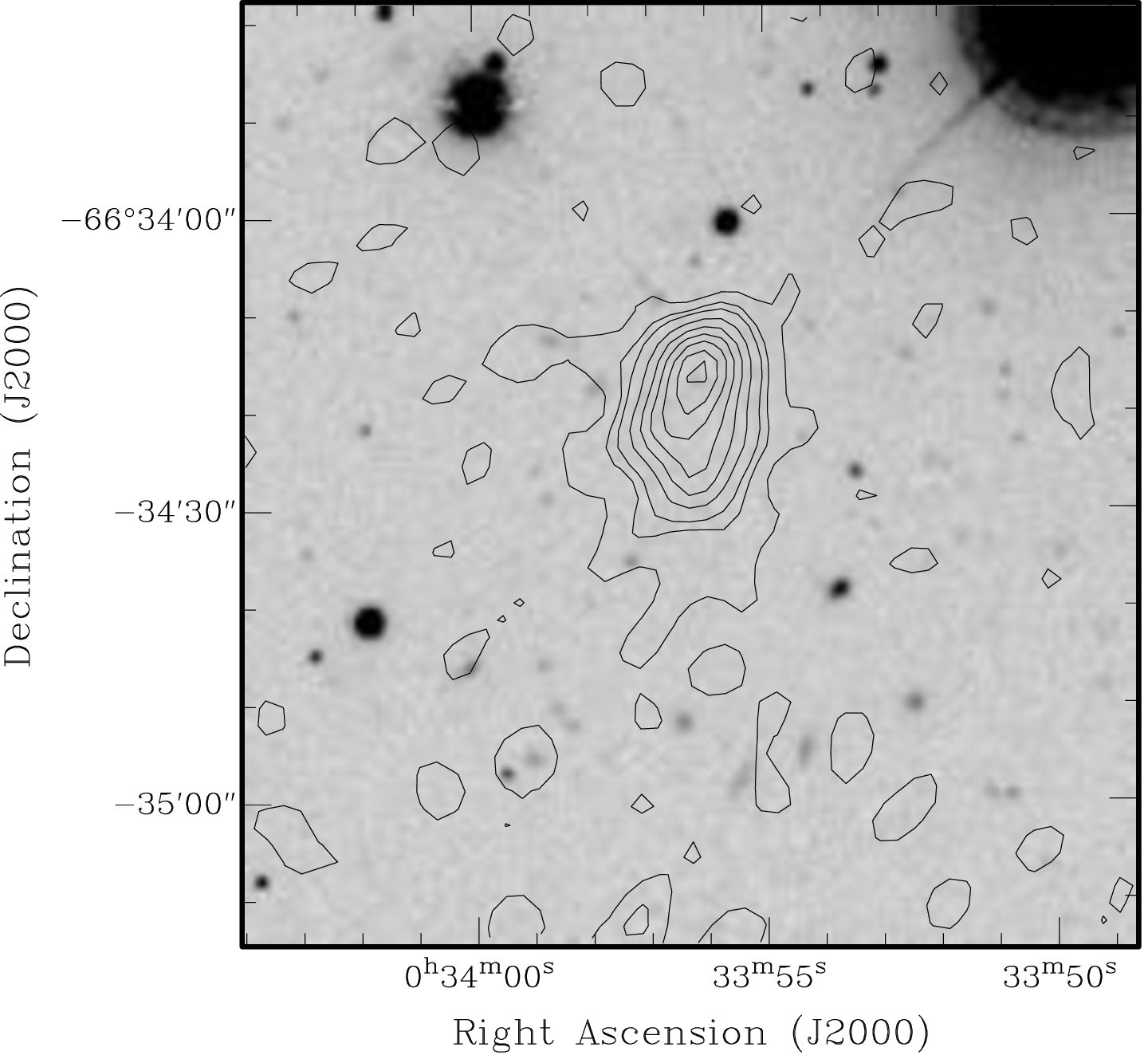} }}
\caption{J0034.0-6639: The left panel shows the full extent of the source at 50" resolution. 
The contour levels are: $10^{-4}$ Jy x 1,  2,  4,  8,  16,  32,  50,  64. 
The right panel shows the region of the inner component in the North lobe at 6" resolution.
The contour levels are: $10^{-4}$ Jy x 1.5,   3,   4,   6,   8,   10,   12, 14, 16.
}
\end{figure*}

\begin{figure*}
\centering
\begin{tabular}{cc}

\begin{minipage}{0.47\linewidth}
\frame{\includegraphics[angle=-90, width=2.8in]{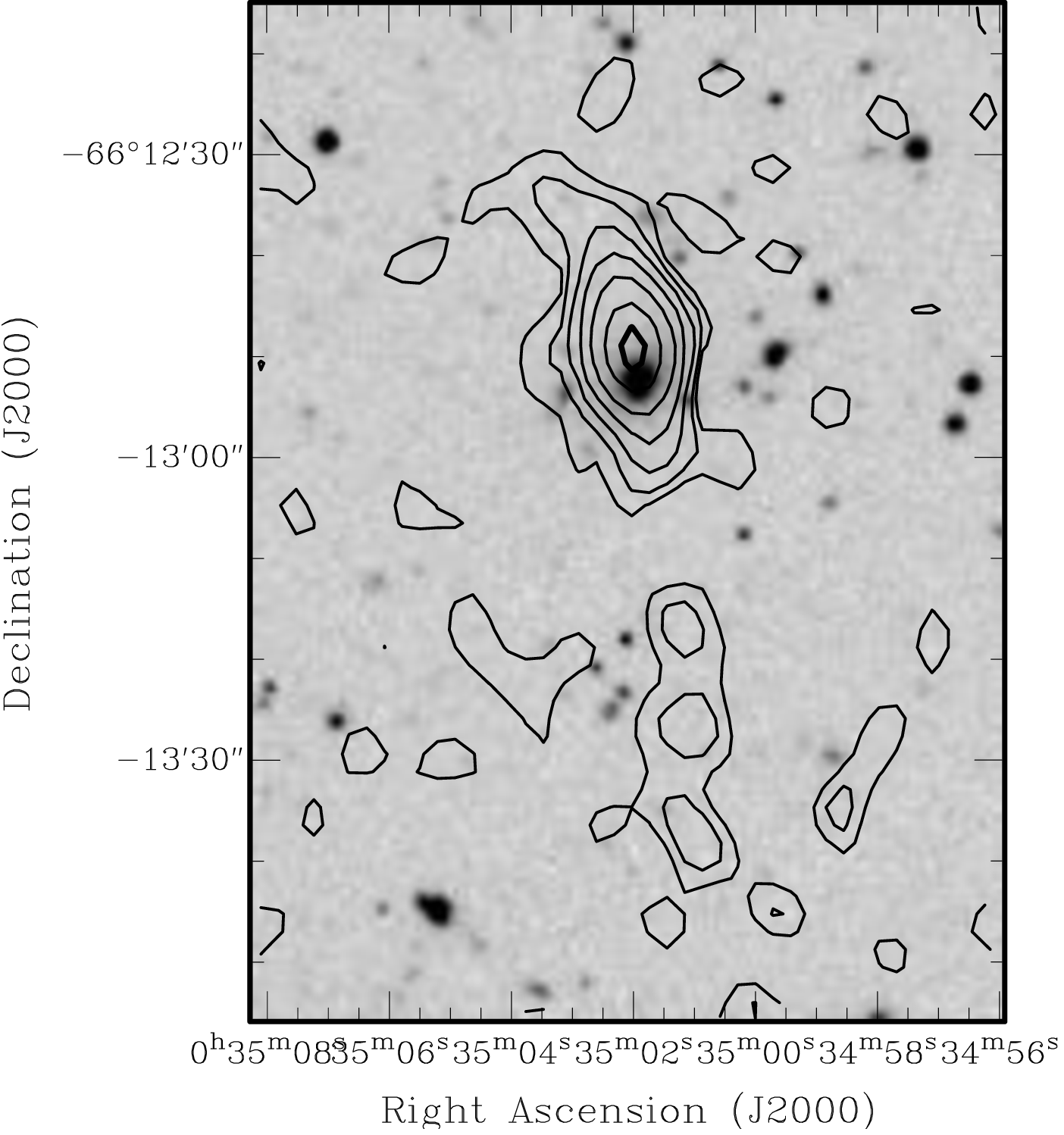}}
\caption{J0035.0-6612: $10^{-4}$ Jy x 1,  2,  4,  8,  16,  32,  50.}
\end{minipage}
&
\begin{minipage}{0.47\linewidth}
\frame{\includegraphics[angle=-90, width=2.8in]{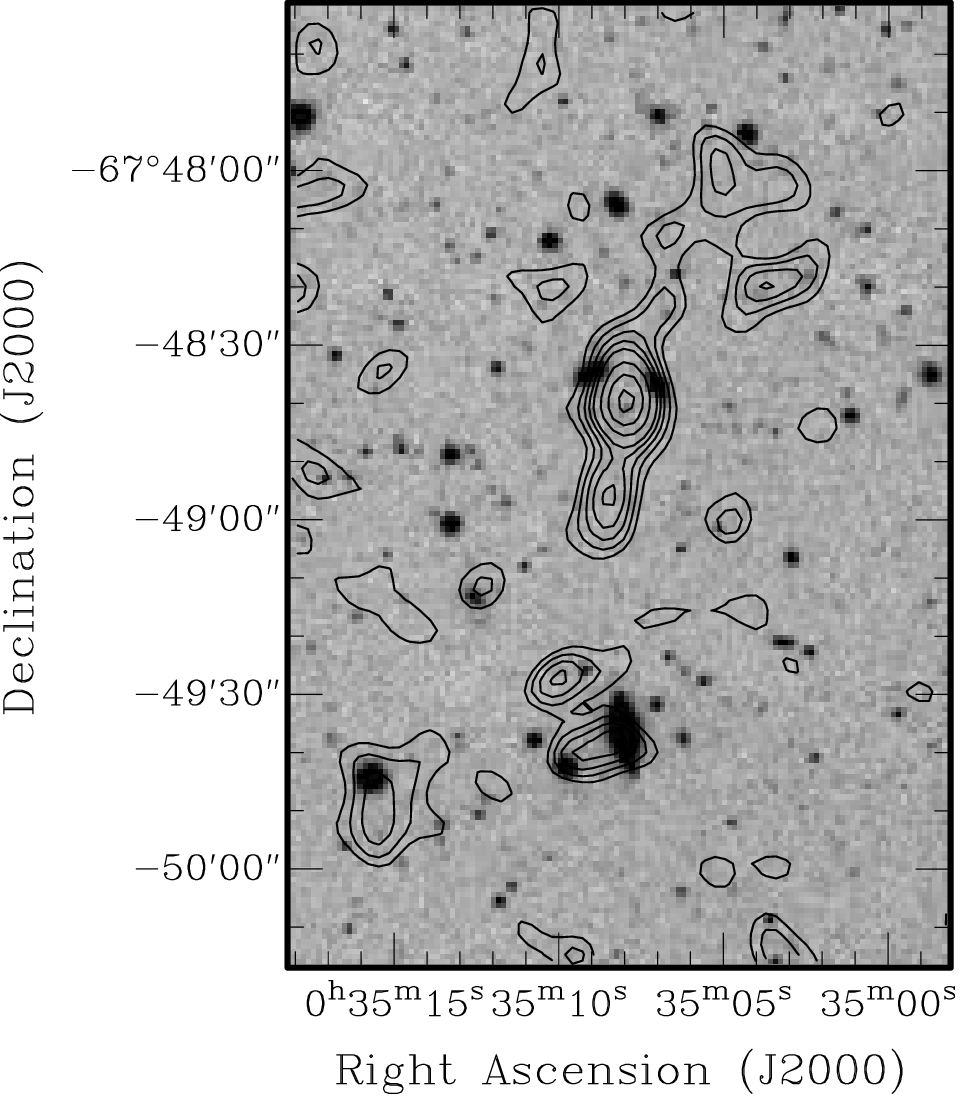}}
\caption{J0035.1-6748: $10^{-4}$ Jy x 1,  1.5,  2,  2.5,  3,  4,  5,  6. Resolution of the image is 10".} 
\end{minipage}
\\
\end{tabular}
\end{figure*}

\clearpage

\begin{figure*}
\centering
\begin{tabular}{cc}
\begin{minipage}{0.47\linewidth}
\frame{\includegraphics[angle=-90, width=2.8in]{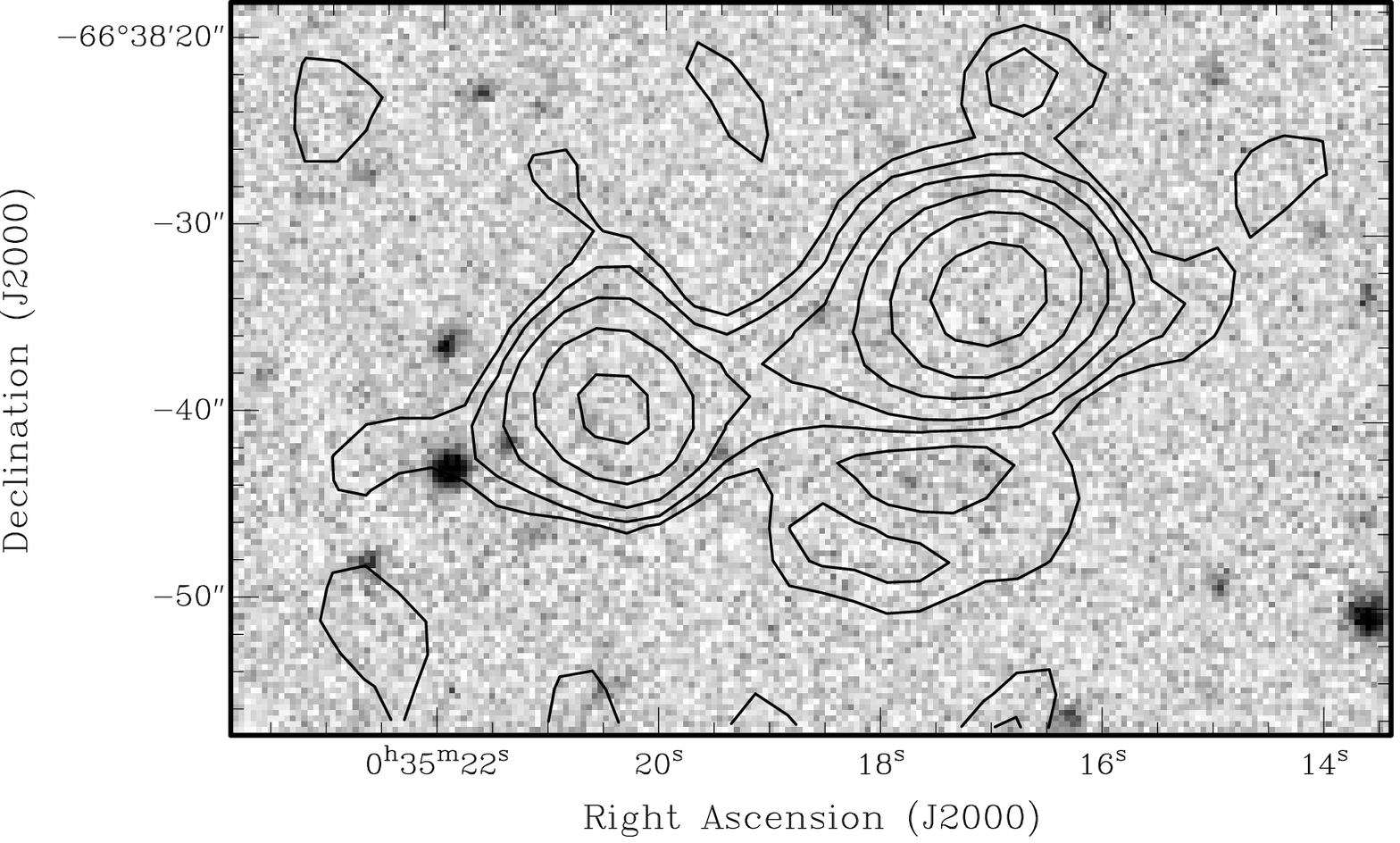}}
\caption{J0035.2-6638: $10^{-4}$ Jy x 1,  2,  4,  8,  16,  32.} 
\end{minipage}
&
\begin{minipage}{0.47\linewidth}
\frame{\includegraphics[angle=-90, width=2.8in] {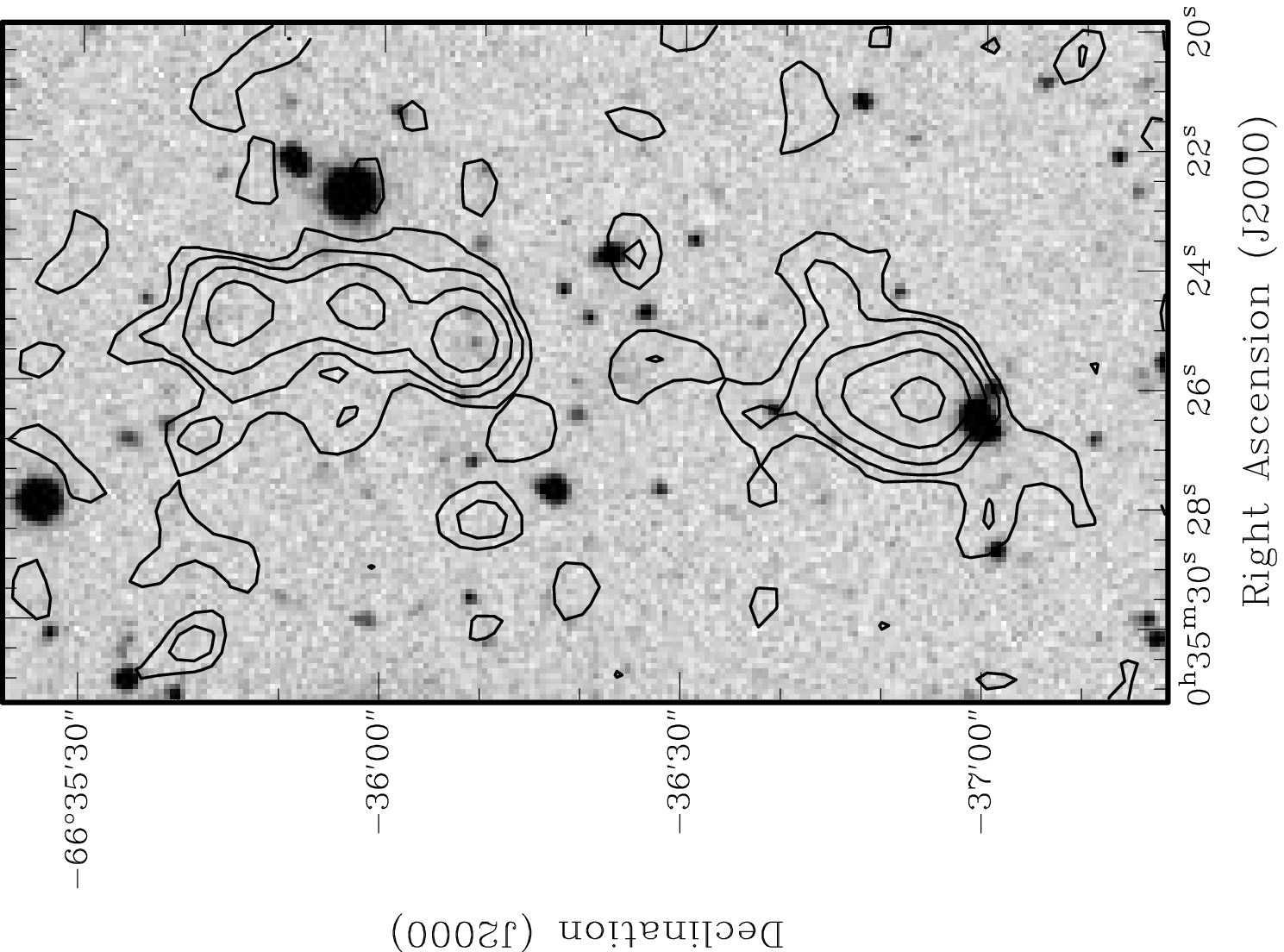}}
\caption{J0035.4-6636: $10^{-4}$ Jy x 1,  2,  4,  8,  16.} 
\end{minipage}
\\
\end{tabular}
\end{figure*}

\begin{figure*}
\centering
\begin{tabular}{l}
\begin{minipage}{0.47\linewidth}
\frame{\includegraphics[angle=-90, width=2.8in]{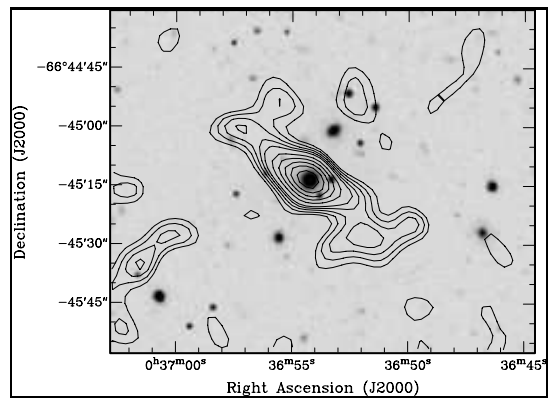}}
\caption{J0036.9-6645: 10"; $10^{-4}$ Jy x 1,  1.5,  2,  2.5,  3,  4,  5,  6,  7,  8. } 
\end{minipage}
\\
\end{tabular}
\end{figure*}

\clearpage


\begin{figure*}
\centering
\begin{tabular}{cc}
\begin{minipage}{0.47\linewidth}
\frame{\includegraphics[angle=-90, width=2.8in]{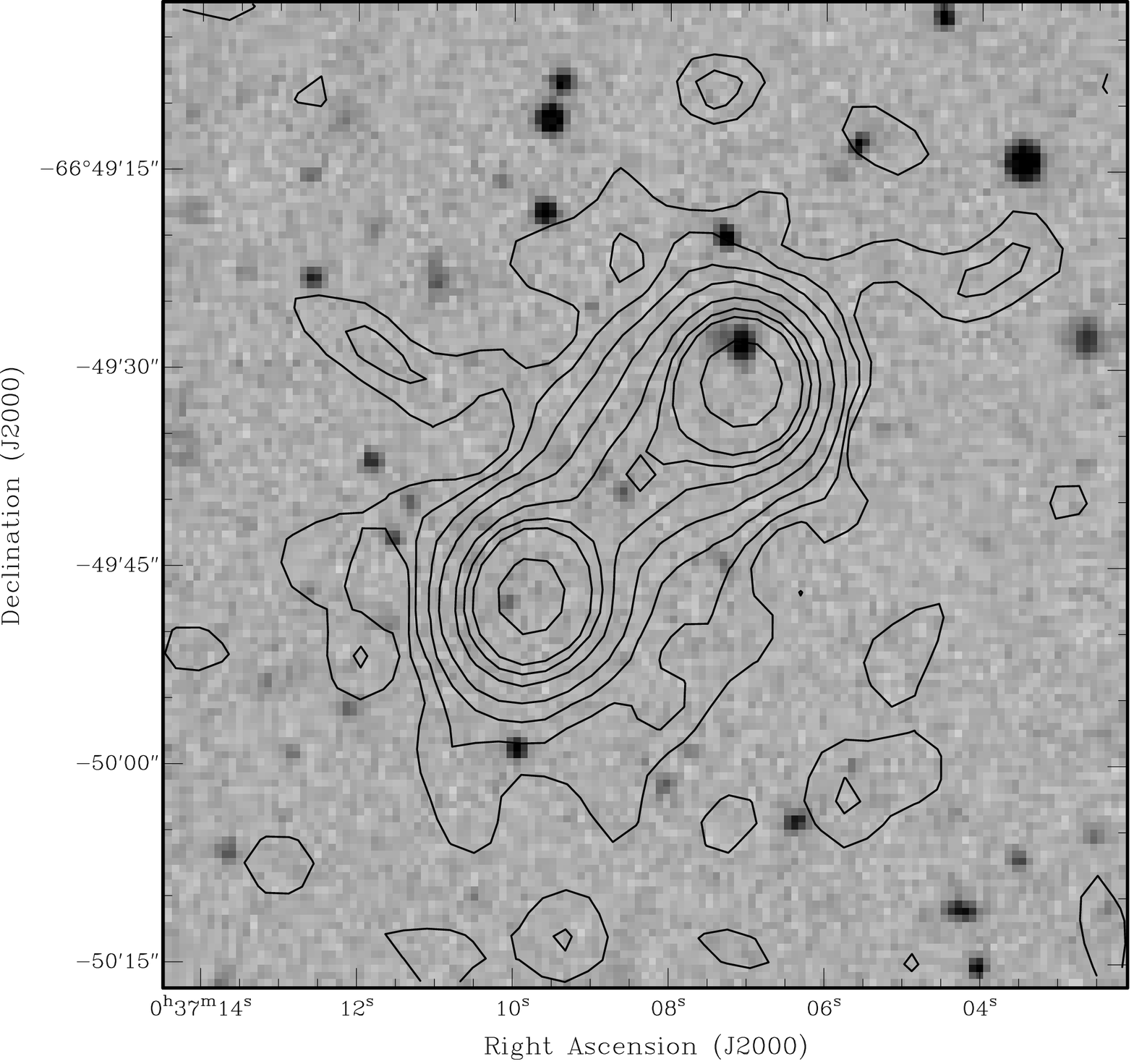}}
\caption{J0037.1-6649: $10^{-4}$ Jy x 1,  2,  4,  8,  16,  24,  32,  64.}
\end{minipage}
&
\begin{minipage}{0.47\linewidth}
\frame{\includegraphics[angle=-90, width=2.8in]{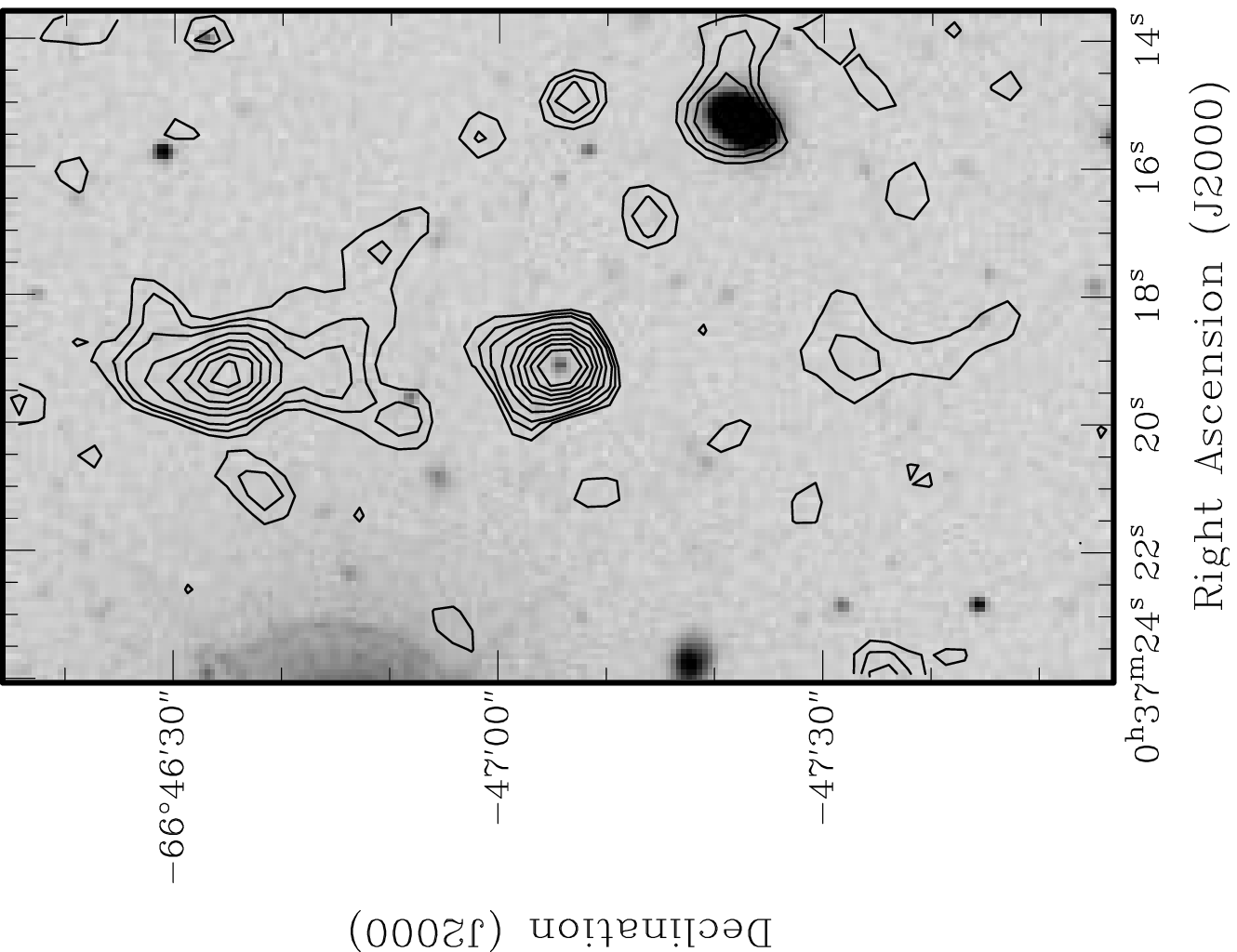}}
\caption{J0037.3-6647: $10^{-4}$ Jy x 1,  1.5,  2,  3,  4,  5,  6,  7,  8.} 
\end{minipage}
\\
\end{tabular}
\end{figure*}

\begin{figure*}
\centering
\begin{tabular}{cc}
\begin{minipage}{0.47\linewidth}
\frame{\includegraphics[angle=-90, width=2.8in]{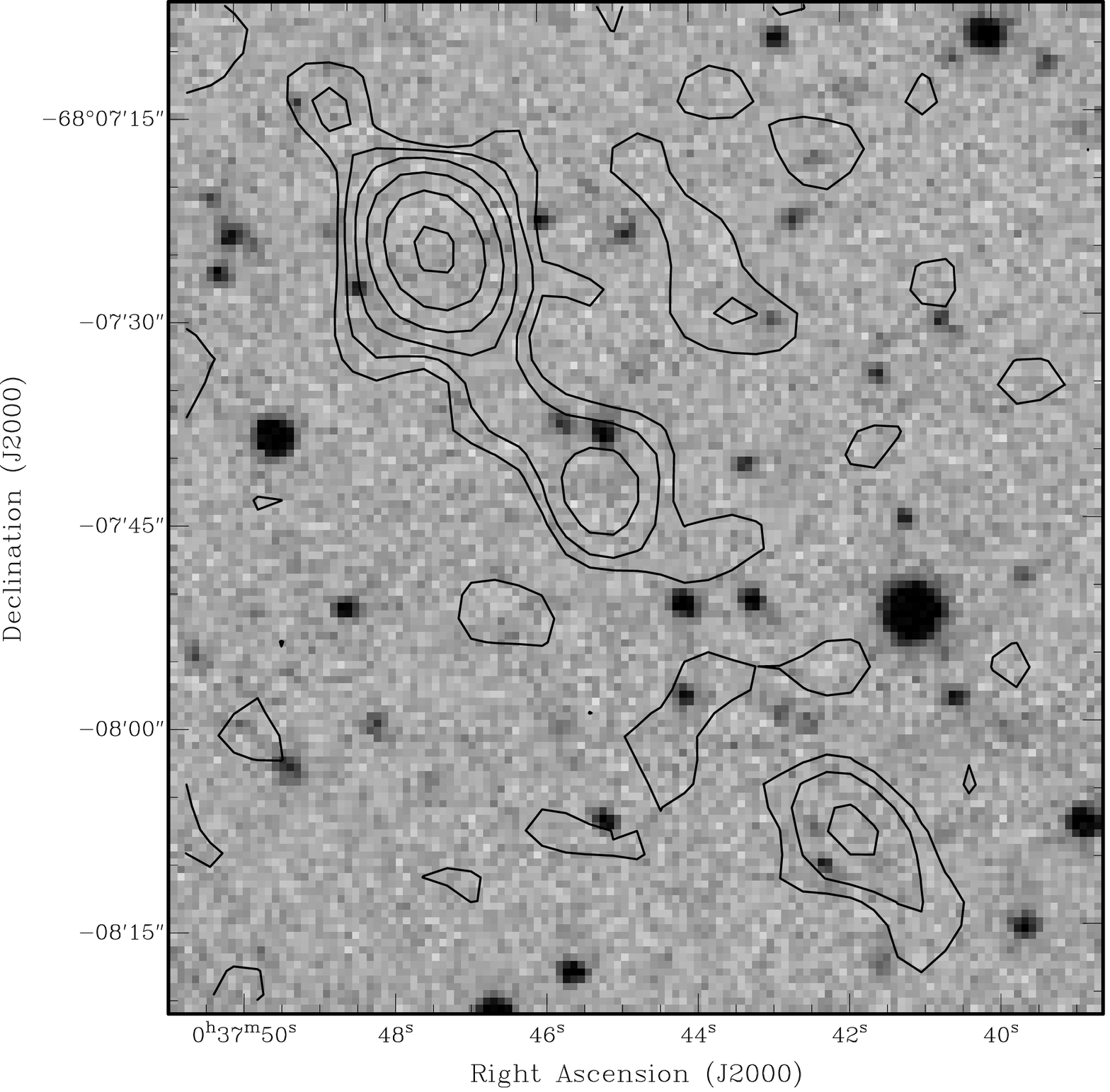}}
\caption{J0037.7-6807: $10^{-4}$ Jy x 1,  2,  4,  8,  16,  32.} 
\end{minipage}
&
\begin{minipage}{0.47\linewidth}
\frame{\includegraphics[angle=-90, width=2.8in] {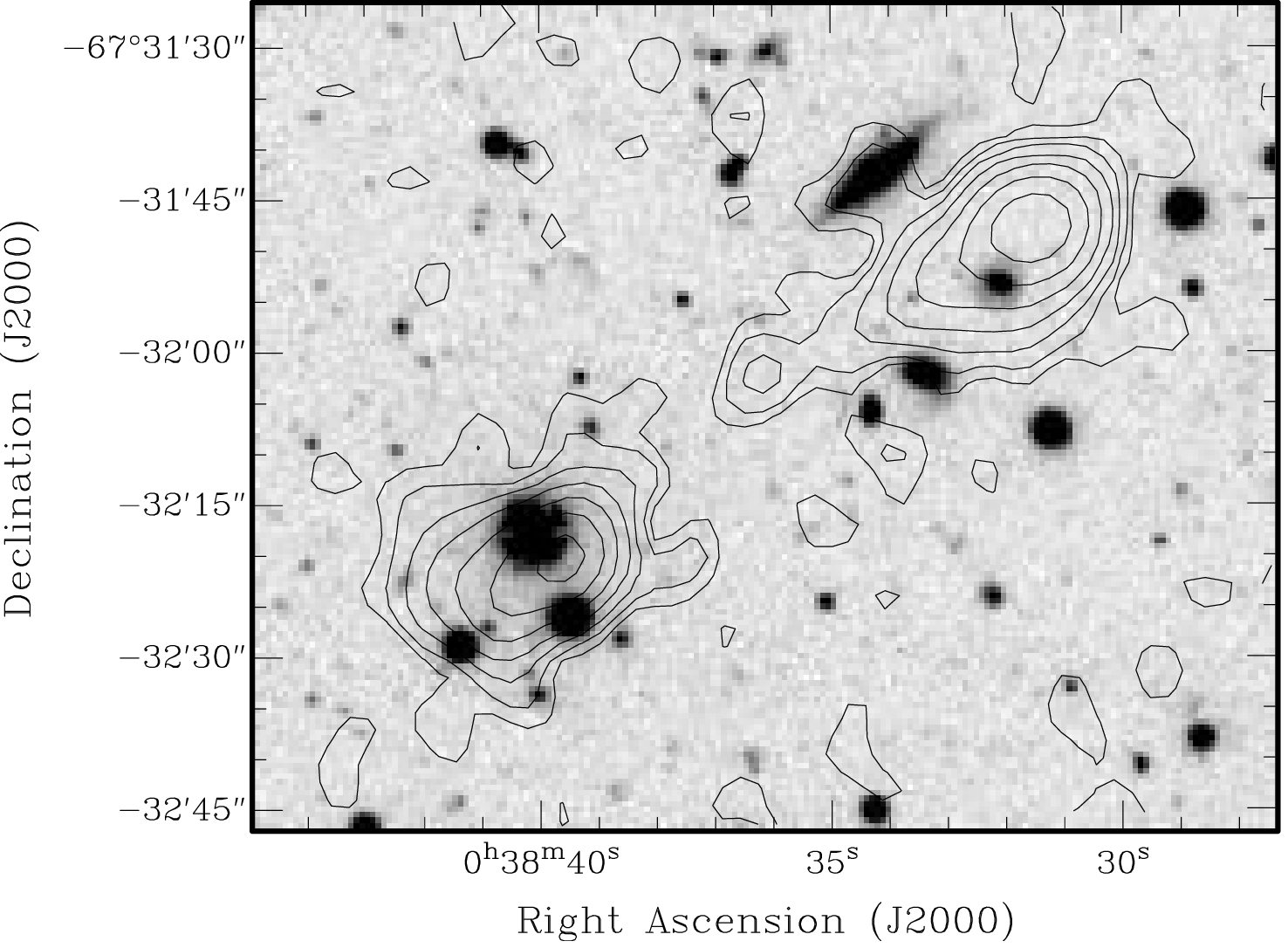}}
\caption{J0038.6-6732: $10^{-4}$ Jy x 1,  2,  4,  8,  16,  32,  64.} 
\end{minipage}
\\
\end{tabular}
\end{figure*}

\begin{figure*}
\centering
\begin{tabular}{cc}
\begin{minipage}{0.47\linewidth}
\frame{\includegraphics[angle=-90, width=2.8in]{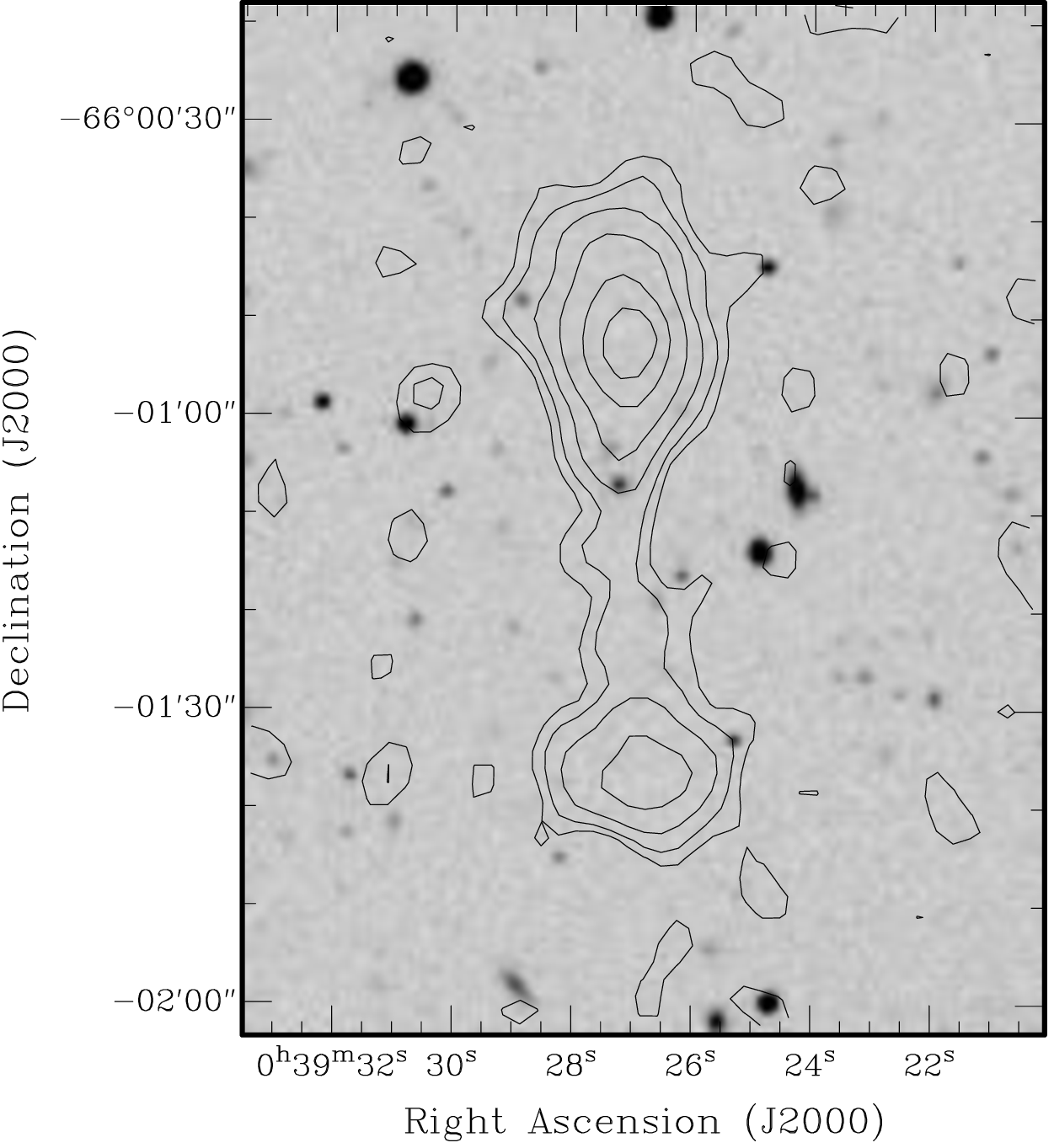}}
\caption{J0039.4-6601: $10^{-4}$ Jy x 1,  2,  4,  8,  16,  24.}
\end{minipage}
&
\begin{minipage}{0.47\linewidth}
\frame{\includegraphics[angle=-90, width=2.8in]{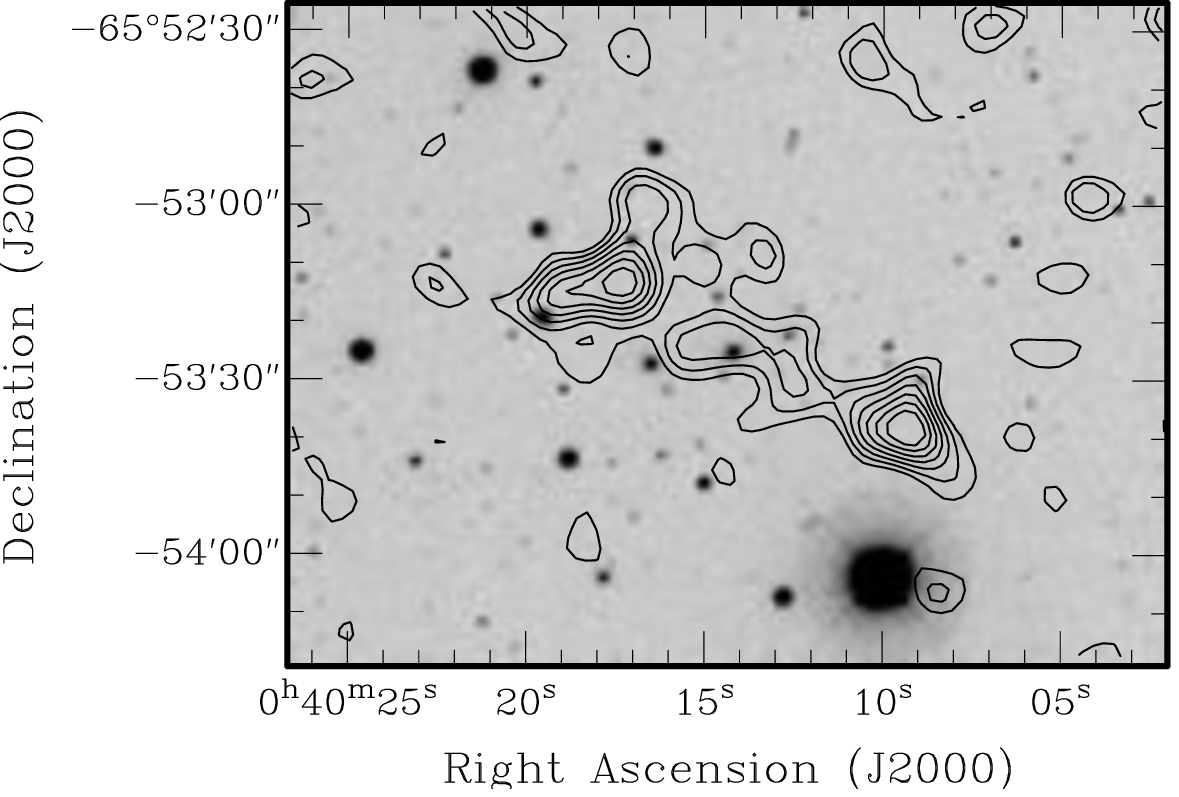}}
\caption{J0040.2-6553: $10^{-4}$ Jy x 1,  1.5,  2,  2.5,  3,  3.5,  4. } 
\end{minipage}
\\
\end{tabular}
\end{figure*}

\begin{figure*}[ht]
\centering
\mbox{\subfigure{\includegraphics[angle=-90, width=2.8in]{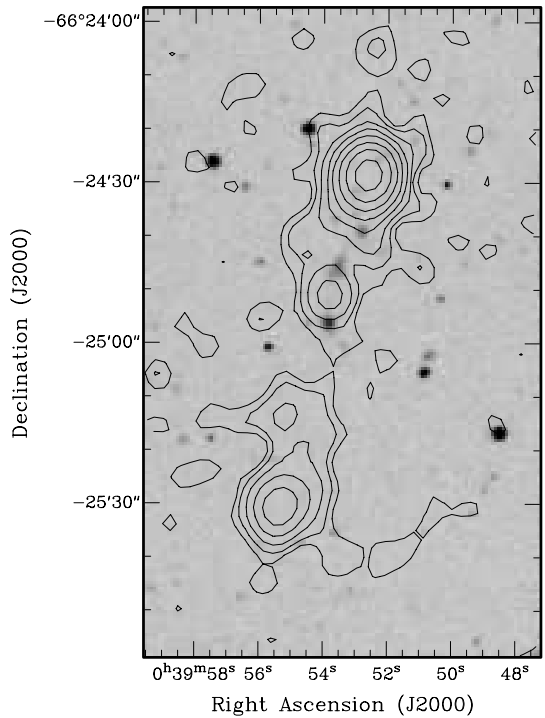}}\quad
\subfigure{\includegraphics[angle=-90, width=2.8in]{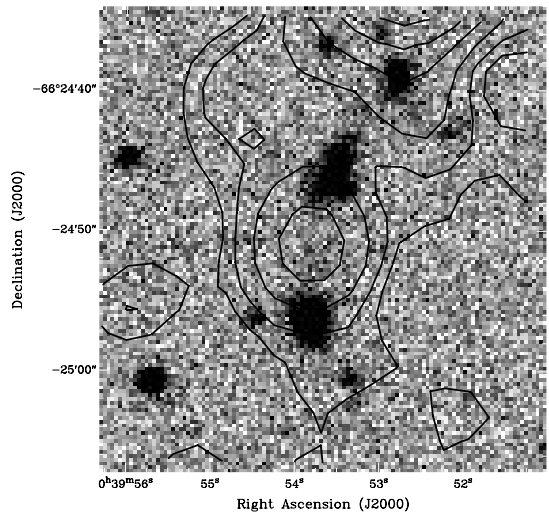}}}
\caption{ J0039.8-6624: The left panel shows the full extent of the source at 6" resolution. 
The contour levels are: $10^{-4}$ Jy x 1,  2,  4,  8,  16,  32,  64,  128.
The right panel shows the core region with the faint optical identification.}
\end{figure*}

\begin{figure*}
\centering
\begin{tabular}{l}
\begin{minipage}{0.47\linewidth}
\frame{\includegraphics[angle=-90, width=2.8in]{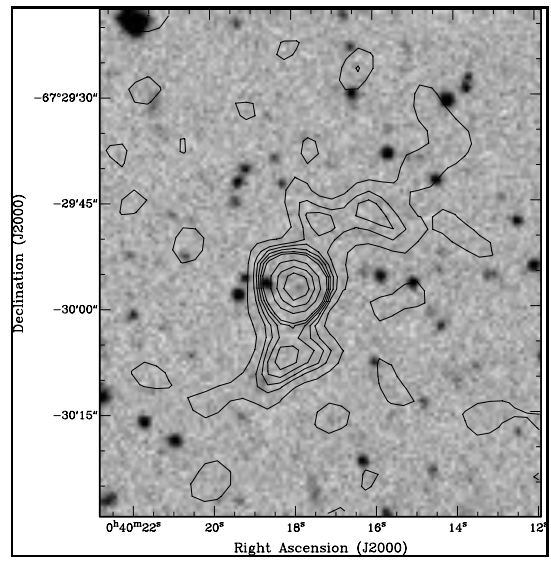}}
\caption{J0040.2-6729: $10^{-4}$ Jy x 1,  2,  2.5,  3,  3.5,  4,  6,  8,  10.} 
\end{minipage}
\\
\end{tabular}
\end{figure*}

\clearpage


\begin{figure*}
\centering
\begin{tabular}{cc}
\begin{minipage}{0.47\linewidth}
\frame{\includegraphics[angle=-90, width=2.8in]{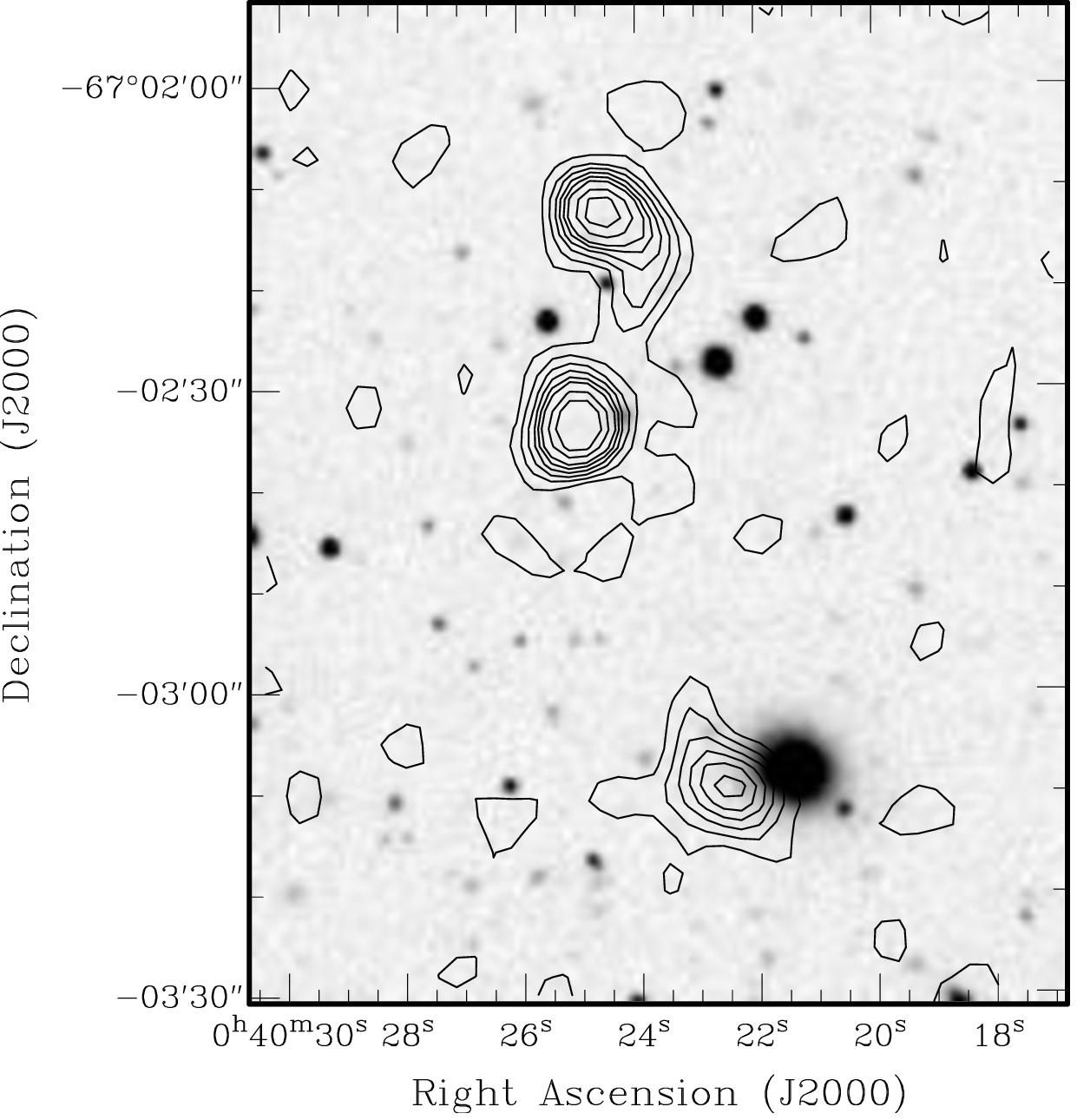}}
\caption{J0040.3-6703: $10^{-4}$ Jy x 1,  2,  3,  4,  5,  6,  8,  10.} 
\end{minipage}
&
\begin{minipage}{0.47\linewidth}
\frame{\includegraphics[angle=-90, width=2.8in] {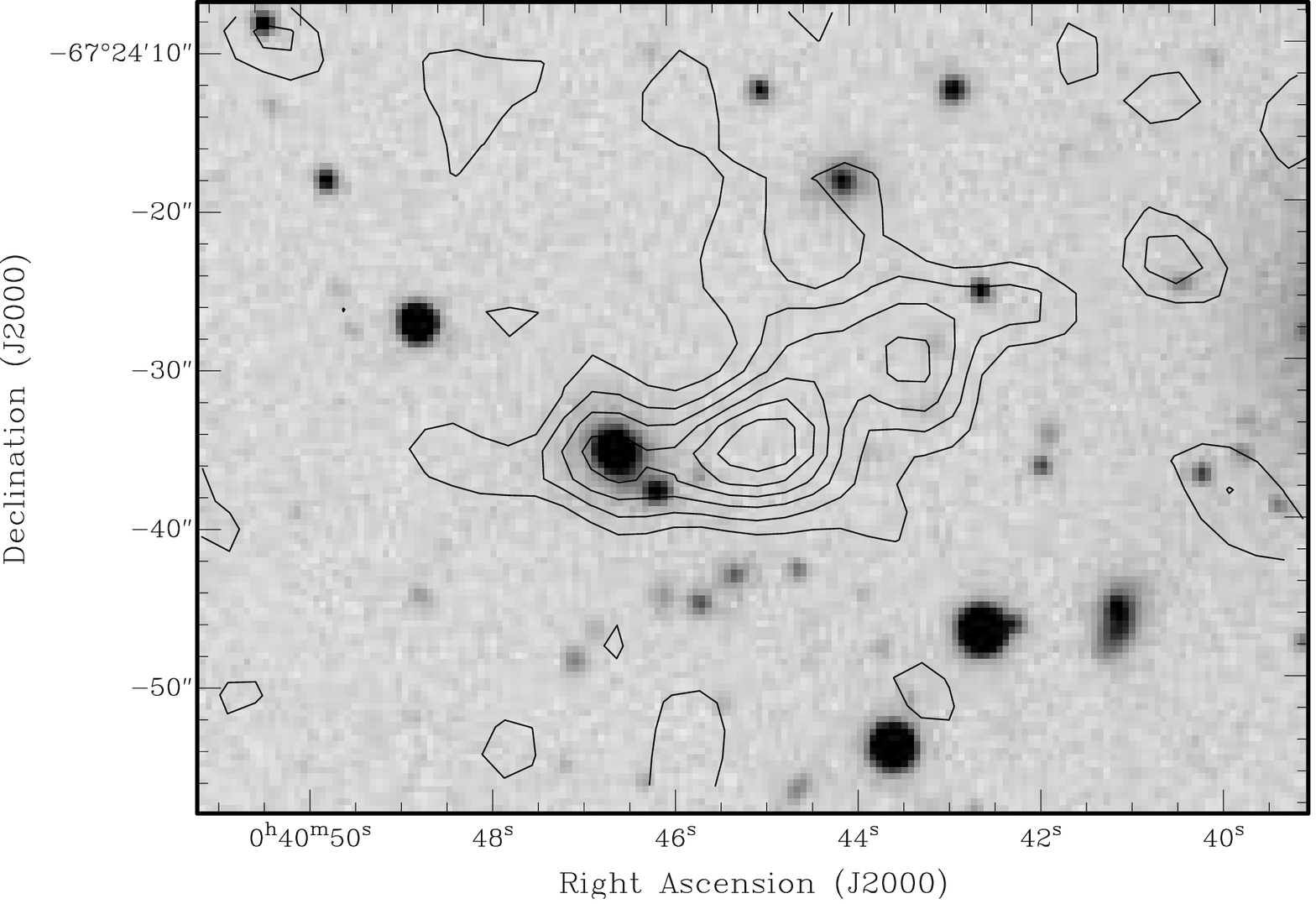}}
\caption{J0040.7-6724: $10^{-4}$ Jy x 1,  2,  3,  4,  5,  6.} 
\end{minipage}
\\
\end{tabular}
\end{figure*}

\begin{figure*}
\centering
\begin{tabular}{cc}
\begin{minipage}{0.47\linewidth}
\frame{\includegraphics[angle=-90, width=2.8in]{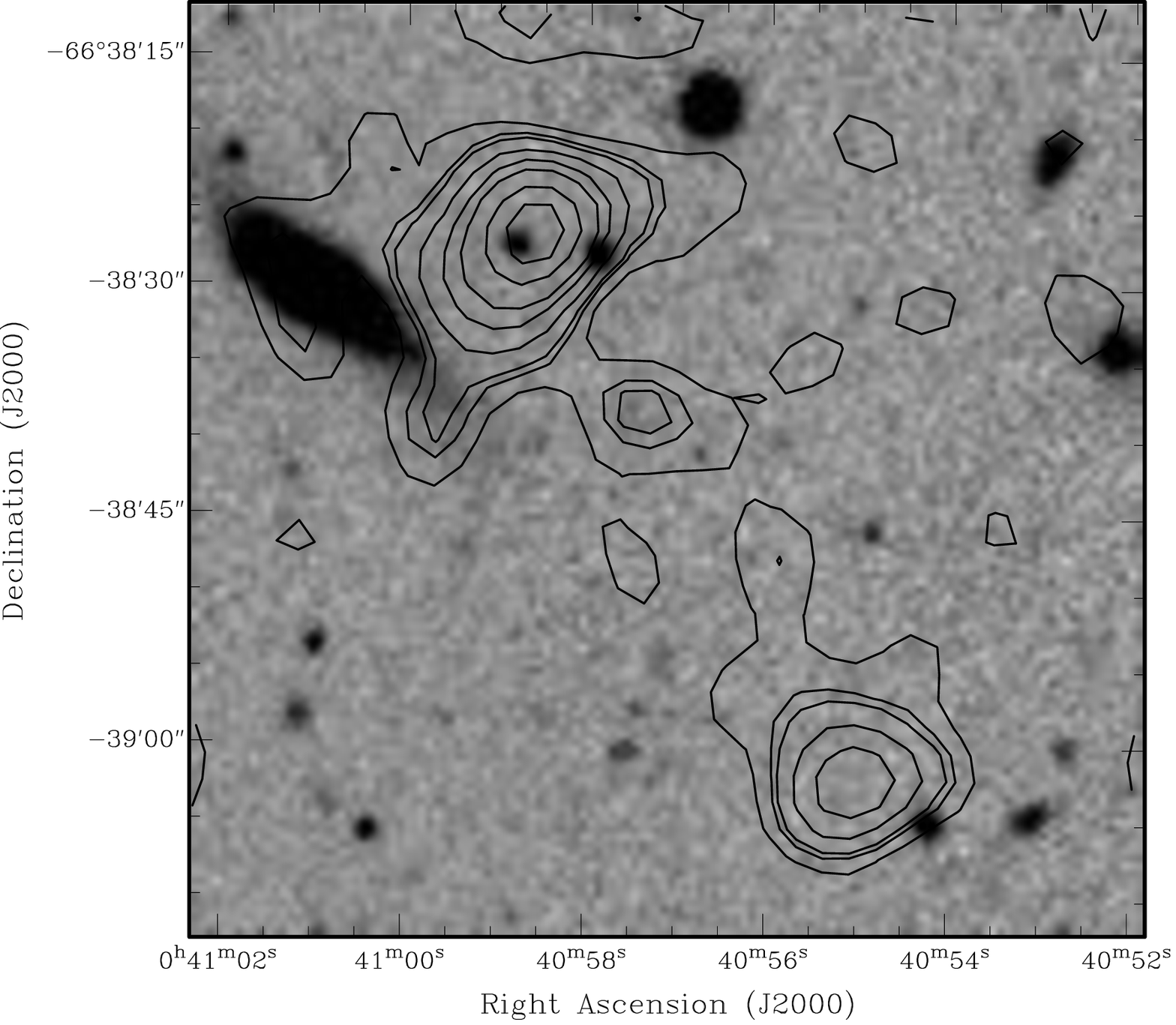}}
\caption{J0040.9-6638: $10^{-4}$ Jy x 1,  2,  2.5,  4,  6,  8,  12,  16,  32,  64.}
\end{minipage}
&
\begin{minipage}{0.47\linewidth}
\frame{\includegraphics[angle=-90, width=2.8in]{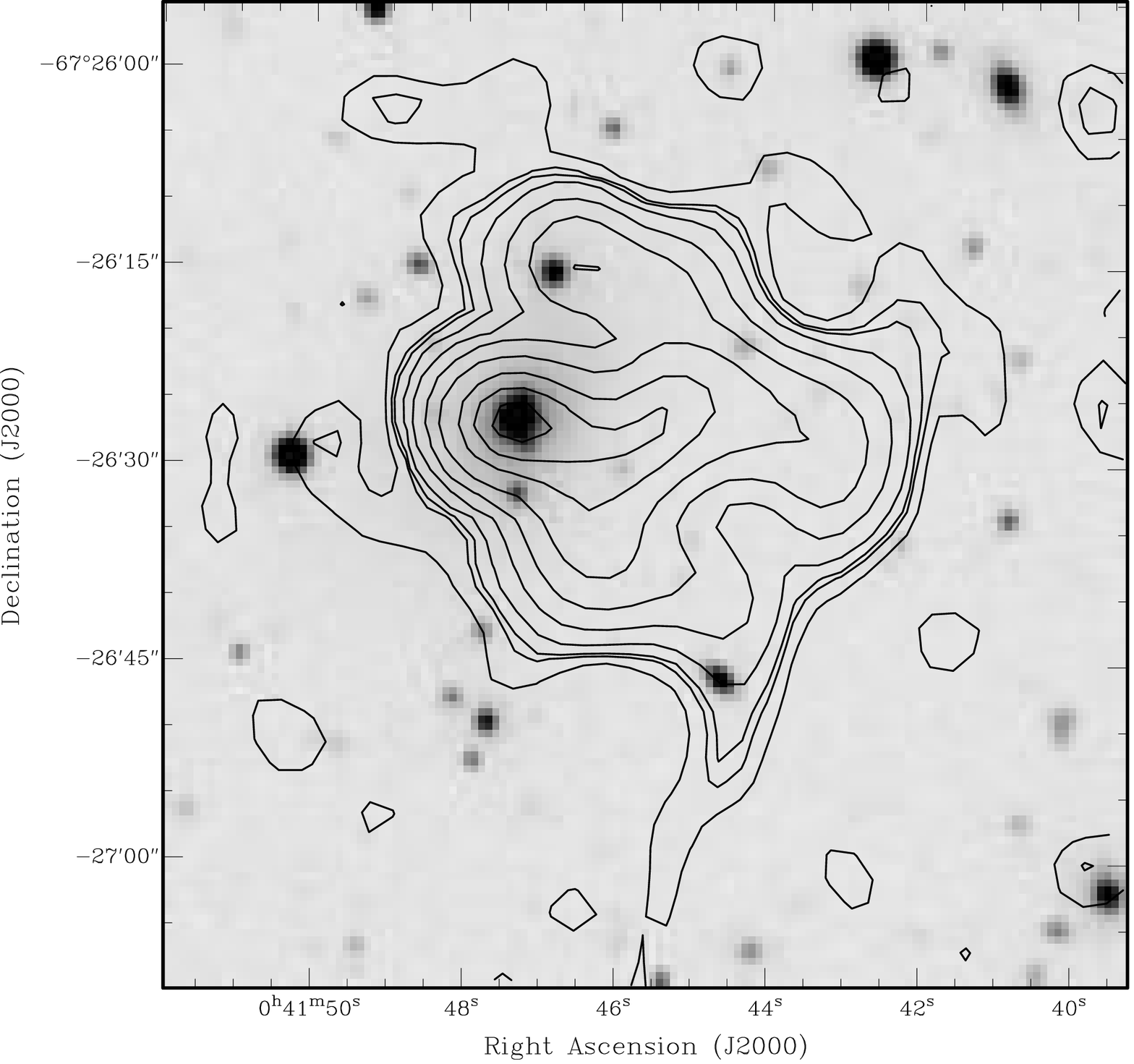}}
\caption{J0041.7-6726: $10^{-4}$ Jy x 1,  2,  2.5,  4,  8,  16,  32,  64,  96, 128.} 
\end{minipage}
\\
\end{tabular}
\end{figure*}

\begin{figure*}
\centering
\begin{tabular}{cc}
\begin{minipage}{0.47\linewidth}
\frame{\includegraphics[angle=-90, width=2.8in]{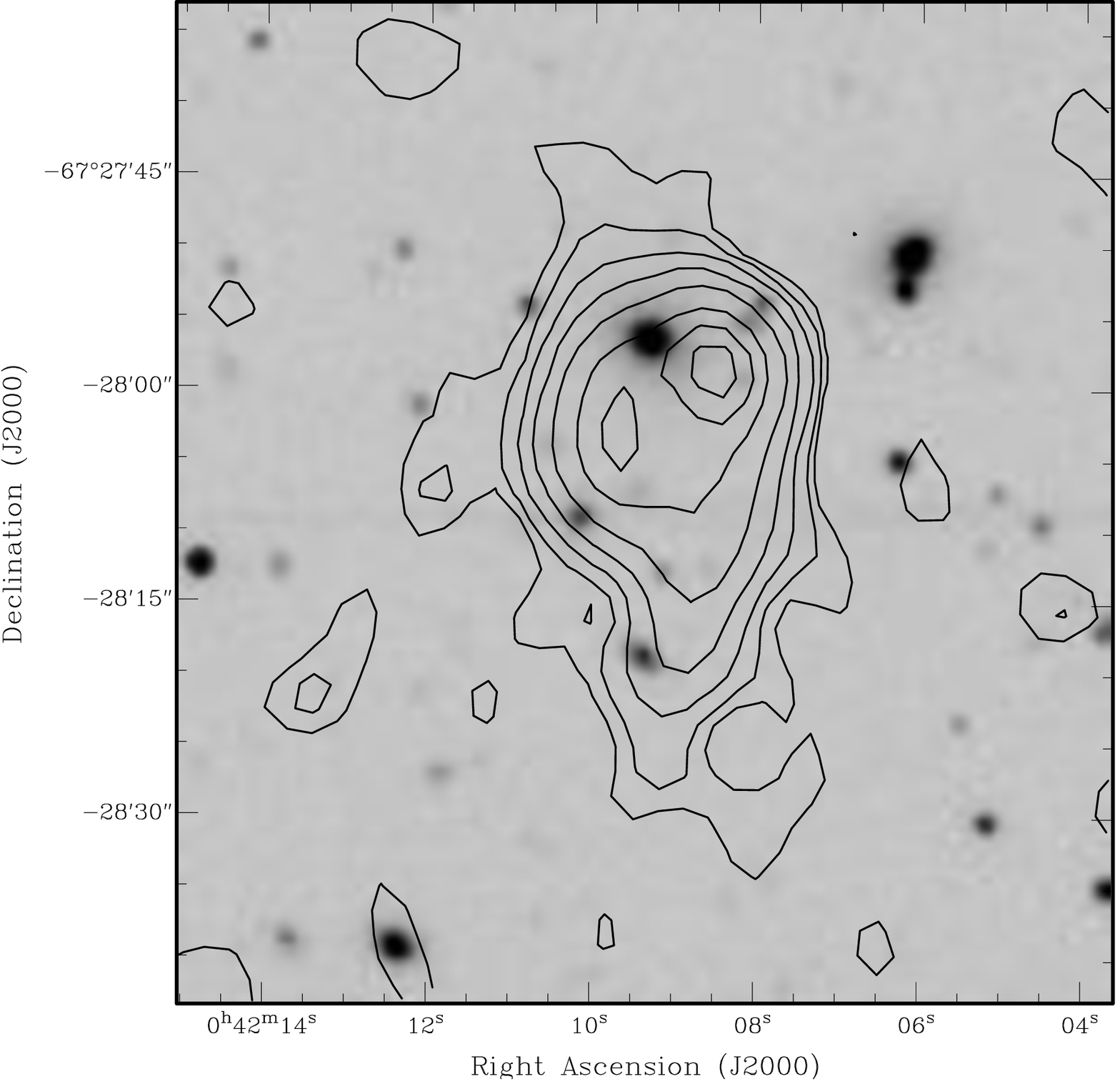}}
\caption{J0042.1-6728: $10^{-4}$ Jy x 1,  2,  4,  8,  16, 32,  48,  64.} 
\end{minipage}
&
\begin{minipage}{0.47\linewidth}
\frame{\includegraphics[angle=-90, width=2.8in] {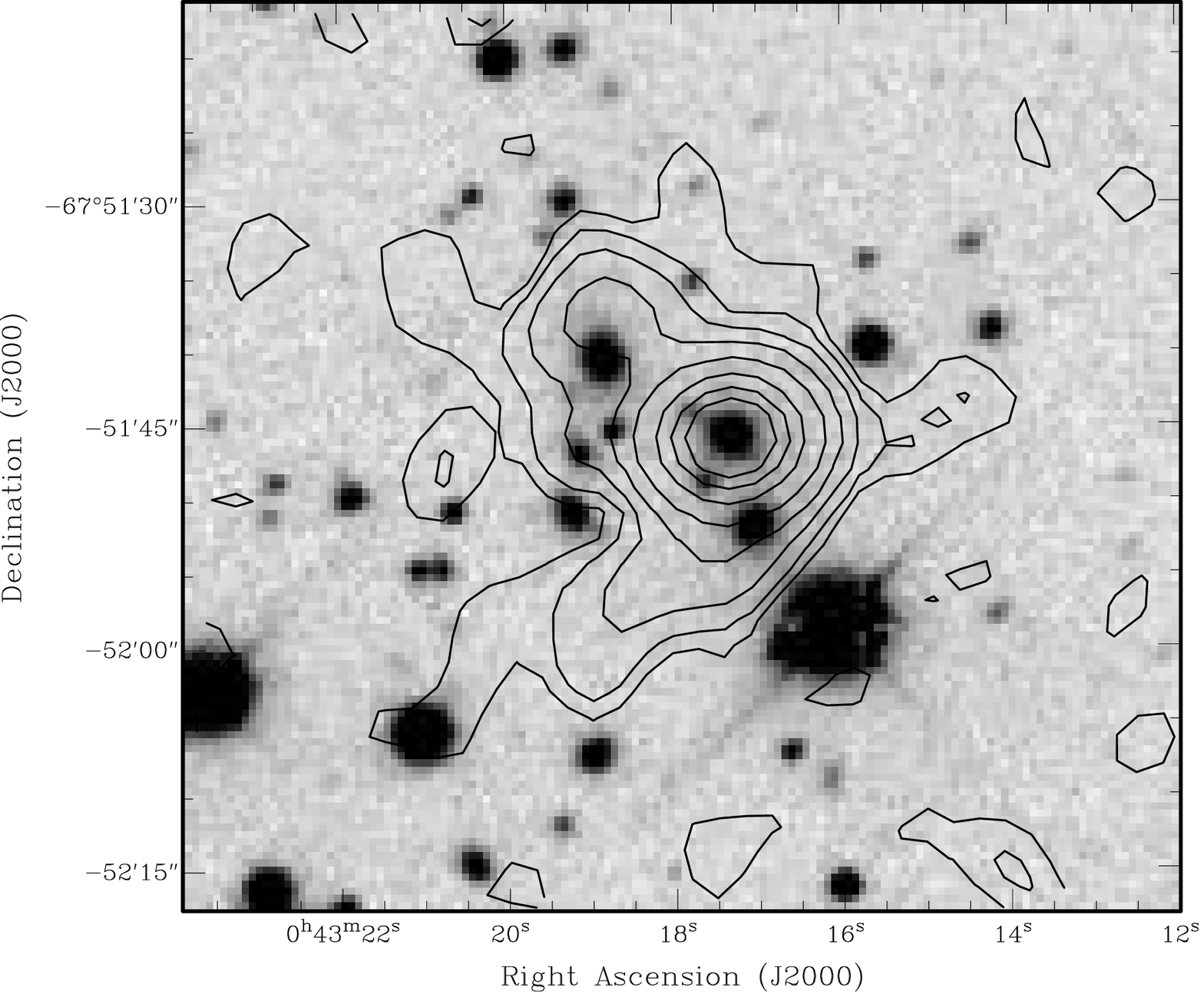}}
\caption{J0043.2-6751: $10^{-4}$ Jy x 1,  2,  4,  8,  16,  32,  48,  64.} 
\end{minipage}
\\
\end{tabular}
\end{figure*}

\clearpage

\begin{figure*}
\centering
\begin{tabular}{cc}
\begin{minipage}{0.47\linewidth}
\frame{\includegraphics[angle=-90, width=2.8in]{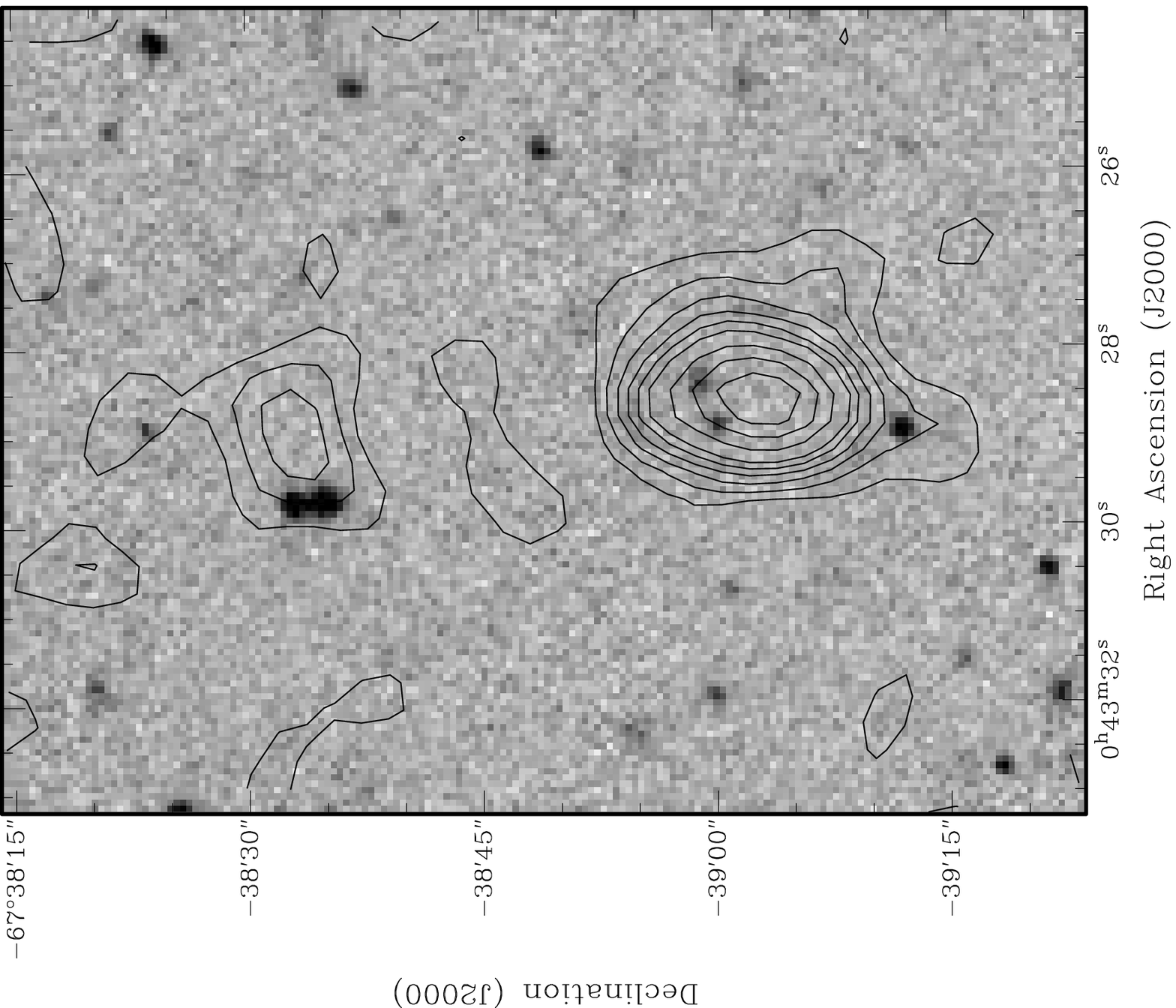}}
\caption{J0043.4-6738: $10^{-4}$ Jy x 1,  2,  3,  4,  5,  6,  8,  10,  12.}
\end{minipage}
&
\begin{minipage}{0.47\linewidth}
\frame{\includegraphics[angle=-90, width=2.8in]{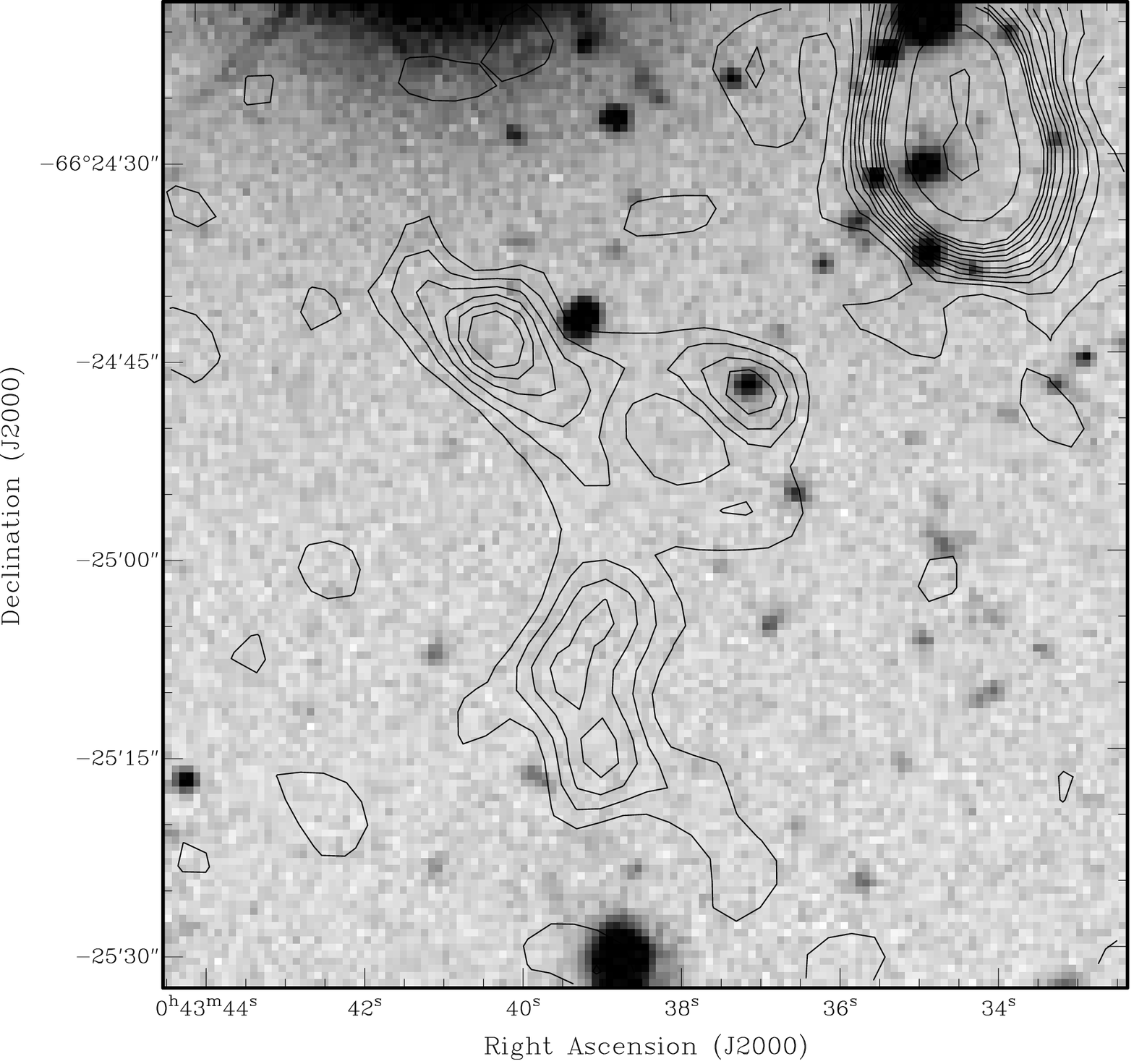}}
\caption{J0043.6-6624: $10^{-4}$ Jy x 1,  2,  3,  4,  5,  6.} 
\end{minipage}
\\
\end{tabular}
\end{figure*}

\begin{figure*}
\centering
\begin{tabular}{cc}
\begin{minipage}{0.47\linewidth}
\frame{\includegraphics[angle=-90, width=2.8in]{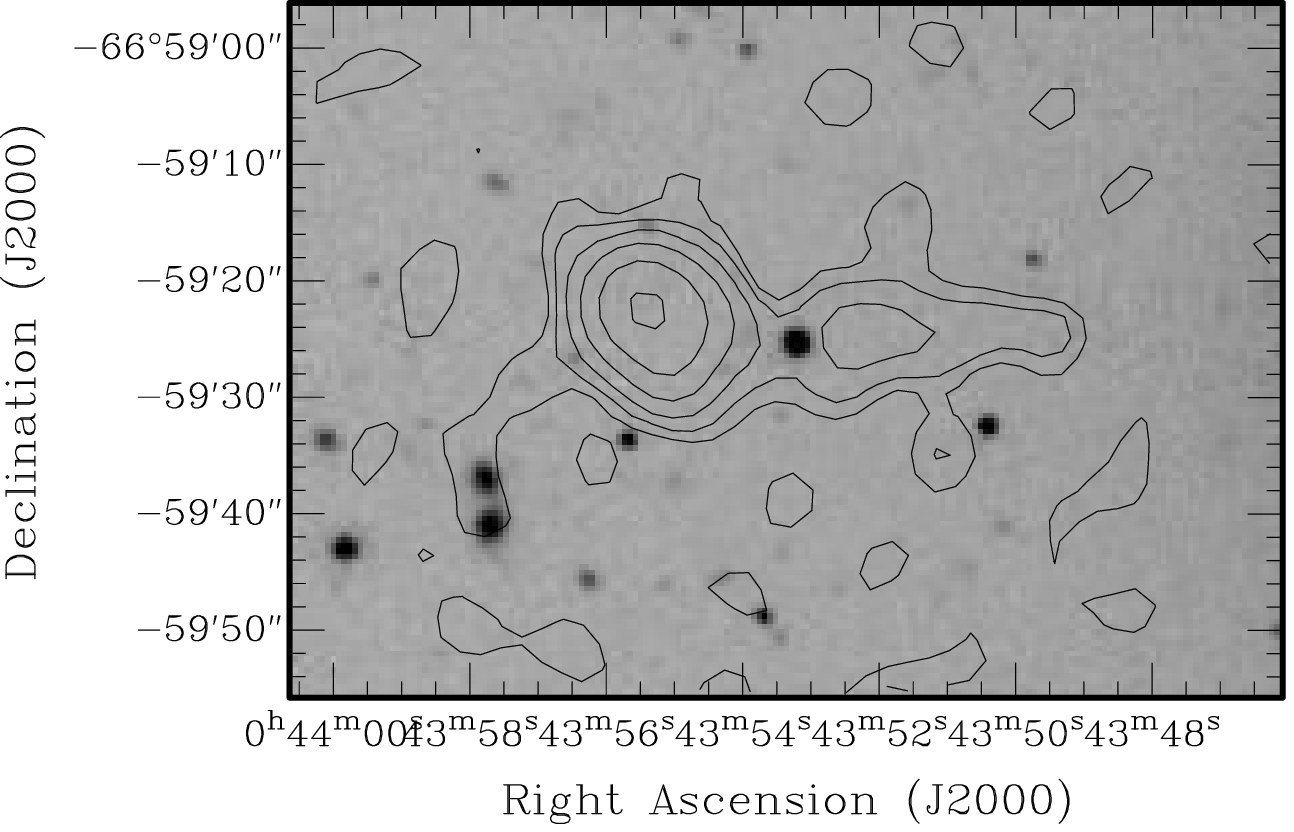}}
\caption{J0043.8-6659: $10^{-4}$ Jy x 1,  2,  4,  8,  16,  32.} 
\end{minipage}
&
\begin{minipage}{0.47\linewidth}
\frame{\includegraphics[angle=-90, width=2.8in] {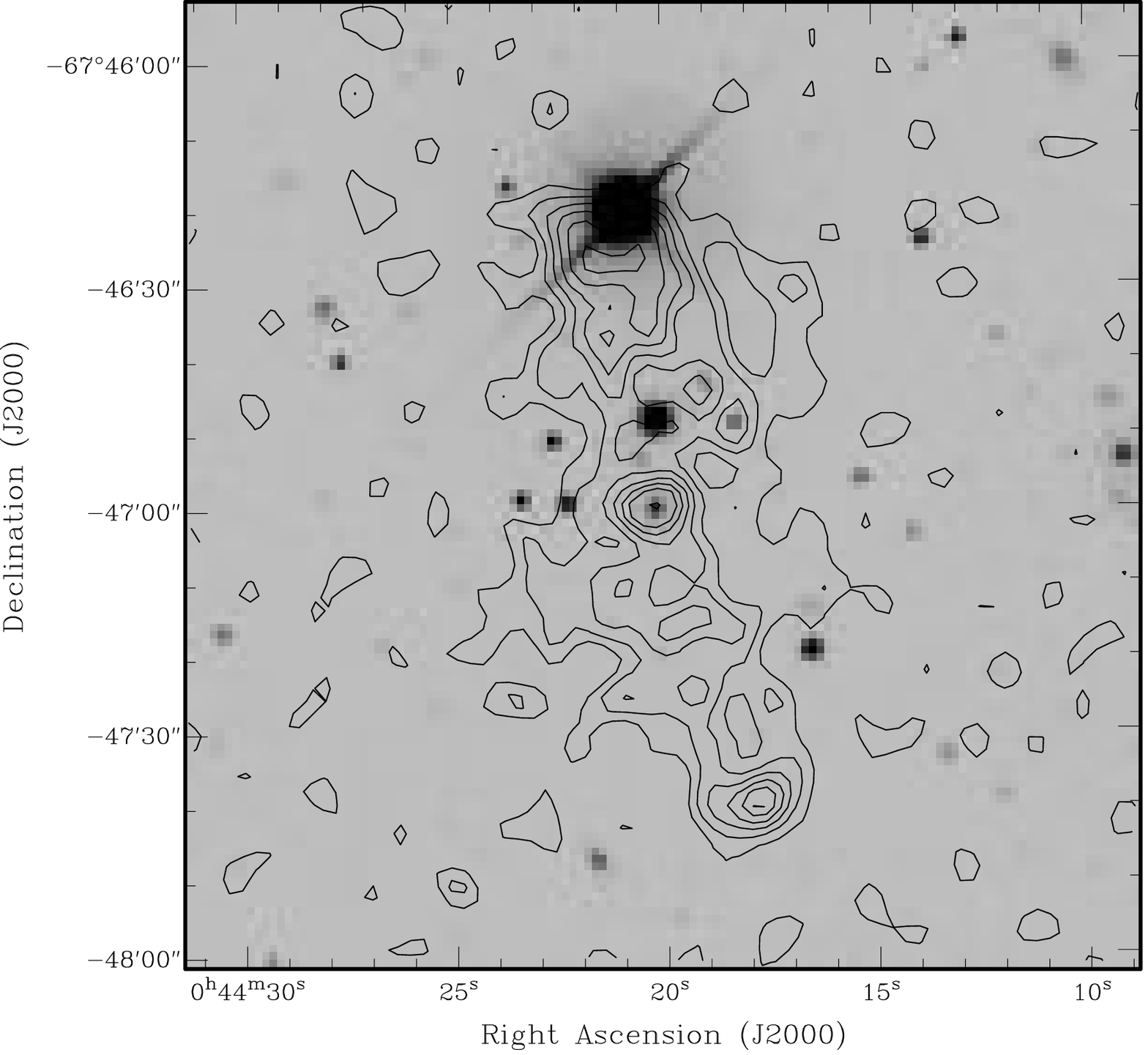}}
\caption{J0044.3-6746: $10^{-4}$ Jy x 1,  2,  3,  4,  5,  6,  8.} 
\end{minipage}
\\
\end{tabular}
\end{figure*}

\begin{figure*}
\centering
\begin{tabular}{cc}
\begin{minipage}{0.47\linewidth}
\frame{\includegraphics[angle=-90, width=2.8in]{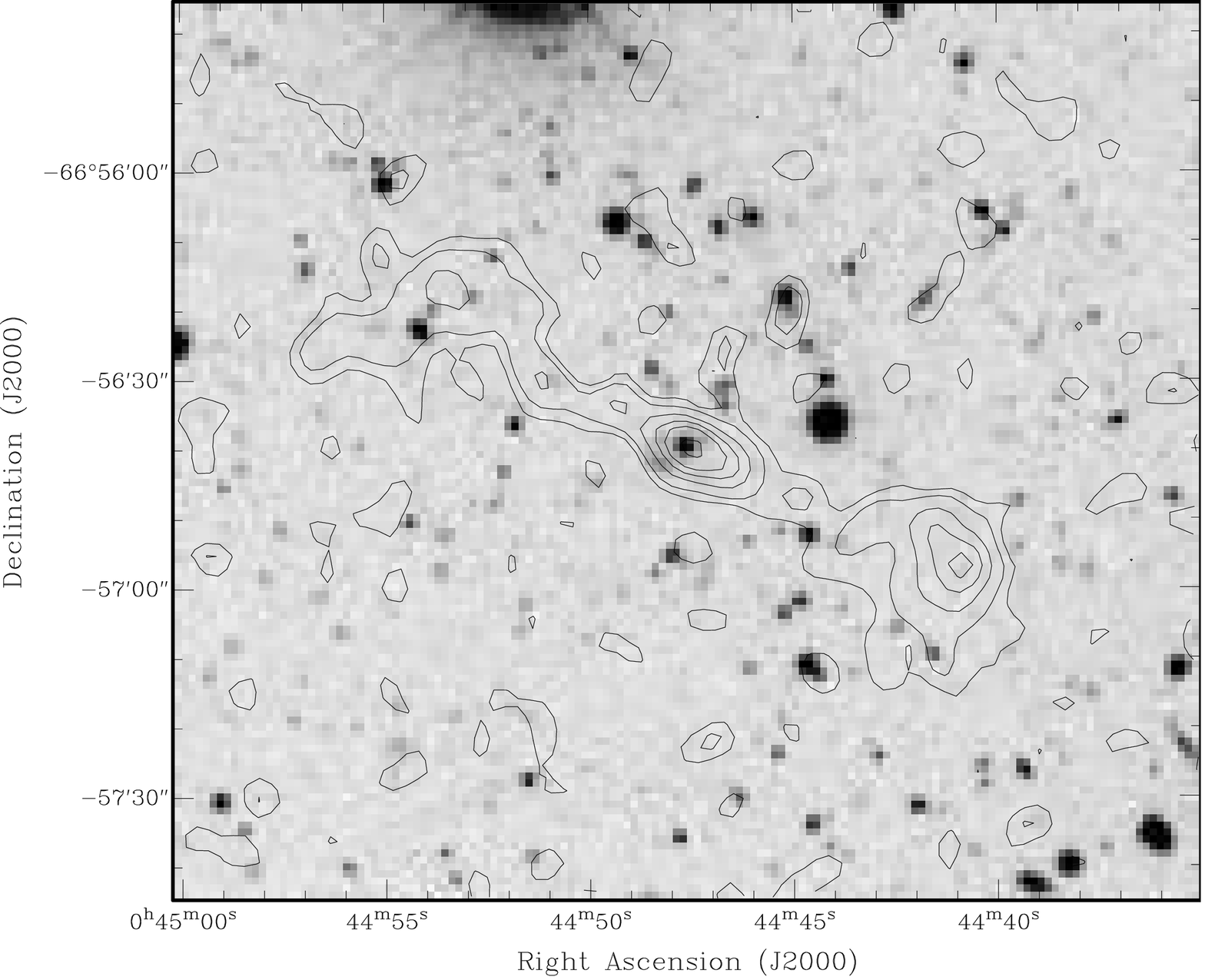}}
\caption{J0044.7-6656: $10^{-4}$ Jy x 1,  2,  4,  6,  8,  12.}
\end{minipage}
&
\begin{minipage}{0.47\linewidth}
\frame{\includegraphics[angle=-90, width=2.8in]{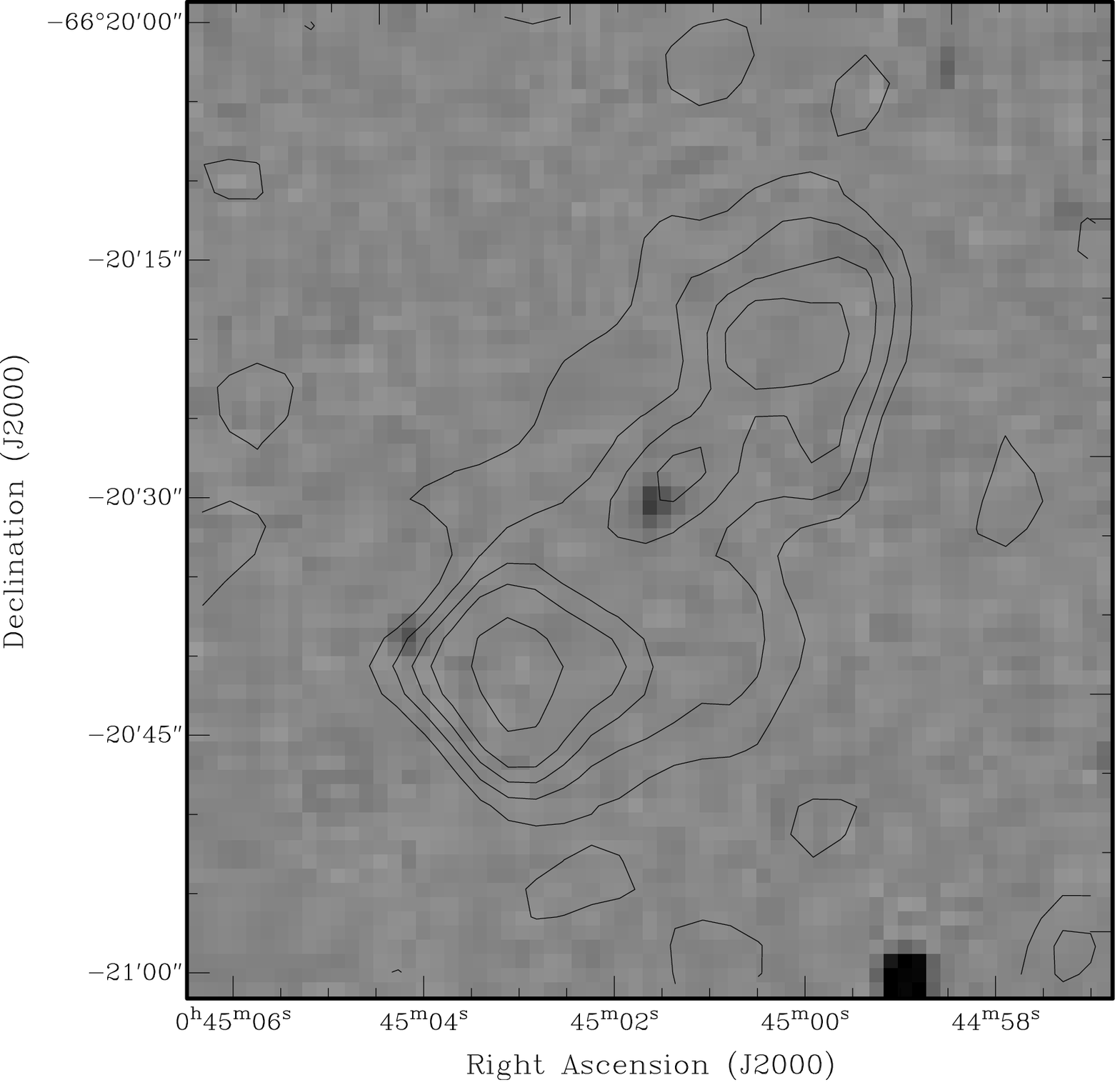}}
\caption{J0045.0-6620: $10^{-4}$ Jy x 1,  2,  3,  4,  6. K-band image is used.} 
\end{minipage}
\\
\end{tabular}
\end{figure*}

\begin{figure*}
\centering
\begin{tabular}{cc}
\begin{minipage}{0.47\linewidth}
\frame{\includegraphics[angle=-90, width=2.8in]{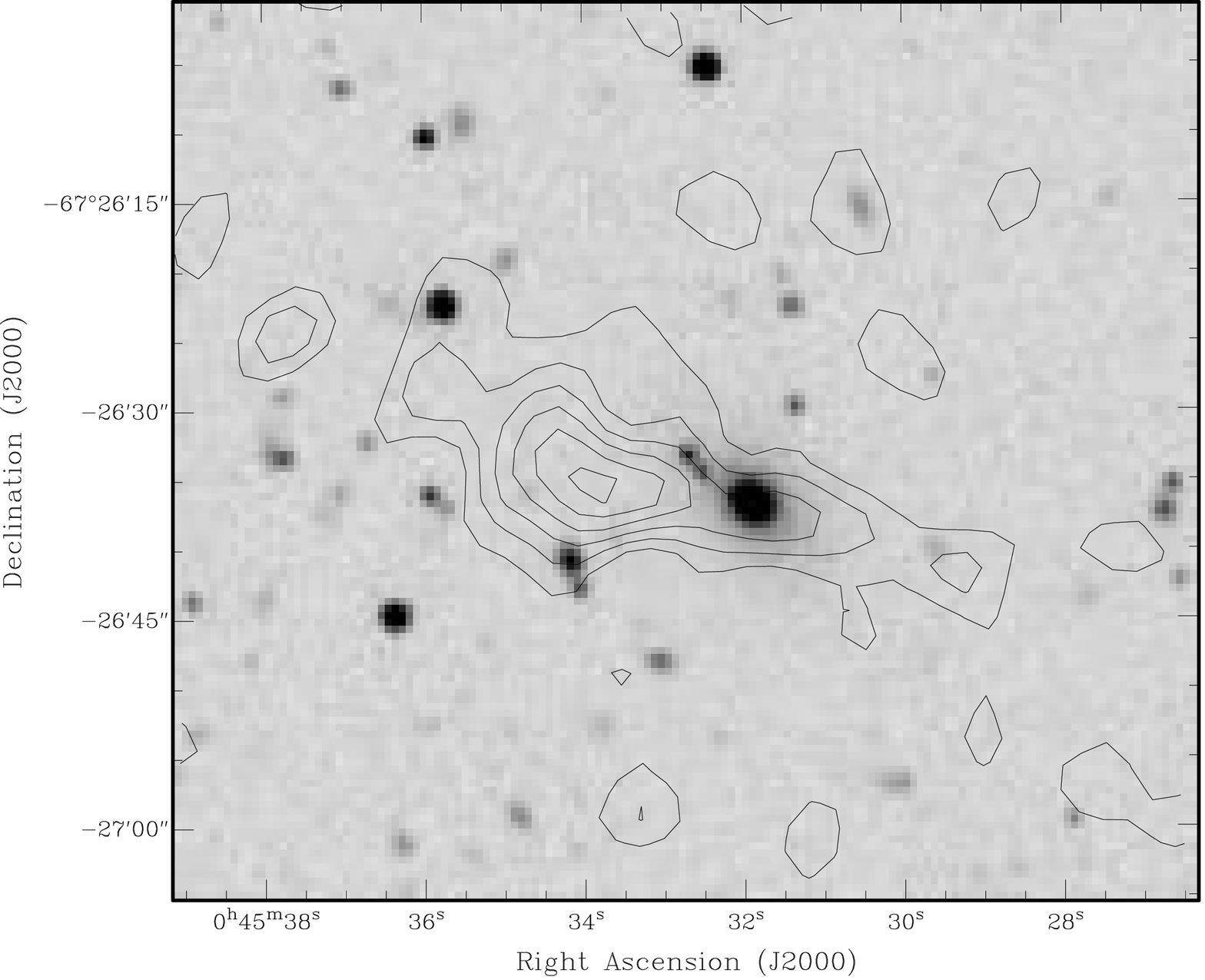}}
\caption{J0045.5-6726: $10^{-4}$ Jy x 1,  2,  3,  4,  5, 6.} 
\end{minipage}
&
\begin{minipage}{0.47\linewidth}
\frame{\includegraphics[angle=-90, width=2.8in] {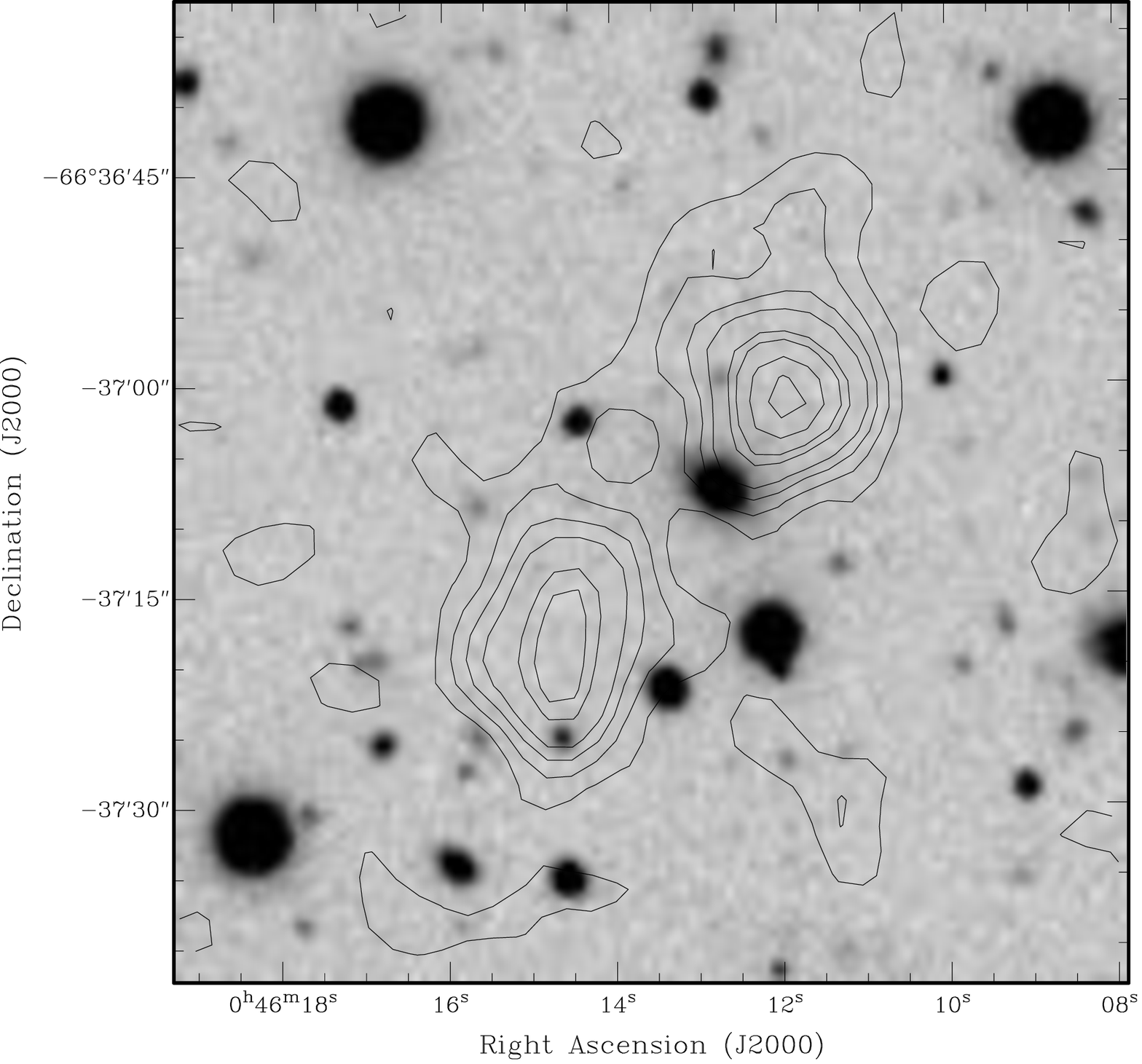}}
\caption{J0046.2-6637: $10^{-4}$ Jy x 1,  2,  3,  4,  6,  7,  9,  11.} 
\end{minipage}
\\
\end{tabular}
\end{figure*}

\begin{figure*}
\centering
\begin{tabular}{l}
\begin{minipage}{0.47\linewidth}
\frame{\includegraphics[angle=-90, width=2.8in]{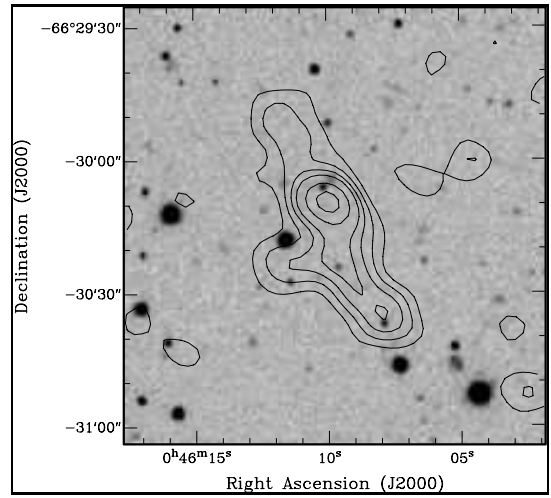}}
\caption{J0046.1-6630: Beam 10"; $10^{-4}$ Jy x 1,  2,  3,  4,  6,  8. } 
\end{minipage}
\\
\end{tabular}
\end{figure*}

\clearpage


\clearpage

\begin{figure*}
\centering
\begin{tabular}{cc}
\begin{minipage}{0.47\linewidth}
\frame{\includegraphics[angle=-90, width=2.8in]{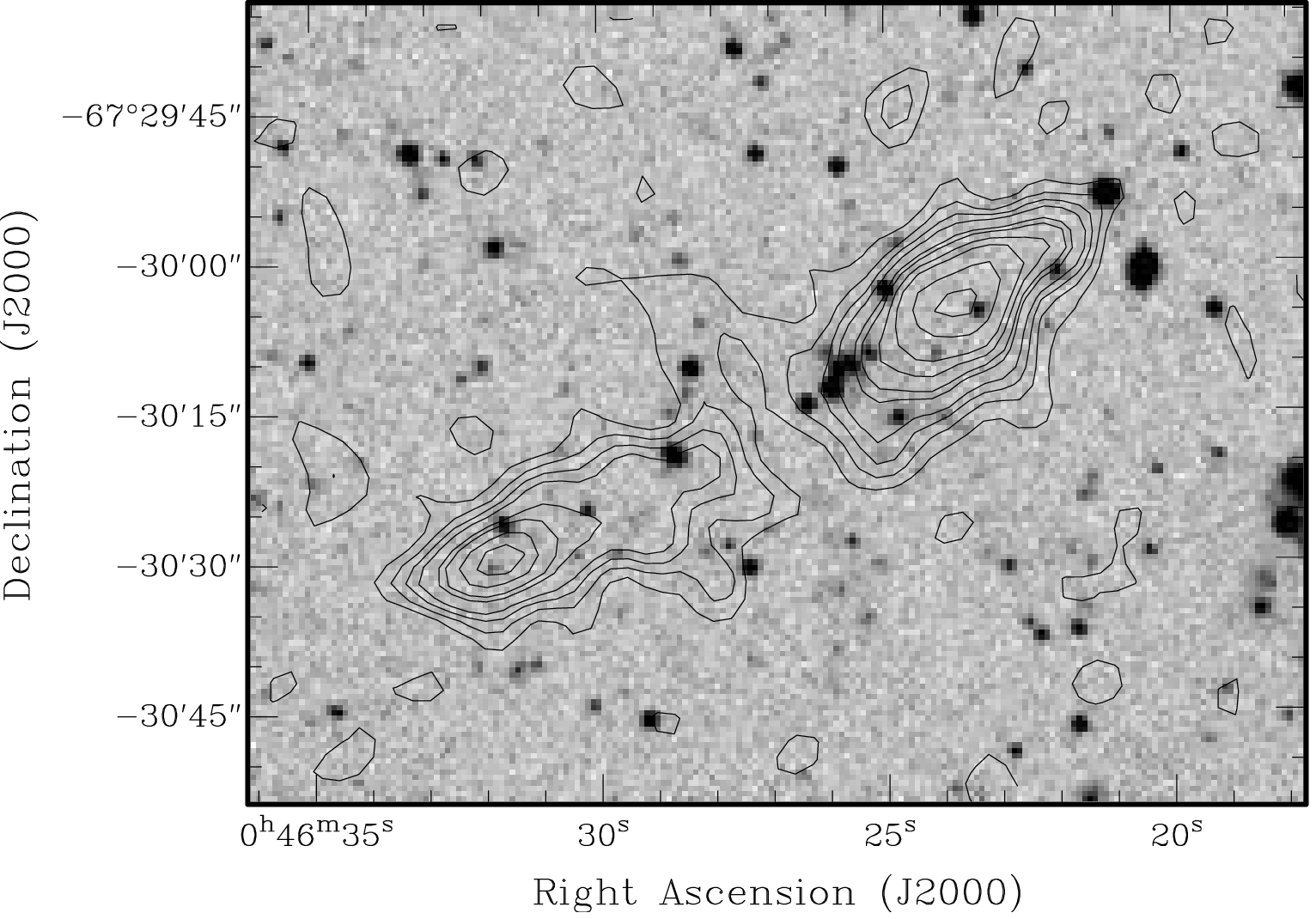}}
\caption{J0046.4-6730: $10^{-4}$ Jy x 1,  2,  3,  4,  6,  7,  9,  11,  16,  21.}
\end{minipage}
&
\begin{minipage}{0.47\linewidth}
\frame{\includegraphics[width=2.8in]{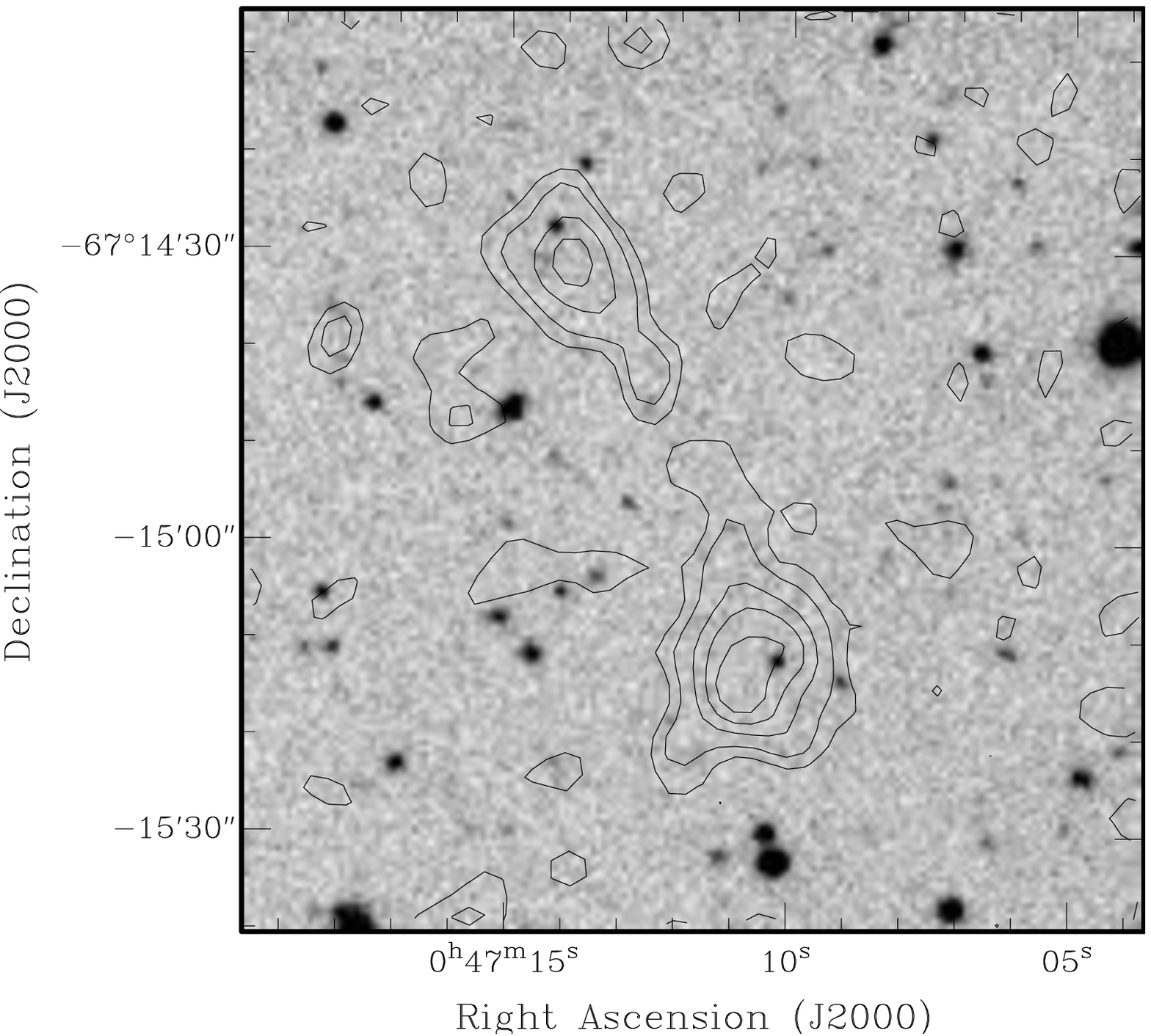}}
\caption{J0047.1-6715: $10^{-4}$ Jy x 1,  2,  4,  6,  8.} 
\end{minipage}
\\
\end{tabular}
\end{figure*}

\begin{figure*}
\centering
\begin{tabular}{cc}
\begin{minipage}{0.47\linewidth}
\frame{\includegraphics[angle=-90, width=2.8in]{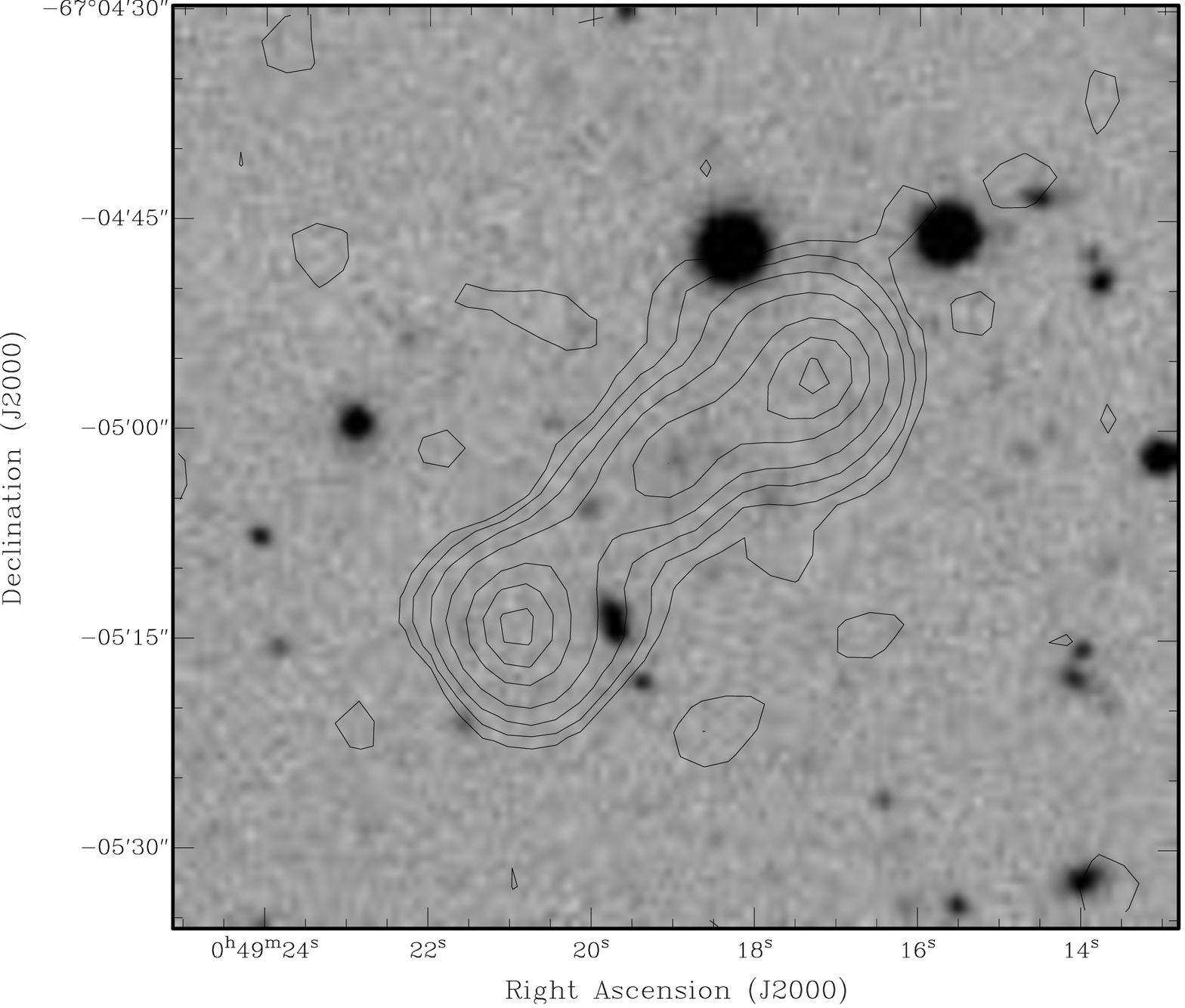}}
\caption{J0049.3-6705: $10^{-4}$ Jy x 1,  2,  4,  8,  16,  24,  32.} 
\end{minipage}
&
\begin{minipage}{0.47\linewidth}
\frame{\includegraphics[angle=-90, width=2.8in]{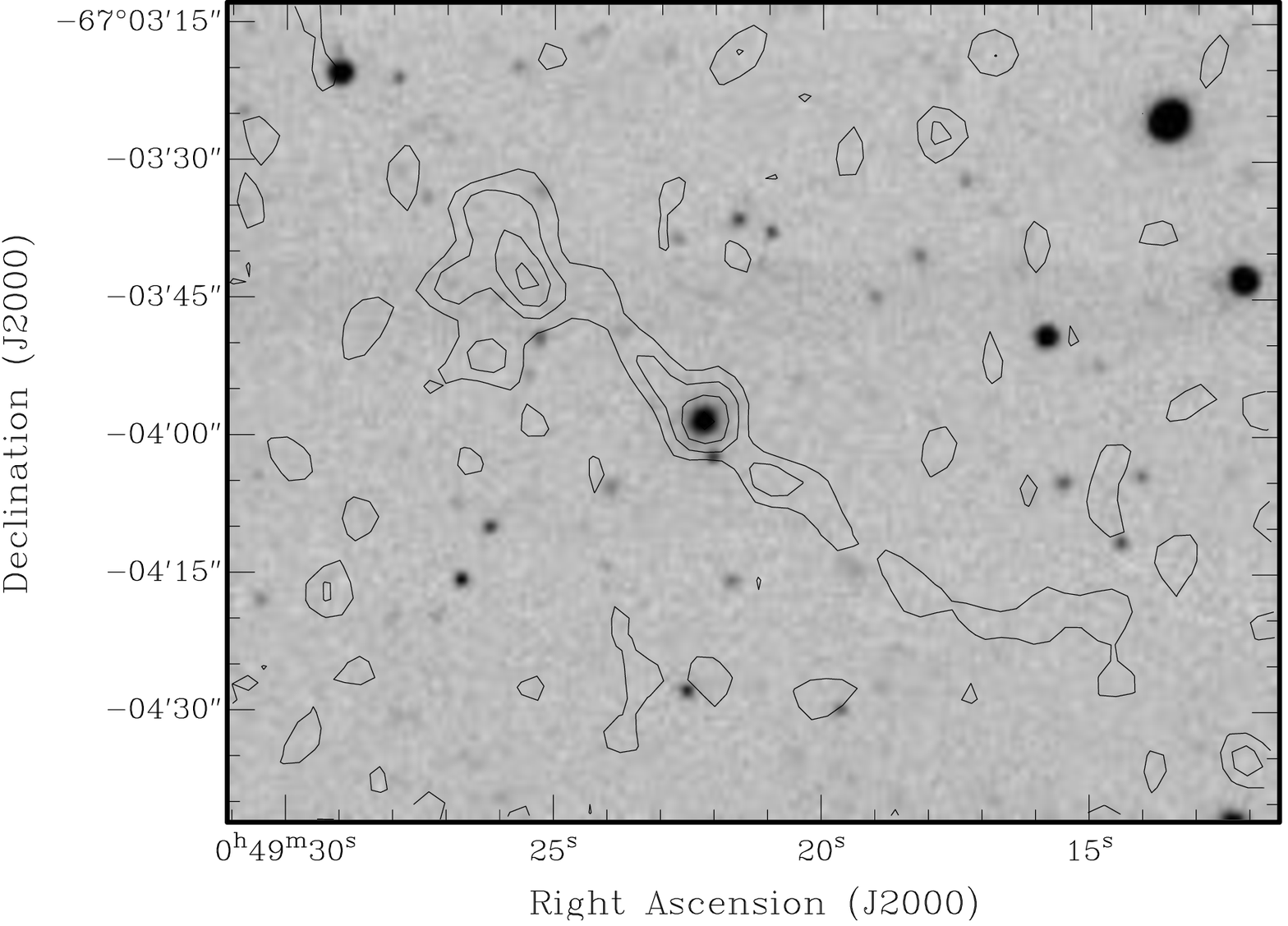}}
\caption{J0049.3-6703: $10^{-4}$ Jy x 1,  2,  3, 4.} 
\end{minipage}
\\
\end{tabular}
\end{figure*}

\clearpage

\begin{figure*}
\centering
\begin{tabular}{cc}
\begin{minipage}{0.47\linewidth}
\frame{\includegraphics[angle=-90, width=2.8in]{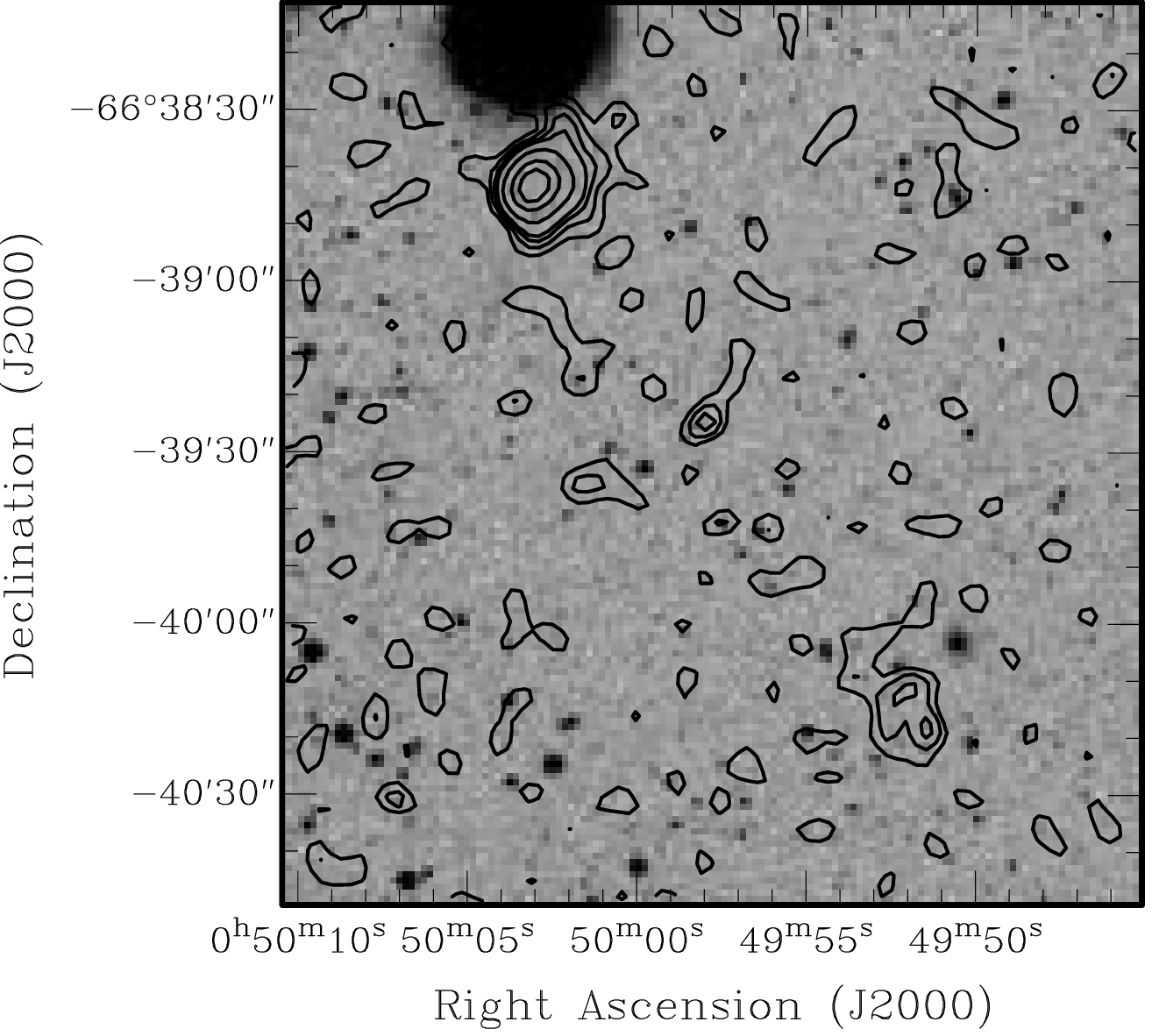}}
\caption{J0049.9-6639: $10^{-4}$ Jy x 1,  2,  3,  4,  8,  16,  24.}
\end{minipage}
&
\begin{minipage}{0.47\linewidth}
\frame{\includegraphics[angle=-90, width=2.8in]{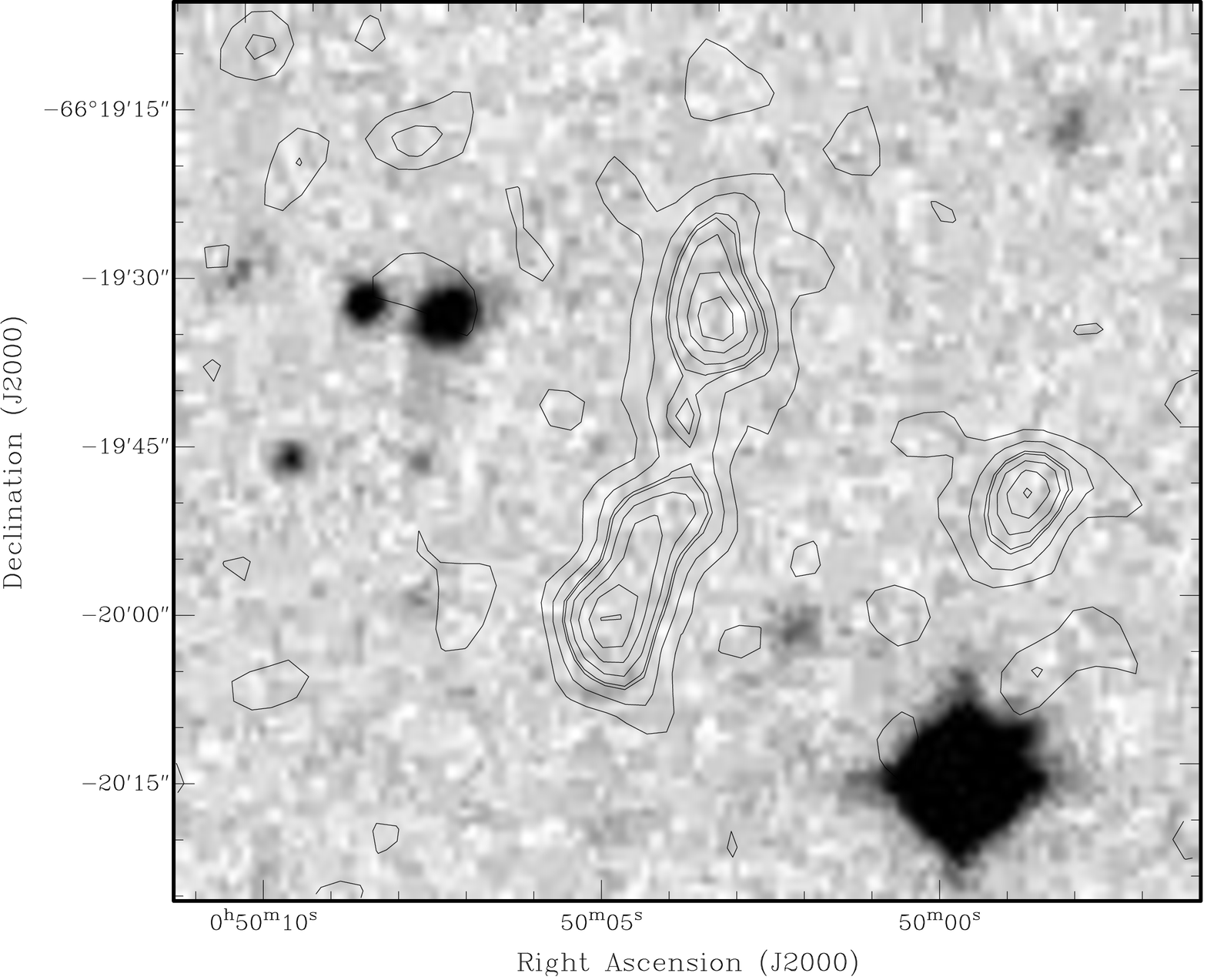}}
\caption{J0050.0-6619: $10^{-4}$ Jy x 1,  2,  3, 3.25, 4,  5,  6, 7. B-band image is used.} 
\end{minipage}
\\
\end{tabular}
\end{figure*}

\begin{figure*}
\centering
\begin{tabular}{cc}
\begin{minipage}{0.47\linewidth}
\frame{\includegraphics[angle=-90, width=2.8in]{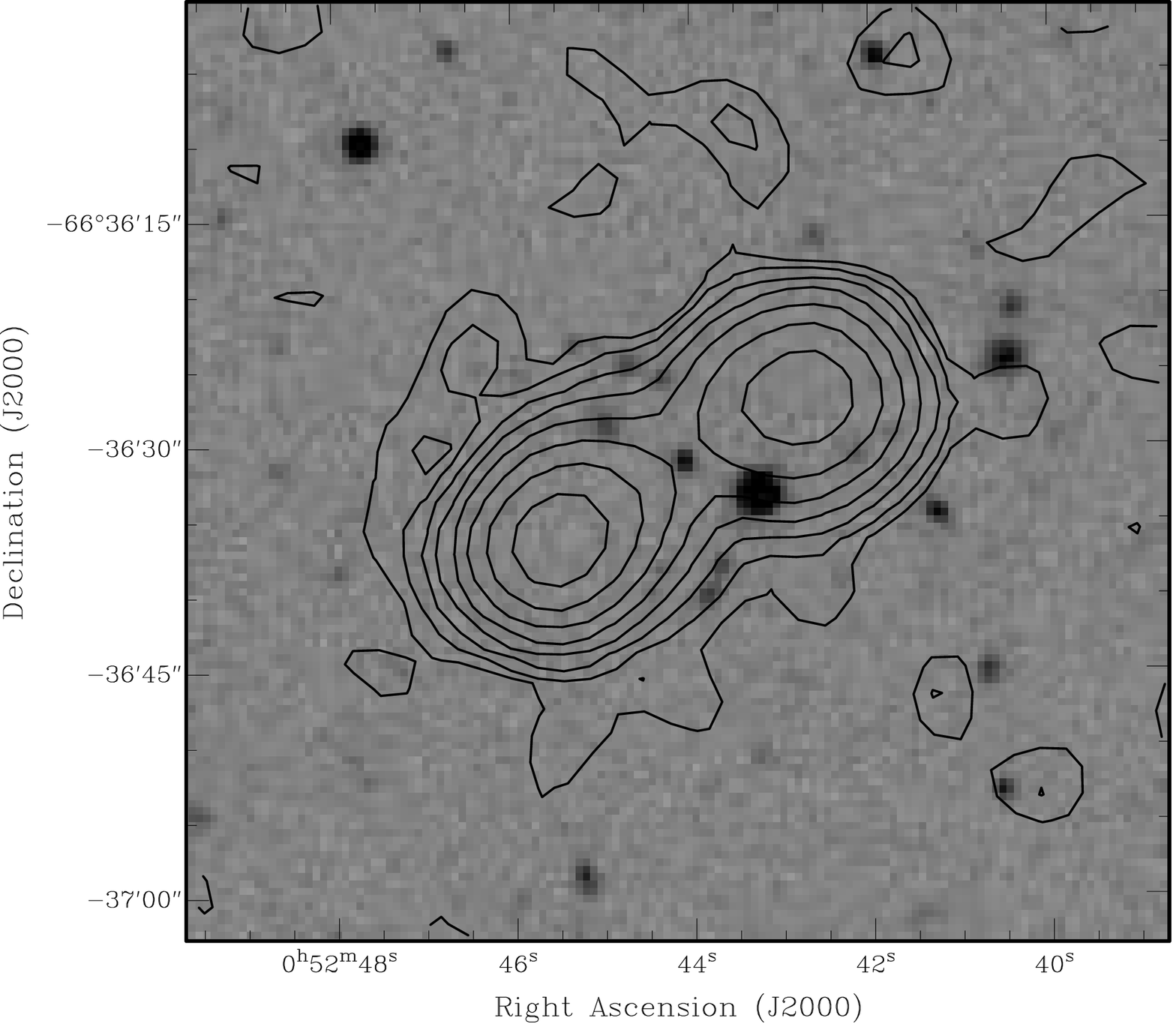}}
\caption{J0052.7-6636: $10^{-4}$ Jy x 1,  2,  4,  8,  16,  32,  64, 128.} 
\end{minipage}
&
\begin{minipage}{0.47\linewidth}
\frame{\includegraphics[angle=-90, width=2.8in]{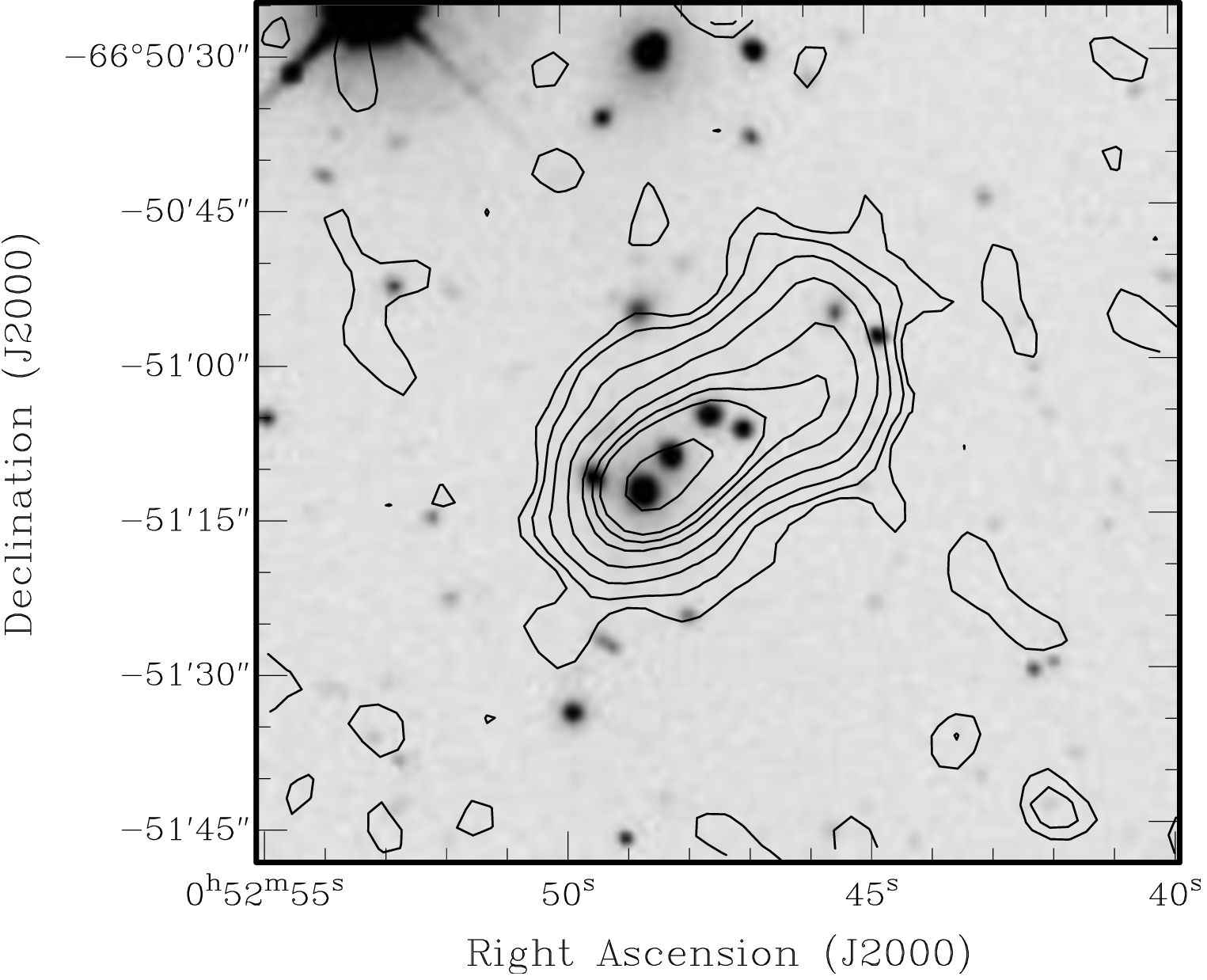}}
\caption{J0052.7-6651: $10^{-4}$ Jy x 1,  2,  4,  8,  16,  24,  32,  64.} 
\end{minipage}
\\
\end{tabular}
\end{figure*}

\begin{figure*}
\centering
\begin{tabular}{cc}
\begin{minipage}{0.47\linewidth}
\frame{\includegraphics[angle=-90, width=2.8in]{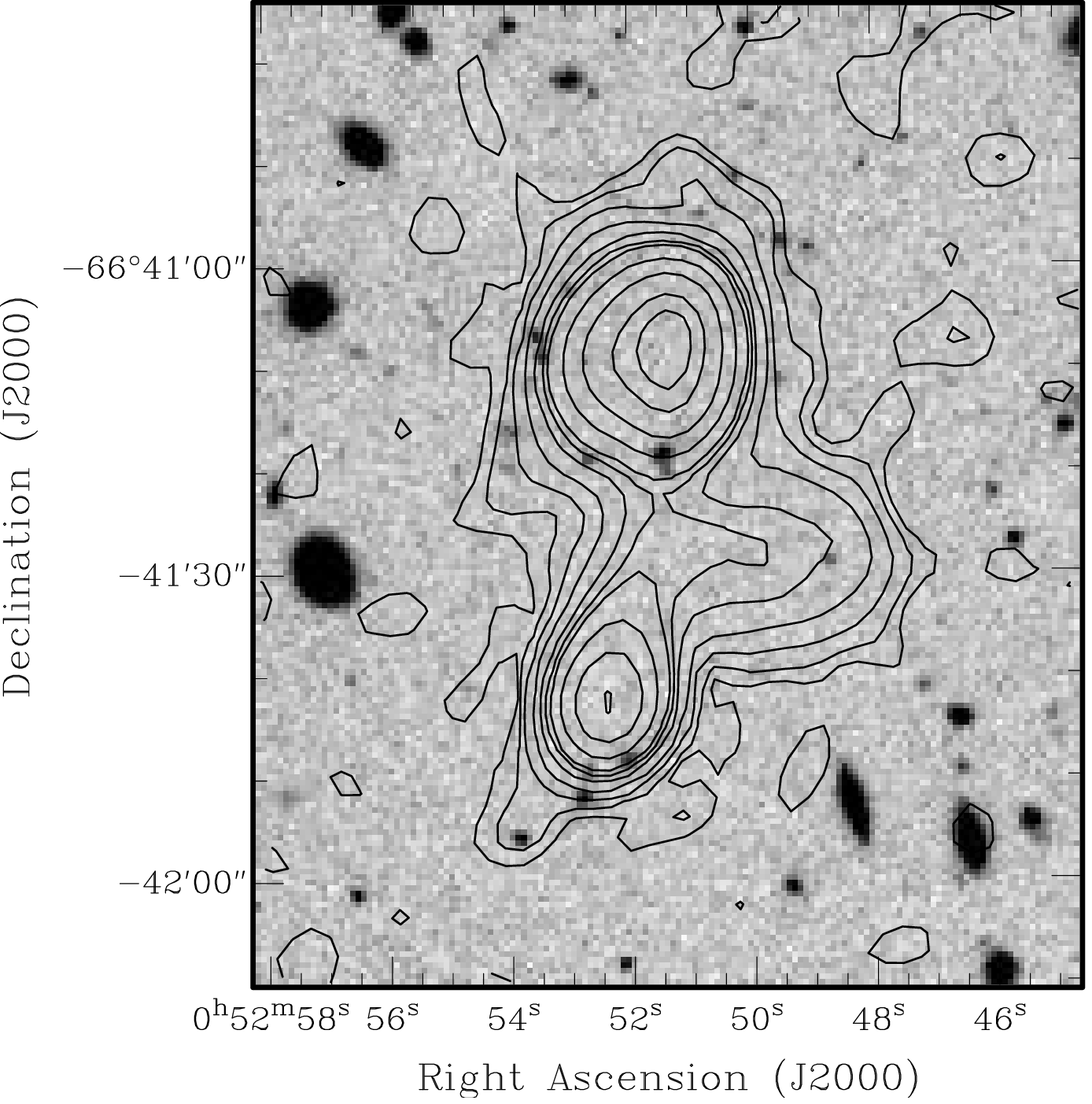}}
\caption{J0052.8-6641: $10^{-4}$ Jy x 1,  2,  4,  8,  16,  24,  32,  64, 128, 256, 360.}
\end{minipage}
&
\begin{minipage}{0.47\linewidth}
\frame{\includegraphics[angle=-90, width=2.8in]{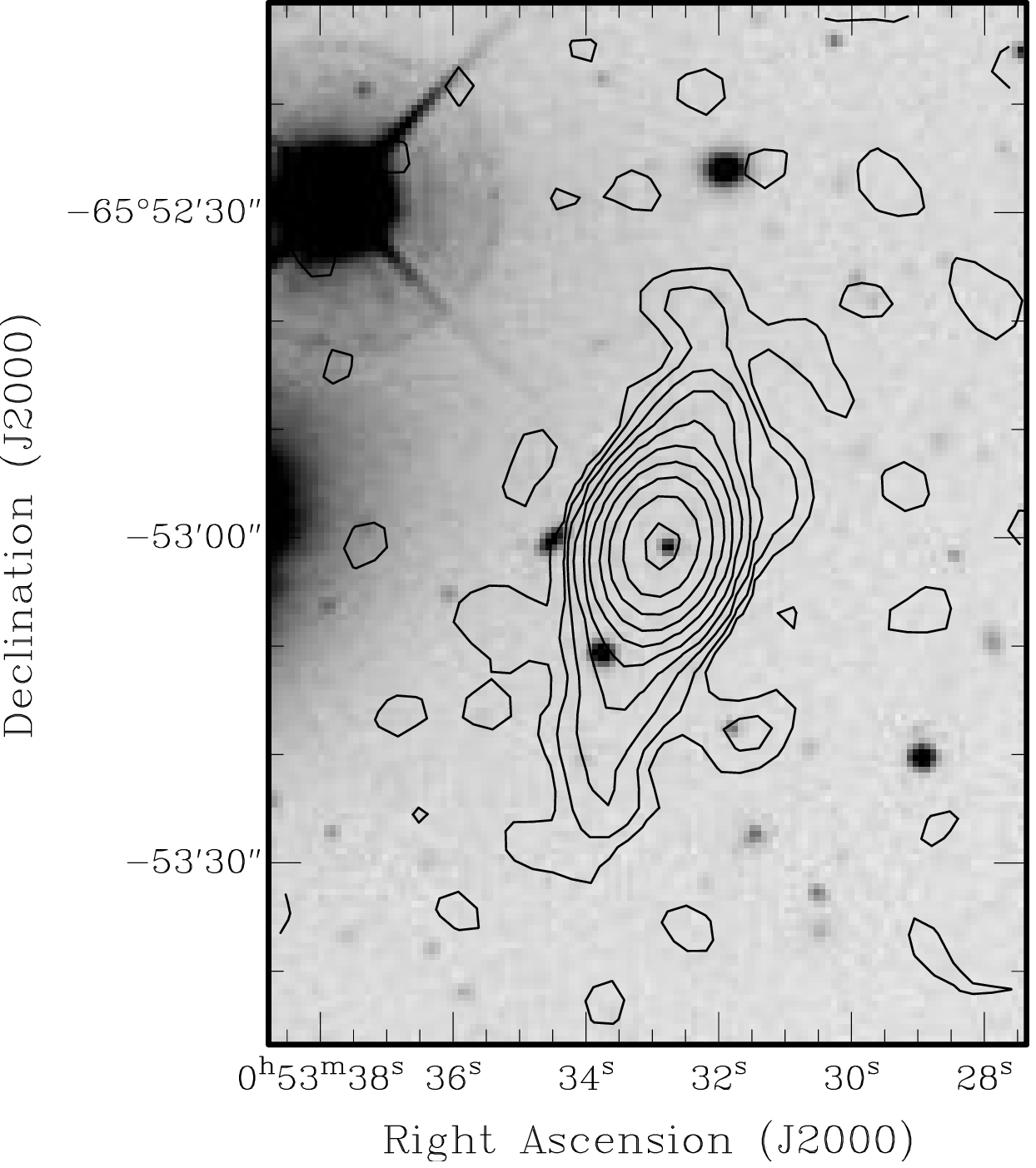}}
\caption{J0053.5-6553: $10^{-4}$ Jy x 1,  2,  4,  8,  16,  32,  64, 128, 256, 512.} 
\end{minipage}
\\
\end{tabular}
\end{figure*}

\begin{figure*}
\centering
\begin{tabular}{cc}
\begin{minipage}{0.47\linewidth}
\frame{\includegraphics[angle=-90, width=2.8in]{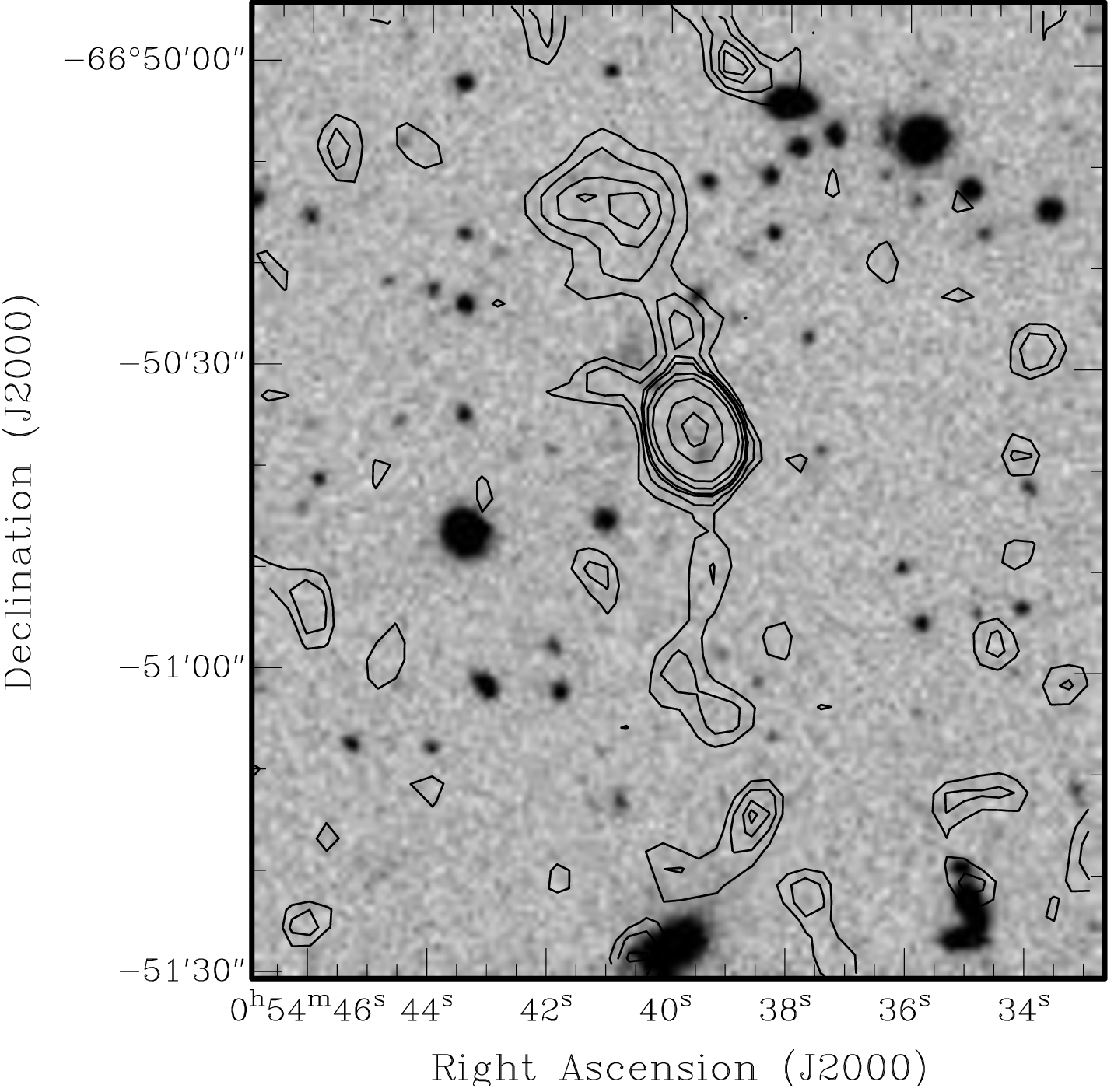}}
\caption{J0054.6-6650: $10^{-4}$ Jy x 1,  1.5,  2,  2.3,  3,  4,  8,  12.} 
\end{minipage}
&
\begin{minipage}{0.47\linewidth}
\frame{\includegraphics[angle=-90, width=2.8in] {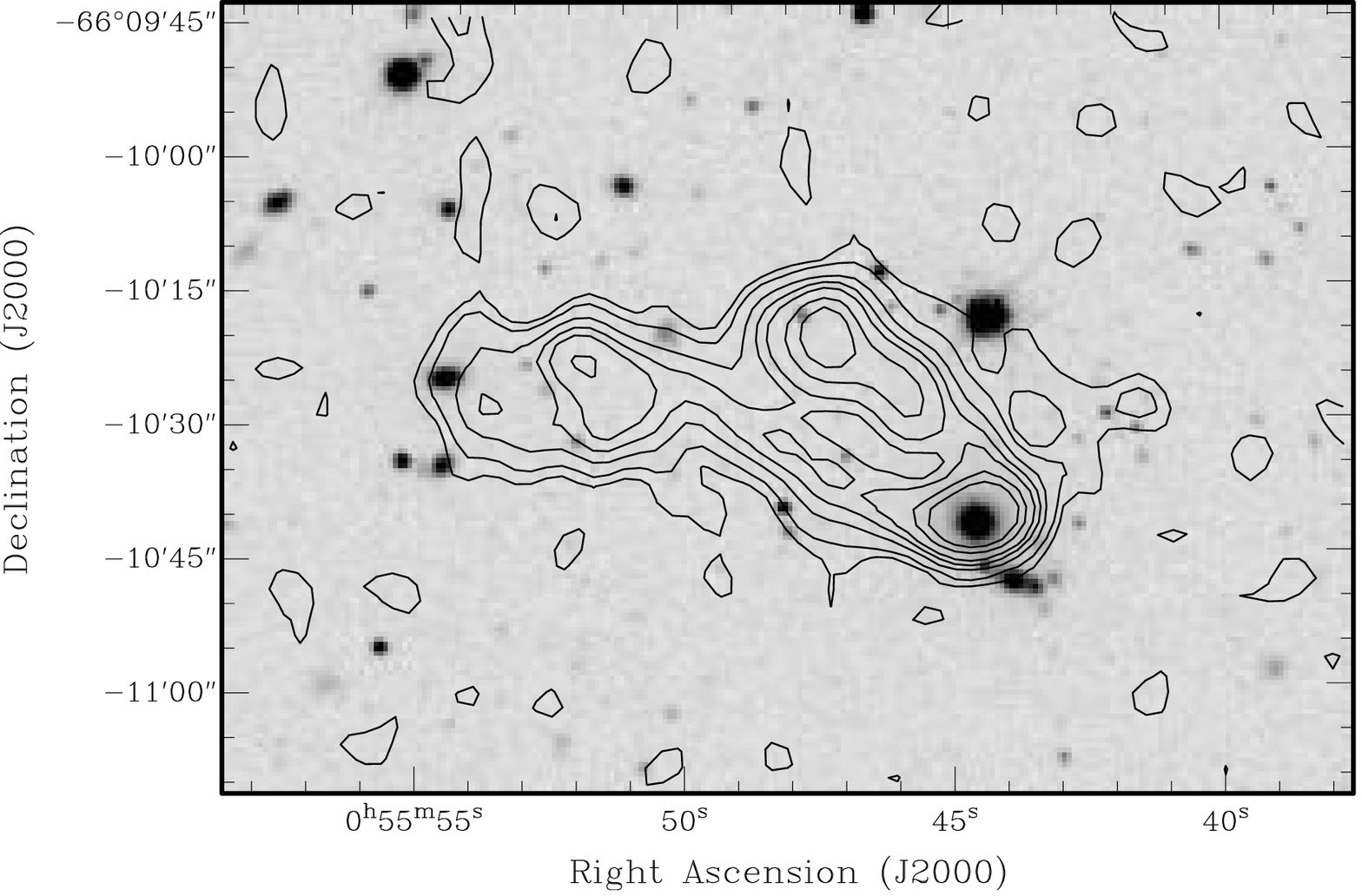}}
\caption{J0055.7-6610: $10^{-4}$ Jy x 1,  2,  4,  6,  8,  12,  16,  32,  64.} 
\end{minipage}
\\
\end{tabular}
\end{figure*}

\begin{figure*}
\centering
\begin{tabular}{cc}
\begin{minipage}{0.47\linewidth}
\frame{\includegraphics[angle=-90, width=2.8in]{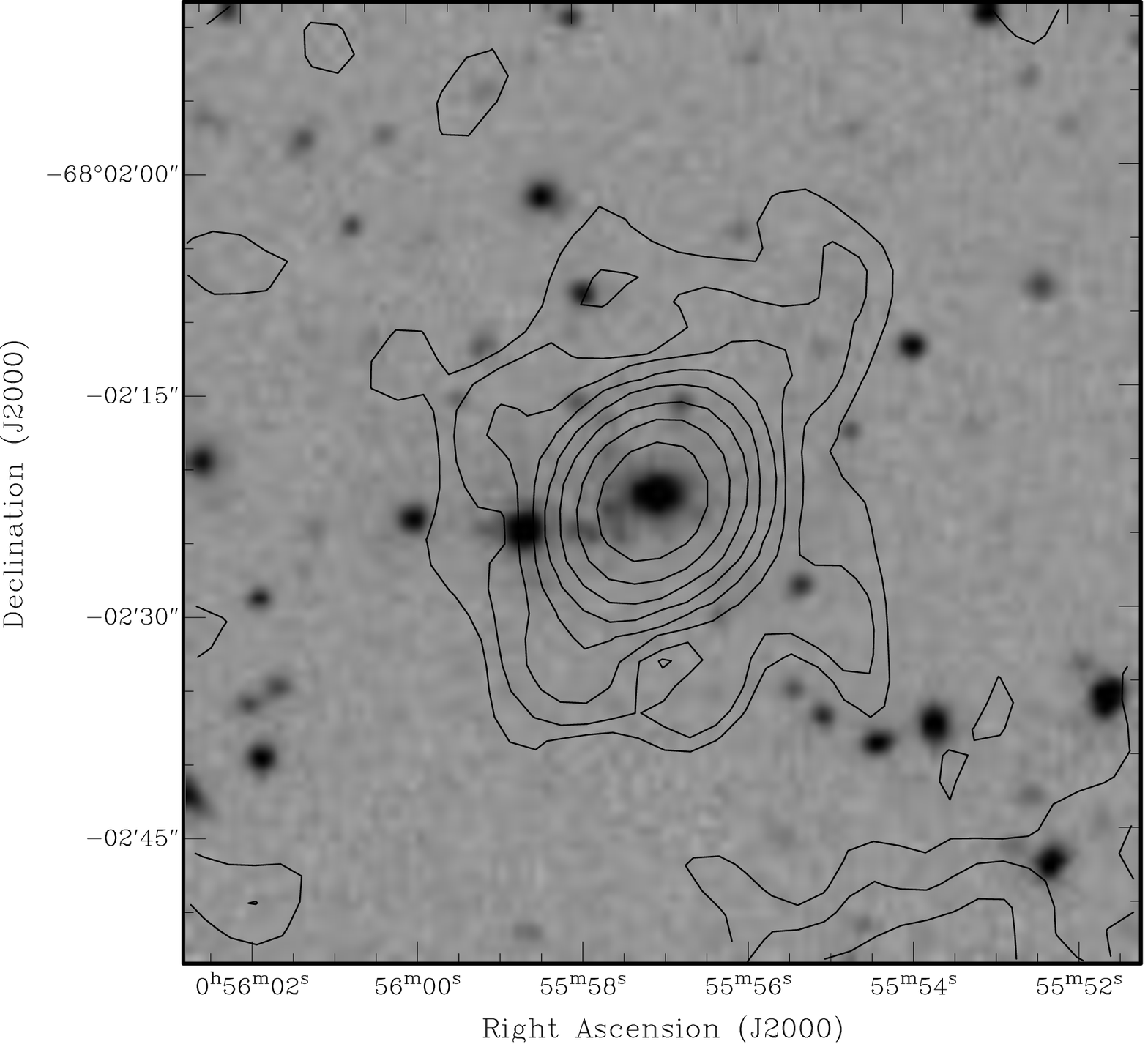}}
\caption{J0055.9-6802: $10^{-4}$ Jy x 1,  2,  4,  8,  16,  32,  64, 128.}
\end{minipage}
&
\begin{minipage}{0.47\linewidth}
\frame{\includegraphics[angle=-90, width=2.8in]{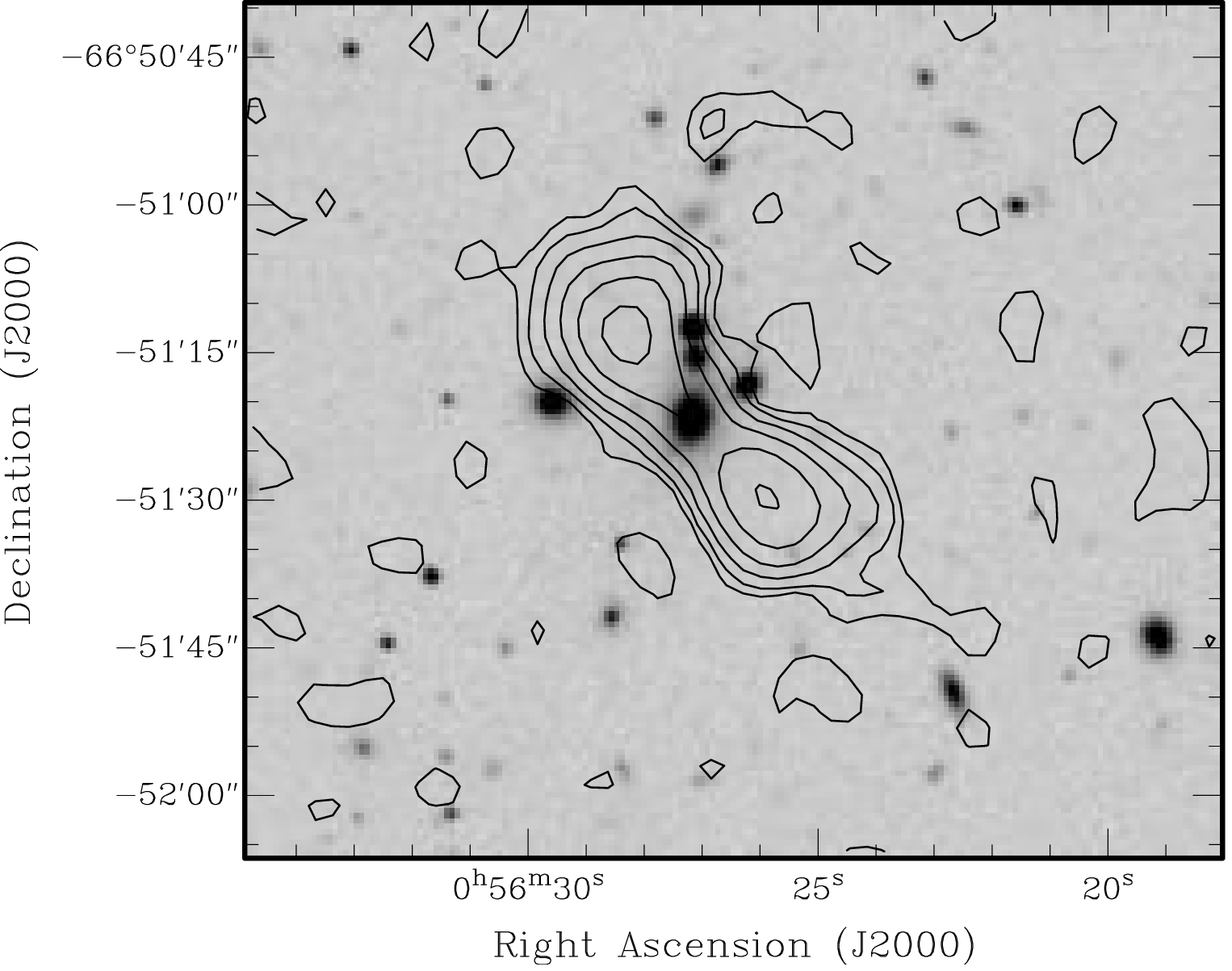}}
\caption{J0056.4-6651: $10^{-4}$ Jy x 1,  2,  4,  8,  16,  32.} 
\end{minipage}
\\
\end{tabular}
\end{figure*}

\begin{figure*}
\centering
\begin{tabular}{cc}
\begin{minipage}{0.47\linewidth}
\frame{\includegraphics[angle=-90, width=2.8in]{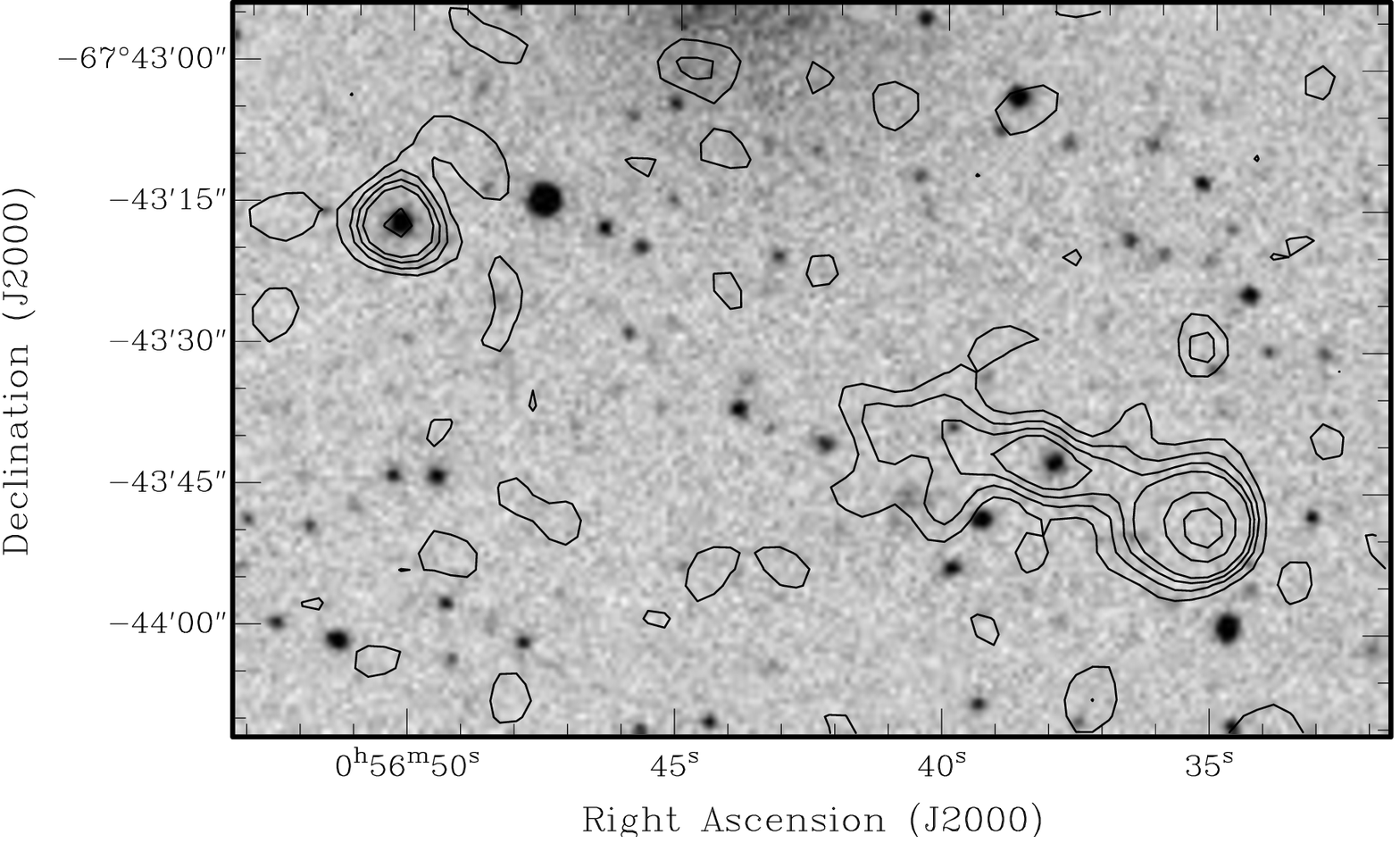}}
\caption{J0056.6-6743: $10^{-4}$ Jy x 1,  2,  3,  4,  8,  12.} 
\end{minipage}
&
\begin{minipage}{0.47\linewidth}
\frame{\includegraphics[angle=-90, width=2.8in] {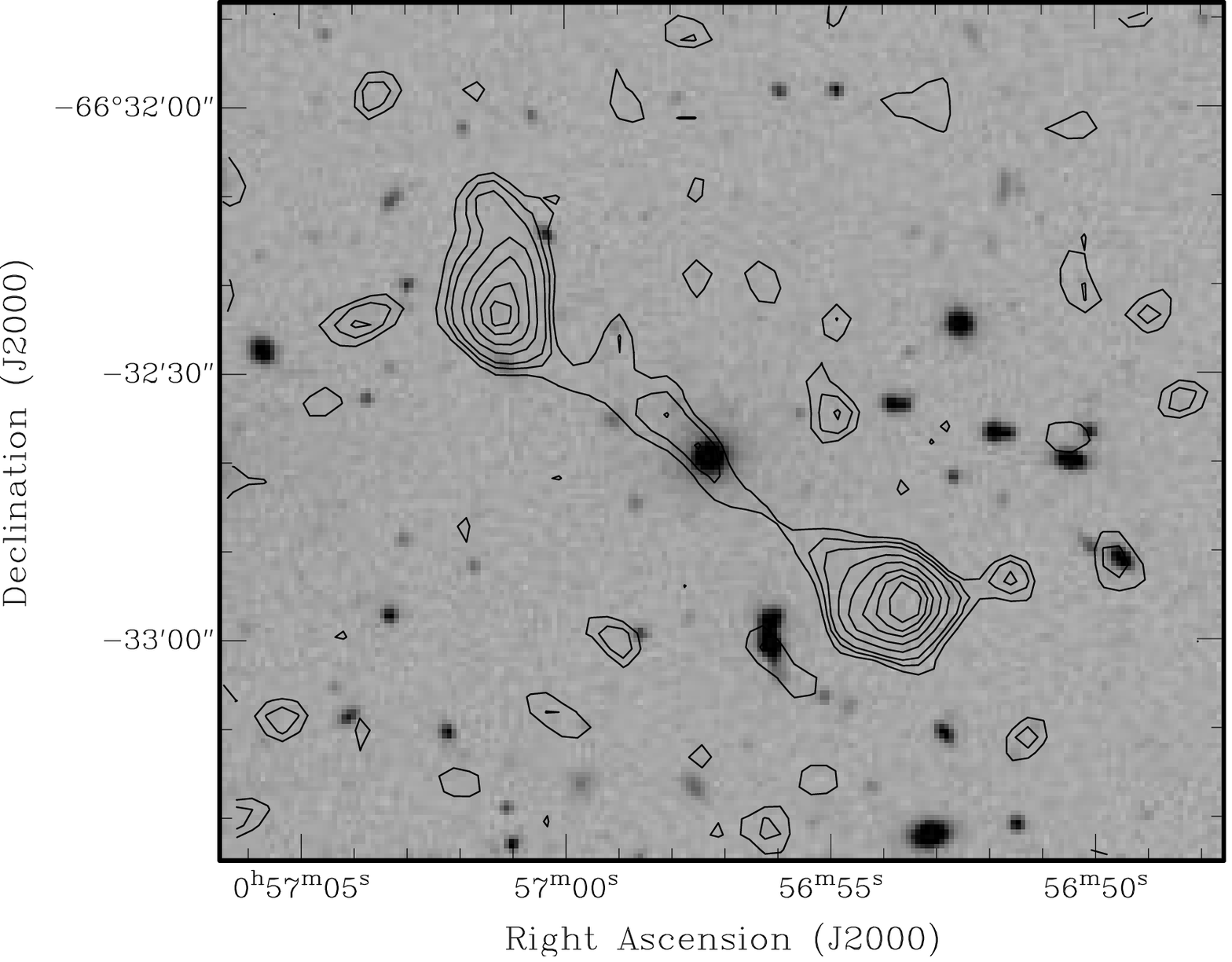}}
\caption{J0056.9-6632: $10^{-4}$ Jy x 1, 1.5, 2, 3, 4, 5, 5.7, 6.5.} 
\end{minipage}
\\
\end{tabular}
\end{figure*}

\begin{figure*}
\centering
\begin{tabular}{cc}
\begin{minipage}{0.47\linewidth}
\frame{\includegraphics[angle=-90, width=2.8in]{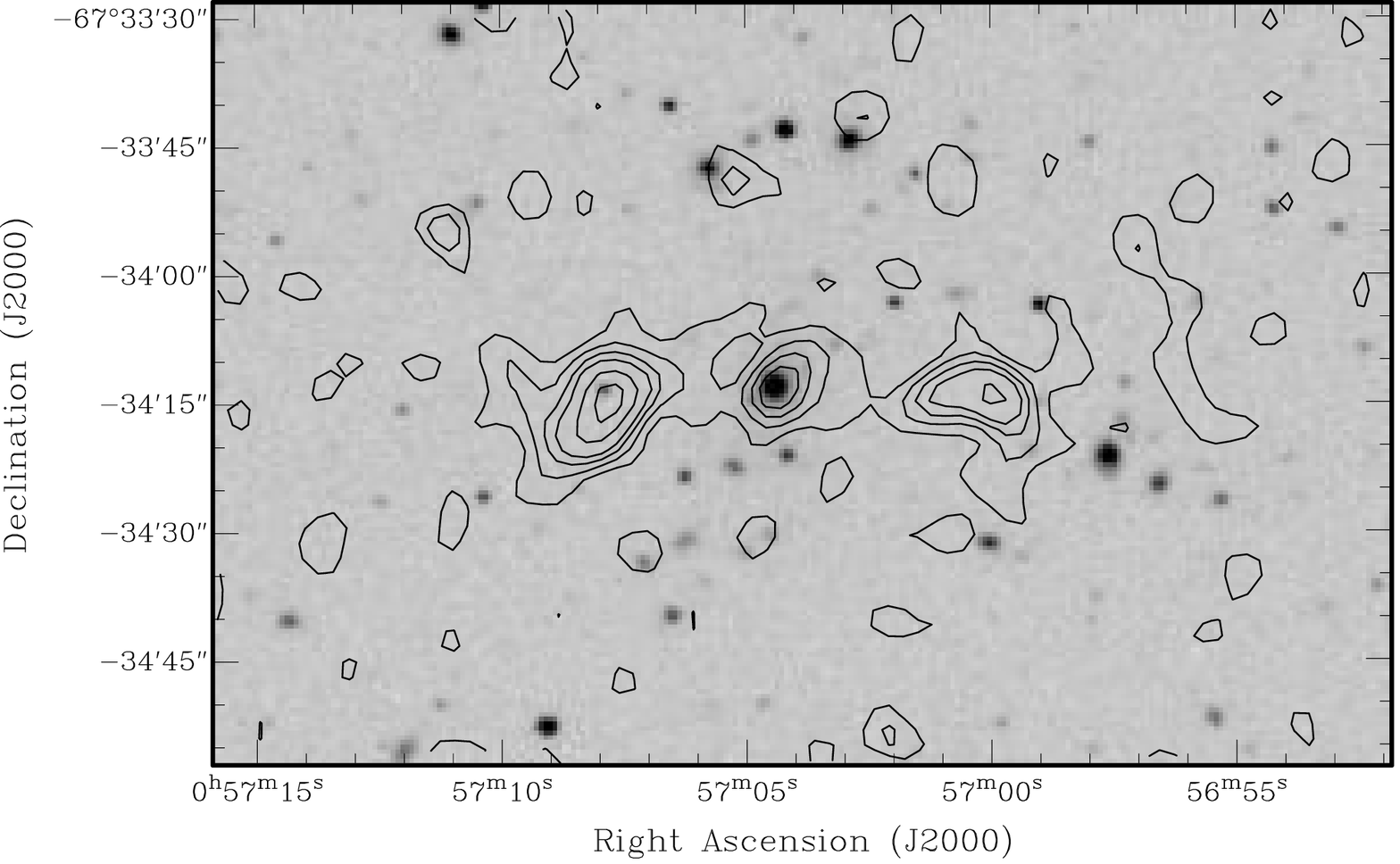}}
\caption{J0057.0-6734: $10^{-4}$ Jy x 1,  2,  3,  4,  6,  8.}
\end{minipage}
&
\begin{minipage}{0.47\linewidth}
\frame{\includegraphics[angle=-90, width=2.8in]{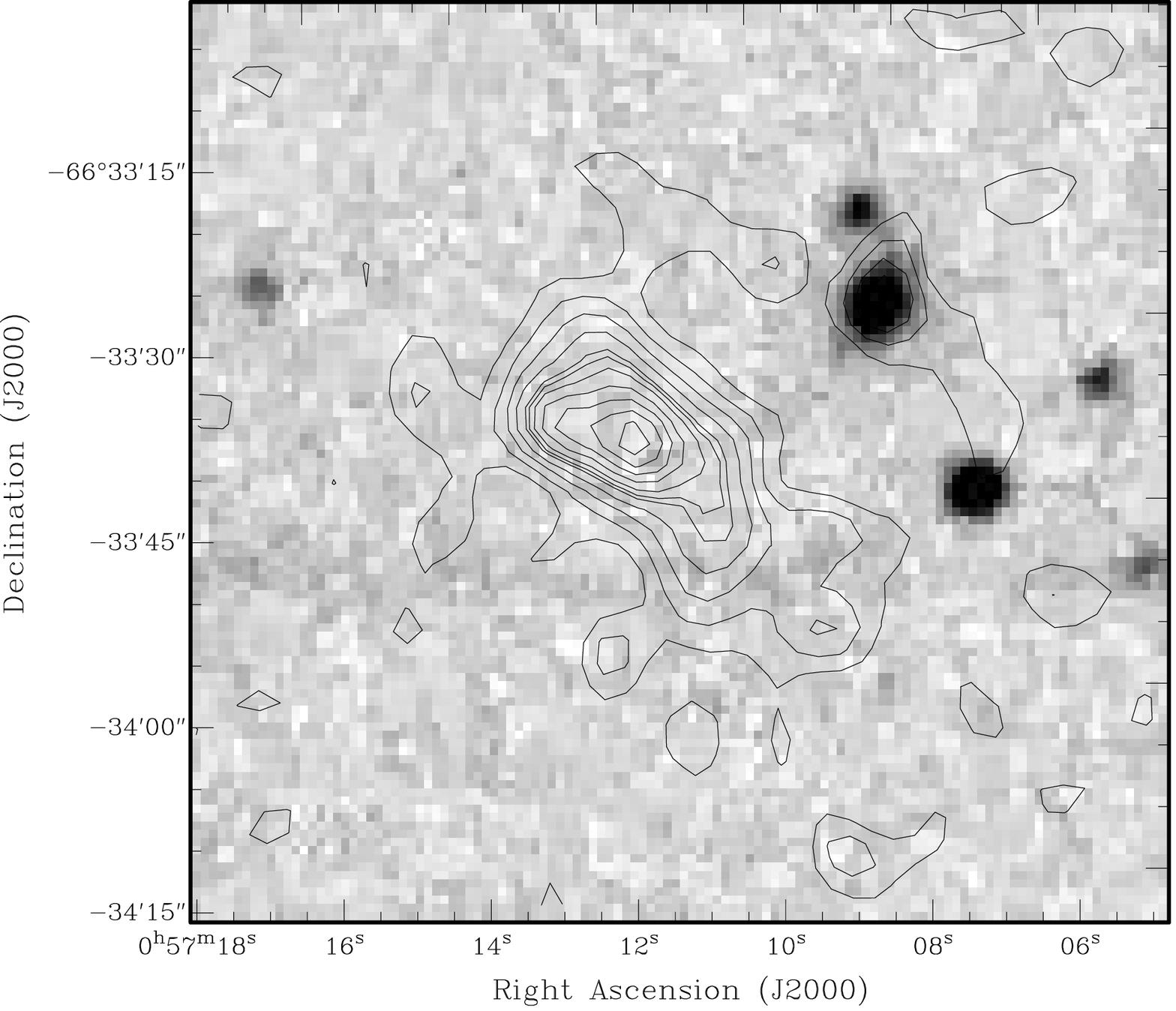}}
\caption{J0057.1-6633: $10^{-4}$ Jy x 1,  2,  3,  4,  5,  5.4,  6, 7, 8, 9, 10. R band image is used.} 
\end{minipage}
\\
\end{tabular}
\end{figure*}

\begin{figure*}
\centering
\begin{tabular}{l}
\begin{minipage}{0.47\linewidth}
\frame{\includegraphics[angle=-90, width=2.8in]{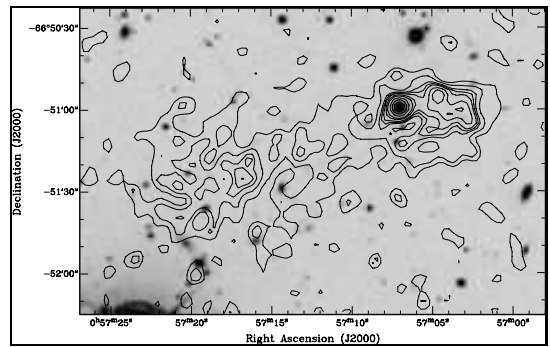}}
\caption{J0057.2-6651: $10^{-4}$ Jy x 1,  2,  3,  3.5,  4,  5,  6,  8,  10,  12.} 
\end{minipage}
\\
\end{tabular}
\end{figure*}

\clearpage



\begin{figure*}
\centering
\begin{tabular}{cc}
\begin{minipage}{0.47\linewidth}
\frame{\includegraphics[angle=-90, width=2.8in]{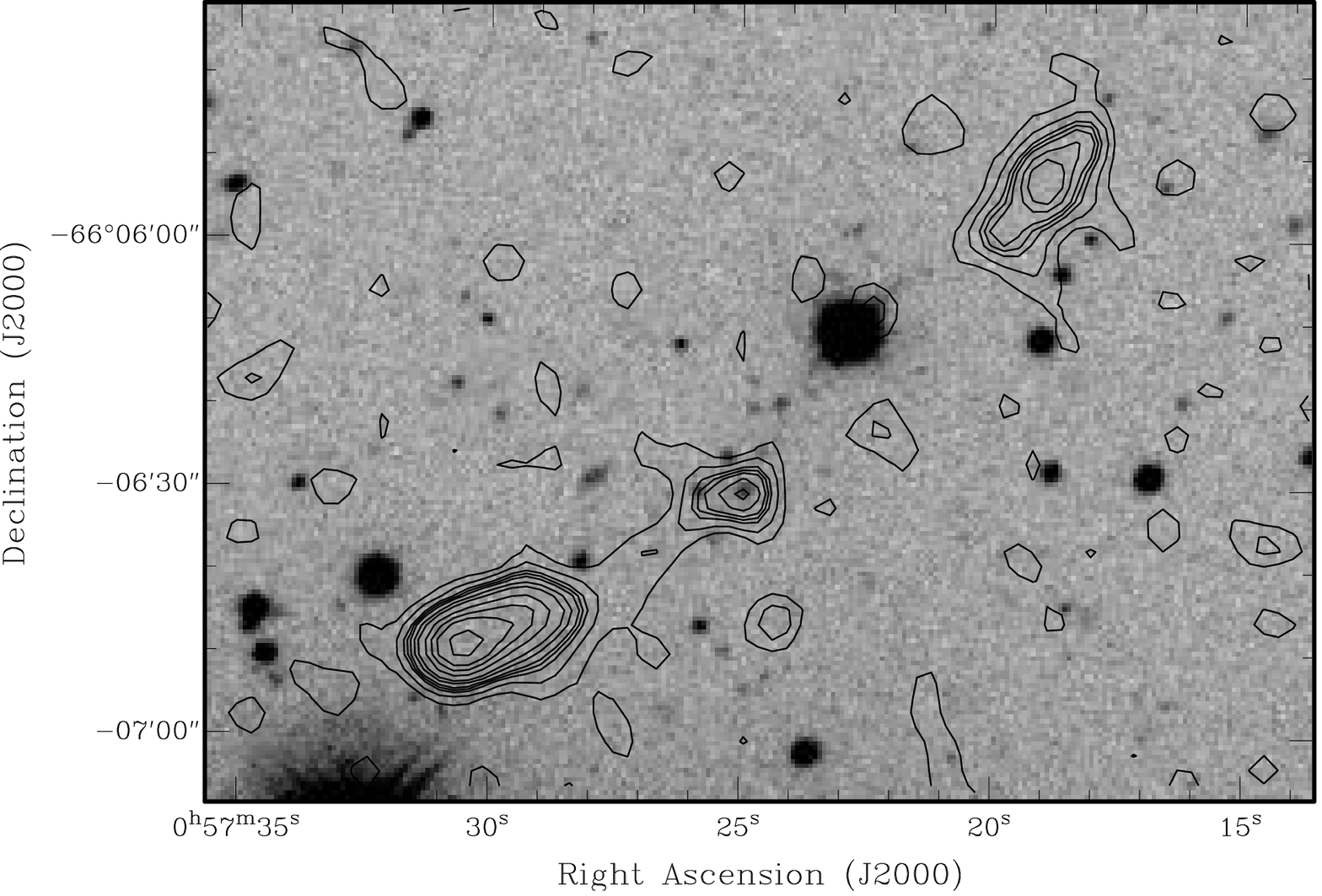}}
\caption{J0057.4-6606: $10^{-4}$ Jy x 1,  2,  3,  3.5,  4,  5,  6,  8,  10,  12,  16.} 
\end{minipage}
&
\begin{minipage}{0.47\linewidth}
\frame{\includegraphics[angle=-90, width=2.8in] {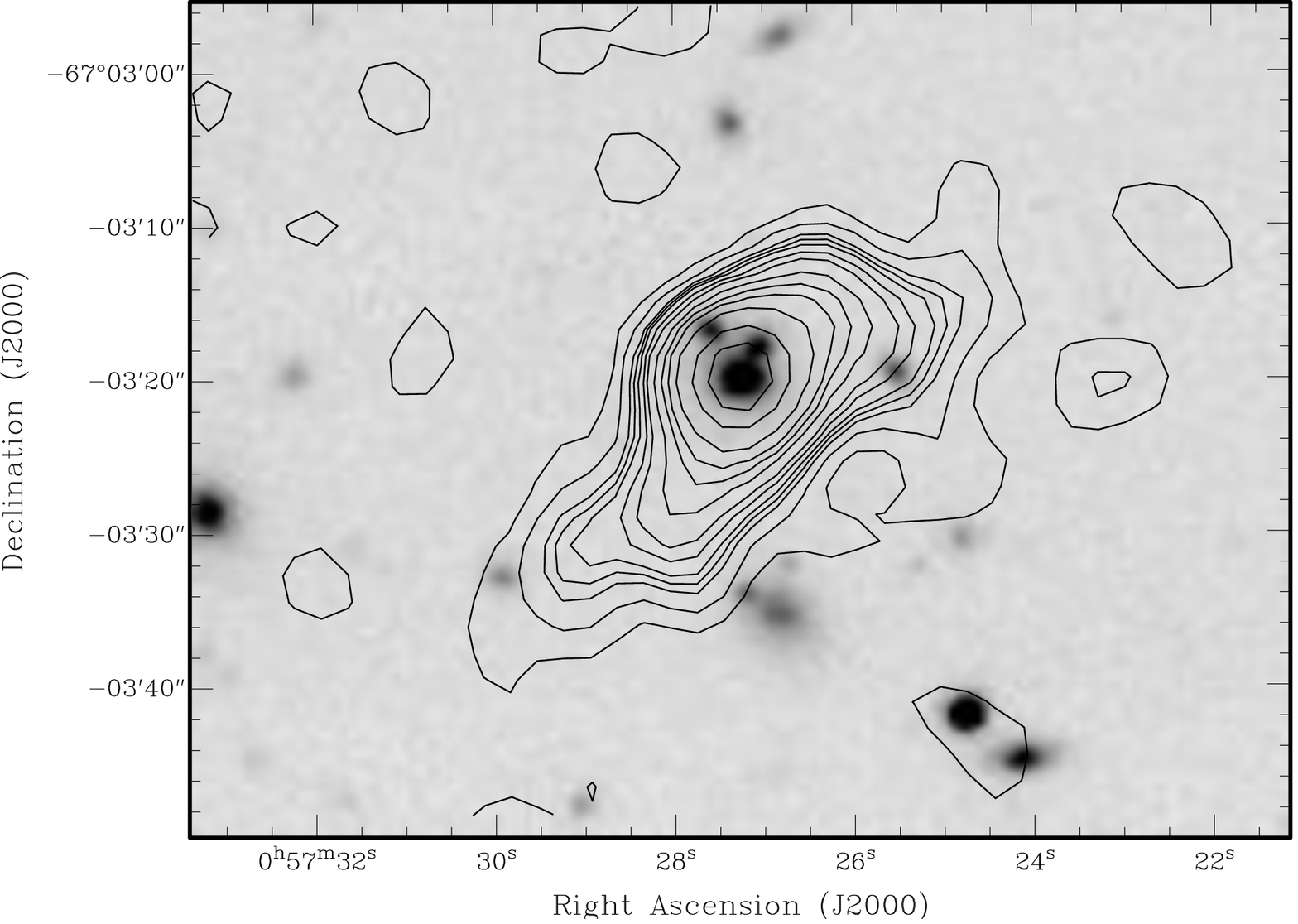}}
\caption{J0057.4-6703: $10^{-4}$ Jy x 1,  2,  3,  3.5,  4,  5,  6,  8,  10,  12,  16,  24,  32.} 
\end{minipage}
\\
\end{tabular}
\end{figure*}

\begin{figure*}
\centering
\begin{tabular}{cc}
\begin{minipage}{0.47\linewidth}
\frame{\includegraphics[angle=-90, width=2.8in]{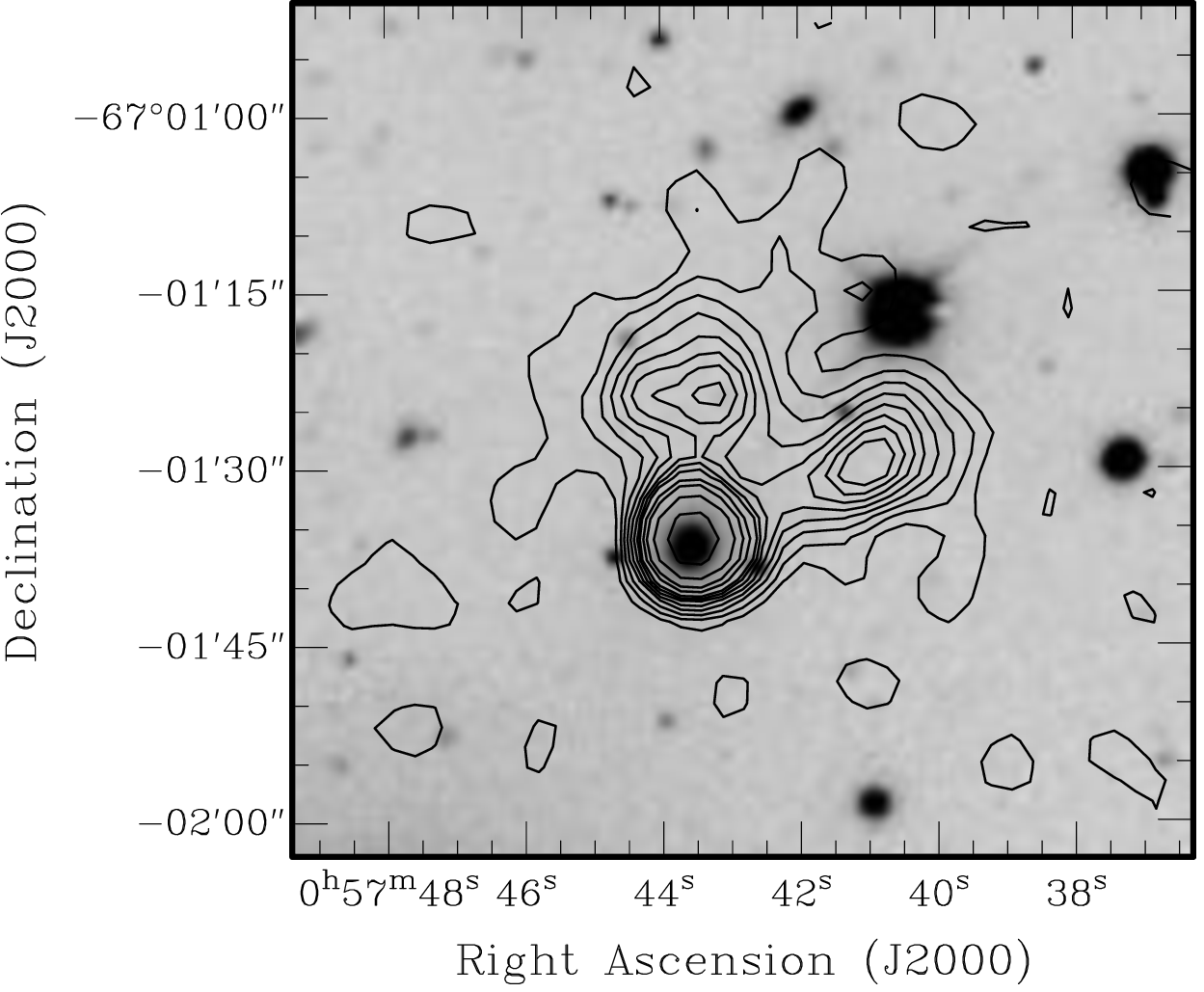}}
\caption{J0057.7-6701: $10^{-4}$ Jy x 1,  2,  3,  4,  5,  6,  7,  8,  10,  12,  16,  24.}
\end{minipage}
&
\begin{minipage}{0.47\linewidth}
\frame{\includegraphics[angle=-90, width=2.8in]{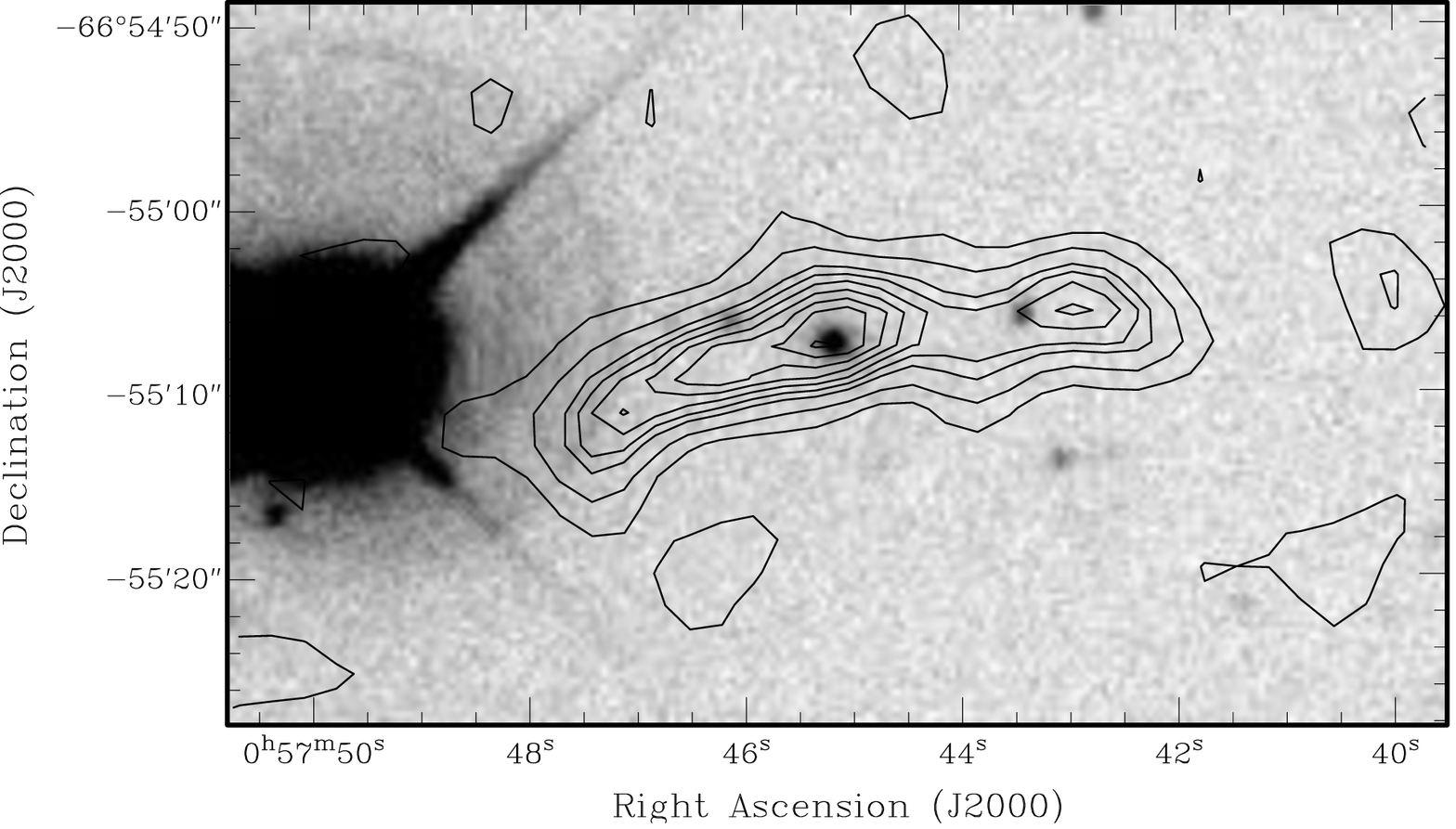}}
\caption{J0057.7-6655: $10^{-4}$ Jy x 1,  2,  3,  3.5,  4,  4.5,  5,  5.5,  6,.} 
\end{minipage}
\\
\end{tabular}
\end{figure*}

\begin{figure*}
\centering
\begin{tabular}{cc}
\begin{minipage}{0.47\linewidth}
\frame{\includegraphics[angle=-90, width=2.8in]{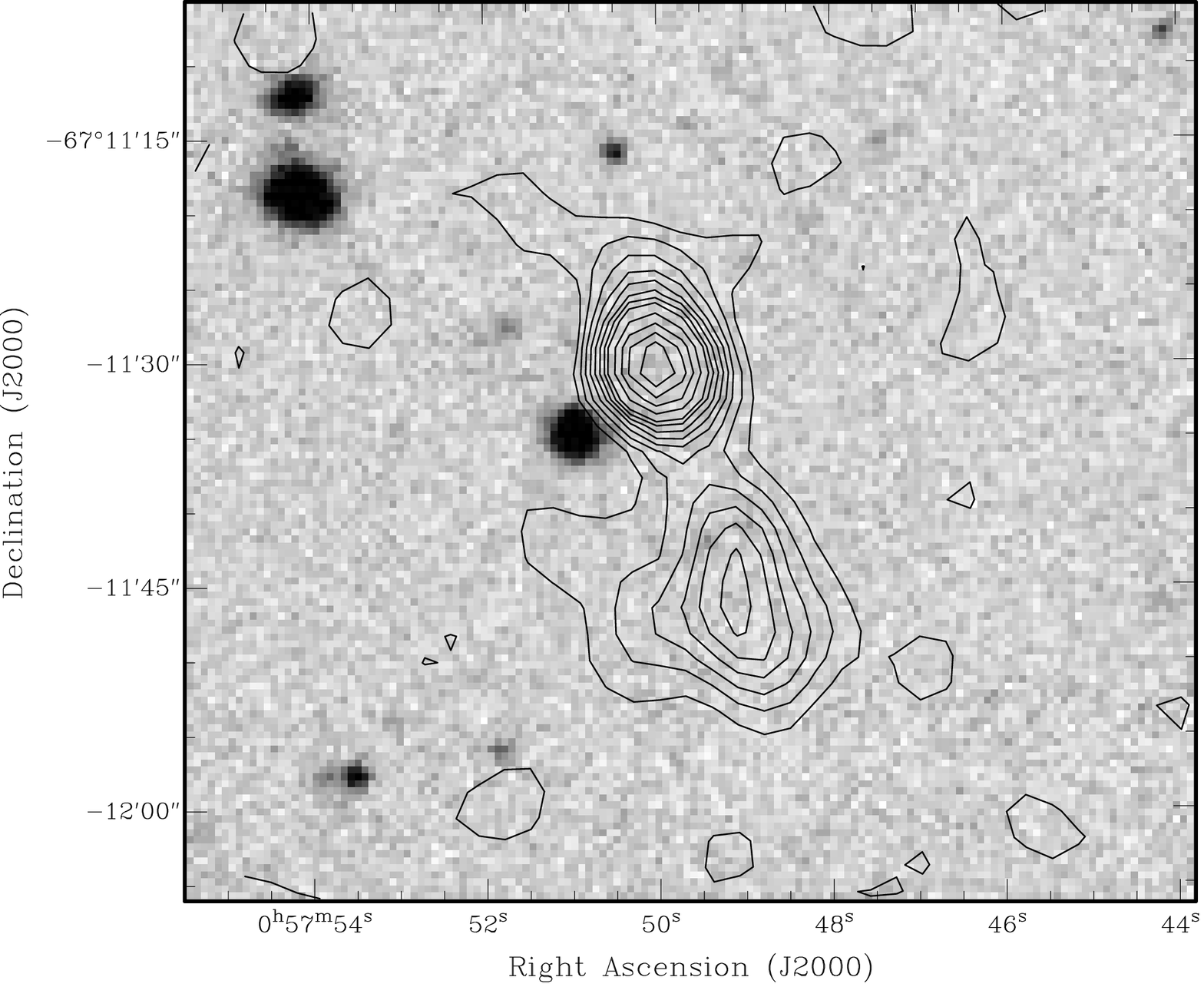}}
\caption{J0057.8-6711: $10^{-4}$ Jy x 1,  2,  3,  4,  5,  6,  7,  8,  10,  12,  14,  16.} 
\end{minipage}
&
\begin{minipage}{0.47\linewidth}
\frame{\includegraphics[angle=-90, width=2.8in] {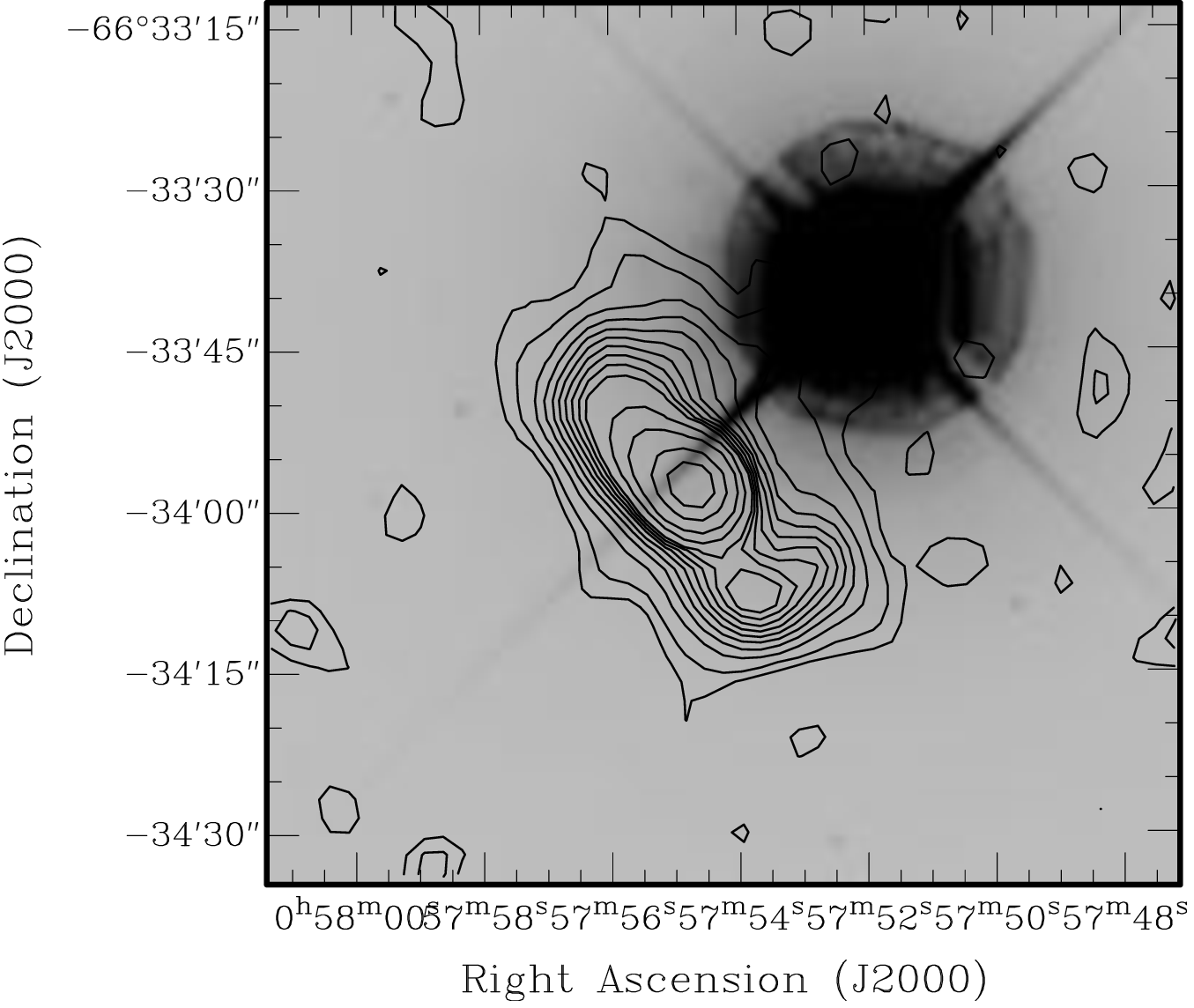}}
\caption{J0057.9-6633: $10^{-4}$ Jy x 1,  2,  3,  4,  5,  6,  7,  8,  
10,  12,  14,  16,  18,  24,  32,  48,  64.} 
\end{minipage}
\\
\end{tabular}
\end{figure*}

\begin{figure*}
\centering
\begin{tabular}{cc}
\begin{minipage}{0.47\linewidth}
\frame{\includegraphics[angle=-90, width=2.8in]{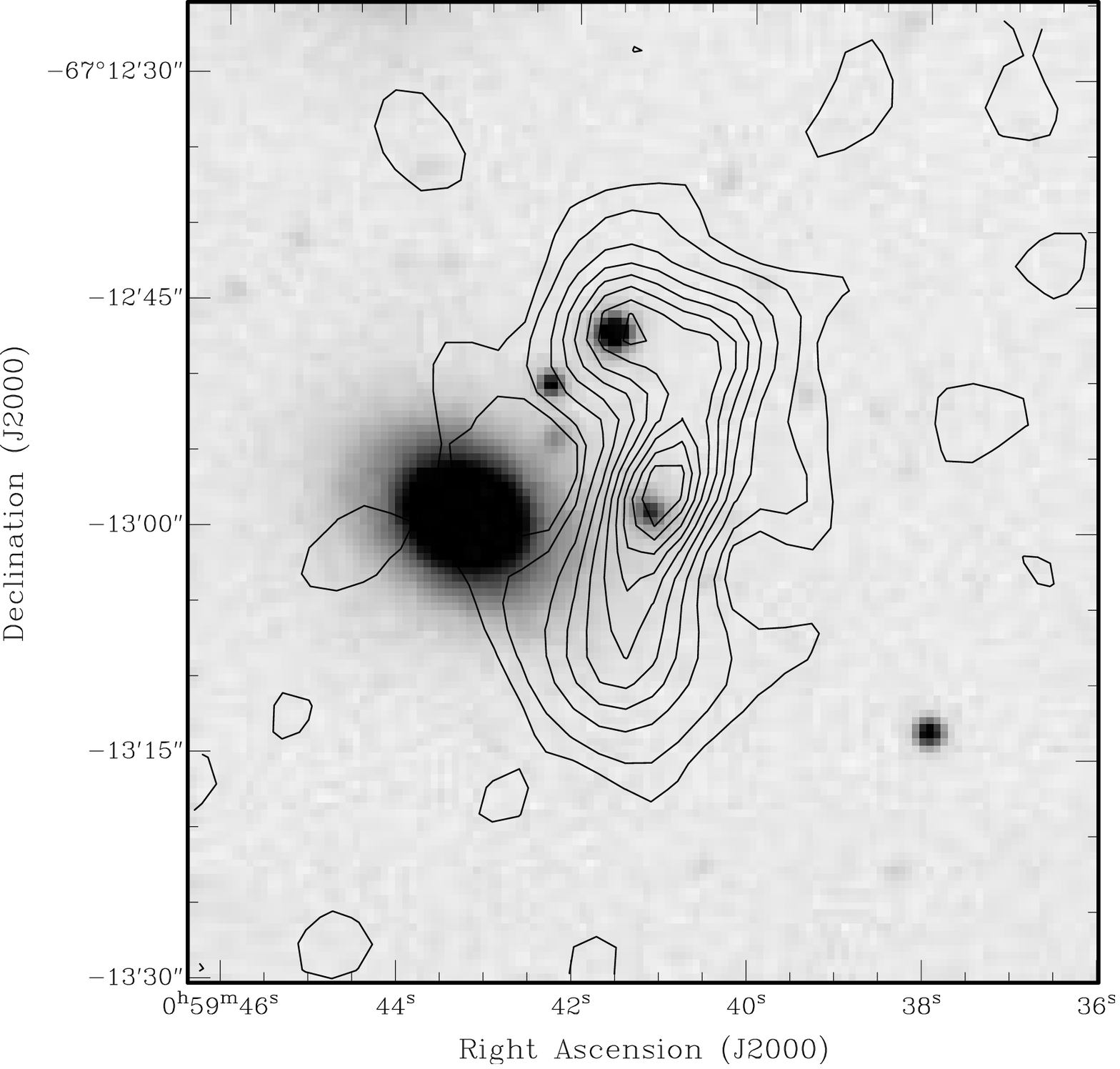}}
\caption{J0059.6-6712: $10^{-4}$ Jy x 1,  2,  4,  6,  8,  10,  12,  14,  16,  18.}
\end{minipage}
&
\begin{minipage}{0.47\linewidth}
\frame{\includegraphics[angle=-90, width=2.8in]{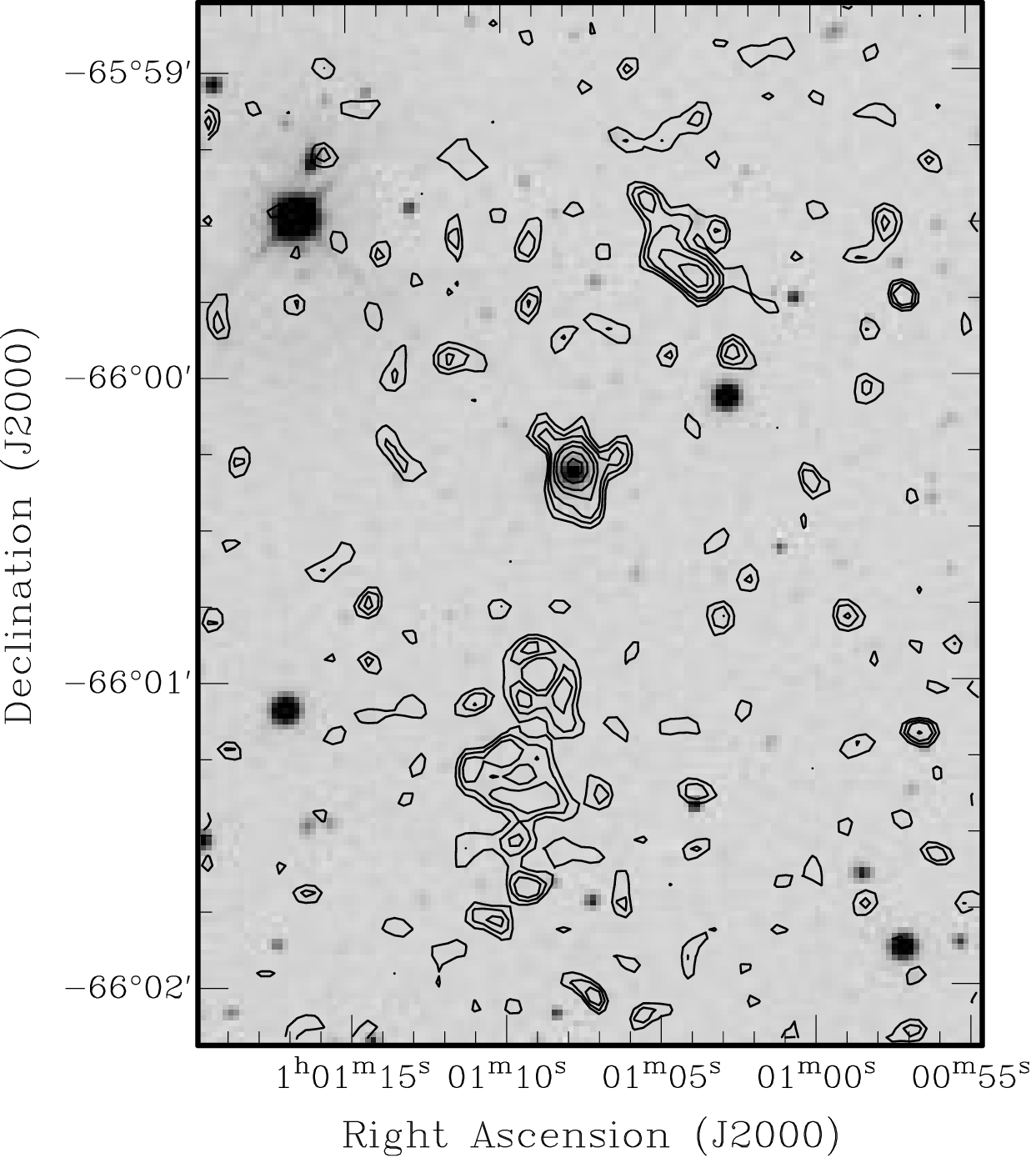}}
\caption{J0101.1-6600: $10^{-4}$ Jy x 1,  1.5,  2,  3,  4,  6,  8.} 
\end{minipage}
\\
\end{tabular}
\end{figure*}

\begin{figure*}
\centering
\begin{tabular}{cc}
\begin{minipage}{0.47\linewidth}
\frame{\includegraphics[angle=-90, width=2.8in]{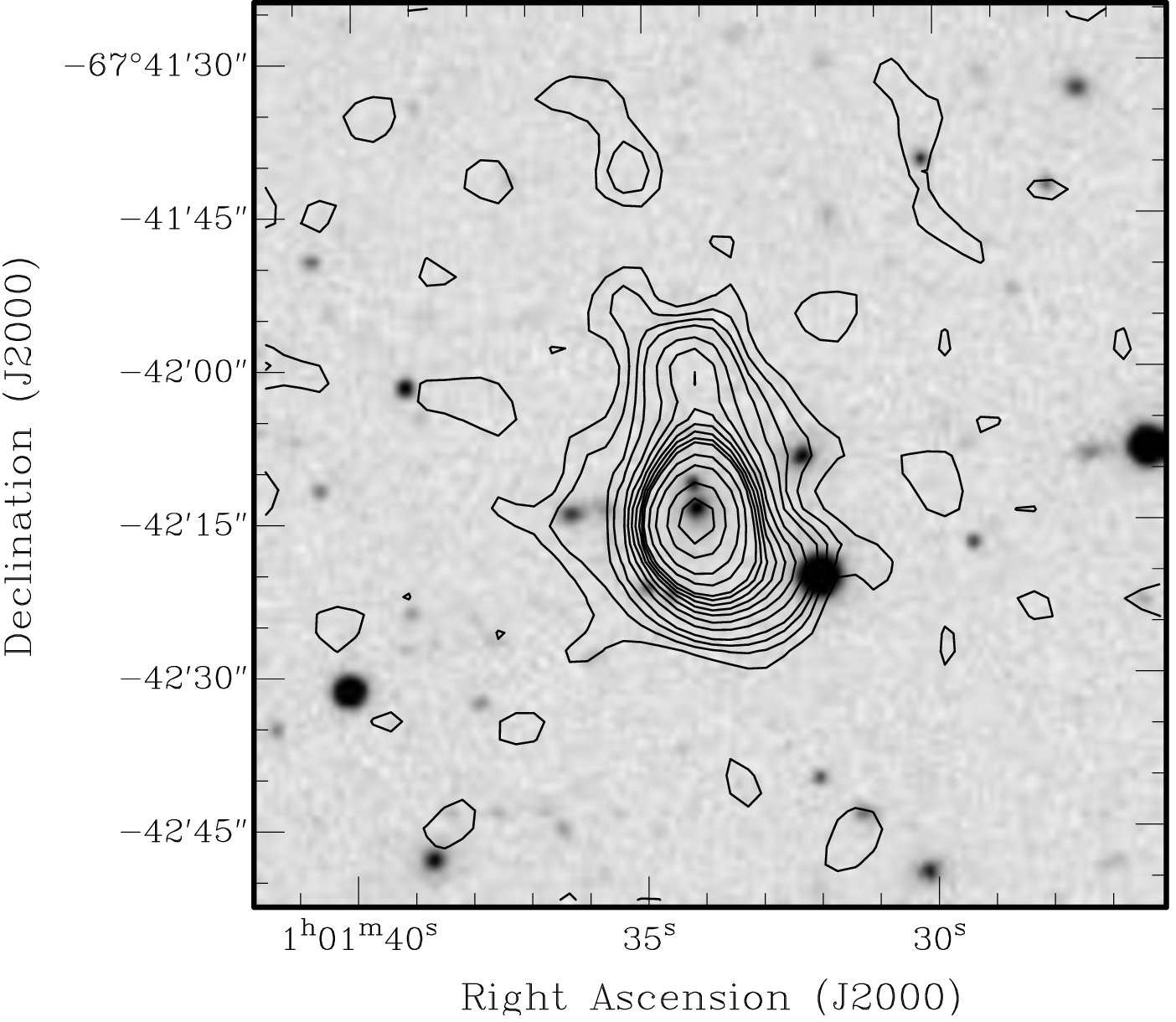}}
\caption{J0101.5-6742: $10^{-4}$ Jy x 1,  2,  3,  4,  6,  8,  10,  12,  14,  16,  18,  24,  32,  48,  64.} 
\end{minipage}
&
\begin{minipage}{0.47\linewidth}
\frame{\includegraphics[angle=-90, width=2.8in] {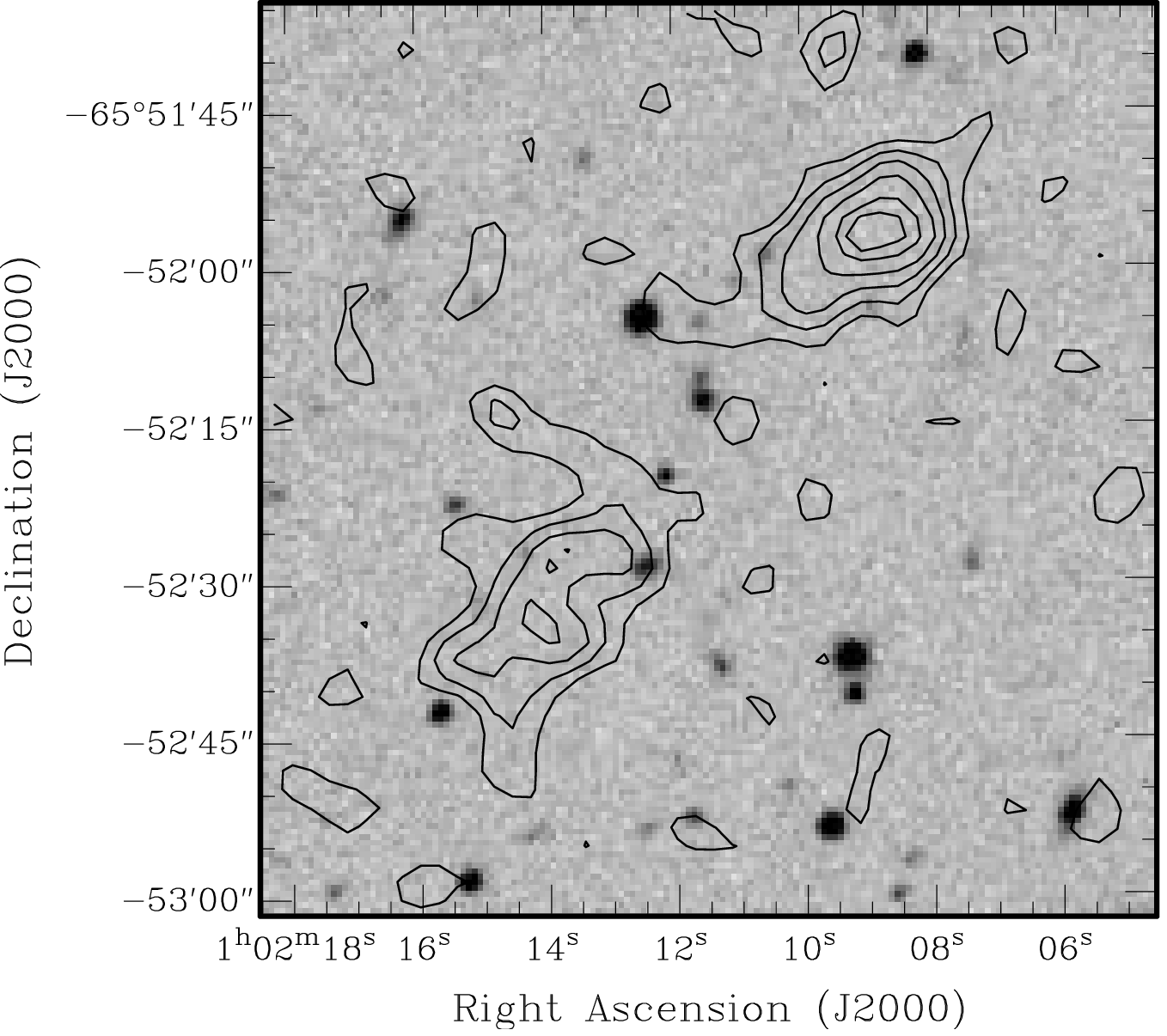}}
\caption{J0102.1-6552: $10^{-4}$ Jy x 1,  2,  3,  4,  5,  5.5.} 
\end{minipage}
\\
\end{tabular}
\end{figure*}

\begin{figure*}
\centering
\begin{tabular}{cc}
\begin{minipage}{0.47\linewidth}
\frame{\includegraphics[angle=-90, width=2.8in]{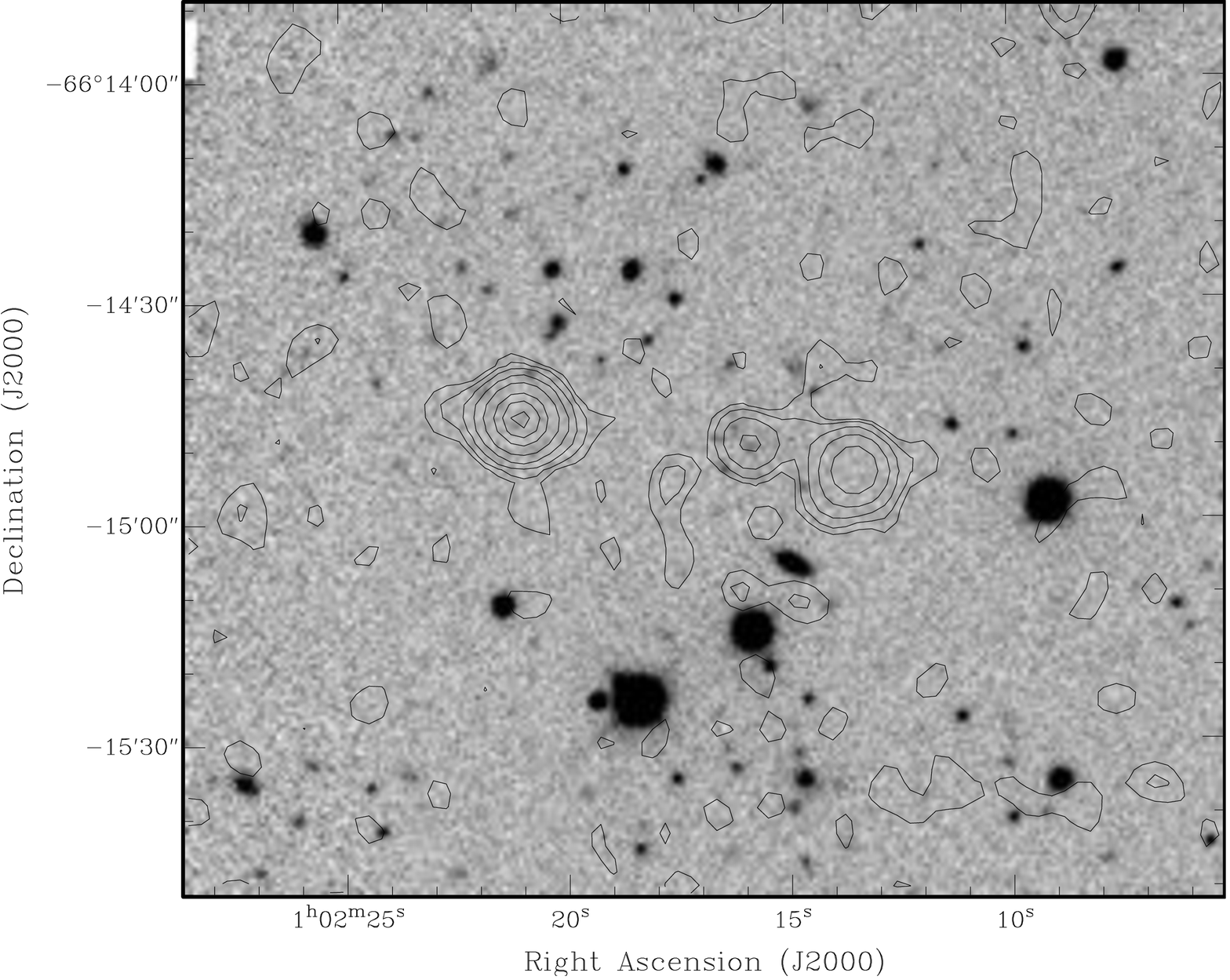}}
\caption{J0102.3-6614: $10^{-4}$ Jy x 1,  2,  4,  8,  16,  32,  48,  64.}
\end{minipage}
&
\begin{minipage}{0.47\linewidth}
\frame{\includegraphics[angle=-90, width=2.8in]{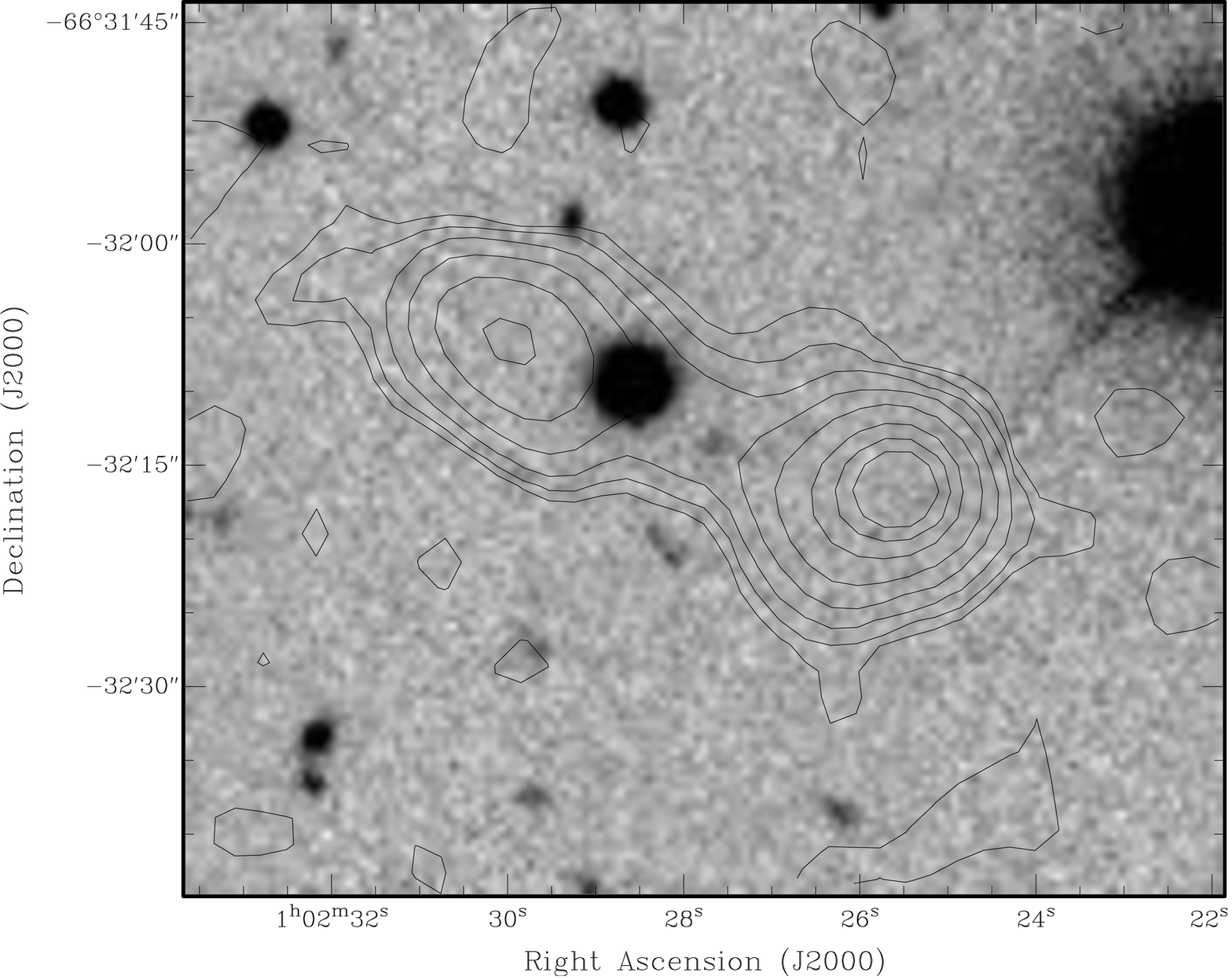}}
\caption{J0102.4-6632: $10^{-4}$ Jy x 1,  2,  4,  8,  16,  32,  48,  64.} 
\end{minipage}
\\
\end{tabular}
\end{figure*}

\begin{figure*}
\centering
\begin{tabular}{cc}
\begin{minipage}{0.47\linewidth}
\frame{\includegraphics[angle=-90, width=2.8in]{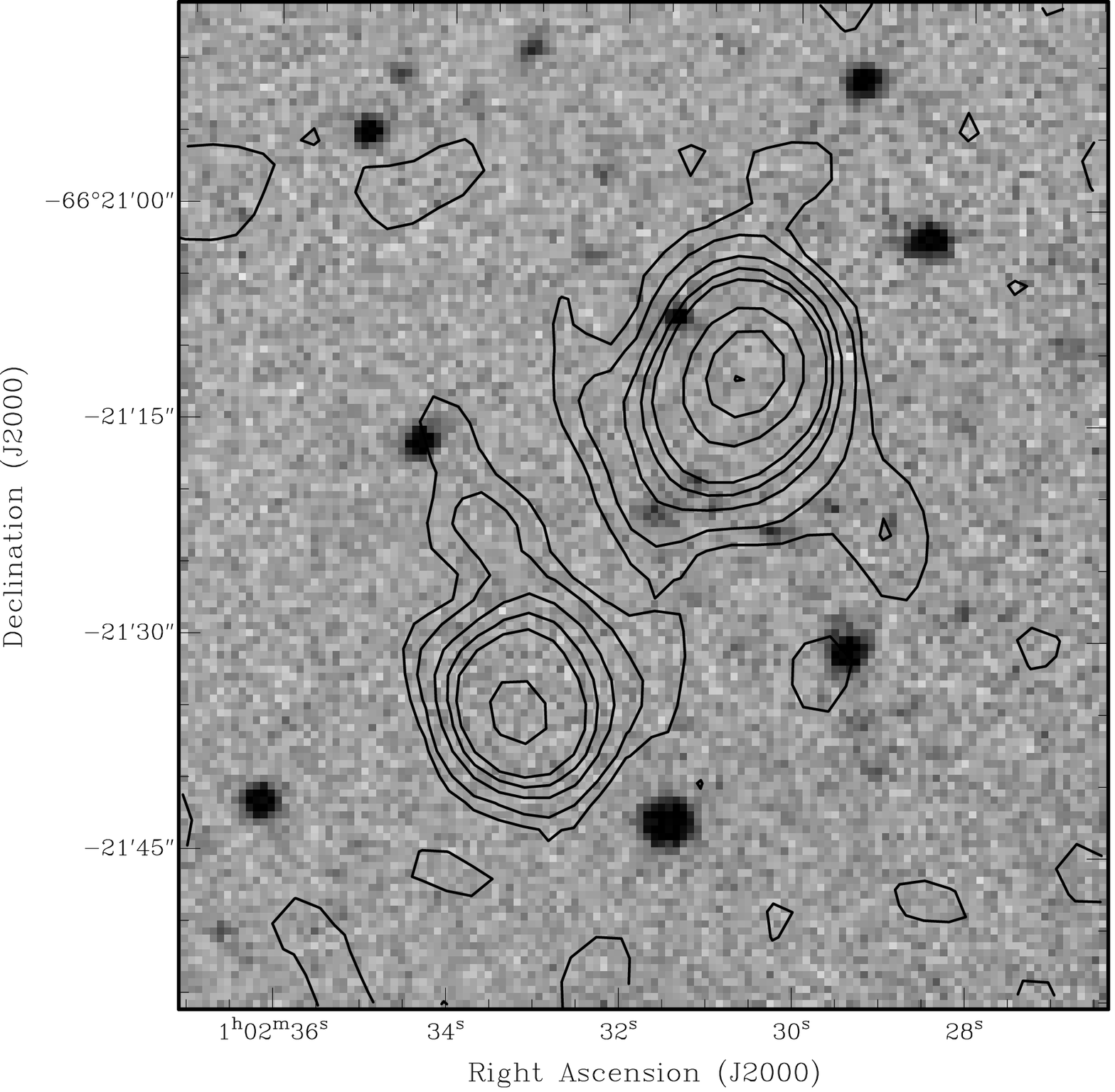}}
\caption{J0102.5-6621: $10^{-4}$ Jy x 1,  2,  4,  6,  8,  16,  24,  32.} 
\end{minipage}
&
\begin{minipage}{0.47\linewidth}
\frame{\includegraphics[angle=-90, width=2.8in] {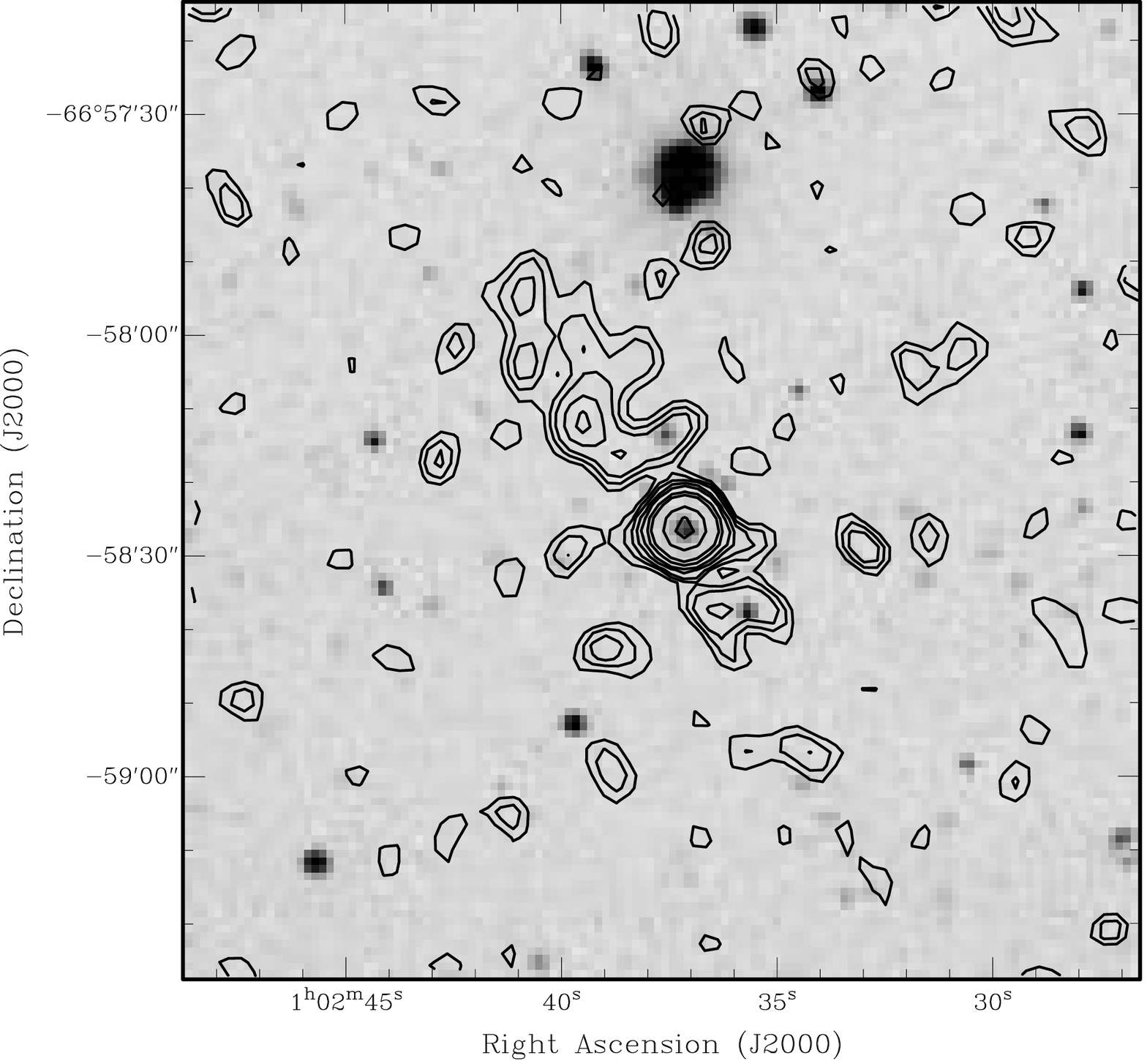}}
\caption{J0102.6-6658: $10^{-4}$ Jy x 1,  1.5,  2,  3,  4,  6,  8,  16,  24.} 
\end{minipage}
\\
\end{tabular}
\end{figure*}

\begin{figure*}
\centering
\begin{tabular}{cc}
\begin{minipage}{0.47\linewidth}
\frame{\includegraphics[angle=-90, width=2.8in]{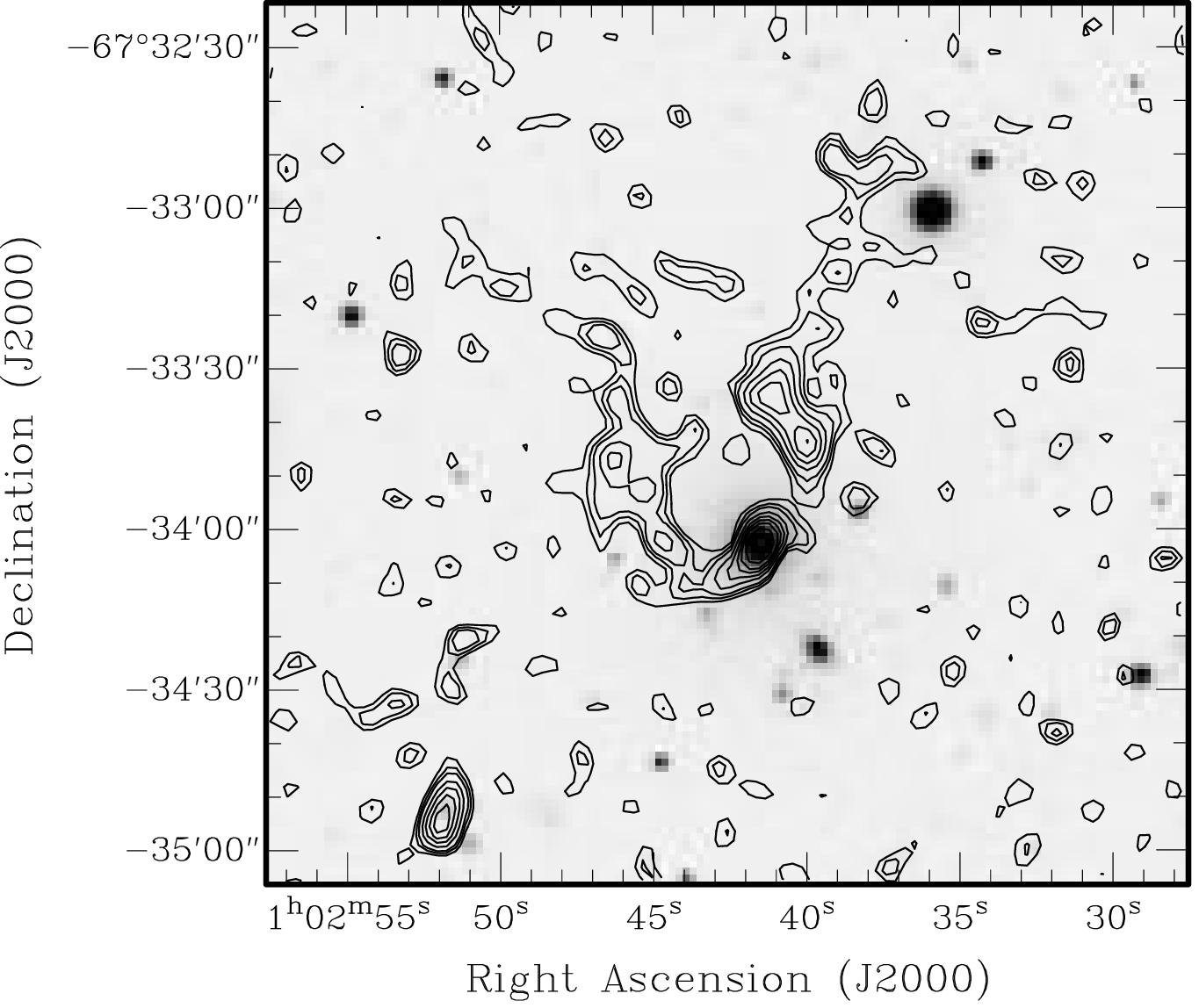}}
\caption{J0102.6-6734: $10^{-4}$ Jy x 1,  1.5,  2,  3,  4,  5,  6,  8.}
\end{minipage}
&
\begin{minipage}{0.47\linewidth}
\frame{\includegraphics[angle=-90, width=2.8in]{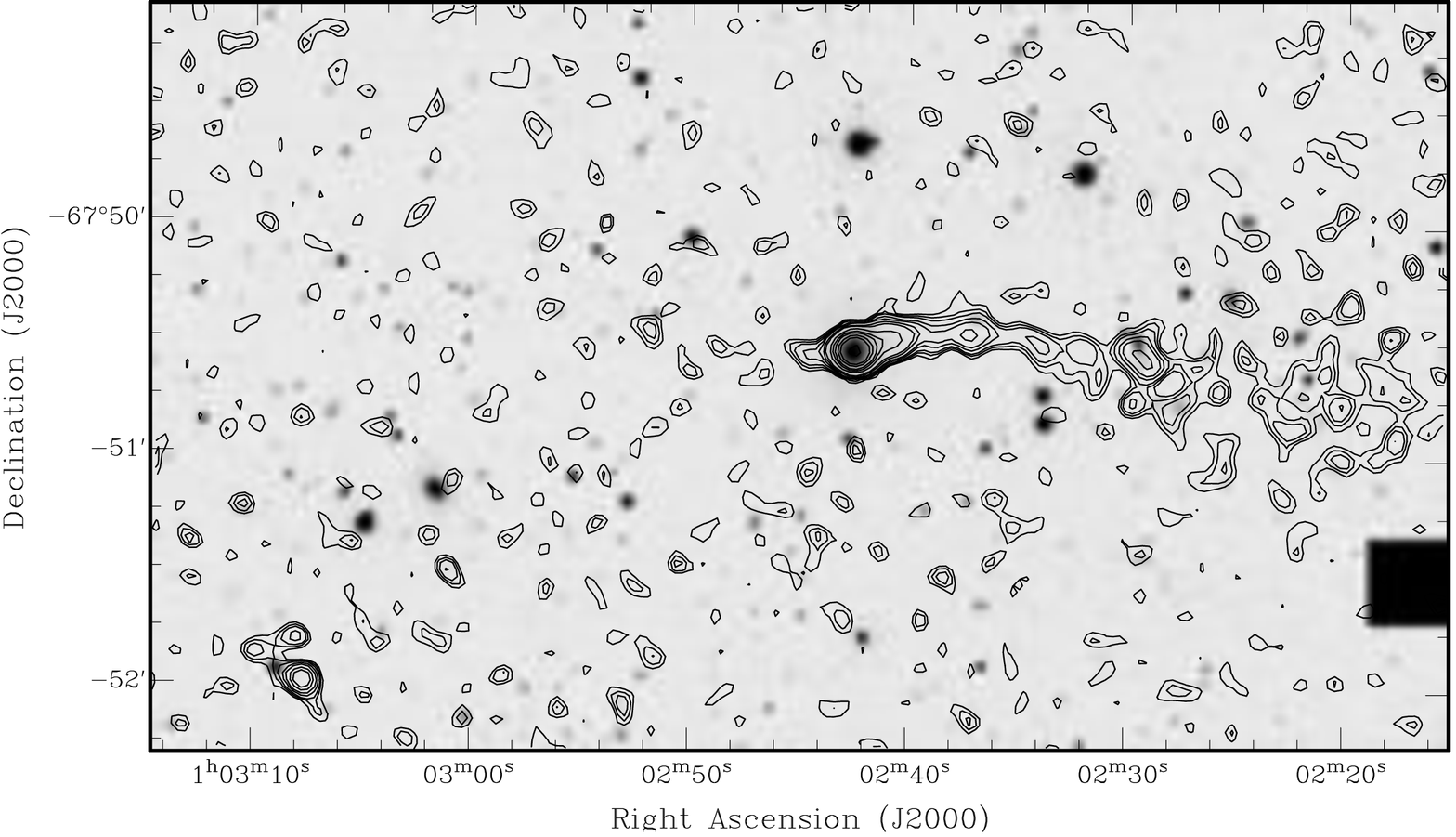}}
\caption{J0102.6-6750: $10^{-4}$ Jy x 1,  1.5,  2,  3,  4,  6,  8,  16,  24,  32,  48,  64.} 
\end{minipage}
\\
\\
\end{tabular}
\end{figure*}

\begin{figure*}
\centering
\begin{tabular}{cc}
\begin{minipage}{0.47\linewidth}
\frame{\includegraphics[angle=-90, width=2.8in]{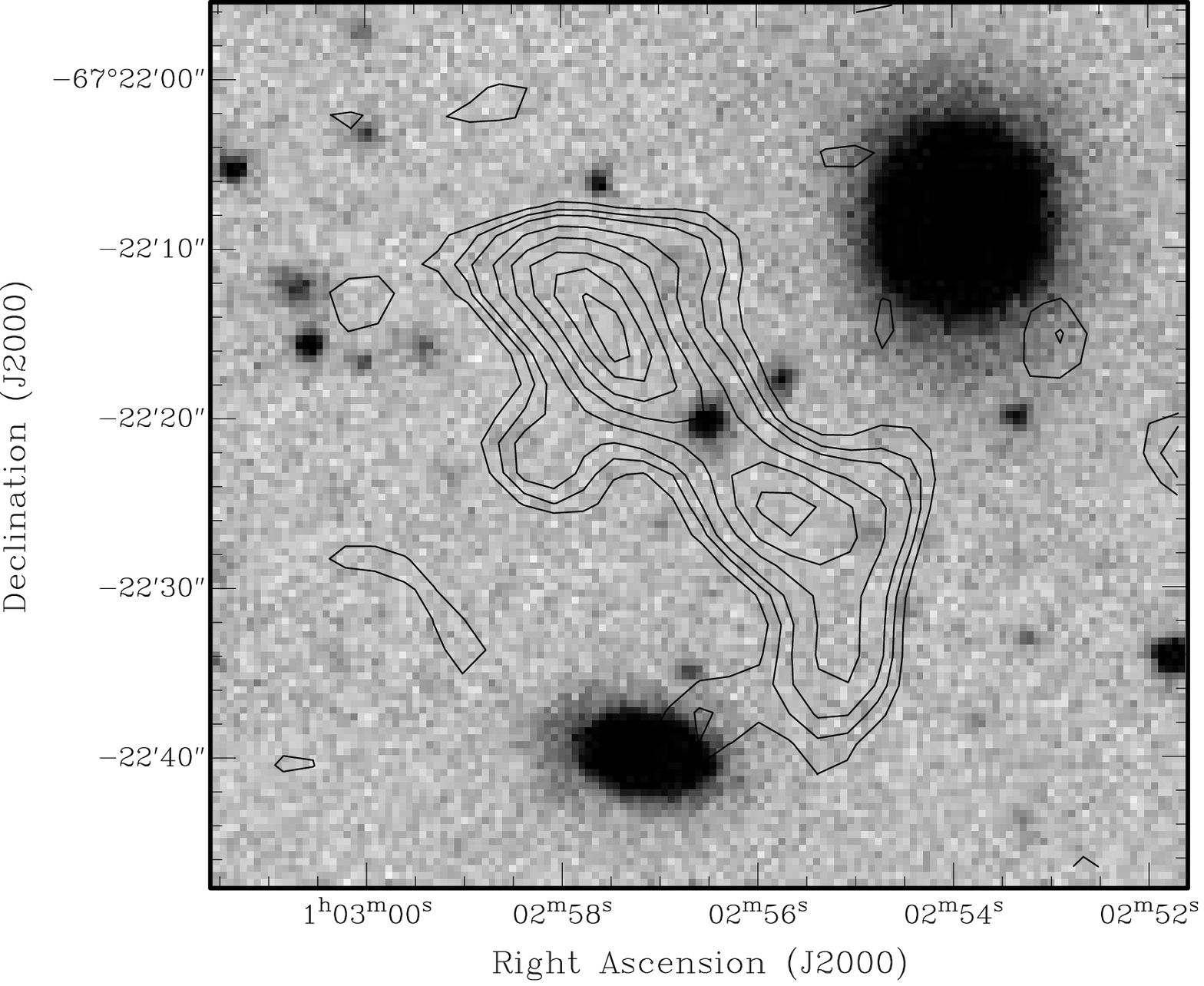}}
\caption{J0102.9-6722: $10^{-4}$ Jy x 1,  1.5,  2,  3,  4,  5,  6,  6.7.} 
\end{minipage}
&
\begin{minipage}{0.47\linewidth}
\frame{\includegraphics[angle=-90, width=2.8in] {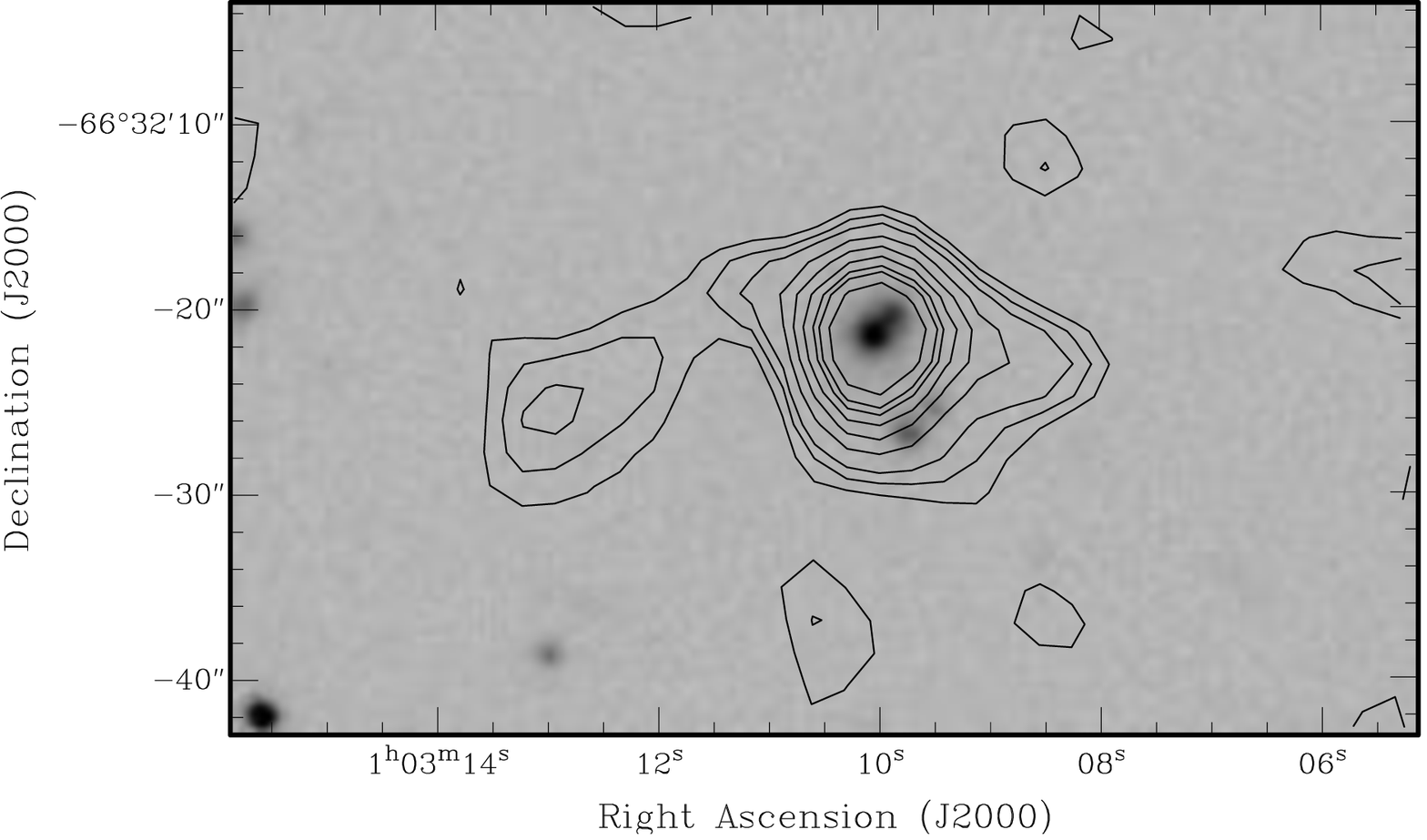}}
\caption{J0103.1-6632: $10^{-4}$ Jy x 1,  1.5,  2,  3,  4,  5,  6,  6.7,  8.} 
\end{minipage}
\\
\end{tabular}
\end{figure*}

\begin{figure*}
\centering
\begin{tabular}{cc}
\begin{minipage}{0.47\linewidth}
\frame{\includegraphics[angle=-90, width=2.8in]{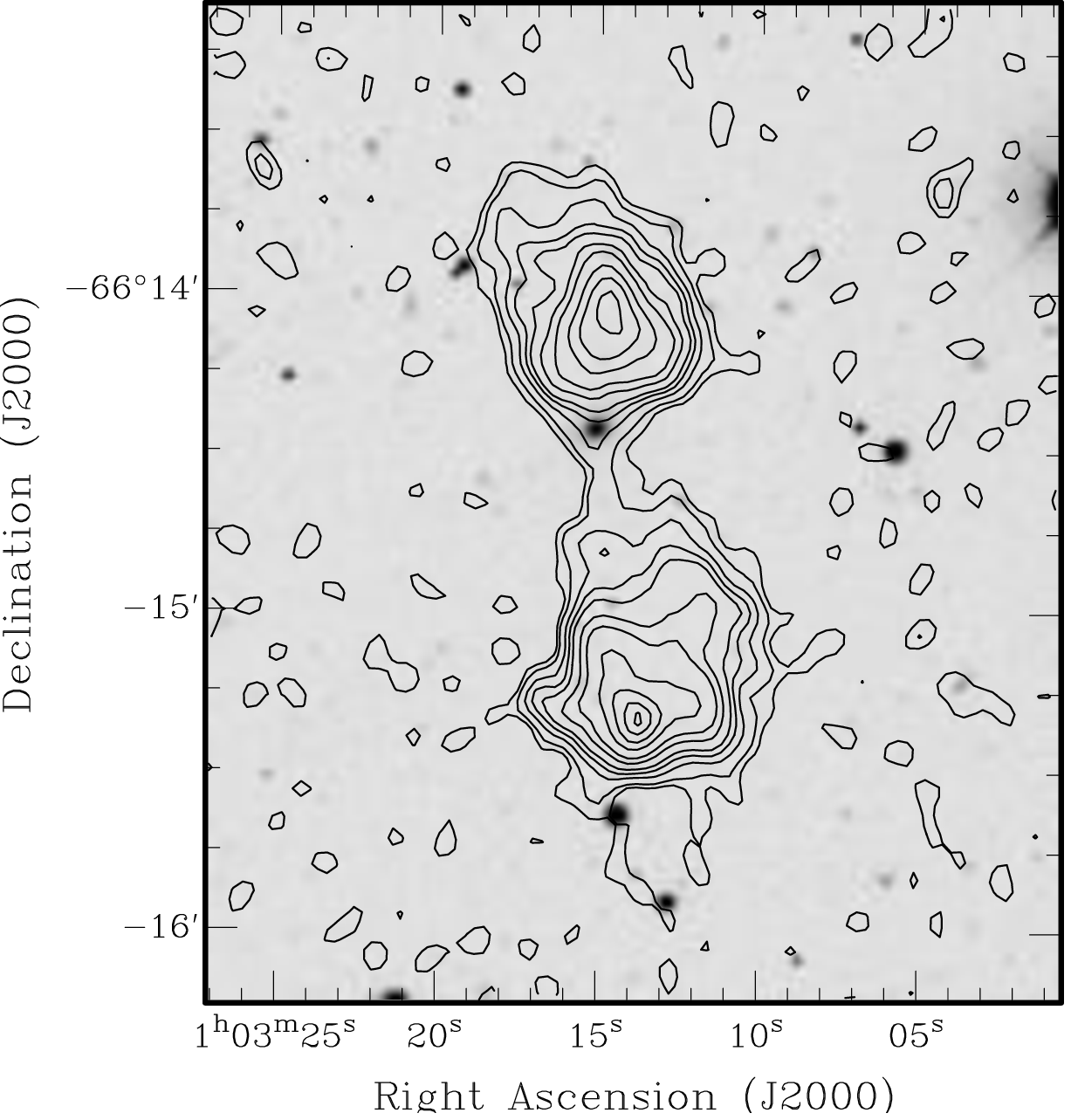}}
\caption{J0103.2-6614: $10^{-4}$ Jy x 1,  2,  4,  6,  8,  12,  16,  24,  32,  41,  48.}
\end{minipage}
&
\begin{minipage}{0.47\linewidth}
\frame{\includegraphics[angle=-90, width=2.8in]{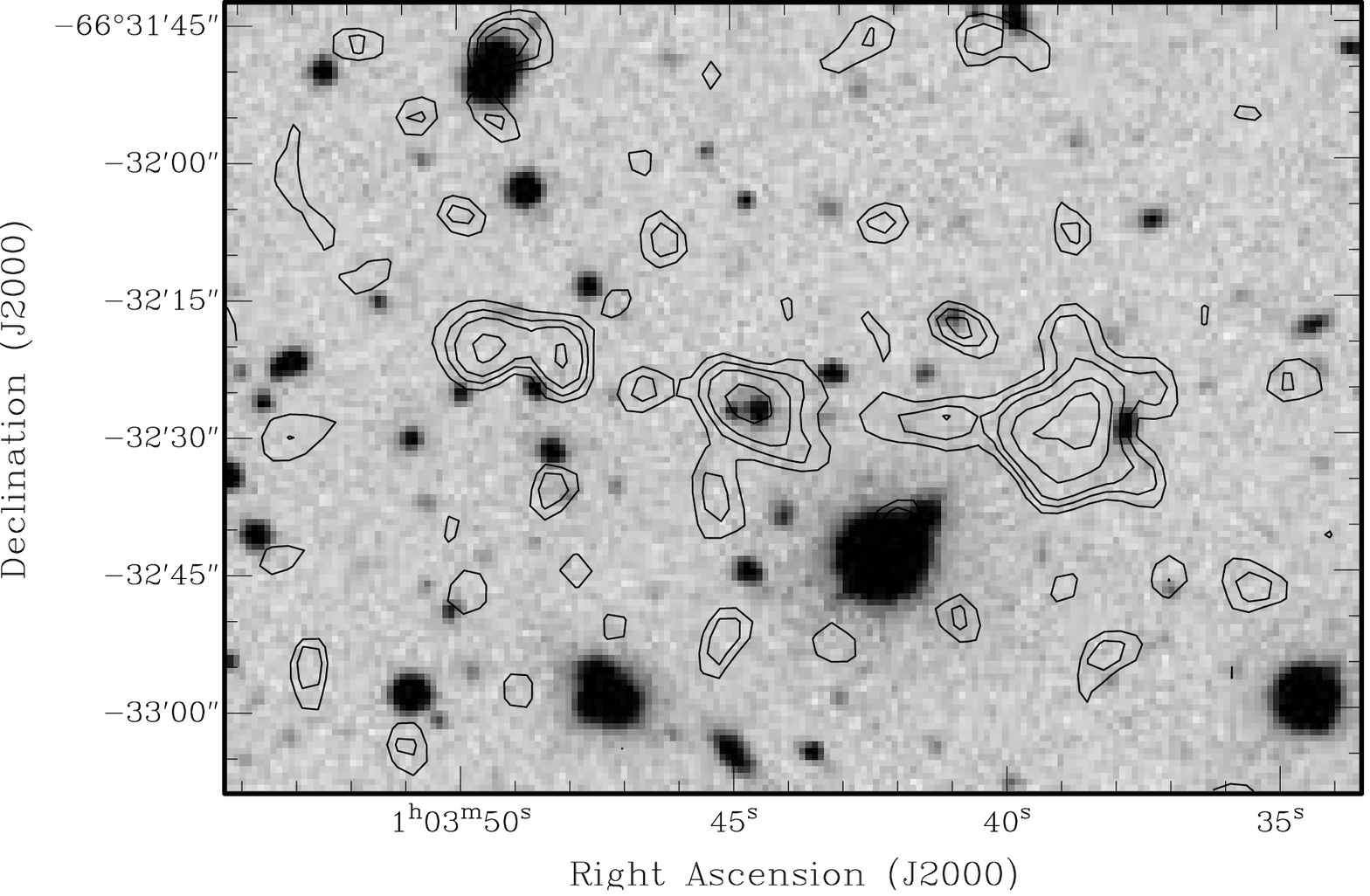}}
\caption{J0103.7-6632: $10^{-4}$ Jy x 1,  1.5,  2,  3,  4.} 
\end{minipage}
\\
\end{tabular}
\end{figure*}

\begin{figure*}
\centering
\begin{tabular}{cc}
\begin{minipage}{0.47\linewidth}
\frame{\includegraphics[angle=-90, width=2.8in]{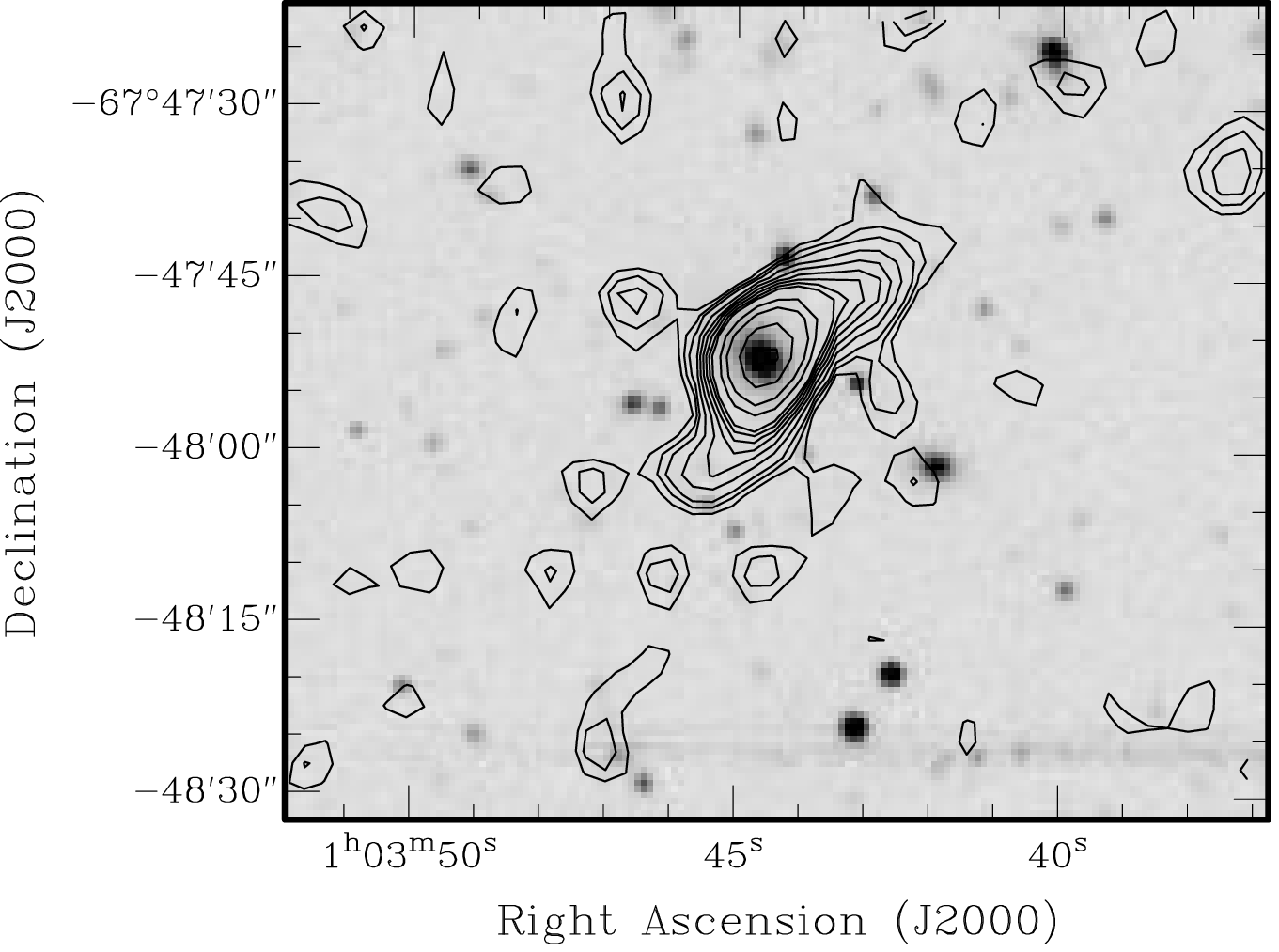}}
\caption{J0103.7-6747: $10^{-4}$ Jy x 1,  1.5,  2,  3,  4,  5,  6,  7,  8,  12,  16,  24.} 
\end{minipage}
&
\begin{minipage}{0.47\linewidth}
\frame{\includegraphics[angle=-90, width=2.8in] {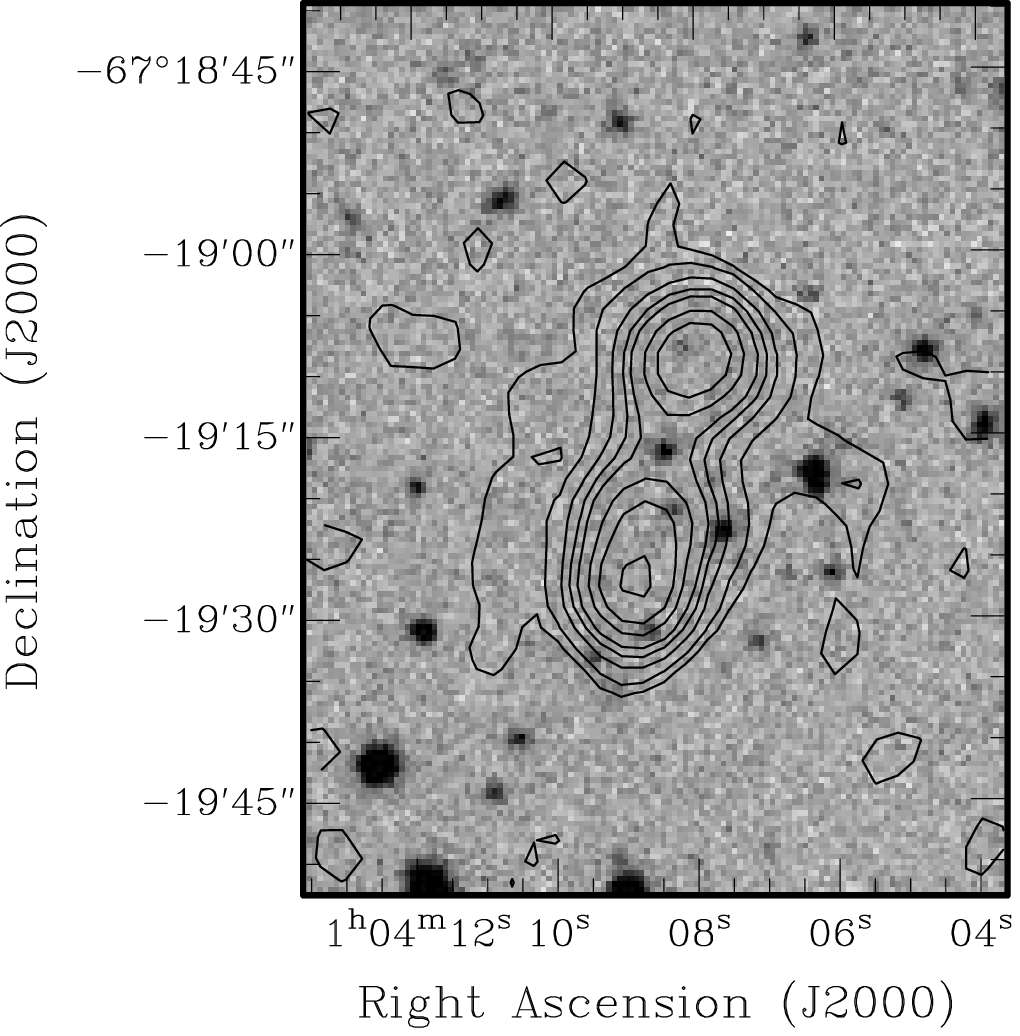}}
\caption{J0104.1-6719: $10^{-4}$ Jy x 1,  2,  4,  6,  8,  12,  16,  24.} 
\end{minipage}
\\
\end{tabular}
\end{figure*}

\begin{figure*}
\centering
\begin{tabular}{cc}
\begin{minipage}{0.47\linewidth}
\frame{\includegraphics[angle=-90, width=2.8in]{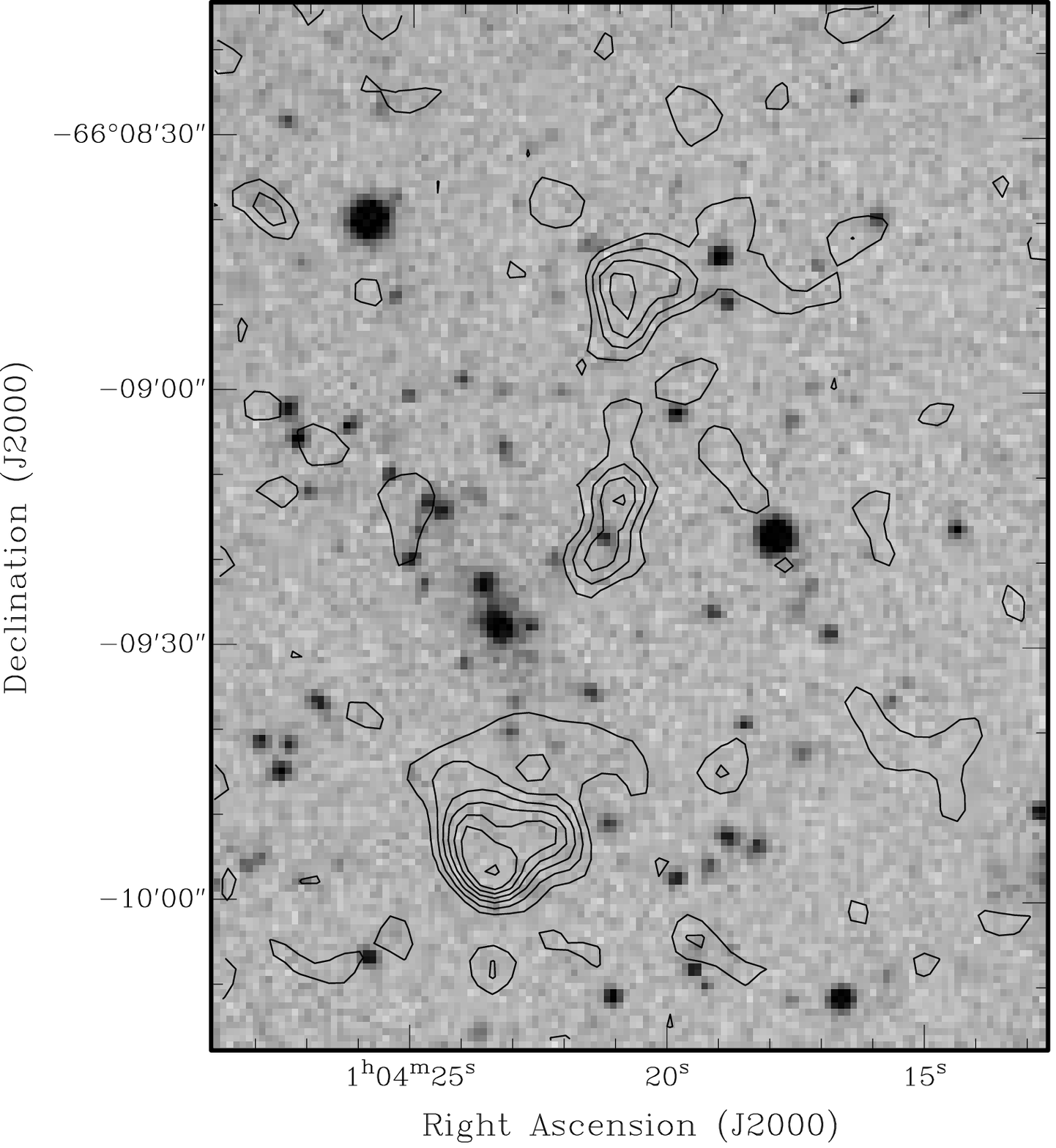}}
\caption{J0104.3-6609: $10^{-4}$ Jy x 1,  2,  3,  4,  5,  6,  8.}
\end{minipage}
&
\begin{minipage}{0.47\linewidth}
\frame{\includegraphics[angle=-90, width=2.8in]{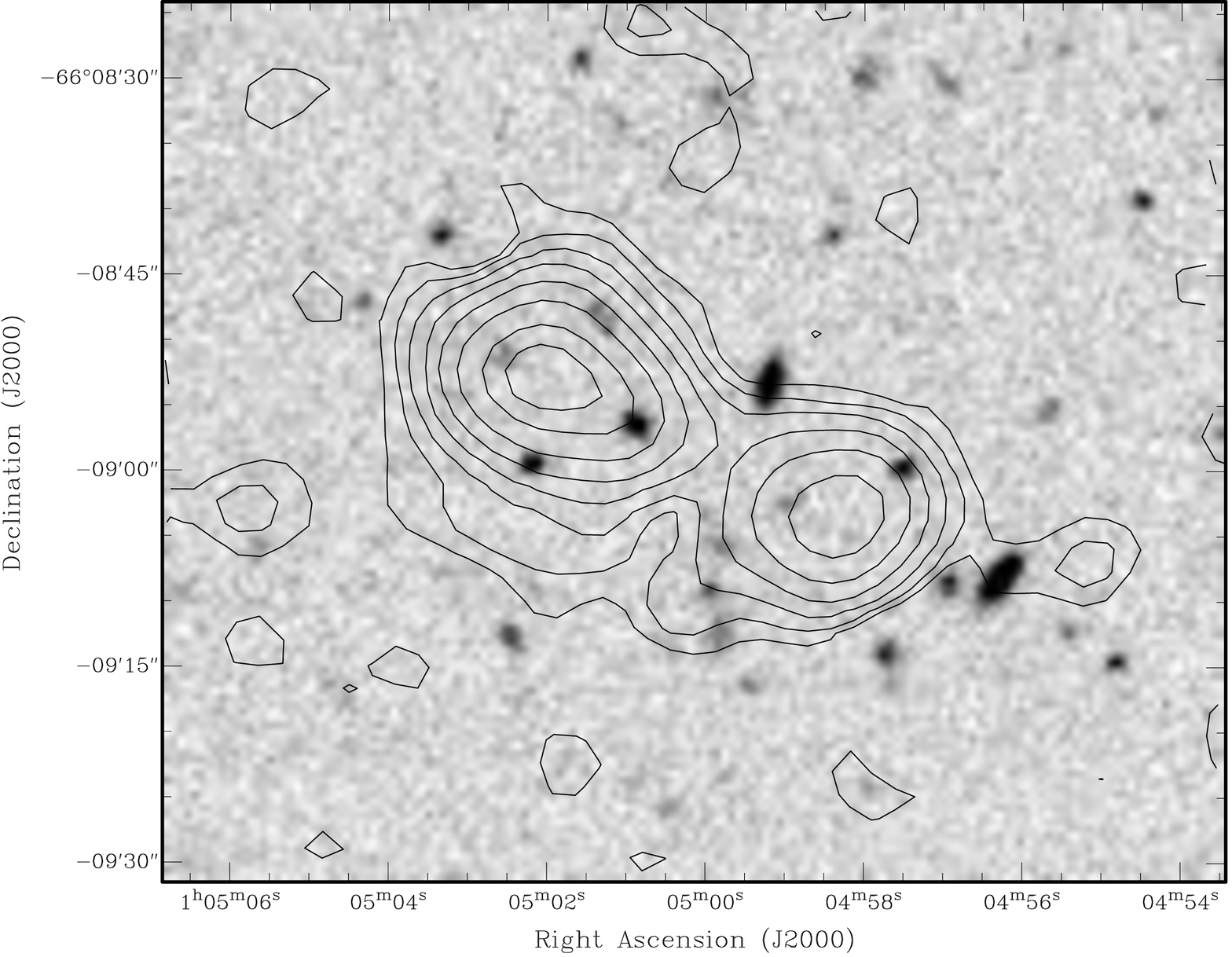}}
\caption{J0105.0-6608: $10^{-4}$ Jy x 1,  2,  4,  8,  16,  32,  64,  96.} 
\end{minipage}
\\
\end{tabular}
\end{figure*}

\begin{figure*}[ht]
\centering
\mbox{\subfigure{\includegraphics[angle=-90, width=2.8in]{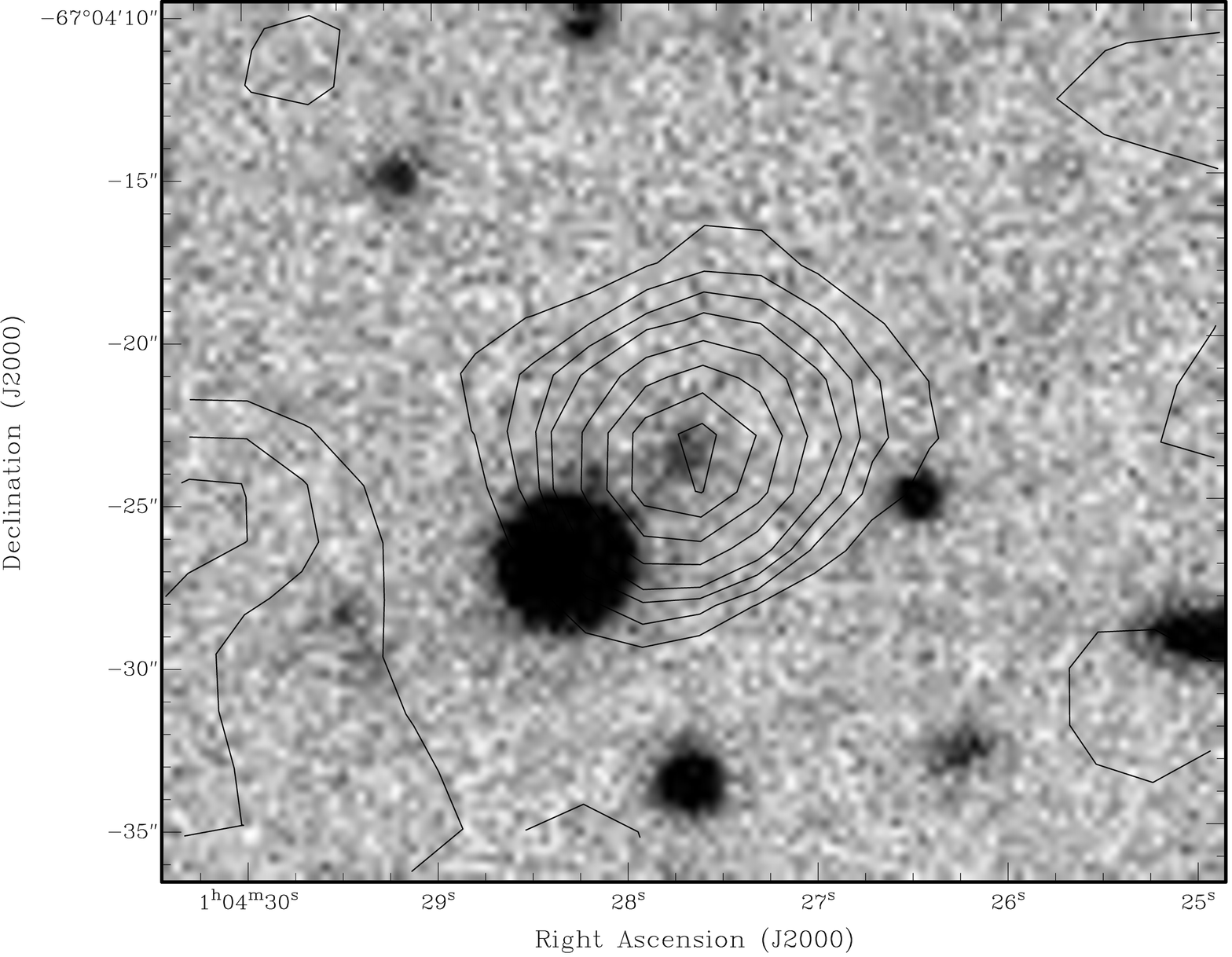}}\quad
\subfigure{\includegraphics[angle=-90, width=2.8in]{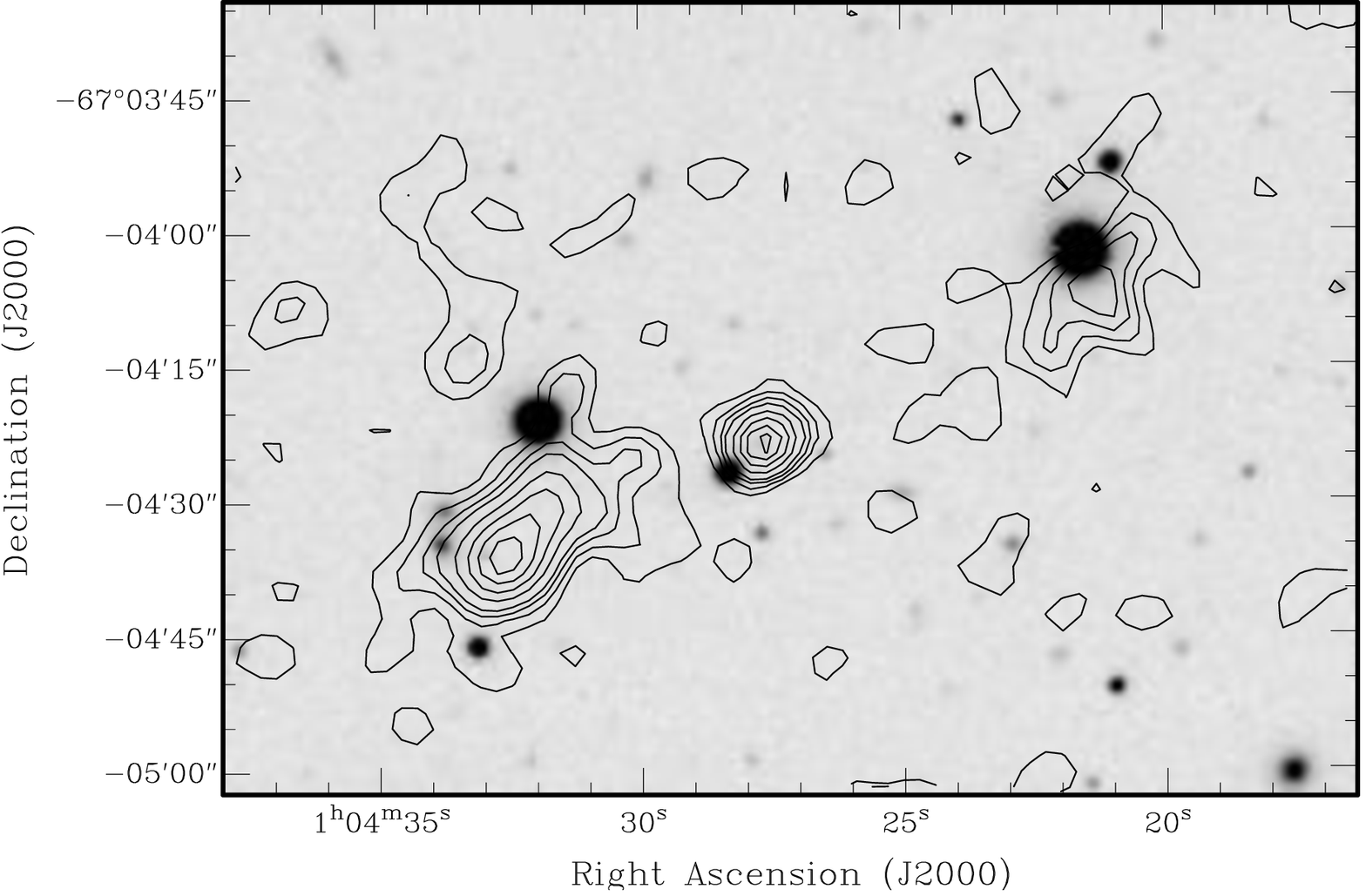}}}
\caption{J0104.4-6704: The left panel shows the core region at 6" resolution; 
the host galaxy is clearly seen. At right is the image of the full angular extent of the source.
The contour levels are: $10^{-4}$ Jy x 1,  2,  3,  4,  6,  8,  10,  12.}
\end{figure*}

\begin{figure*}[ht]
\centering
\mbox{\subfigure{\includegraphics[angle=-90, width=2.8in]{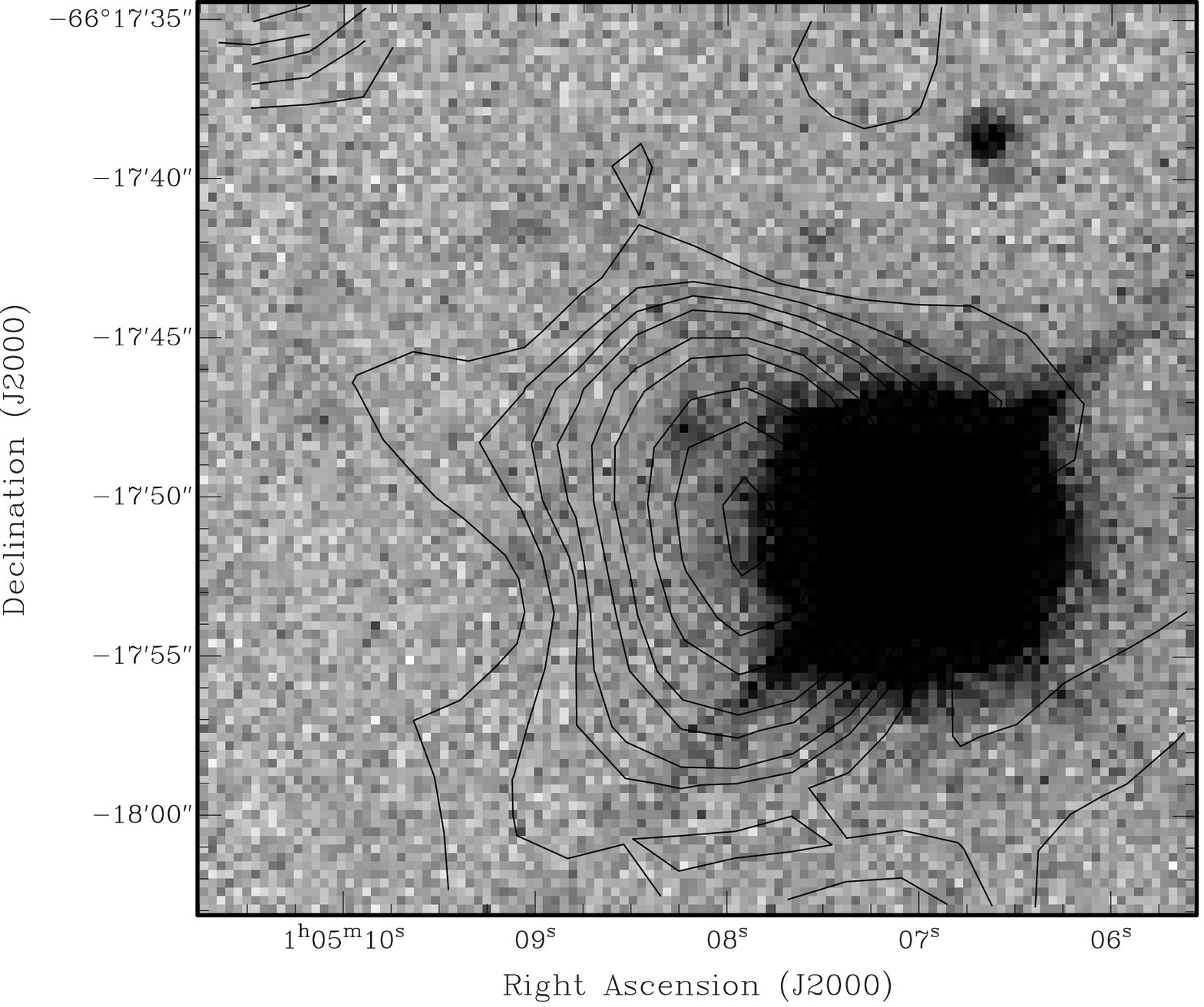}}\quad
\subfigure{\includegraphics[angle=-90, width=2.8in]{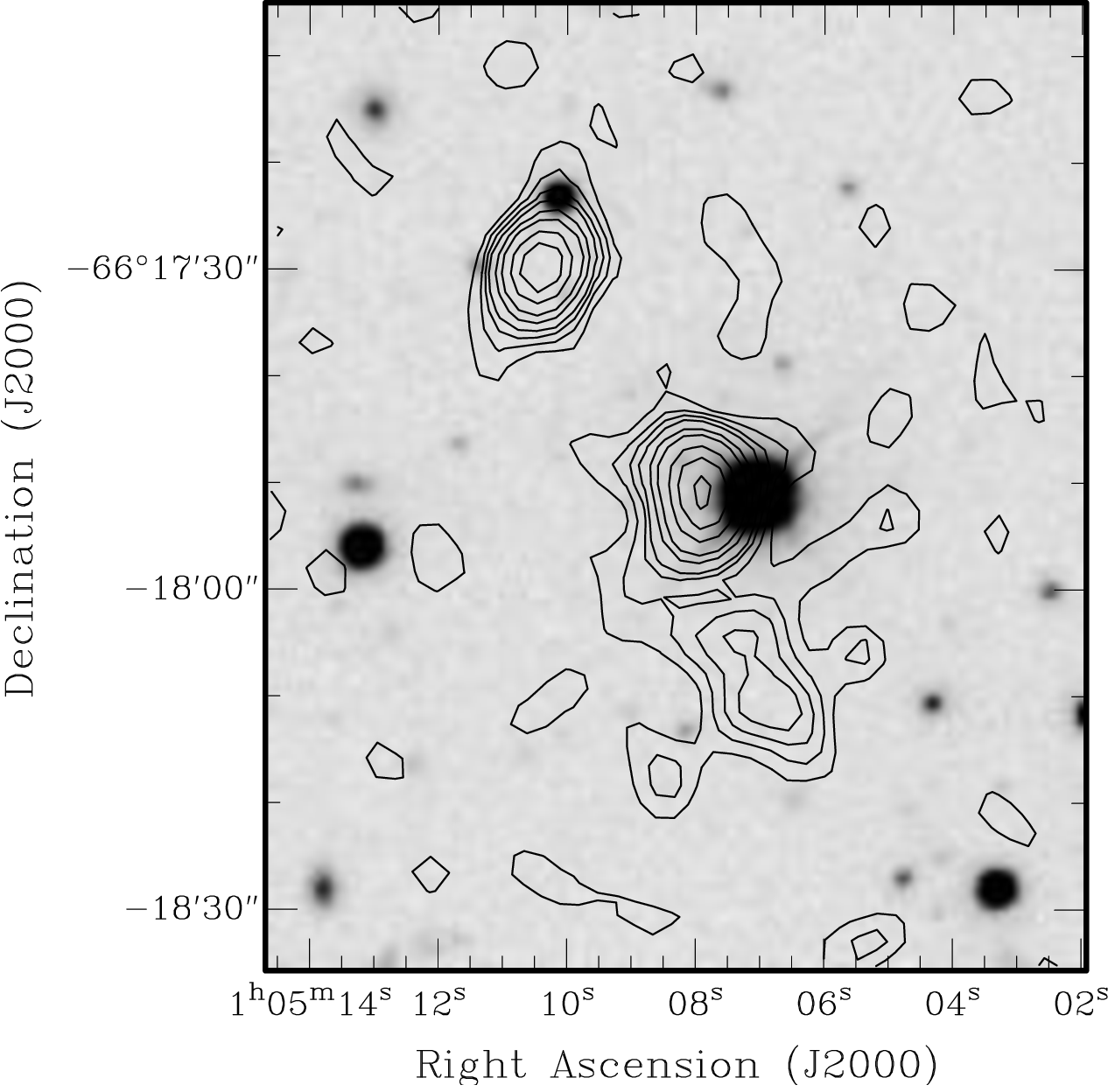}}}
\caption{J0105.1-6617: The left panel shows the core region at 6" resolution; 
the candidate host galaxy is seen. At right is the image of the full angular extent of the source.
The contour levels are: $10^{-4}$ Jy x 1,  2,  3,  4,  6,  8,  12,  16,  21.}
\end{figure*}

\begin{figure*}
\centering
\begin{tabular}{cc}
\begin{minipage}{0.47\linewidth}
\frame{\includegraphics[angle=-90, width=2.8in]{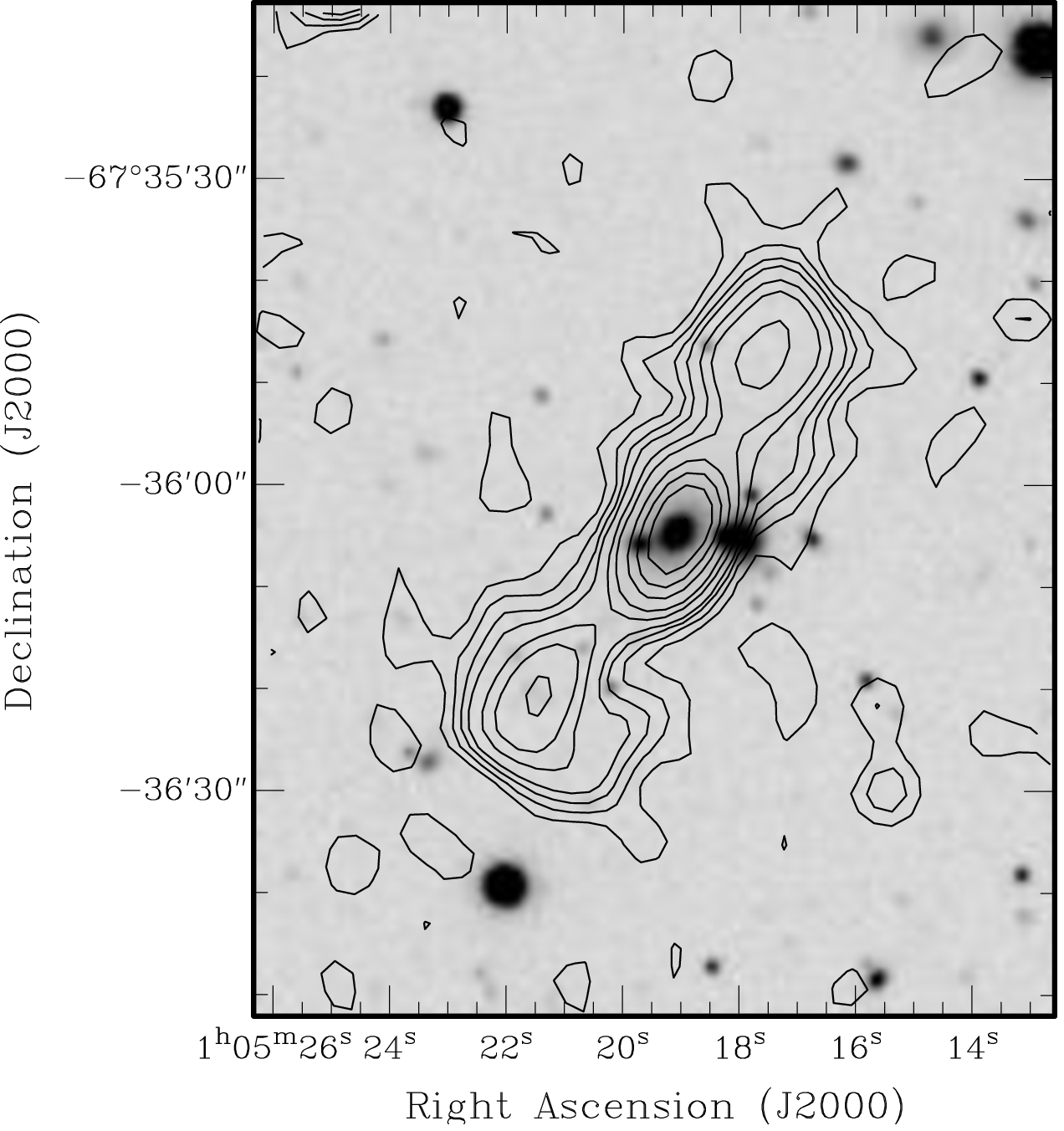}}
\caption{J0105.3-6736: $10^{-4}$ Jy x 1,  2,  3,  4,  6,  8,  12,  16,  21.} 
\end{minipage}
&
\begin{minipage}{0.47\linewidth}
\frame{\includegraphics[angle=-90, width=2.8in] {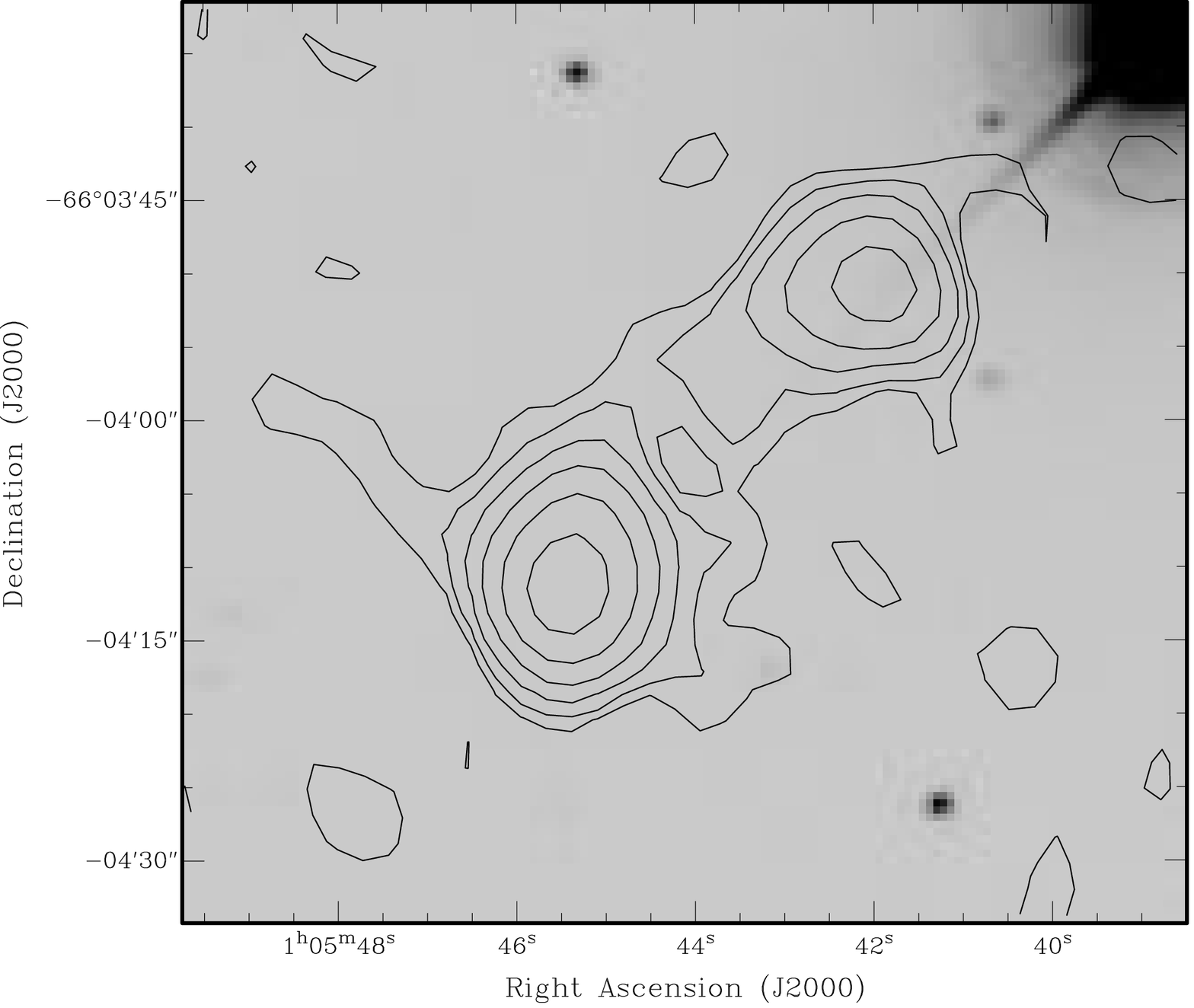}}
\caption{J0105.7-6604: $10^{-4}$ Jy x 1,  2,  4,  8,  16,  32.} 
\end{minipage}
\\
\end{tabular}
\end{figure*}

\begin{figure*}[ht]
\centering
\mbox{\subfigure{\includegraphics[angle=-90, width=2.8in]{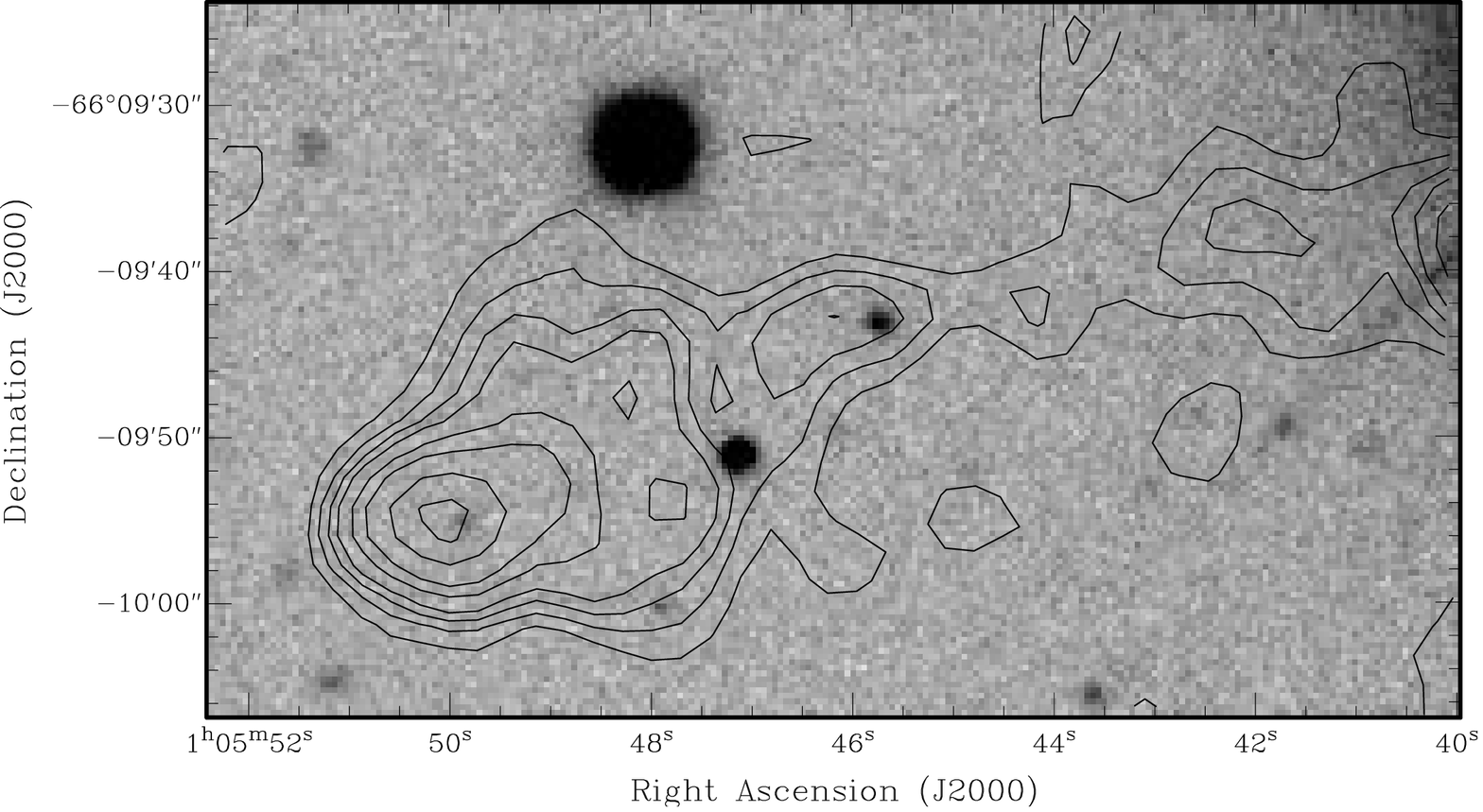}}\quad
\subfigure{\includegraphics[angle=-90, width=2.8in]{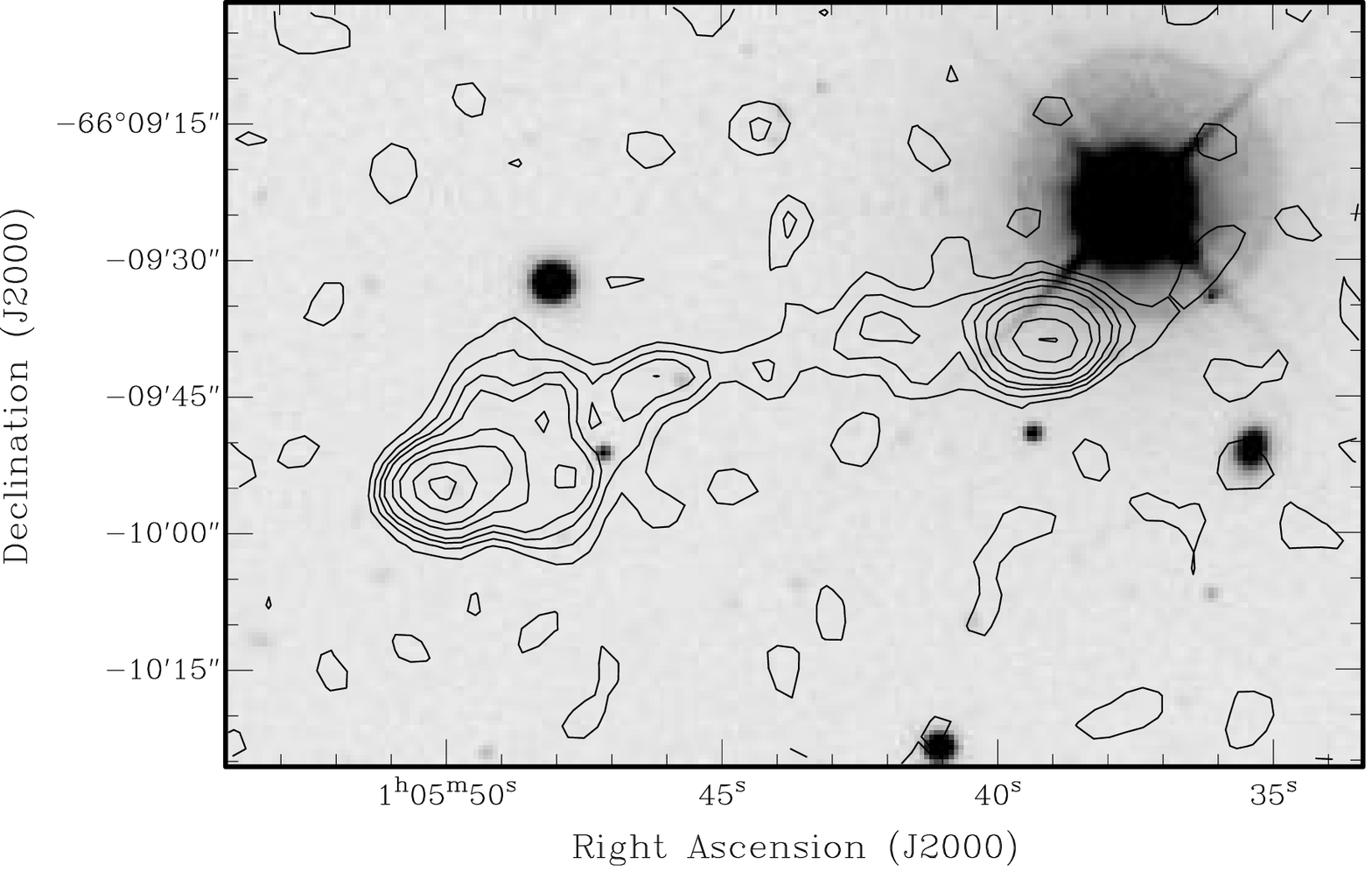}}}
\caption{J0105.7-6609: The left panel shows the core region at 6" resolution; 
the host galaxy is clearly seen. At right is the image of the full angular extent of the source.
The contour levels are: $10^{-4}$ Jy x 1,  2,  3,  4,  6,  8,  12,  16.}
\end{figure*}

\begin{figure*}
\centering
\begin{tabular}{cc}
\begin{minipage}{0.47\linewidth}
\frame{\includegraphics[angle=-90, width=2.8in]{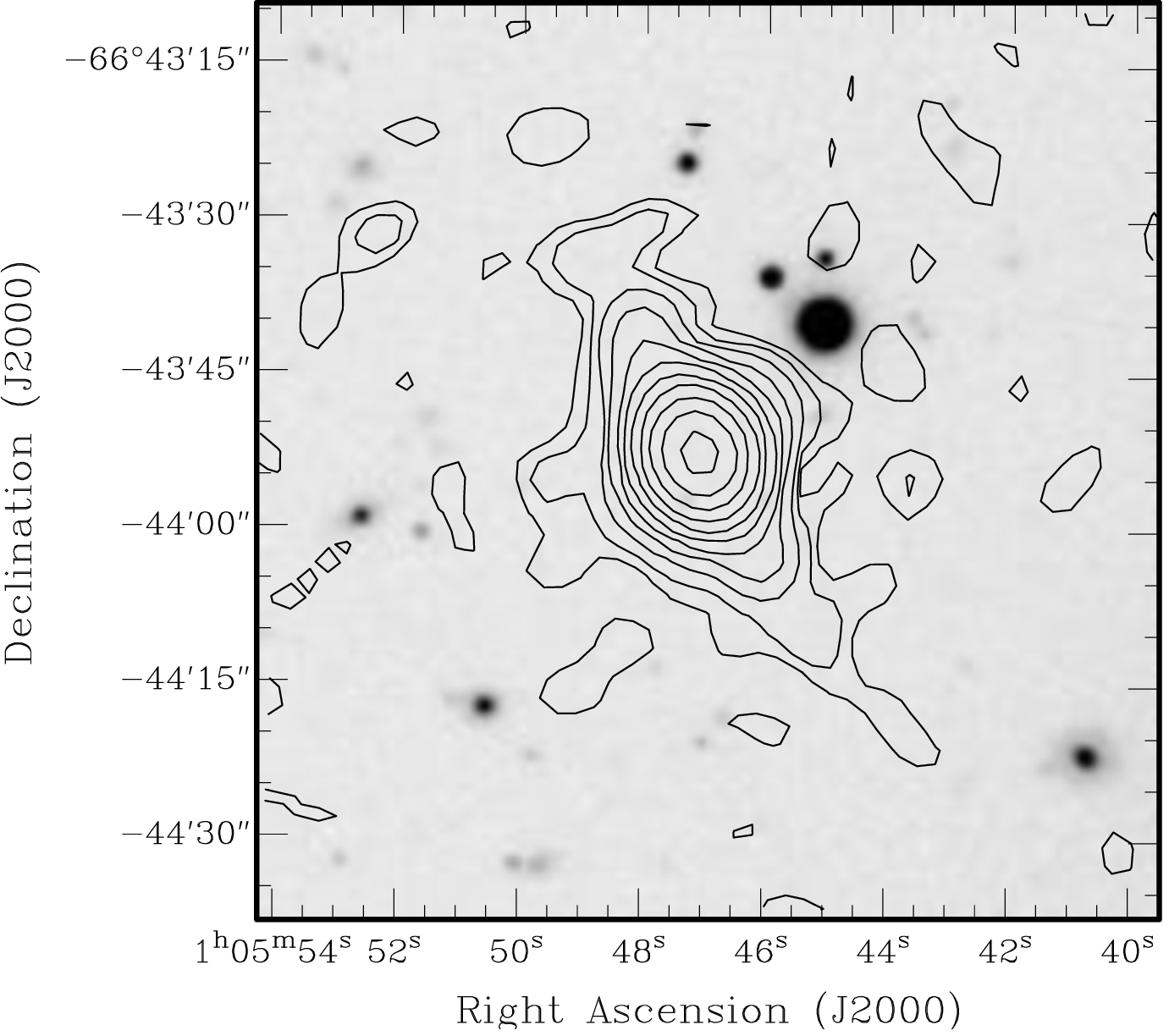}}
\caption{J0105.7-6643: $10^{-4}$ Jy x 1,  
2,  4,  8,  16,  32,  64, 128, 256, 512, 1024.}
\end{minipage}
&
\begin{minipage}{0.47\linewidth}
\frame{\includegraphics[angle=-90, width=2.8in]{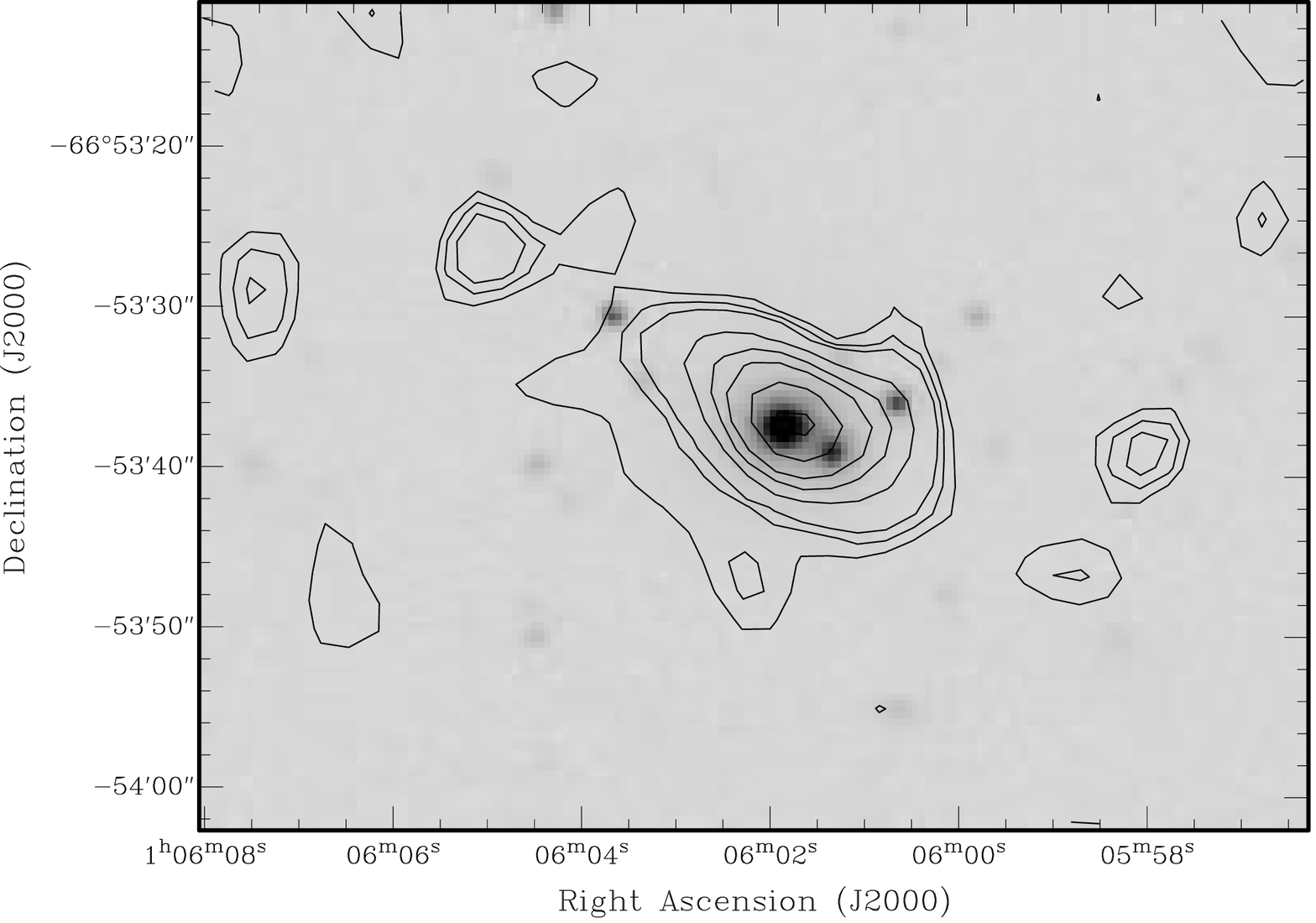}}
\caption{J0106.0-6653: $10^{-4}$ Jy x 1,  1.5,  2,  4,  6,  8, 12, 16.} 
\end{minipage}
\\
\end{tabular}
\end{figure*}

\clearpage

\begin{figure*}
\centering
\begin{tabular}{cc}
\begin{minipage}{0.47\linewidth}
\frame{\includegraphics[angle=-90, width=2.8in]{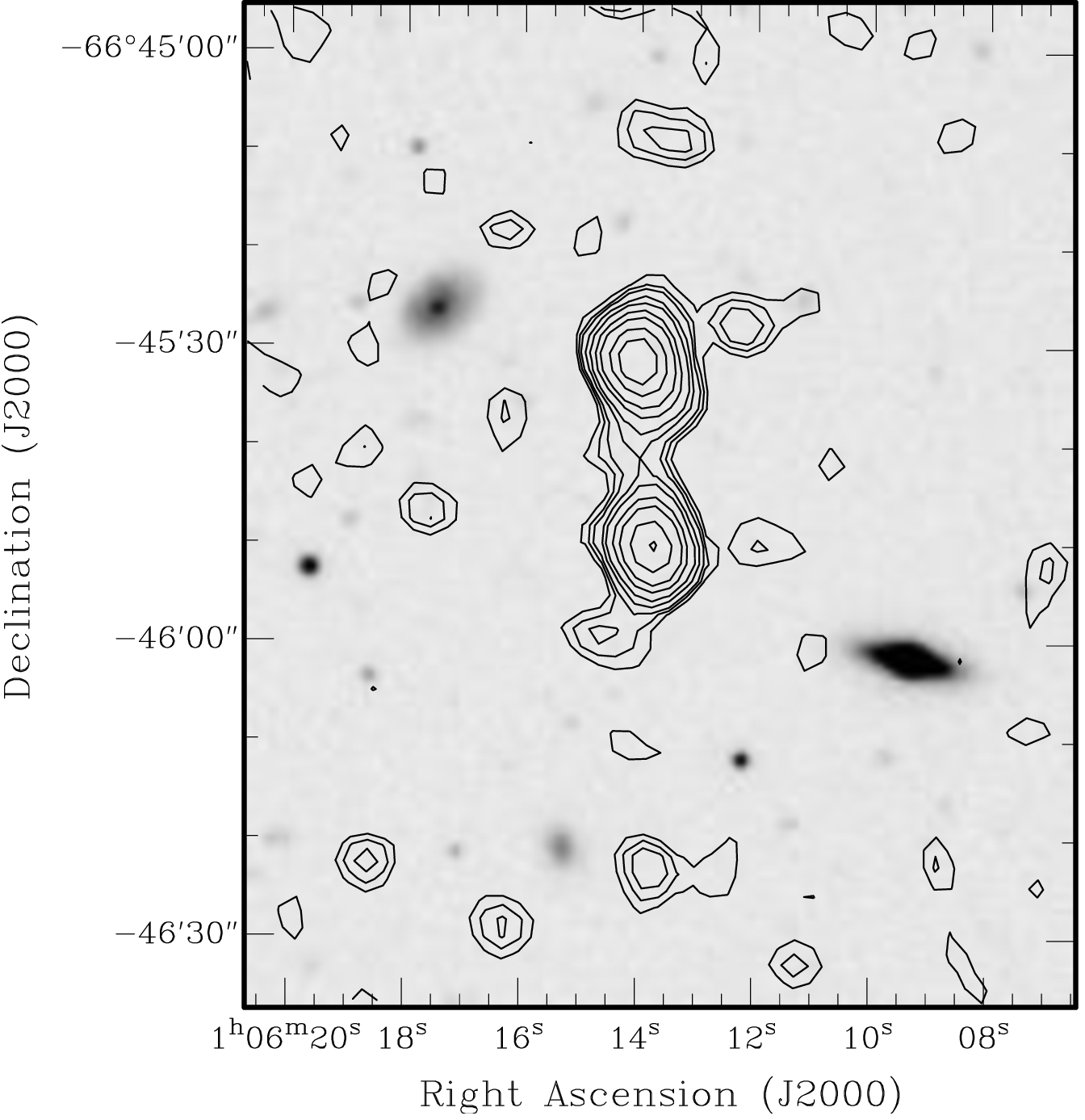}}
\caption{J0106.2-6645: $10^{-4}$ Jy x 1,  1.5,  2,  3,  4,  
6,  8,  12,  16.} 
\end{minipage}
&
\begin{minipage}{0.47\linewidth}
\frame{\includegraphics[angle=-90, width=2.8in] {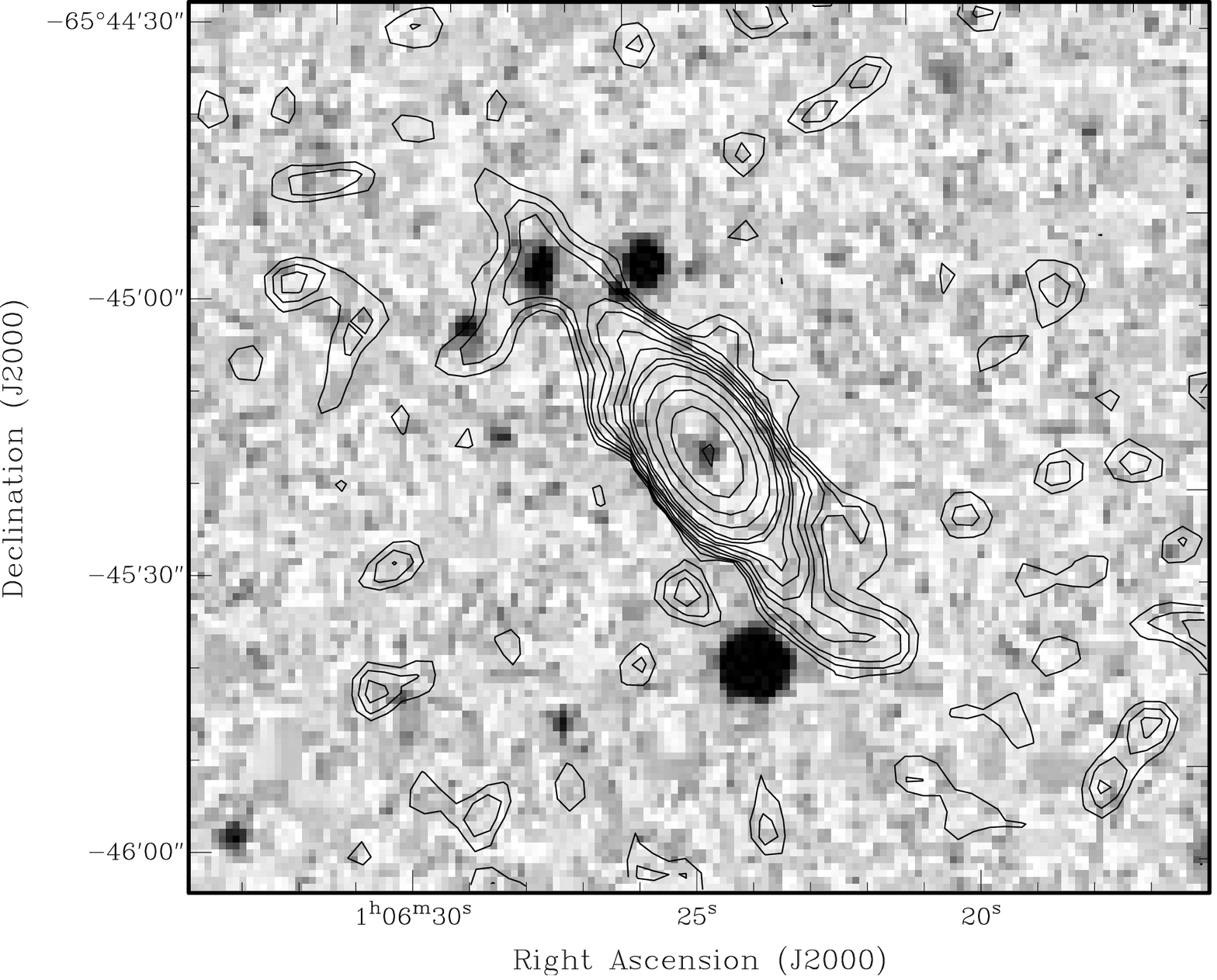}}
\caption{J0106.2-6545: $10^{-4}$ Jy x 1,  1.5,  2,  3,  4,  6,  8,  
12,  16,  32,  64, 128, 256. B-band image is used.} 
\end{minipage}
\\
\end{tabular}
\end{figure*}

\begin{figure*}
\centering
\begin{tabular}{cc}
\begin{minipage}{0.47\linewidth}
\frame{\includegraphics[angle=-90, width=2.8in]{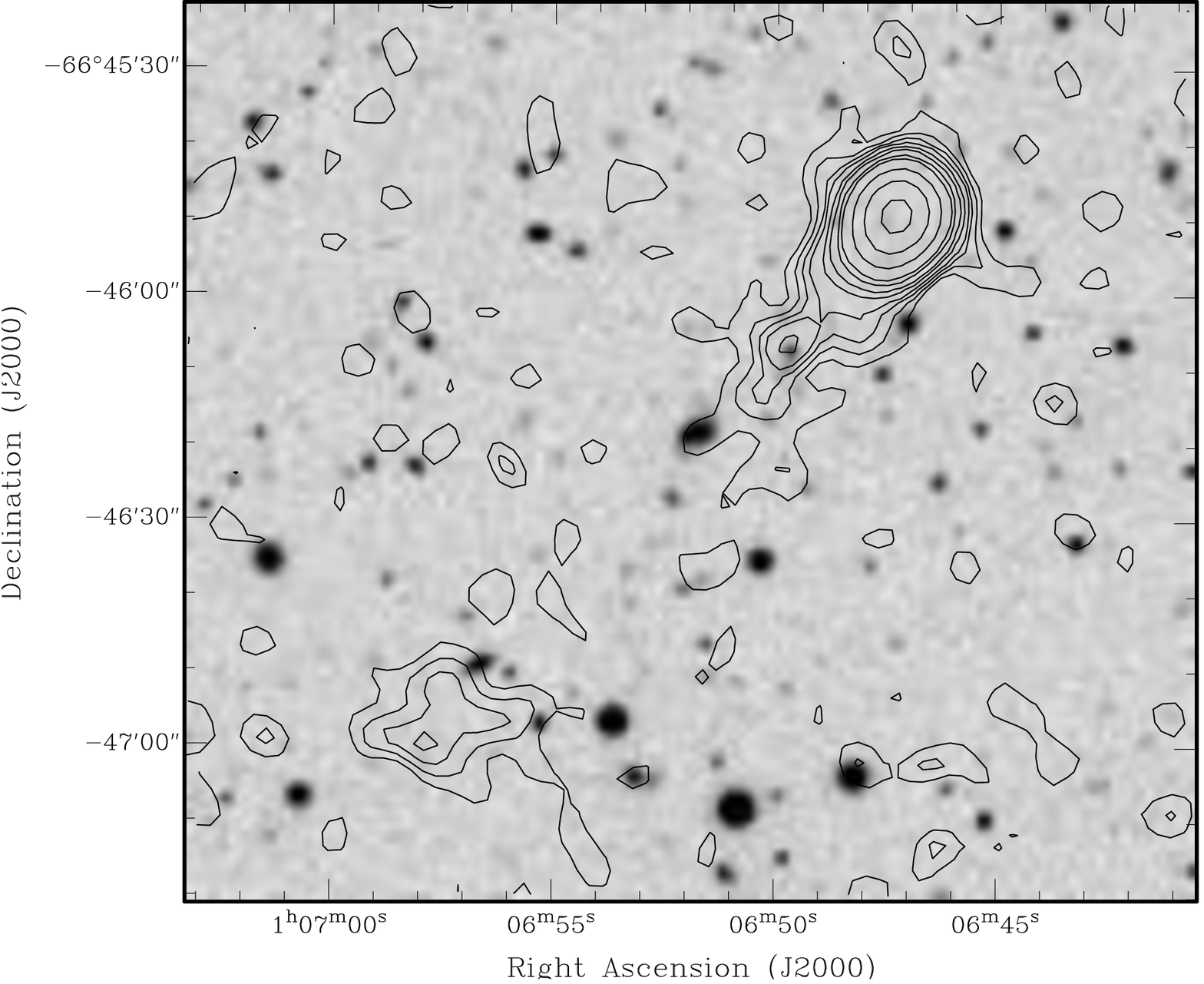}}
\caption{J0106.8-6645: $10^{-4}$ Jy x 1,  2,  3,  4,  6,  8,  12,  16,  32,  64, 128.}
\end{minipage}
&
\begin{minipage}{0.47\linewidth}
\frame{\includegraphics[angle=-90, width=2.8in]{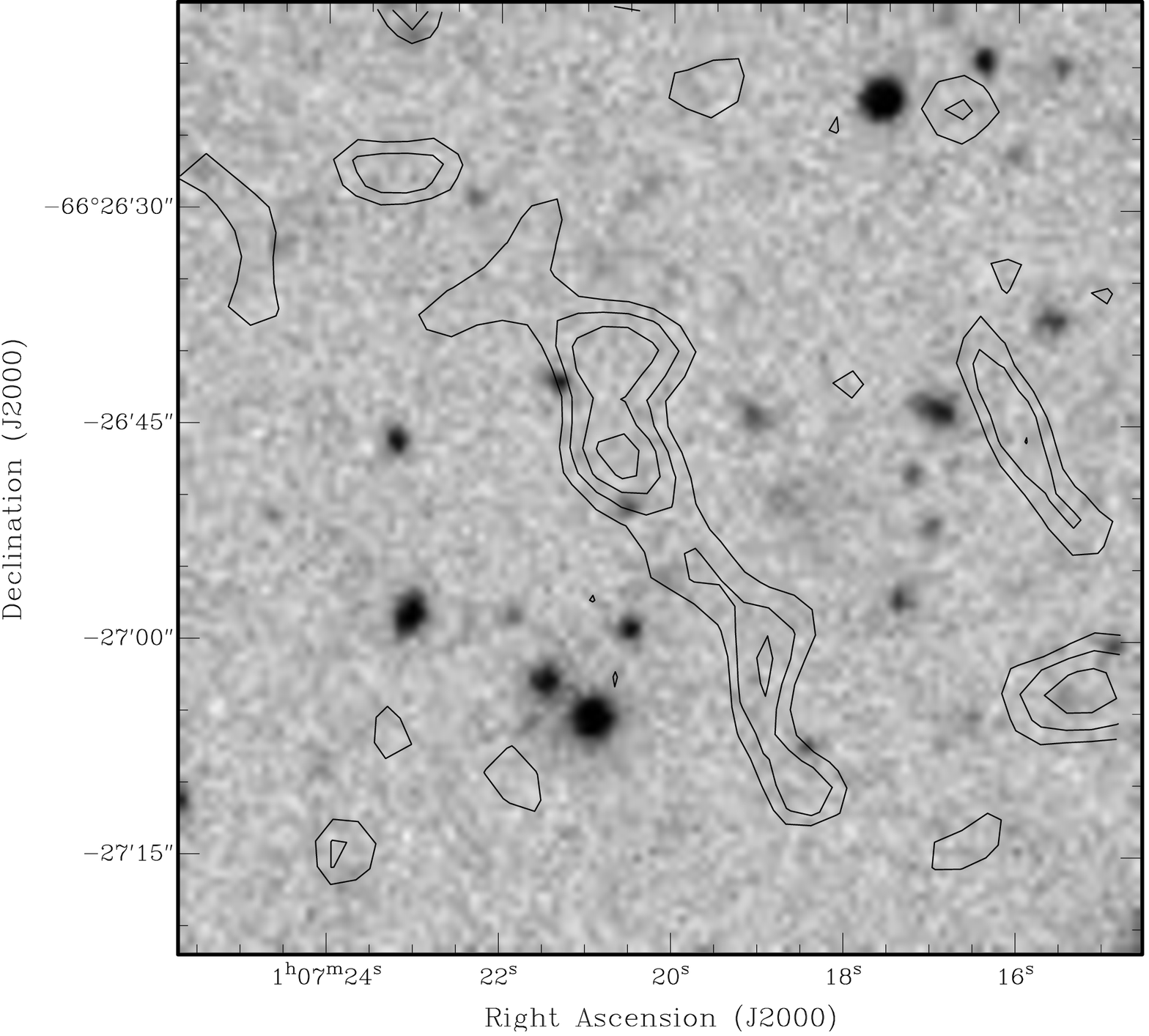}}
\caption{J0107.3-6626: $10^{-4}$ Jy x 1,  1.5,  2,  2.5.} 
\end{minipage}
\\
\end{tabular}
\end{figure*}

\clearpage

\begin{figure*}
\centering
\begin{tabular}{cc}
\begin{minipage}{0.47\linewidth}
\frame{\includegraphics[angle=-90, width=2.8in]{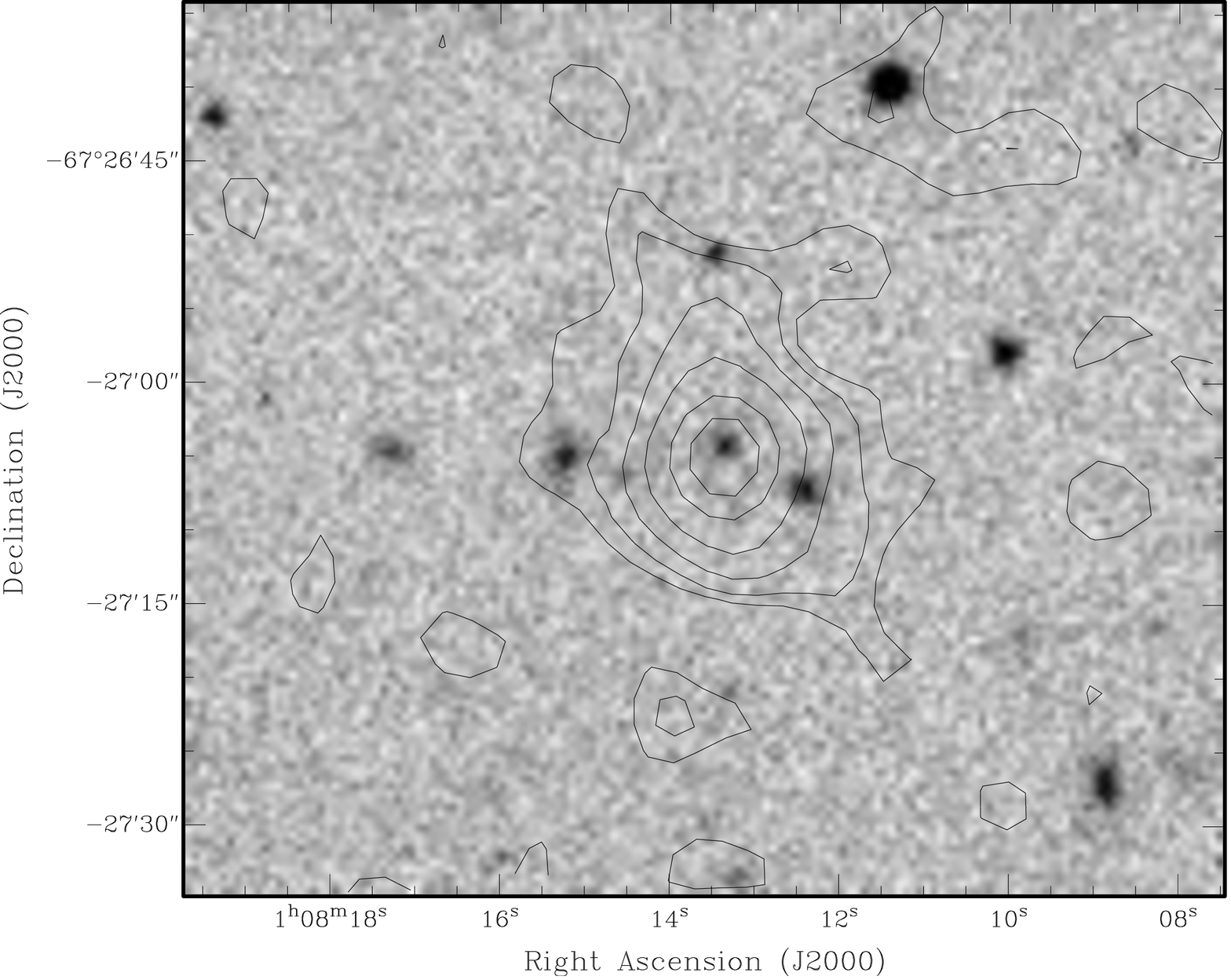}}
\caption{J0108.2-6727: $10^{-4}$ Jy x 1,  2,  4,  8,  16,  24.} 
\end{minipage}
&
\begin{minipage}{0.47\linewidth}
\frame{\includegraphics[angle=-90, width=2.8in]{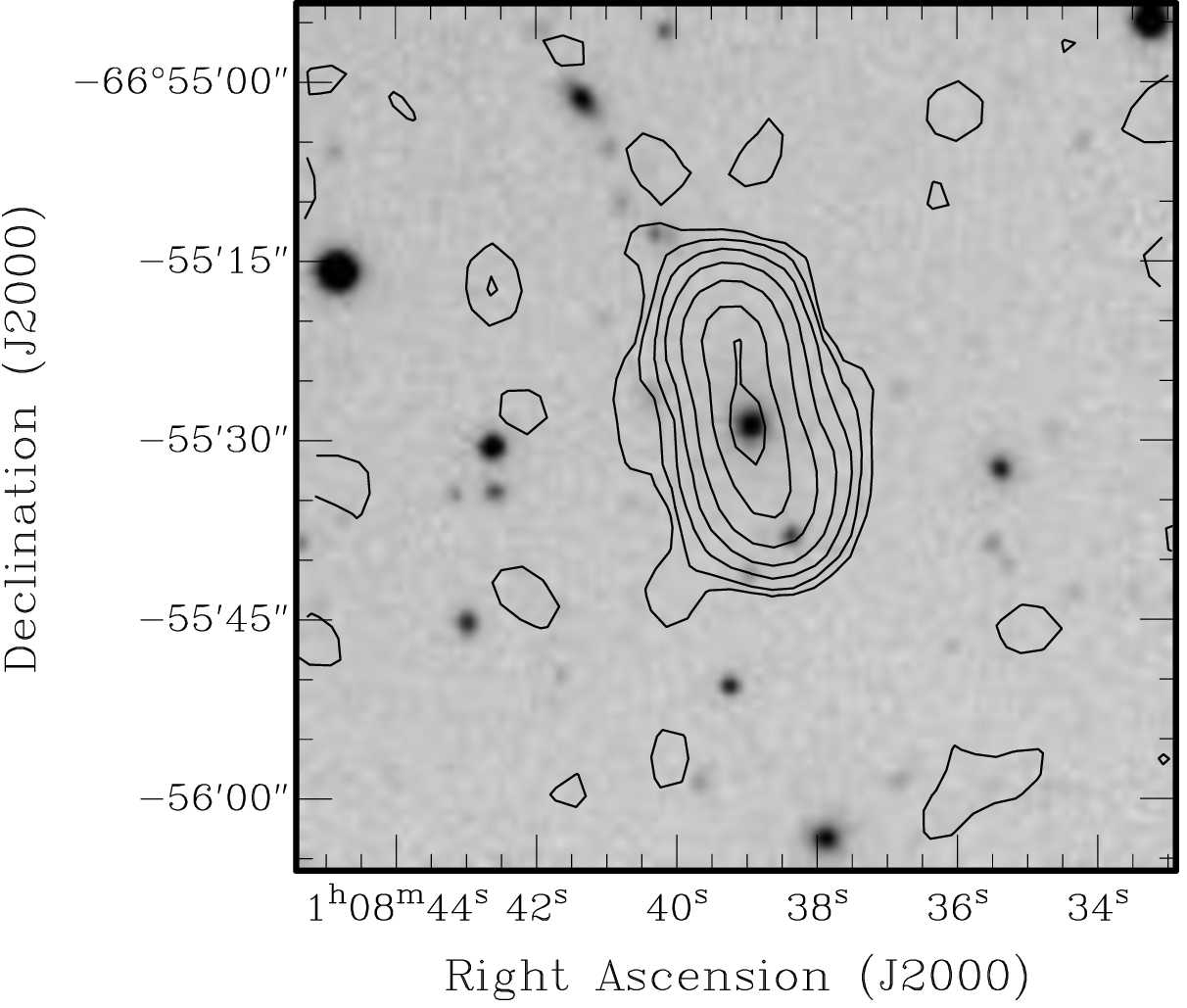}}
\caption{J0108.6-6655: $10^{-4}$ Jy x 1,  2,  4,  8,  16,  32,  48.} 
\end{minipage}
\\
\end{tabular}
\end{figure*}

\begin{figure*}[ht]
\centering
\mbox{\subfigure{\includegraphics[angle=-90, width=2.8in]{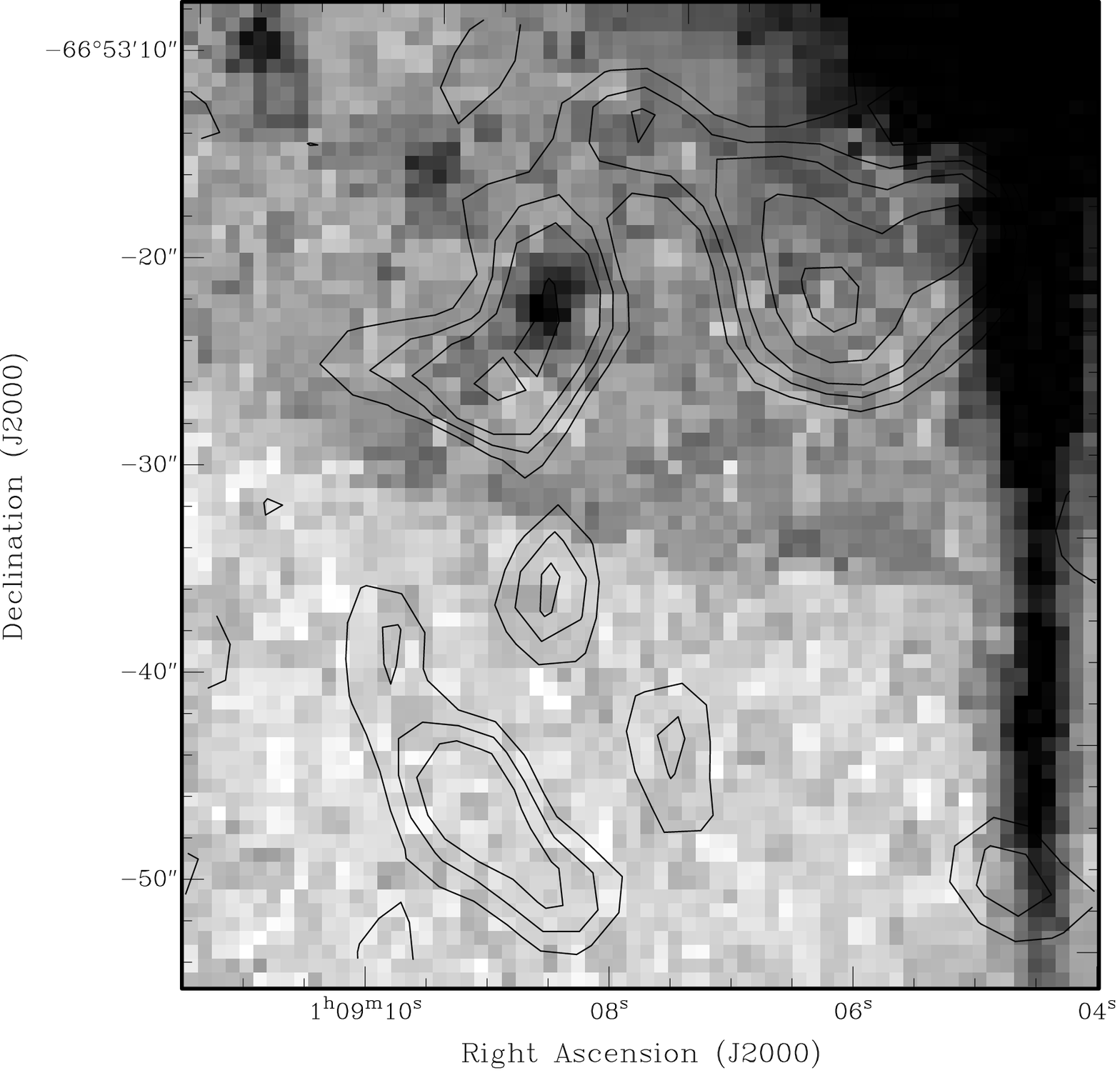}}\quad
\subfigure{\includegraphics[angle=-90, width=2.8in]{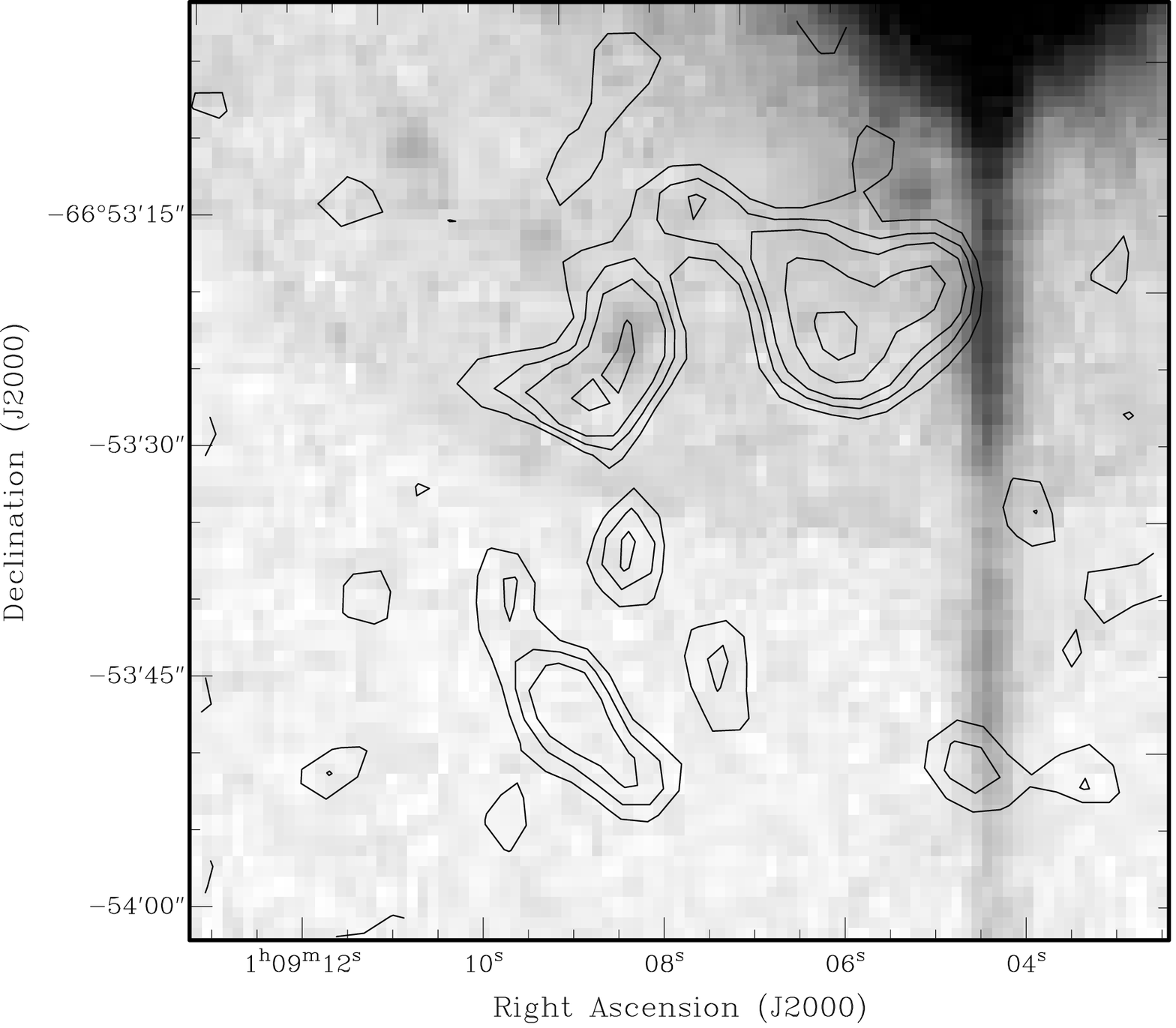}}}
\caption{J0109.1-6653: The left panel shows the core region at 6" resolution; 
the host galaxy is clearly seen. At right is the image of the full angular extent of the source.
The contour levels are: $10^{-4}$ Jy x 1,  1.5,  2,  3,  4. R-band image is used.}
\end{figure*}

\begin{figure*}
\centering
\begin{tabular}{cc}
\begin{minipage}{0.47\linewidth}
\frame{\includegraphics[angle=-90, width=2.8in] {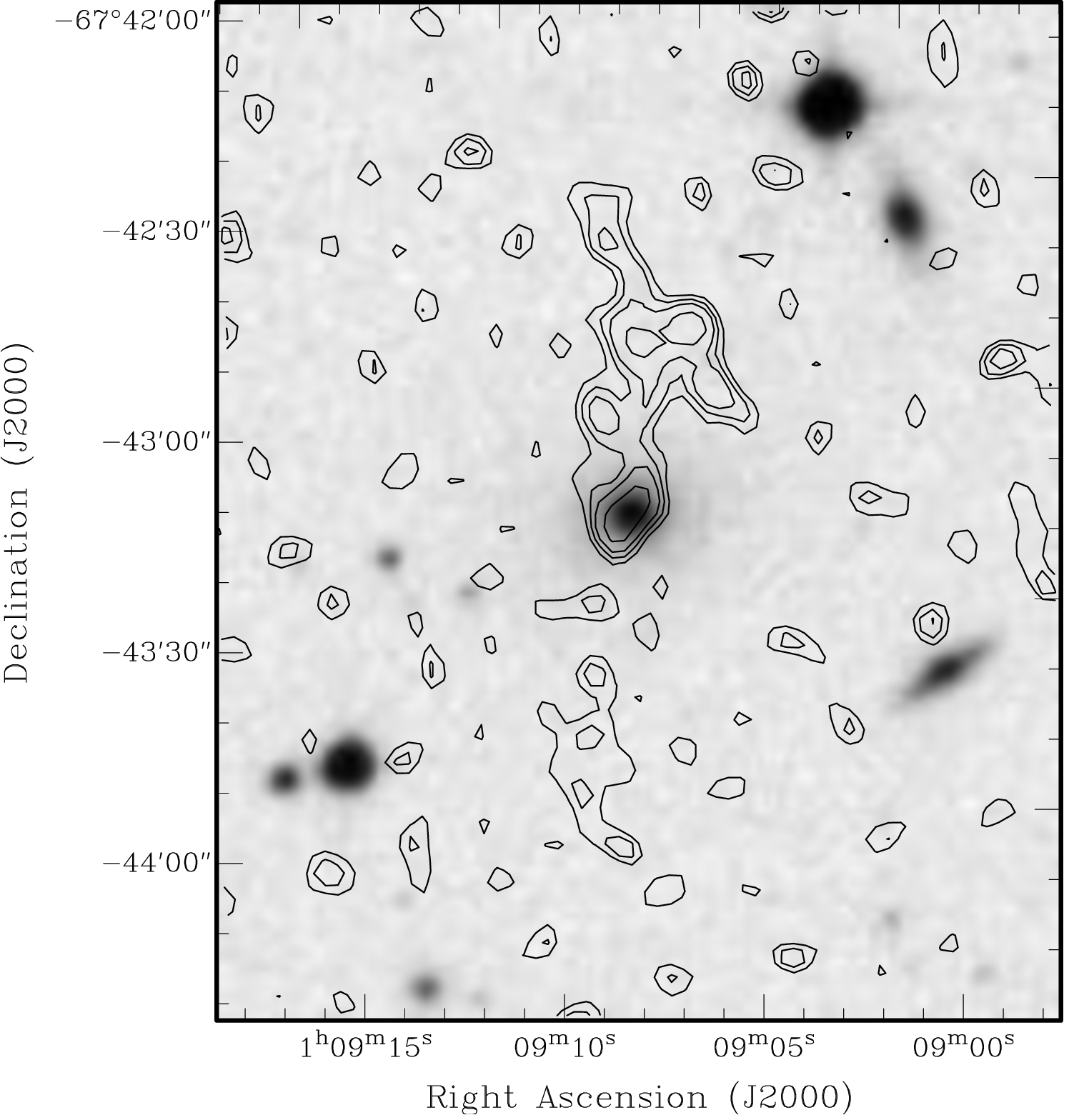}}
\caption{J0109.1-6743: $10^{-4}$ Jy x 1,  1.5,  2,  3. R-band image is used.} 
\end{minipage}
&
\begin{minipage}{0.47\linewidth}
\frame{\includegraphics[angle=-90, width=2.8in]{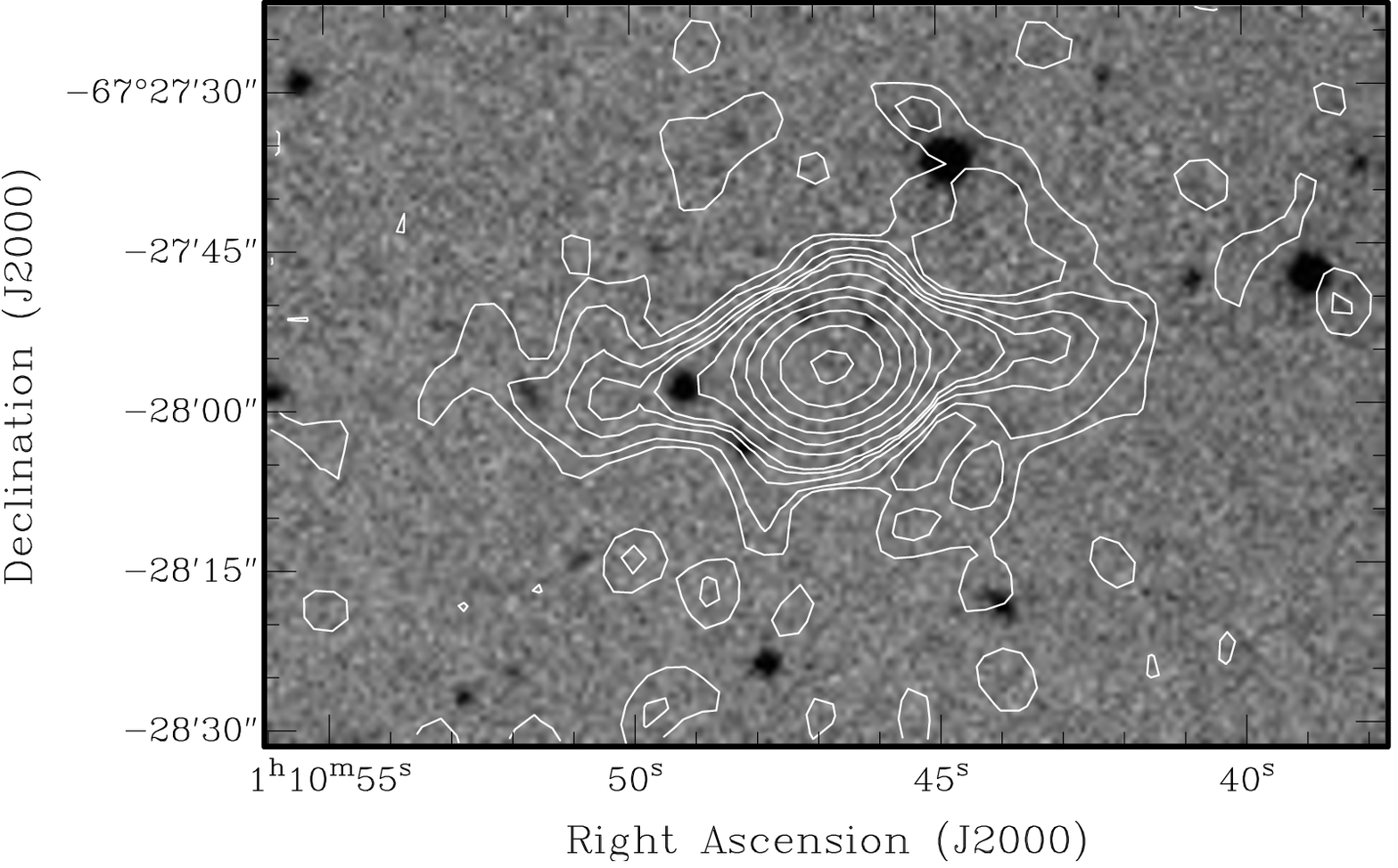}}
\caption{J0110.7-6727: $10^{-4}$ Jy x 1,  2,  4,  6, 8,  16,  32,  64, 128, 
256.}
\end{minipage}
\\
\end{tabular}
\end{figure*}

\begin{figure*}
\centering
\begin{tabular}{cc}
\begin{minipage}{0.47\linewidth}
\frame{\includegraphics[angle=-90, width=2.8in]{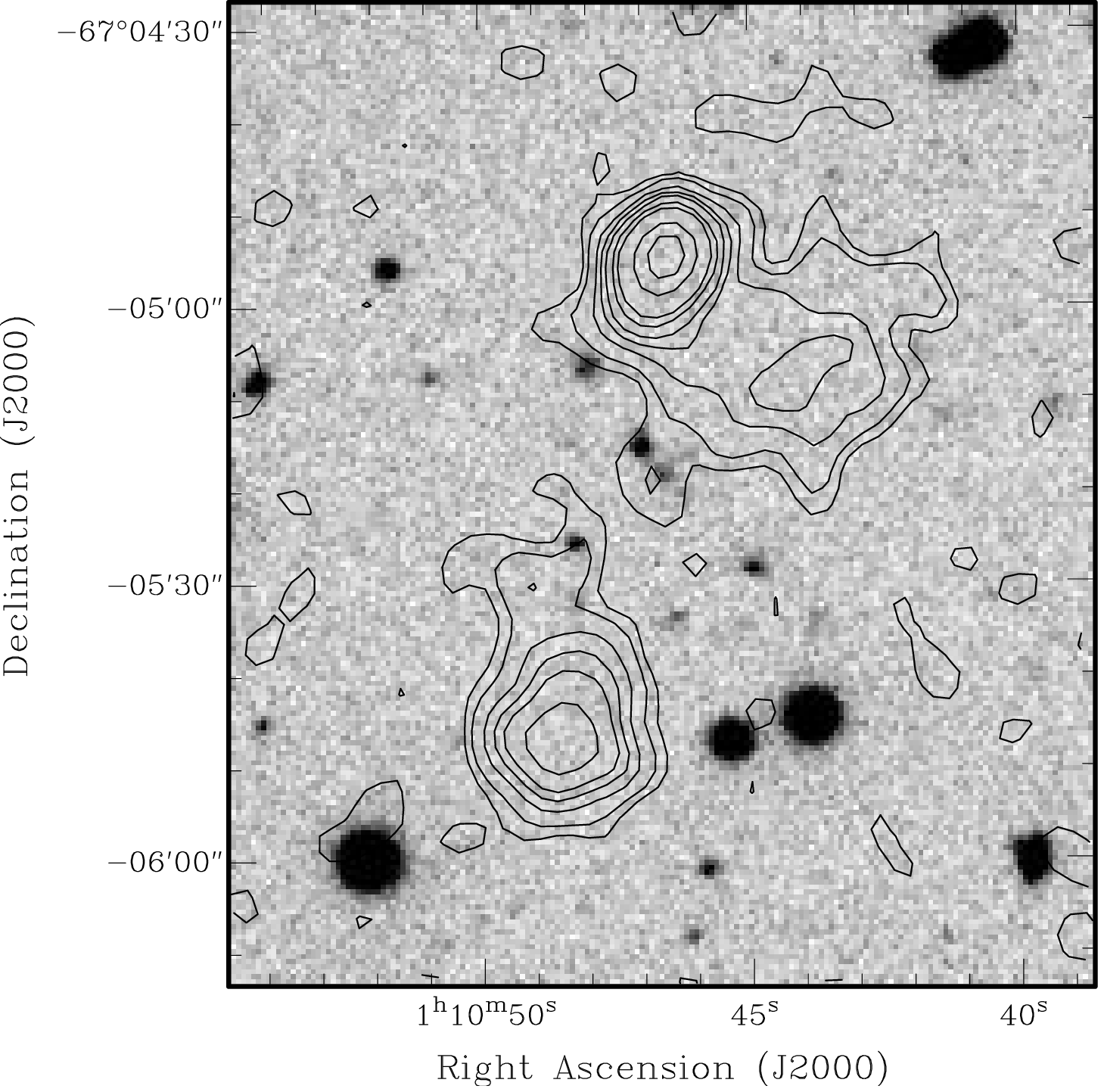}}
\caption{J0110.7-6705: $10^{-4}$ Jy x 1,  2,  4,  6,  8,  12,  16,  32,  48.} 
\end{minipage}
&
\begin{minipage}{0.47\linewidth}
\frame{\includegraphics[angle=-90, width=2.8in]{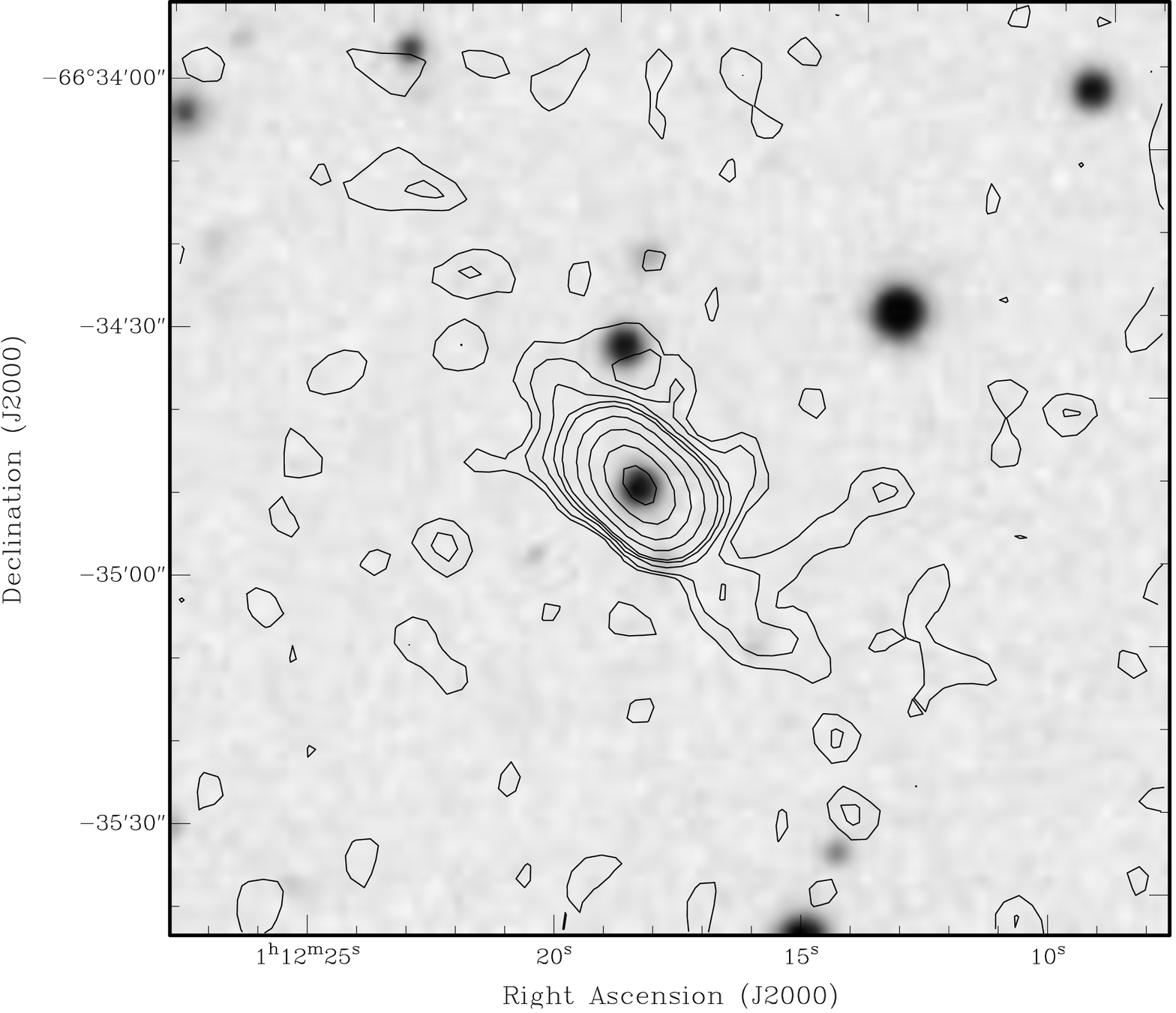}}
\caption{J0112.3-6634: $10^{-4}$ Jy x 1,  2,  4,  6,  8,  16,  32,  64, 110. B-image is used.} 
\end{minipage}
\\
\end{tabular}
\end{figure*}

\begin{figure*}
\centering
\begin{tabular}{l}
\begin{minipage}{0.47\linewidth}
\frame{\includegraphics[angle=-90, width=2.8in]{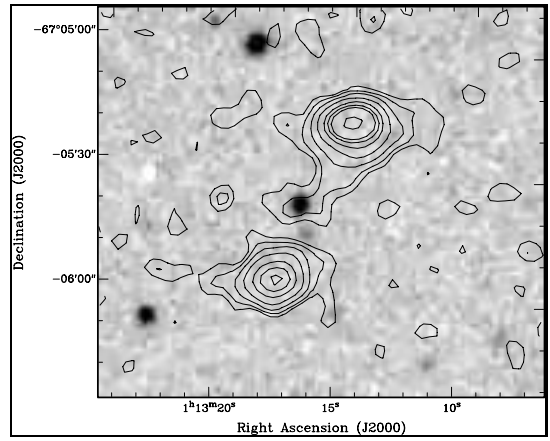}}
\caption{J0113.2-6705: $10^{-4}$ Jy x 1,  2,  4,  8,  16,  24,  32,  64. B-image is used.} 
\end{minipage}
\\
\end{tabular}
\end{figure*}

\clearpage

\end{document}